\documentclass{article}
\usepackage[utf8]{inputenc}
\usepackage{graphicx}
\usepackage{geometry}
\usepackage{authblk}
\usepackage{amsmath}
\usepackage{amssymb}
\usepackage{enumitem}
\usepackage{xcolor}
\usepackage{hyperref}
\usepackage[sorting = none, backend = bibtex, style=numeric-comp]{biblatex}

\title{Comparing Three Generations of D-Wave Quantum Annealers for Minor Embedded Combinatorial Optimization Problems}

\author[1]{Elijah Pelofske
\thanks{Email: epelofske@lanl.gov}}
\affil[1]{Los Alamos National Laboratory}
\date{\vspace{-6ex}}

\begin{document}

\maketitle

\begin{abstract}
Quantum annealing is a novel type of analog computation that aims to use quantum mechanical fluctuations to search for optimal solutions of Ising problems. Quantum annealing in the Transverse Ising model, implemented on D-Wave QPUs, are available as cloud computing resources. In this study we report concise benchmarks across three generations of D-Wave quantum annealers, consisting of four different devices, for the NP-hard discrete combinatorial optimization problems unweighted maximum clique and unweighted maximum cut on random graphs. The Ising, or equivalently QUBO, formulation of these problems do not require auxiliary variables for order reduction, and their overall structure and weights are not highly variable, which makes these problems simple test cases to understand the sampling capability of current D-Wave quantum annealers. All-to-all minor embeddings of size $52$, with relatively uniform chain lengths, are used for a direct comparison across the Chimera, Pegasus, and Zephyr device topologies. A grid-search over annealing times and the minor embedding chain strengths is performed in order to determine the level of reasonable performance for each device and problem type. Experiment metrics that are reported are approximation ratios for non-broken chain samples, chain break proportions, and Time-to-Solution for the maximum clique problem instances. How fairly the quantum annealers sample optimal maximum cliques, for instances which contain multiple maximum cliques, is quantified using entropy of the measured ground state distributions. The newest generation of quantum annealing hardware, which has a Zephyr hardware connectivity, performed the best overall with respect to approximation ratios and chain break frequencies. 
\end{abstract}

\section{Introduction}
\label{section:introduction}

Quantum annealing (QA) is a quantum computational technology that can be viewed as a heuristic version of Adiabatic Quantum Computing (AQC). QA uses quantum fluctuations to search for a minimum variable assignment of a combinatorial optimization problem \cite{Kadowaki_1998, doi:10.1126/science.284.5415.779, morita2008mathematical, das2008colloquium, finnila1994quantum, santoro2006optimization, PhysRevX.4.021041}. Quantum annealing has been physically implemented using superconducting flux qubits in commercial devices by companies such as D-Wave \cite{johnson2011quantum}. D-Wave quantum annealers are available to be used as a cloud compute resource. The topic of benchmarking quantum annealers capabilities in the NISQ era \cite{Preskill_2018, hauke2020perspectives} is a topic of considerable interest \cite{vert2021benchmarking, mcgeoch2015benchmarking, Zaborniak_2021, Willsch_2022, grant2022benchmarking, PhysRevApplied.15.014012, doi:10.7566/JPSJ.90.064001, 9465651, https://doi.org/10.48550/arxiv.2210.04291}. For reviews on quantum annealing, see refs. \cite{Yarkoni_2022, hauke2020perspectives}. Quantum annealing in the transverse field Ising model works by initializing the system in the groundstate of the transverse field Hamiltonian, defined by:

\begin{equation}
    H_\text{initial} = \sum_{i}^n \sigma_i^x
    \label{equation:transverse_ising}
\end{equation}

Where $\sigma_i^x$ is the Pauli matrix for qubit $i$. The system then applies a user specified problem Hamiltonian:

\begin{equation}
    H(t) = A(t)H_\text{initial} + B(t)H_\text{ising}
    \label{equation:QA_eq}
\end{equation}

The total \emph{annealing time} can be chosen by the user. The problem Hamiltonian is defined as:

\begin{equation}
    H_\text{ising} = \sum_{i}^n h_i \sigma_i^z + \sum_{i < j}^n J_{ij} \sigma_i^z \sigma_j^z
    \label{equation:problem_ising}
\end{equation}

Importantly, the user programmed Hamiltonian in eq.~\eqref{equation:problem_ising} is the form of Ising problems, or equivalently Quadratic Unconstrained Binary Optimization problems (QUBOs). NP-hard problems, and in general discrete combinatorial optimization problems, can be formulated as Isings or QUBOs \cite{Lucas_2014}. 

Quantum annealing has been experimentally applied to a variety of problems, including the network shortest path problem \cite{9186612}, support vector machines \cite{WILLSCH2020107006}, charged particle tracking \cite{zlokapa2021charged}, and integer factorization \cite{jiang2018quantum, dridi2017prime, peng2019factoring, warren2019factoring, cryptoeprint:2021/527, https://doi.org/10.48550/arxiv.2005.02268}. There is encouraging evidence that D-Wave quantum annealers can sample combinatorial optimization problems with a competitive accuracy and time cost to classical optimization methods \cite{https://doi.org/10.48550/arxiv.2210.04291}. Quantum annealing has also been used for physics simulation problems such as simulating frustrated magnetic systems \cite{harris2018phase, PRXQuantum.2.030317, Abel_2021, king2021qubit, zhou2021experimental, king2021scaling}. 

\begin{equation}
     H = - \frac{A(s)}{2}  \sum_i^n X_i + \frac{B(s)} {2} \left( \sum_{i}^n h_i Z_i + \sum_{i < j}^n J_{ij} Z_i Z_j \right)
    \label{equation:QA_Hamiltonian}
\end{equation}

Eq.~\eqref{equation:QA_Hamiltonian} defines the Hamiltonian that is programmed on D-Wave quantum annealers, which describes that the annealing is driven by the transverse field Hamiltonian $\sum_i^n X_i$ (this is the same $\sum_{i}^n \sigma_i^x$ term in eq.~\eqref{equation:transverse_ising}). The $h_i$ and $J_{ij}$ terms define the classical Ising model that the user programs (eq.~\eqref{equation:problem_ising}). The functions $A(s)$ and $B(s)$ are system-defined functions that provide the proportion between the transverse field driving Hamiltonian and the programmed problem Hamiltonian, parameterized by the unitless quantity $s \in [0,1]$ where $s=0$ corresponds to maximum-strength transverse field and $s=1$ corresponds to minimum-strength transverse field. $s$ is referred to as the \emph{anneal fraction}. Defining $s$ at each point during the anneal gives an \emph{anneal schedule}. The quantum annealing algorithm is inspired from adiabatic quantum computation and ideally would use purely adiabatic evolution, however current hardware implementations do not allow purely adiabatic evolution to be implemented. The fundamental idea of quantum annealing is to transition the system from an easy-to-prepare groundstate to the groundstate of a Hamiltonian whose groundstate is unknown (or hard to prepare) which can correspond to the optimal solution of a programmed classical spin system that is a discrete optimization problem \cite{santoro2006optimization, Kadowaki_1998}.

The default anneal schedule is a linear interpolation of the anneal schedules - specifically as a function of annealing time, the anneal parameter $s$ is linearly interpolated starting at $s=0$ at $t=0$ and then increasing up to $s=1$ at the end of the anneal. However, modifications to the anneal schedule such as adding pauses, have been shown to be beneficial for improving the sampling capabilities of quantum annealers \cite{PhysRevB.100.024302, PhysRevApplied.11.044083}. Note that the different generations of D-Wave quantum annealers The actual hardware parameters $A(s)$ and $B(s)$ can differ; all of them use the standardized $s$ anneal parameter to make the same anneal schedules comparable and transferable to the other devices. $A(s)$ and $B(s)$ change over the anneal in a non-linear way; it is only the anneal parameter $s$ that is a linear schedule in the default configuration. There are many other parameters of quantum annealing which can be tuned to improve performance, including anneal offsets \cite{yarkoni2019boosting}, spin reversals \cite{PhysRevA.104.012619}, and different chain weight distributions \cite{Barbosa_2021}. Typically though, tuning many additional parameters requires more computation time. For this cross device comparison, we want the parameters to be as simple as possible and therefore most of these parameters will be set to default values. The only things which will be optimized are annealing time and chain strength, which can be optimized for in a reasonably sized gridsearch. 

The goal of this study is to present concise results from sampling two well studied graph optimization problems (unweighted maximum cut and unweighted maximum clique) on random graphs using four different D-Wave quantum annealing Quantum Processing Units (QPUs), spanning three D-Wave hardware topology generations. In particular, the aim of this study is to provide a simple and clear analysis of using the currently available cloud-based quantum annealing processors for solving paradigmatic and well-studied graph optimization problems, using no classical post-processing. 

Maximum Clique \cite{pardalos1994maximum, carraghan1990exact, OSTERGARD2002197, WU2015693} and Maximum Cut \cite{10.5555/640044.640046, commander2009maximum} are both NP-hard problems with applications in network analysis tasks \cite{4811845, 7542525} including bioinformatics \cite{eblen2012maximum}. Both of these problems in general are believe to be hard to approximate efficiently \cite{10.1145/335305.335322, rossman2010average, khot2007optimal}. The (unweighted) Maximum Clique problem asks us to find the largest fully connected subgraph of an input undirected graph $G = (V, E)$ which is comprised of some set of vertices $V$ and edges $E$. The (unweighted) Maximum Cut problem asks us to find a partition of the undirected problem graph $G = (V, E)$ such that the number of edges shared between the two partitions, also referred to as the \emph{cut} number, is maximized.

\begin{table*}[h!]
\begin{center}
\begin{tabular}{ |p{4cm}||p{2cm}|p{1.4cm}|p{1.4cm}|p{3.4cm}|p{2.4cm}| } 
 \hline
 D-Wave QPU chip id & Topology name & Available qubits & Available couplers & $K_{52}$ minor embedding chain lengths \newline (min, mean, max) & Annealing time (min, max) \newline [microseconds] \\ 
 \hline
 \hline
 \texttt{DW\_2000Q\_6} & Chimera $C_{16}$ & 2041 & 5974  & (14, 14, 14) & (1, 2000) \\ 
 \hline
 \texttt{Advantage\_system4.1} & Pegasus $P_{16}$ & 5627 & 40279  & (5, 5.81, 6) & (0.5, 2000) \\ 
 \hline
 \texttt{Advantage\_system6.1} & Pegasus $P_{16}$ & 5616 & 40135  & (5, 5.81, 6) & (0.5, 2000) \\ 
 \hline
 \texttt{Advantage2\_prototype1.1} & Zephyr $Z_{4}$ & 563 & 4790  & (4, 4.98, 6) & (1, 2000) \\ 
 \hline
\end{tabular}
\end{center}
    \caption{D-Wave quantum annealing processor hardware summary including the clique $K_{52}$ minor embedding used for each device. Note that each of these four devices have some hardware defects which cause the available hardware (qubits and couplers) to be smaller than the ideal graph lattice structure. The table order of the devices corresponds to the generation of the device where the top device is the oldest generation and the bottom device is the newest generation of D-Wave quantum annealers. The newer generation of quantum annealers have denser connectivity graphs, therefore allowing the structured minor embeddings to use shorter chains. }
    \label{table:hardware_summary}
\end{table*}

Programming large QUBO or Ising problems onto quantum annealers in such a way that the quantum annealers will be able to sample the optimal solution(s) is challenging. There is evidence that problems which are natively embeddable onto D-Wave hardware could show meaningful speedups for combinatorial optimization \cite{https://doi.org/10.48550/arxiv.2210.04291}, but typically for arbitrarily structured problems minor embeddings are required to encode them onto current quantum annealers \cite{patton2019efficiently, lucas2019hard, boothby2016fast, zbinden2020embedding, pelofske2022parallel}, which have relatively sparse connectivity. Typically, large minor embeddings begin to fail to sample the optimal solutions, largely due to higher chain break proportions, but the exact scaling with respect to minor embedding size is not precisely known. At long chain lengths, chain break proportions become significant. Perhaps even more importantly though, the task of computing a graph minor (e.g. minor embedding) is very computationally intensive in general (e.g. for general graphs), and therefore it is typically advantageous to be able to compute a fixed minor embedding that has an all-to-all connectivity so that one can re-use the same embedding procedure for arbitrarily connected problem graphs (up to the size of the minor embedding). Therefore, in the experiments presented in this article we use a single fixed minor embedding of size $N=52$ for each of the four D-Wave quantum annealers. Beyond developing hardware with denser connectivity, which is being done across the D-wave generations, there are proposed methods for still being able to utilize quantum annealers for problem which are larger than the hardware. For example by employing classical decomposition algorithms, combined with parallel quantum annealing \cite{pelofske2022parallel, https://doi.org/10.48550/arxiv.2205.12165}, it is possible to reliably sample the optimal solution(s) for large minor embedded maximum clique problems \cite{https://doi.org/10.48550/arxiv.2205.12165}, where otherwise the minor embedded QUBO or Ising would fail to sample the optimal solution. The aim of this research is to investigate medium sized problem instances, $52$ node graphs, whose minor embeddings will fit onto all four D-Wave device topologies, and whose solution quality will not drastically suffer due to the effects of large minor embedded problems. 

The intention is for these results to inform users of D-Wave quantum annealers on what (simple) annealing parameters are most useful, and what approximation ratios can be expected for moderately sized minor embedded combinatorial optimization problems. Specifically, the results presented here will demonstrate to readers three points:

\begin{enumerate}[noitemsep]
    \item How the different hardware generations of D-Wave quantum annealers compare against each other for moderately sized minor embedded graph optimization problems. While there is a consistent improvement across the generations of devices, there are nuances in their performance on the maximum clique QUBOs and maximum cut Isings of random graphs which show that there is a consistent improvement of the devices sampling capability over the hardware generations. 
    \item How the parameters of annealing time and chain strength impact the solution quality. This includes not only the approximation ratio for sampling the optimal solutions, but also the rate of chain breaks that comes with each device and parameter combination. 
    \item What modern D-Wave quantum annealers can, with a reasonable amount of simple parameter tuning, realistically sample from combinatorial optimization problems and when they fail to sample optimal solutions. 
\end{enumerate}

Section~\ref{section:methods} defines the minor embeddings which are used on the four D-Wave devices, and defines the problem instances which will be sampled. Section~\ref{section:results} presents the results of sampling these test instances; the results are primarily approximation ratios and chain break proportions. Fair sampling and properties of how chains break is also commented on. Section~\ref{section:conclusions} concludes with what the results indicate in regards to how the four quantum annealers sample minor embedded combinatorial optimization problems. The figures in this article were generated using matplotlib \cite{thomas_a_caswell_2021_5194481, Hunter:2007} and dwave-networkx \footnote{\url{https://github.com/dwavesystems/dwave-networkx}}  \cite{hagberg2008exploring} in Python 3. All data associated with this article is publicly available \cite{elijah_pelofske_2023_7535790}.

%%%%%%%%%%%%%%%%%%%%%%%%%%%%%%%%%%%%%%%%%%%%%%%%%%%%%%%%%%%%%%%%%
%%%%%%%%%%%%%%%%%%%%%%%%%%%%%%%%%%%%%%%%%%%%%%%%%%%%%%%%%%%%%%%%%
%%%%%%%%%%%%%%%%%%%%%%%%%%%%%%%%%%%%%%%%%%%%%%%%%%%%%%%%%%%%%%%%%
\section{Methods}
\label{section:methods}

One of the primary obstacles with encoding a problem onto a quantum annealer is that the quantum annealer has a fixed hardware graph structure which is not very dense. Ideally one would want a connectivity which is all-to-all, meaning that any given problem no matter its connectivity could be programmed onto the quantum annealer as long as its number of variables was less than or equal to the number of physical qubits. When manufacturing these devices it is much easier to create hardware graphs which are sparsely connected. There is a method of using the sparse connectivity so as to embed a logical graph onto the hardware - that is \emph{minor embedding} \cite{patton2019efficiently, lucas2019hard, boothby2016fast, zbinden2020embedding}. Minor embeddings are very useful in this case for allowing dense combinatorial optimization problems to be mapped onto quantum annealers so as to evaluate how effective quantum annealing could be at sampling solutions to those combinatorial optimization problems. Each logical variable is encoded as a \emph{chain}, of physical qubits which are coupled together with ferromagnetic couplers so that they will be penalized for not being in the same state. The standard method is to generate an all-to-all minor embedding because this allows for embedding re-use for arbitrarily structured problems as long as the embedding size can fit the given problem \cite{PRXQuantum.2.040322}. This alleviates the task of re-computing a minor embedding for every problem, which would incur a large overhead, at the expense of possibly poorer minor embeddings (i.e., longer chain lengths). One of the results of minor embedding however, is a decrease in the capability of the annealing process to sample the optimal solutions of the specified problem \cite{PhysRevResearch.2.023020, Marshall_2022}. One of the problems that comes about is the need to handle cases where the physical chains of qubits which represent the logical variable state disagree on what the variable state is (e.g., a broken chain) - it is common in these cases to use a chain break resolution algorithm, such as simple majority voting to repair the chain \cite{Venturelli_2015, 9325394}. However, this chain repair can cause a distortion of what the measured state was. In particular it is not always clear in a broken chain what the ``intended'' state was. Moreover, as minor embedding sizes increase the probability of sampling the optimal solutions of the problem decrease largely because of the ferromagnetic coupling strengths dominating the programmable energy range allowed on the chip \cite{PhysRevResearch.2.023020, osti_1498001, Marshall_2022}.

Note that tailoring minor embedding for sparse problem instances can result in significantly fewer hardware components (qubits and couplers) being used compared to fully connected minor embeddings - and also in the process reducing the chain lengths used in the embeddings. For especially sparse problem graphs this is a potentially useful encoding to use for sampling such problem instances on the hardware. However, for problem instances with varying connectivity structure and density, making use of the fixed fully connected minor embeddings specifically because of the computation time required to find a minor embedding for each individual problem instance, and also because of the potentially poor quality of the embedding (namely, there being chains with significantly varying lengths), is a good approach.

Table~\ref{table:hardware_summary} shows the hardware summary of the D-Wave devices which will be used for testing. There are two parameters which we will vary in a simple grid search. Those parameters are \emph{annealing time} and \emph{chain strength}. The chain strengths for the Maximum Clique problems are varied over $0.5, 1, 2, \dots, 11, 12$. The chain strengths for the Maximum cut problems the chain strengths are varied over $0.5, 1, 2, 3, \dots, 18, 19, 20, 25, 30, \dots 45, 50$. The magnitude of chain strengths for Maximum Cut were set to be greater than what was used for the Maximum Clique QUBOs because it was found that good Maximum Cut solutions were only found at comparatively larger chain strengths. This seems to be the case because the Maximum Cut formulation is setting antiferromagnetic coupling terms on the quadratic coefficients between logical variables (see Section~\ref{section:Maximum_cut_Ising}) - thus partitioning the graph into two sets of nodes whose cut value is maximized. The consequence of this is that the chains themselves are more prone to breaking because of the strong antiferromagnetic terms. Therefore, larger chain strengths are needed in order for the computation to carry out successfully with fewer chain breaks. However, because of the precision limitations for encoding weights on the quantum annealers larger chain strengths (for example on the order of 100) would cause the problem coefficients to be lost in noise. Therefore, a maximum chain strength of $50$ was chosen as a reasonably large chain strength that would make sense to program for minor embedded problems of this size. 

For both Maximum Clique and Maximum Cut problems, annealing times are varied over the minimum annealing time possible on the device, which is either $1$ or $0.5$ microseconds, and $10, 100, 1000, 2000$ microseconds. Each experiment used exactly $1000$ anneal-readout cycles so that the number of samples is held constant (note that because of varying annealing time the total compute time is therefore not held constant across the various experiments). \texttt{readout\_thermalization} and \texttt{programming\_thermalization} time were set to $0$ microseconds so as to quantify the QPU capabilities when all inter-sample thermalization is removed. All other quantum annealing parameters are set to default, for example the annealing schedule is linearly interpolated between the anneal fractions $s=0$ and $s=1$. The clique minor-embeddings were computed using the D-Wave SDK \emph{find clique embedding} \footnote{\url{https://docs.ocean.dwavesys.com/en/stable/docs_minorminer/source/reference/generated/minorminer.busclique.busgraph_cache.find_clique_embedding.html}} method which finds clique embeddings that minimize the chainlengths for the given size and the three different hardware graphs\cite{boothby2016fast, boothby2020next} (note that a similar idea is template based minor-embedding \cite{serra2021templatebased}). Figure~\ref{fig:minor_embedding} in Appendix~\ref{section:appendix_minor_embeddings} shows the exact structure of these fixed minor embeddings on the four different QPU's. These four minor embeddings are the only minor embeddings used in all of the presented experiments. These clique embeddings are much more structured compared to random minor embeddings \cite{pelofske2022parallel}, and give the programmed problems more balanced and shorter chain lengths. For these fixed hardware graphs, all-to-all minor embeddings can be computed efficiently \cite{boothby2016fast, boothby2020next} even though in the general case the problem of minor embedding is NP-hard \cite{https://doi.org/10.48550/arxiv.1406.2741}. These structured and efficiently computable minor embedding algorithms could only find a clique embedding for the \texttt{Advantage2\_prototype1.1} device up to $52$ nodes, which combined with the fact that increasing minor embedding sizes causes decreased solution quality meant that $N=52$ problem sizes would allow for a reasonable direct cross-platform comparison of the four quantum annealers. The relatively uniform chain lengths provided by these structured clique embedding algorithms \cite{boothby2016fast} are important since logical chains that are ferromagnetically bound can lead an anneal path that is different from the rest of the problems, if the chain lengths are not uniform, and if the chains are not calibrated at a low level. Chains can be very finely calibrated using anneal offsets, flux bias offsets, and tuning the programmed linear and quadratic terms such that the effective weights are what was intended \cite{King_2022, King_2023_5000q} - in this case we do not perform such low level calibrations and treat the D-Wave devices as black-box optimization algorithms which we wish to evaluate.

The NP-hard problems unweighted maximum clique and unweighted maximum cut were selected as simple test instances for two reasons.

\begin{enumerate}[noitemsep]
    \item They do not contain higher order terms. Higher order terms introduce additional complexity especially for embedding onto quantum annealers because it becomes necessary to perform order reduction to map the higher order problem to a quadratic problem with auxiliary variables where the ground state solutions match that of the original problem \cite{VALIANTE2021108102, 5444874, pelofske2022quantum, jiang2018quantum}. 
    \item The problem coefficients do not range in scale dramatically. Because of the precision limits on D-Wave quantum annealers, encoding problems which have coefficients that are very different in size will cause the smaller coefficients to be effectively set to $0$ because the devices require the programmed coefficients to be in a strict range. 
\end{enumerate}

\begin{equation}
    C(x) = \sum_{(i,j) \in E} (2z_i - 1)\cdot (2z_j -1)
    \label{equation:cplex_binary_formulation}
\end{equation}

\begin{figure*}[h!]
    \centering
    \includegraphics[width=0.35\textwidth]{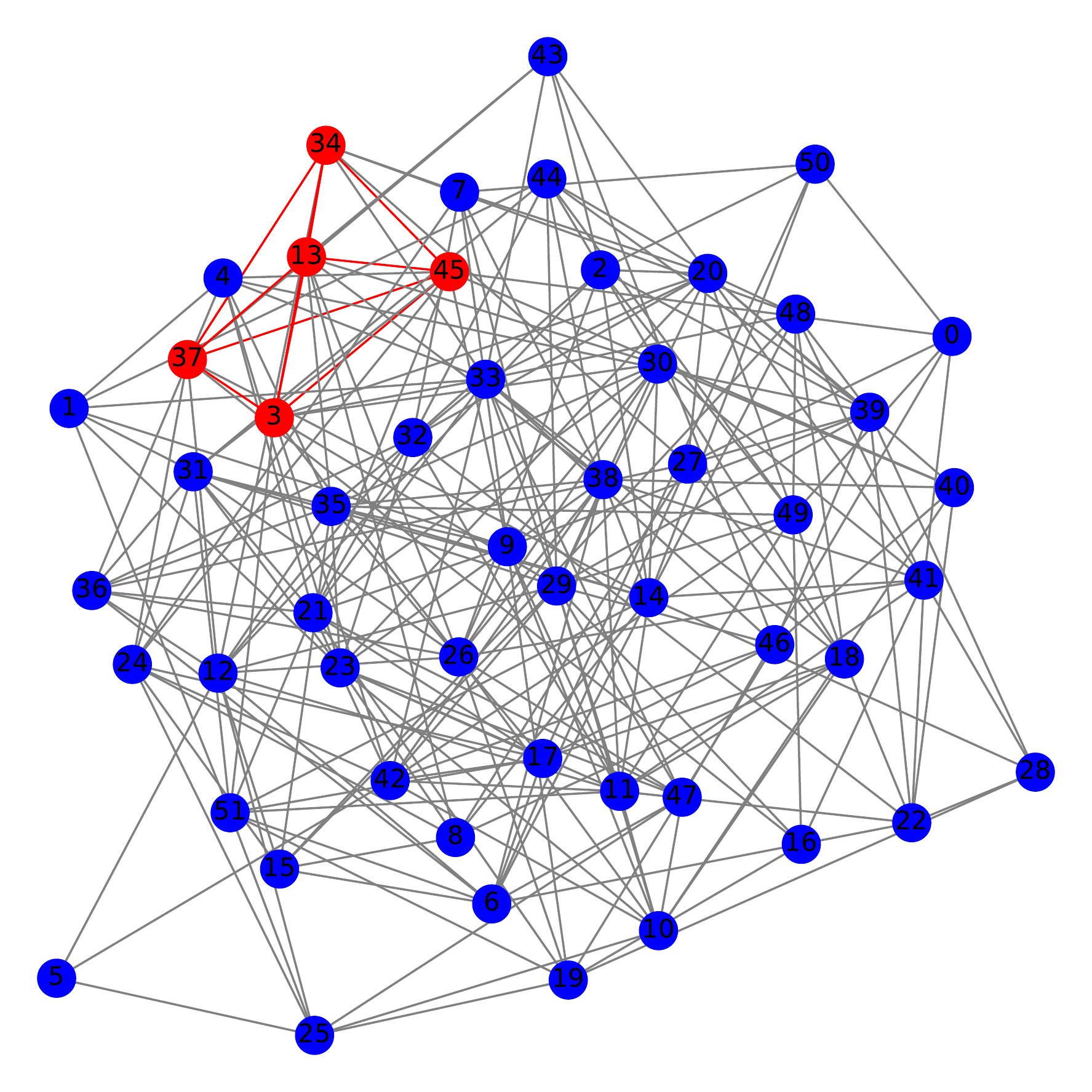}
    \includegraphics[width=0.35\textwidth]{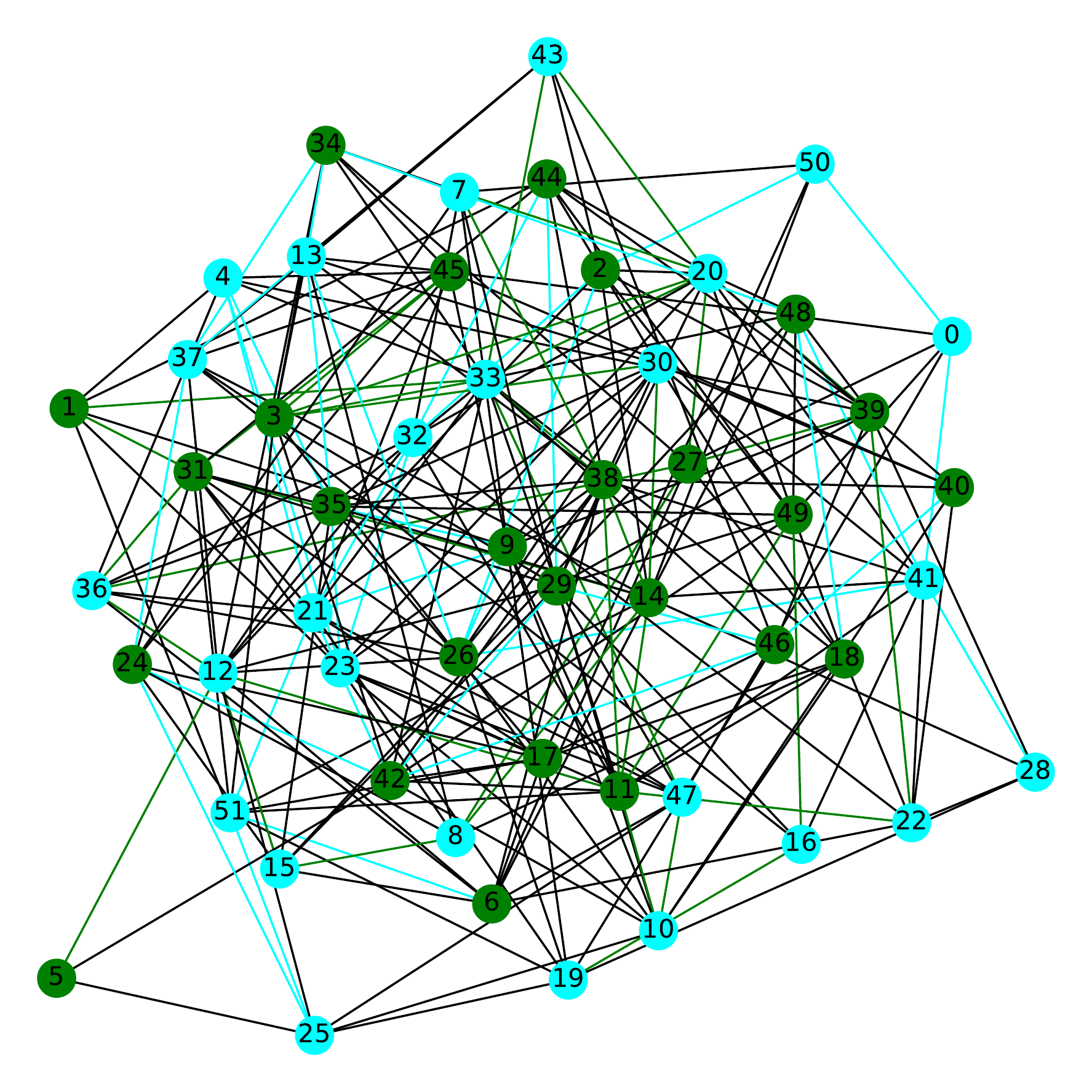}
    \caption{Example random $52$ node graph with $254$ edges. The left plot shows the single Maximum Clique of the graph, which has size $5$. The right plot shows a Maximum Cut partition of the same graph, where the two partition node and edges are green and cyan, and the shared edges between the bi-partition are black. }
    \label{fig:graph_renderings_maxclique_maxcut}
\end{figure*}

The optimal maximum clique solutions are computed using networkx \cite{hagberg2008exploring}. The optimal maximum cut solutions were computed using a Mixed Integer Quadratic Programming (MIQP) formulation of maximum cut with boolean variables using the CPLEX \cite{cplexv12} solver. To adapt the spin variables of the Maximum Cut Ising to boolean variables, eq.~\eqref{equation:cplex_binary_formulation} is used, where $z_i$ and $z_j$ are the arbitrary spin variables in the original Ising model. Eq.~\eqref{equation:cplex_binary_formulation} is defined simply by the edgeset $E$ for any graph $G = (V, E)$. The CPLEX solver does not compute multiple ground states, if they exist, (beyond the trivial maximum cut symmetry), whereas the networkx solver does produce all degenerate Maximum Clique solutions for the given graph. Figure~\ref{fig:graph_renderings_maxclique_maxcut} shows a visual representation of the Maximum Clique and Maximum Cut of an example random graph, drawn using networkx \cite{hagberg2008exploring} with fruchterman reingold layout \cite{fruchterman1991graph}.

The test problems are Erd{\H o}s--R{\'e}nyi random graphs~\cite{ErdosRenyi1960}, generated with density parameter uniformly drawn from the range $(0.05, 0.95)$. All graphs have exactly $52$ nodes. The graphs are also specifically constructed such that there are no degree zero nodes. $200$ random graphs will be sampled for their Maximum Cliques, and $150$ random graphs will be sampled for the Maximum Cuts. 

The primary metric of interest that will be reported is the mean approximation ratio of the $1000$ samples computed for the given problem instance (and annealing time and chain strength). In particular, the mean approximation ratio is measured as the mean of all of the approximation ratios for all of the samples (these measures thus show the mean of the entire sampled distribution). The approximation ratio is the ratio of the objective value of a given sample divided by the optimal objective value possible for a specific graph. In this case, the optimal objective function values would be the Maximum Clique or the Maximum Cut. For the results shown in Section~\ref{section:results}, no un-embedding algorithm is used when computing the mean approximation ratio of the $1000$ samples. In particular, if an anneal has any broken chain then its objective function is computed as $0$, and will be used in the computation of the mean approximation ratio. Chain repair (un-embedding) algorithms are not used in this study so as to perform a direct cross-device comparison without classical post-processing algorithms complicating such an analysis. Un-embedding post-processing algorithms can additionally complicate runtime measurements since the classical compute time would need to be factored in -- therefore the direct discard of broken chains simplifies the compute time measurements. A mean approximation ratio of $1$ is the best sampling rate possible, and a mean approximation ratio of $0$ means that the sampler does not ever find optimal solutions. The other metric that will be reported is the proportion of the $1000$ anneal-readout cycles which contained any broken chains. A good parameter choice of annealing time and chain strength should result in approximation ratios close to, or equal to, $1$, and chain break proportions which are very close to, or equal to, $0$ (without the classical post-processing of an un-embedding algorithm). 

Next, Section~\ref{section:Maximum_cut_Ising} will briefly define the Maximum Cut Ising formulation that will be used, and Section~\ref{section:Maximum_clique_QUBO} will briefly define the Maximum Clique QUBO formulation that will be used. The D-Wave devices can be equivalently programmed to sample Ising or QUBO formulated problems. 

\begin{table*}[h!]
\begin{center}
\begin{tabular}{ |p{4.1cm}||p{2.8cm}|p{2.5cm}| } 
 \hline
 Chip ID & Maximum Clique & Maximum cut \\ 
 \hline
 \hline
 \texttt{DW\_2000Q\_6} & $187/200$ & $40/150$ \\ 
 \hline
 \texttt{Advantage\_system4.1} & $200/200$ & $68/150$ \\
 \hline
 \texttt{Advantage\_system6.1} & $200/200$ & $63/150$ \\
 \hline
 \texttt{Advantage2\_prototype1.1} & $194/200$ & $58/150$  \\
 \hline
 \hline
\end{tabular}
\end{center}
\caption{Optimal solution sampling success rate for the maximum clique and maximum cut problems applied to the random test graphs. Here the success rate is quantified purely in terms of whether \emph{any} sample, without an unembedding algorithm applied, across the varied chain strength and annealing time executions was in the ground state (e.g. it is an optimal solution). Importantly, all maximum clique problems were able to be solved to optimality, in contrast to the maximum cut problem instances where not all problems were solved to optimality. The other interesting observation is that the two quantum annealers with Pegasus hardware graphs performed the best at obtaining optimal maximum cut solutions, compared to the other two devices.  }
\label{table:optimal_solution_table}
\end{table*}

\subsection{Maximum Cut Ising Model}
\label{section:Maximum_cut_Ising}
Eq.~\eqref{equation:maxcut_ising} describes the Ising formulation for the maximum cut problem \cite{8123654} on an undirected graph $G = (V, E)$. $V$ is the set of vertices, also referred to as nodes, in the graph which are connected by the set of edges $E$. The maximum cut Hamiltonian formulation gives a connectivity graph that is the same interaction graph as the problem graph instance. Ising problems have variables assignments $x_i \in \{+1, -1\}$, and in the case of eq.~\eqref{equation:maxcut_ising} the vector $x$ is describing the variable assignments for each variable. The important fact is that by minimizing the objective function of eq.~\eqref{equation:maxcut_ising}, the maximum cut variable assignments for any graph $G$ can be found. Which of the two variable states each variable is assigned in the lowest energy solution determines which partition that corresponding node is in. 

\begin{equation}
    H(x) = \sum_{(u, v) \in E} x_u v_v
    \label{equation:maxcut_ising}
\end{equation}

\subsection{Maximum Clique QUBO}
\label{section:Maximum_clique_QUBO}
Eq.~\eqref{equation:maxclique_QUBO} describes the QUBO formulation for the maximum clique problem \cite{chapuis2019finding, pelofske2019solving, 8123654, Lucas_2014, https://doi.org/10.48550/arxiv.1801.08653, 10.1145/3075564.3075575} that will be used in the experiments, which defines a Hamiltonian interaction connectivity graph that is the complement of the graph problem instance. 

\begin{equation}
    H(x) = -A \sum_{i \in V} x_i + B \sum_{(u, v) \in \overline{E}} x_u x_v
    \label{equation:maxclique_QUBO}
\end{equation}

Setting $A=1$, $B=2$ ensures the minimum of eq.~\eqref{equation:maxclique_QUBO} corresponds to the maximum clique on the undirected graph $G = (V, E)$ \cite{chapuis2019finding, pelofske2019solving}. $\overline{E}$ denotes the complement of the edgeset of the graph $G$. QUBO problems have variable assignments $x_i \in \{+1, 0\}$; the vector $x$ in eq.~\eqref{equation:maxclique_QUBO} is a set of binary variable assignments. Those variable assignments are then evaluated by eq.~\eqref{equation:maxclique_QUBO}. That objective function is minimized when the variables which represent the nodes in the graph $G$ which form a maximum clique are assigned $1$, and all other variables are assigned $0$.

%%%%%%%%%%%%%%%%%%%%%%%%%%%%%%%%%%%%%%%%%%%%%%%%%%%%%%%%%%%%%%%%%
%%%%%%%%%%%%%%%%%%%%%%%%%%%%%%%%%%%%%%%%%%%%%%%%%%%%%%%%%%%%%%%%%
%%%%%%%%%%%%%%%%%%%%%%%%%%%%%%%%%%%%%%%%%%%%%%%%%%%%%%%%%%%%%%%%%
\section{Results}
\label{section:results}

This section presents the experimental results which compare the four D-Wave quantum annealers. Section~\ref{section:results_maximum_clique} details the Maximum Clique results, Section~\ref{section:results_maximum_cut} details the Maximum Cut results, Section~\ref{section:results_fair_sampling} briefly examines how fairly degenerate ground states, specifically of some of the Maximum Clique problems, are sampled, and lastly Section~\ref{section:results_broken_chain_properties} briefly notes consistent properties of chain breaks. 

The first question to consider is whether the quantum annealers were able to sample the optimal solutions of the combinatorial optimization problems. At the problem size of $52$ variables, these test problems are well out of the trivial range of combinatorial optimization problems where randomly sampling the search space with thousands of samples would find the optimal solution (this is the case for very small problem sizes that have on the order of $10$ variables or less). Table~\ref{table:optimal_solution_table} shows for what devices and what problems the optimal solutions were able to be computed. 

Note that for all of the results presented in this section, for each parameter and setting that is changed (including annealing times, problem instance, QPU, and chain strength), that problem instance is sampled on the D-Wave hardware using a fixed number of samples - specifically $1,000$ anneal-readout cycles. 

\begin{figure*}[h!]
    \centering
    \includegraphics[width=0.24\textwidth]{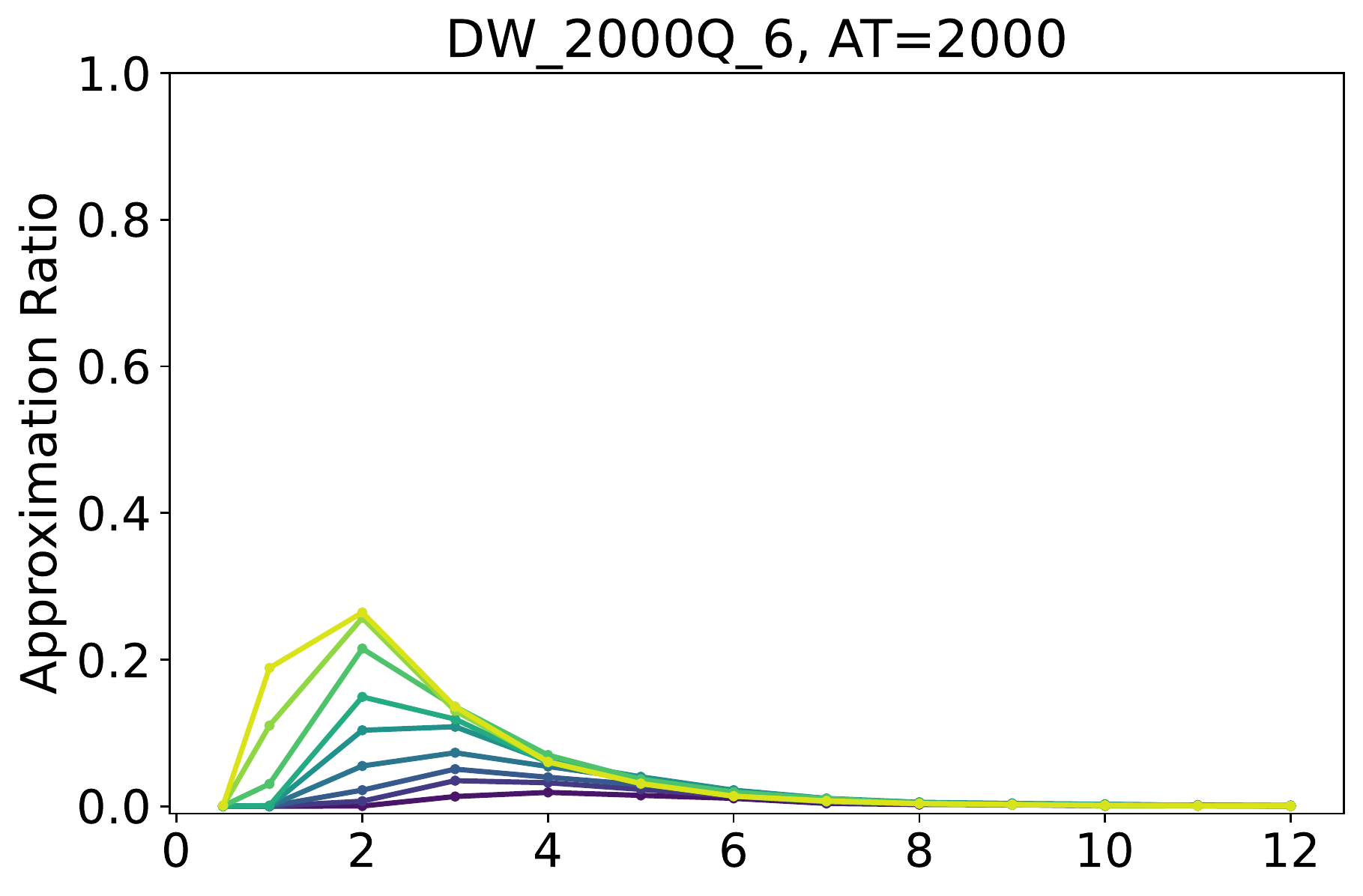}
    \includegraphics[width=0.24\textwidth]{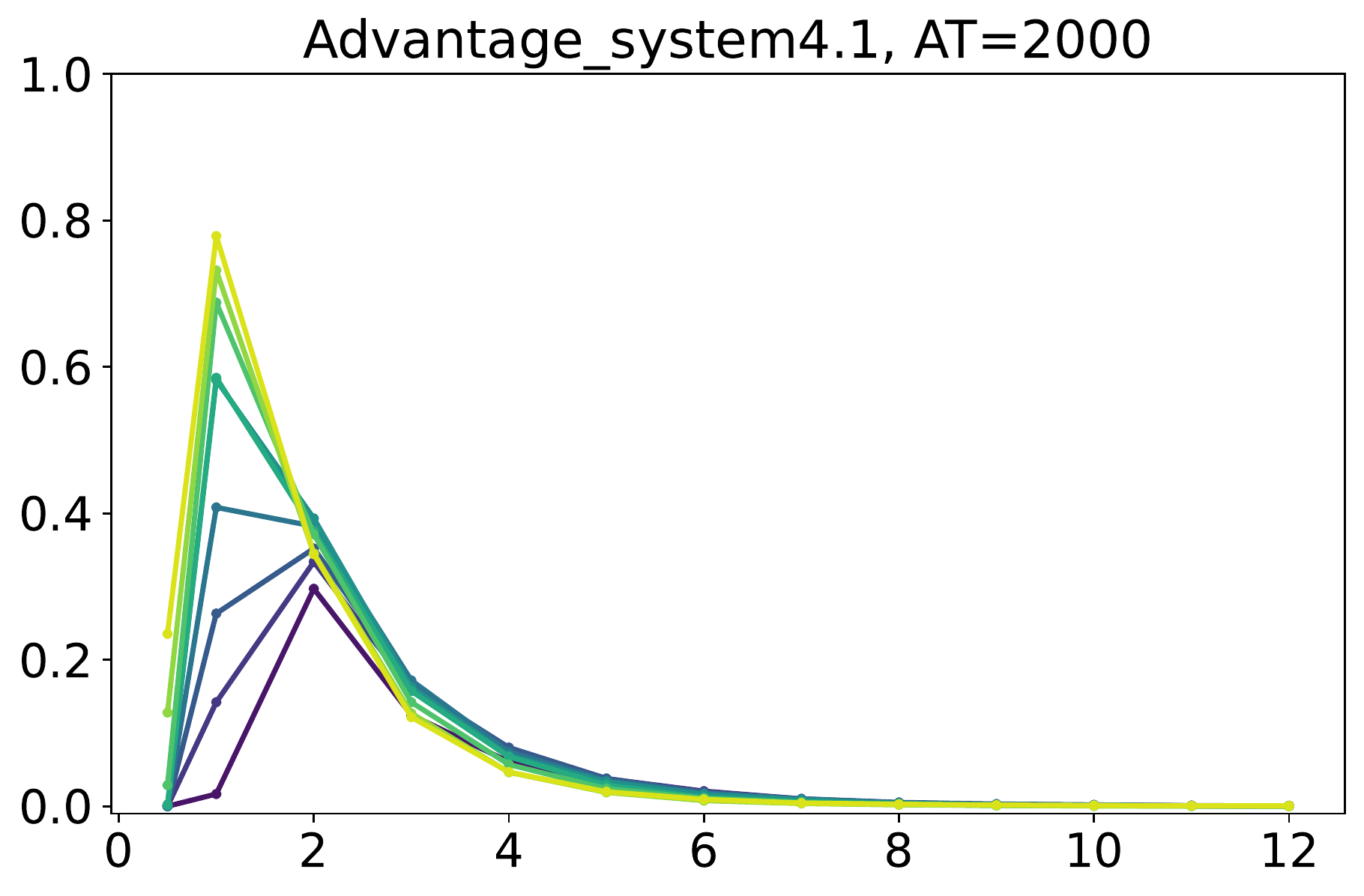}
    \includegraphics[width=0.24\textwidth]{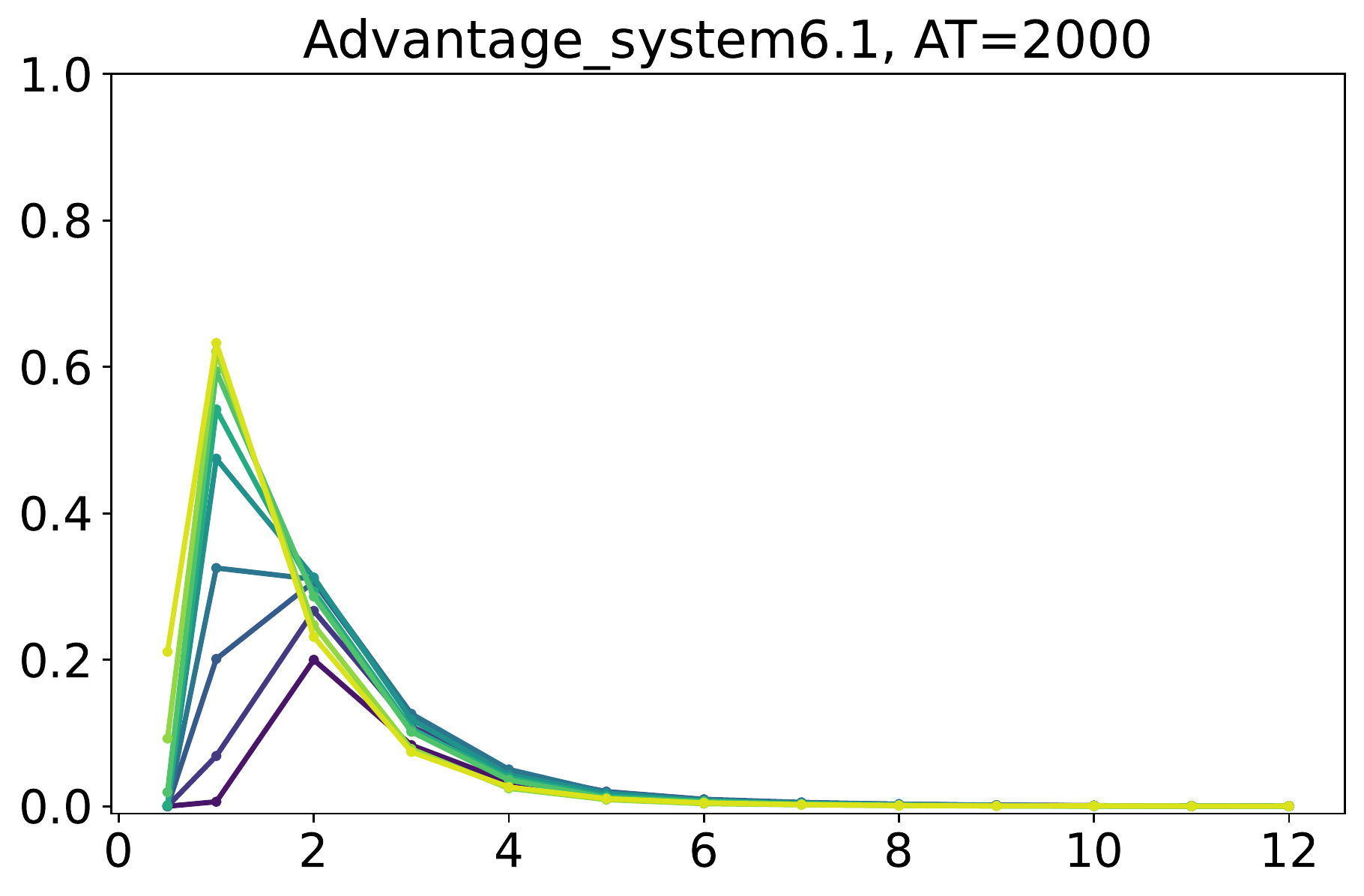}
    \includegraphics[width=0.24\textwidth]{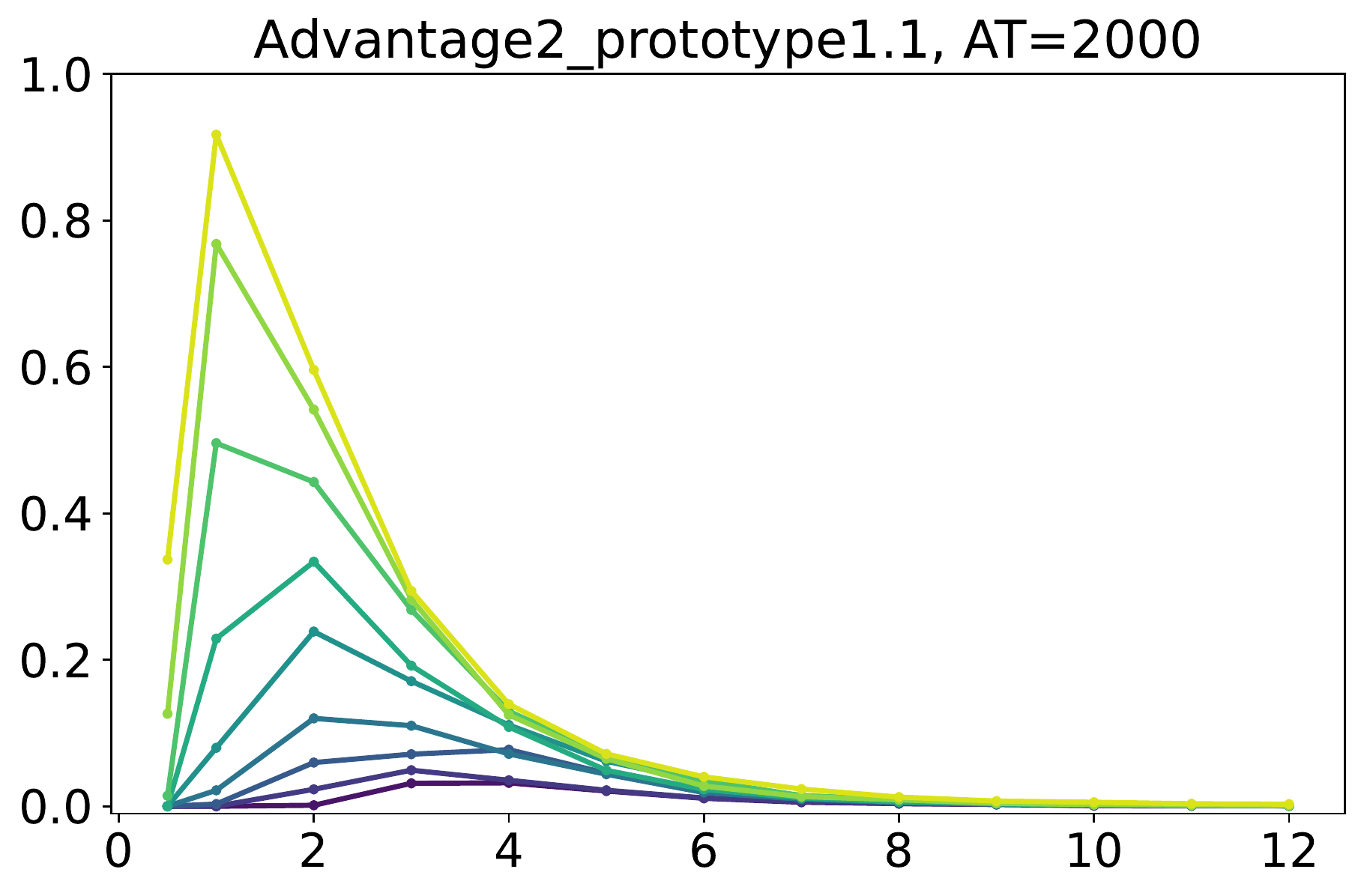}\\
    \includegraphics[width=0.24\textwidth]{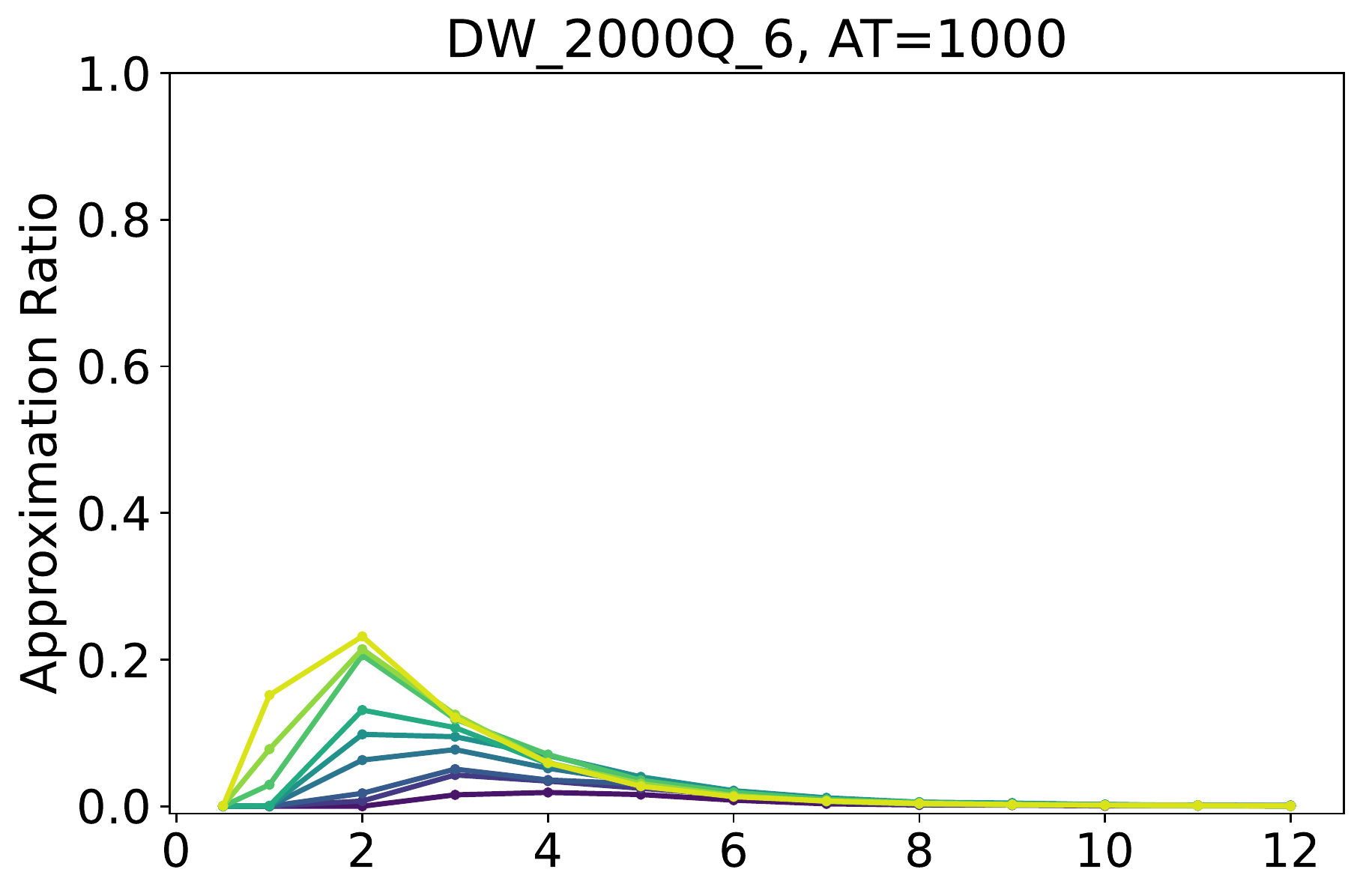}
    \includegraphics[width=0.24\textwidth]{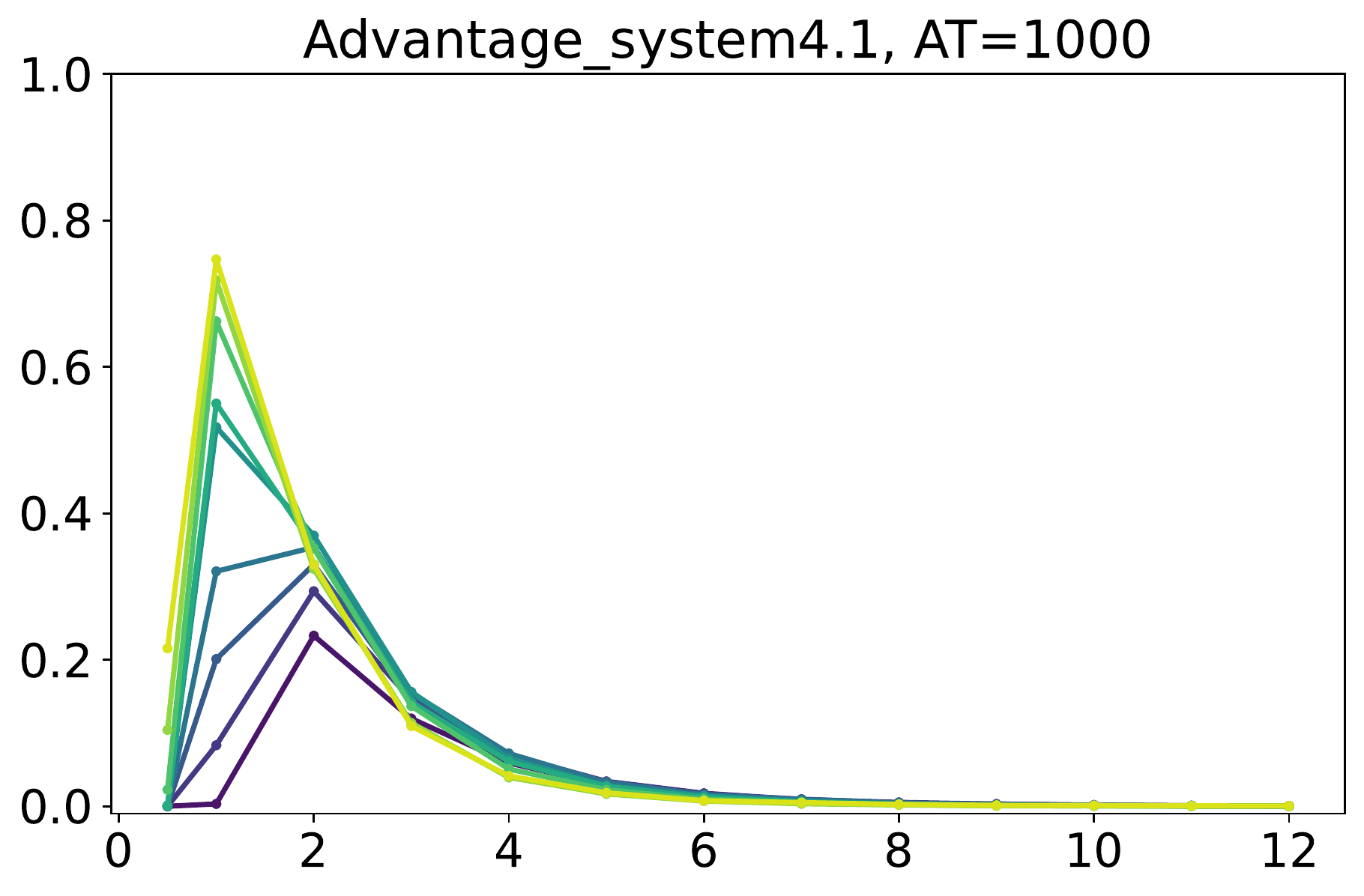}
    \includegraphics[width=0.24\textwidth]{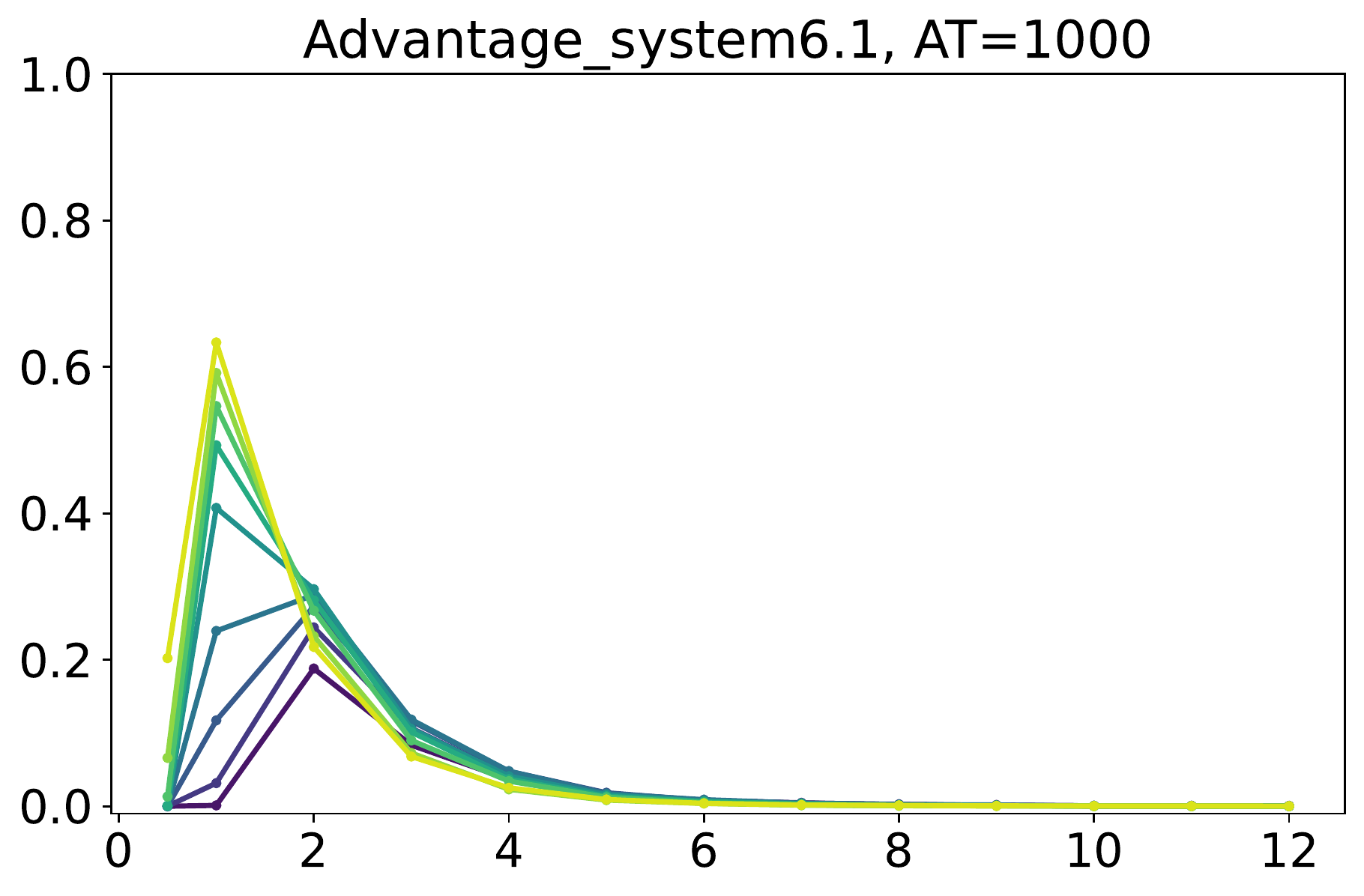}
    \includegraphics[width=0.24\textwidth]{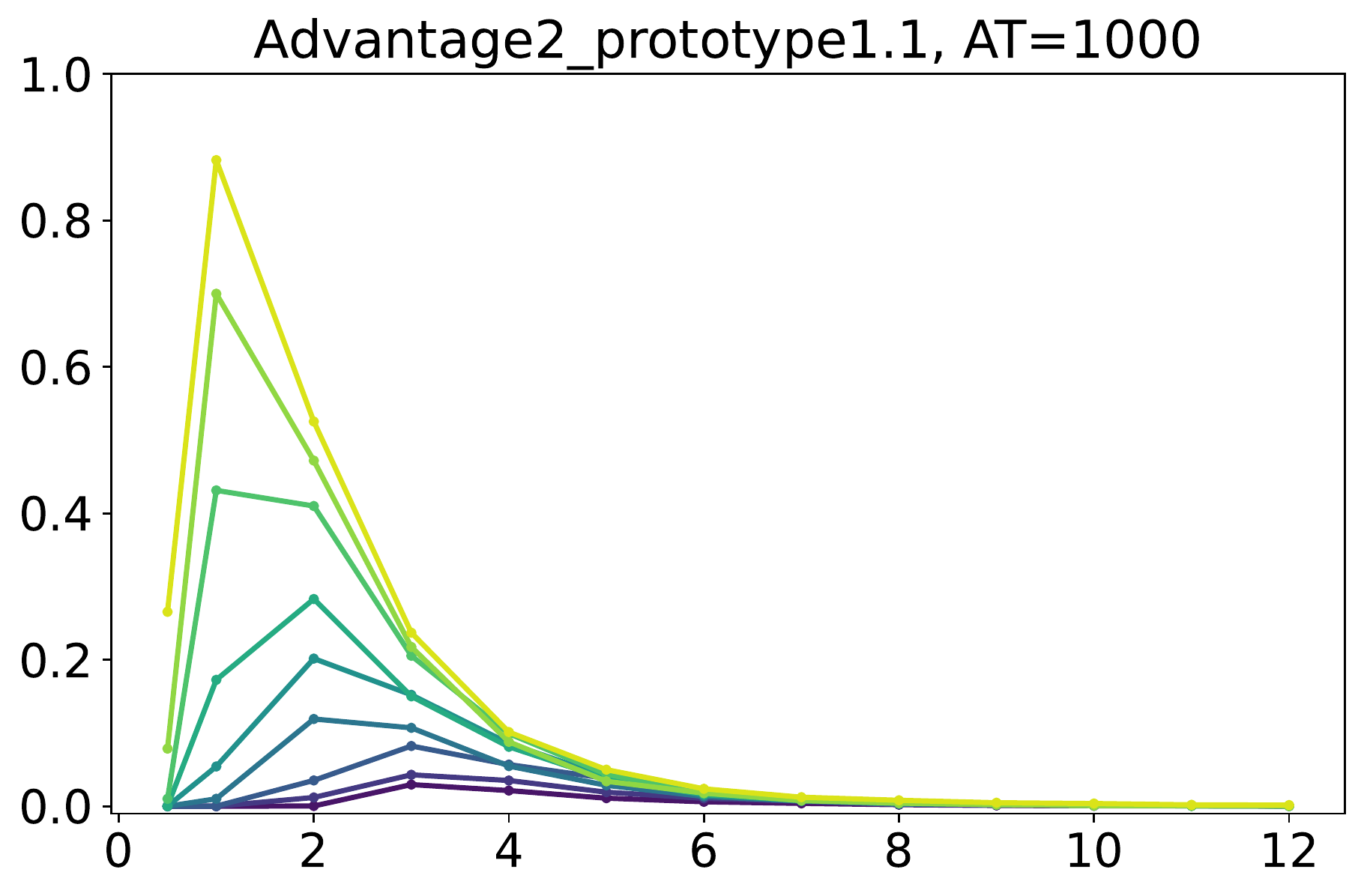}\\
    \includegraphics[width=0.24\textwidth]{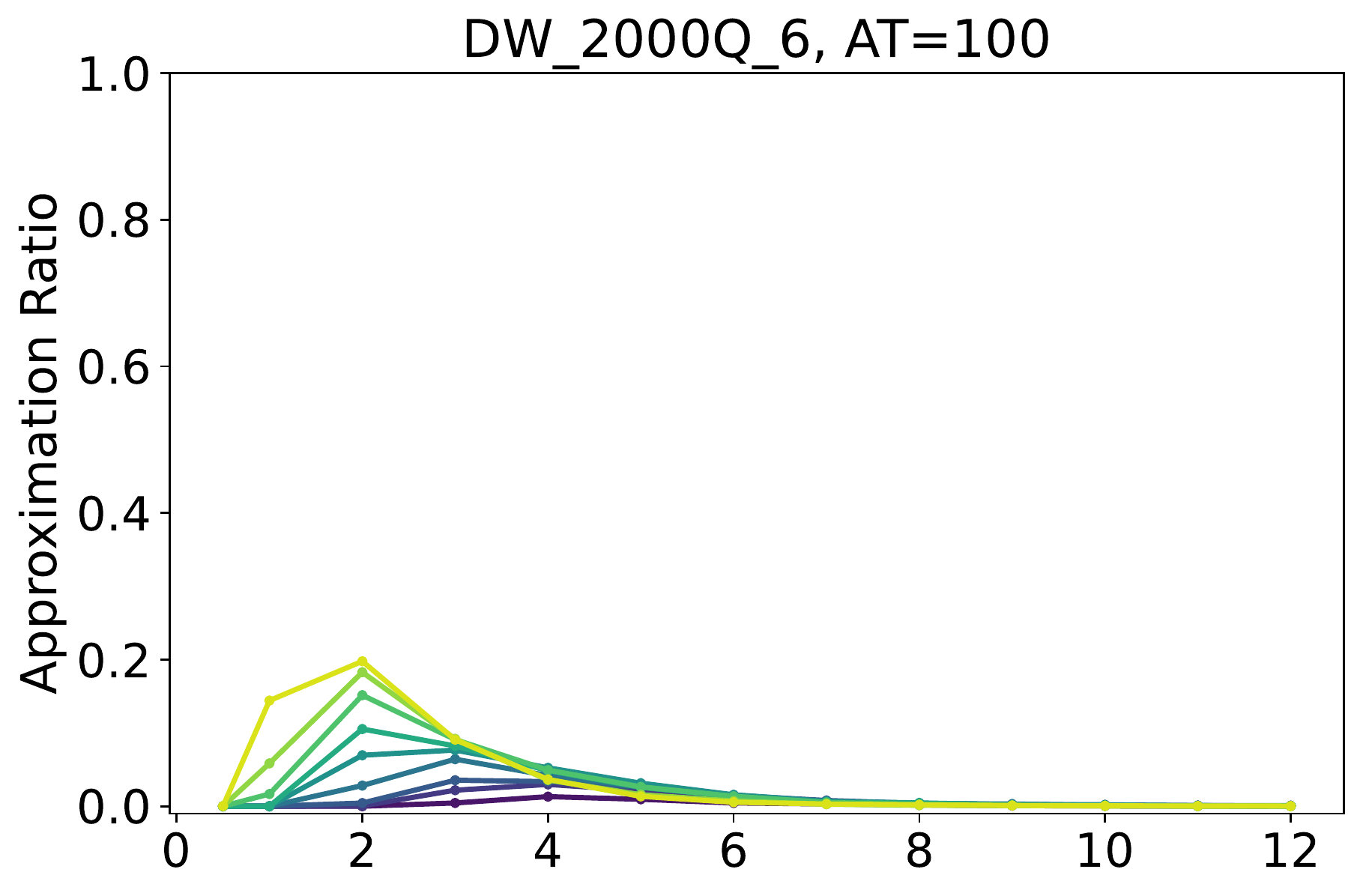}
    \includegraphics[width=0.24\textwidth]{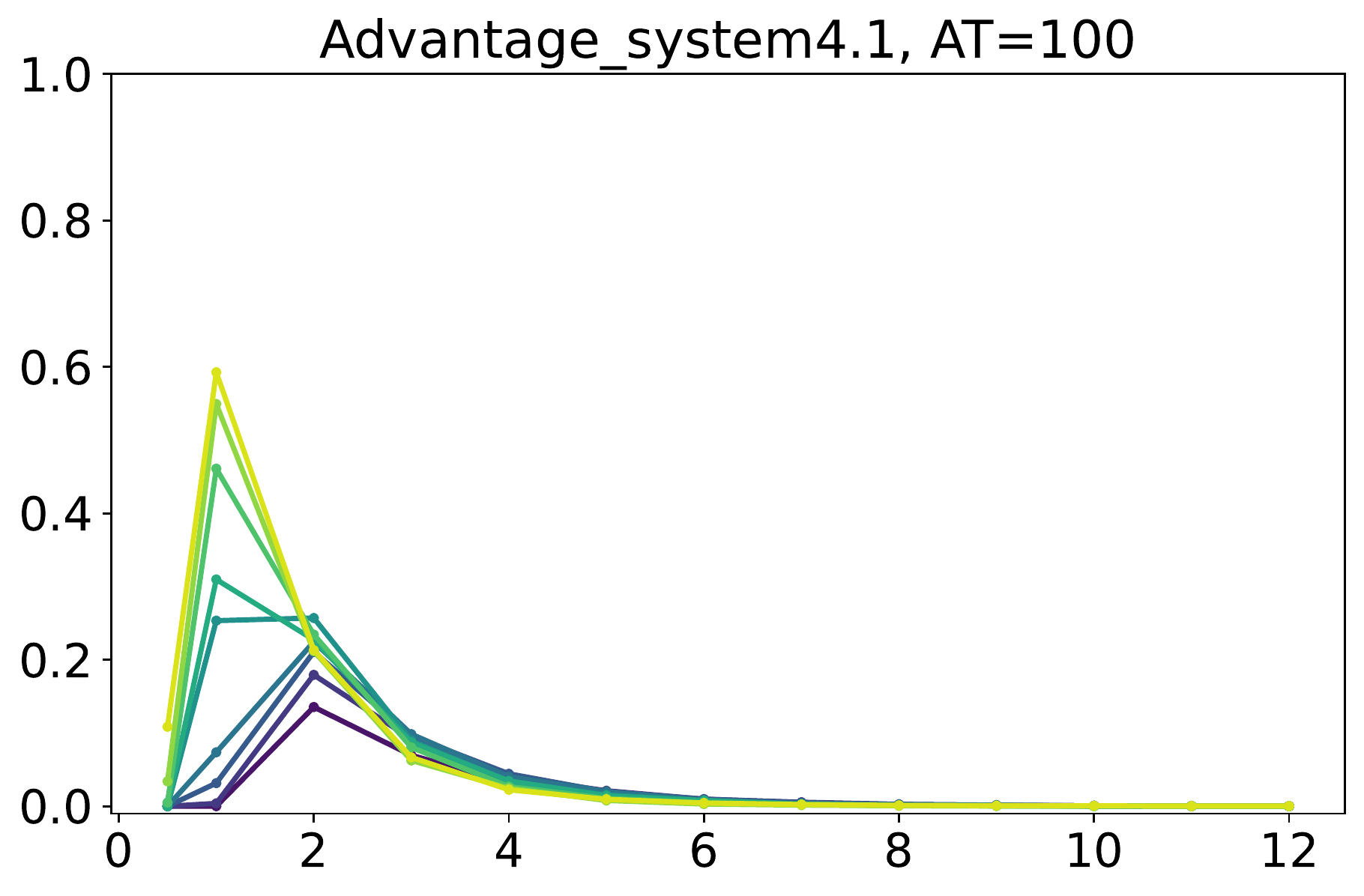}
    \includegraphics[width=0.24\textwidth]{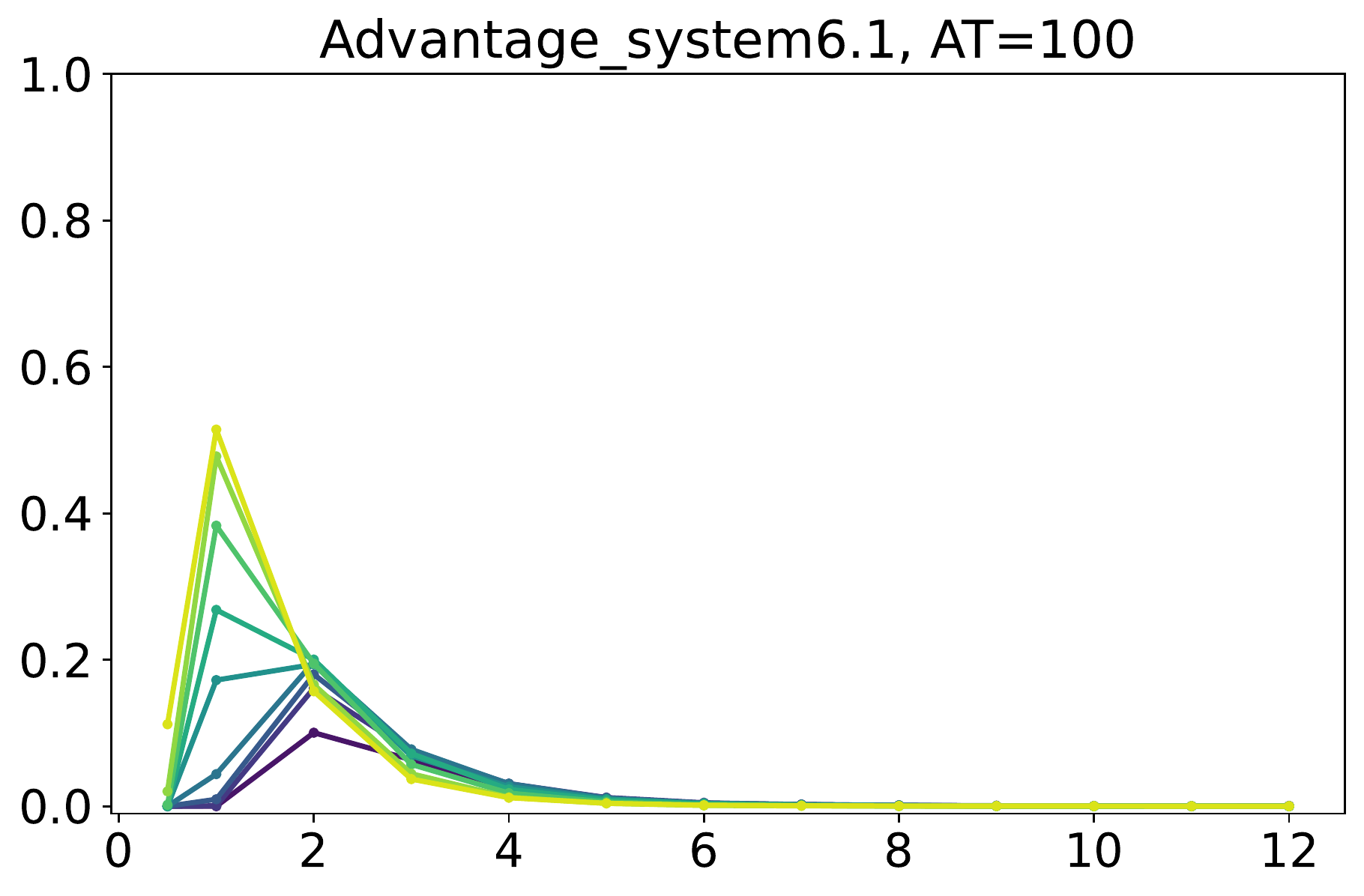}
    \includegraphics[width=0.24\textwidth]{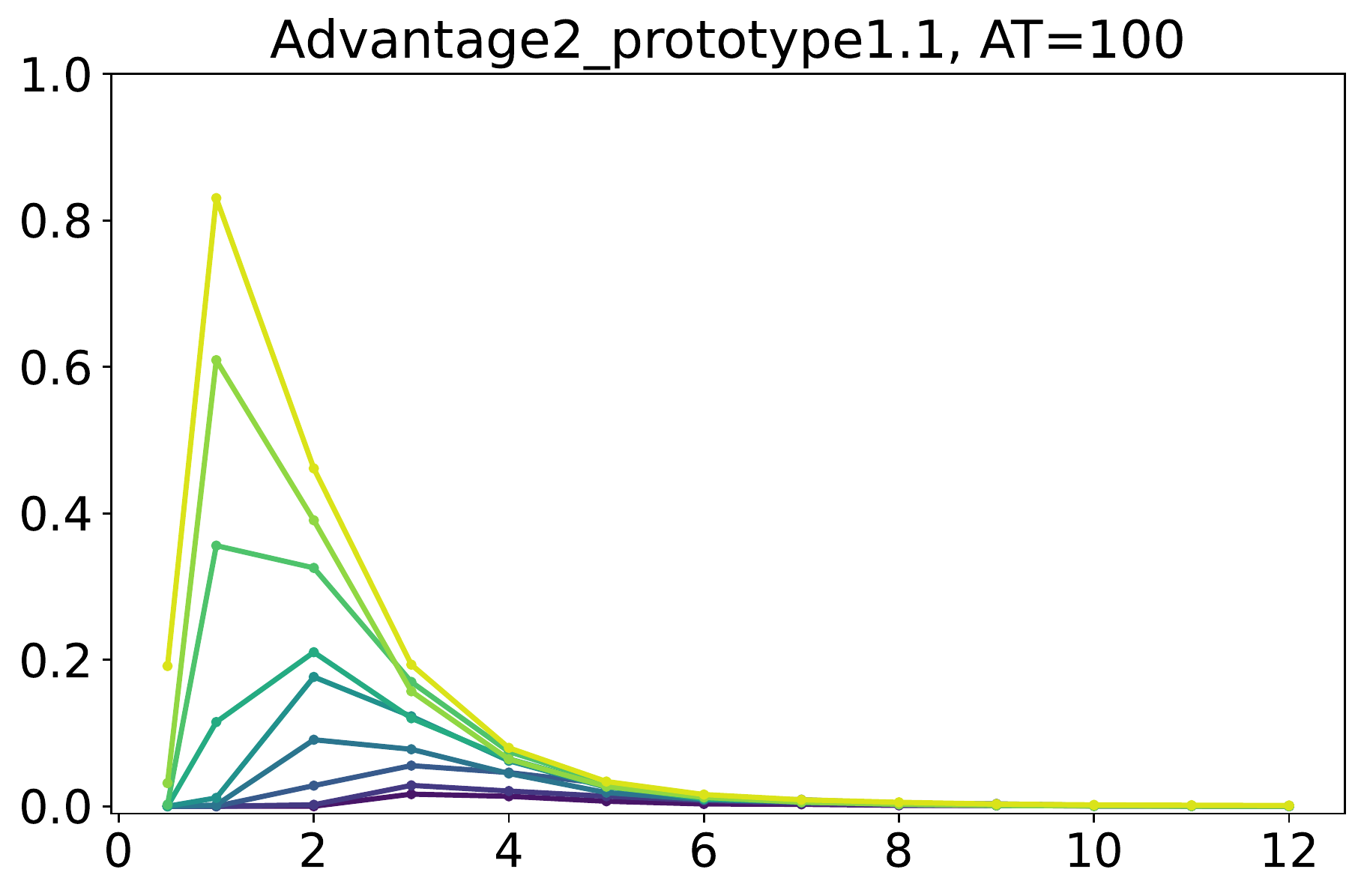}\\
    \includegraphics[width=0.24\textwidth]{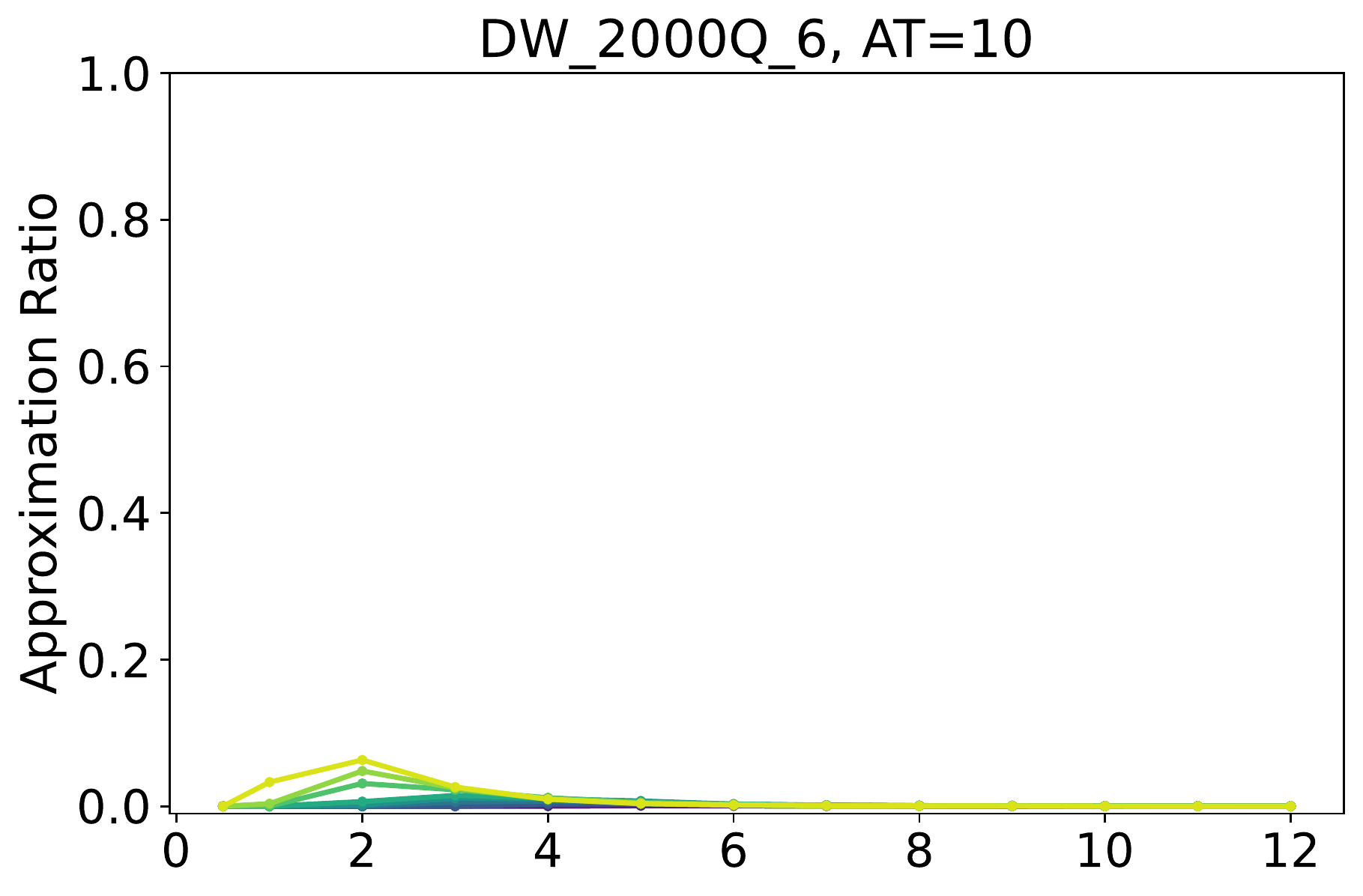}
    \includegraphics[width=0.24\textwidth]{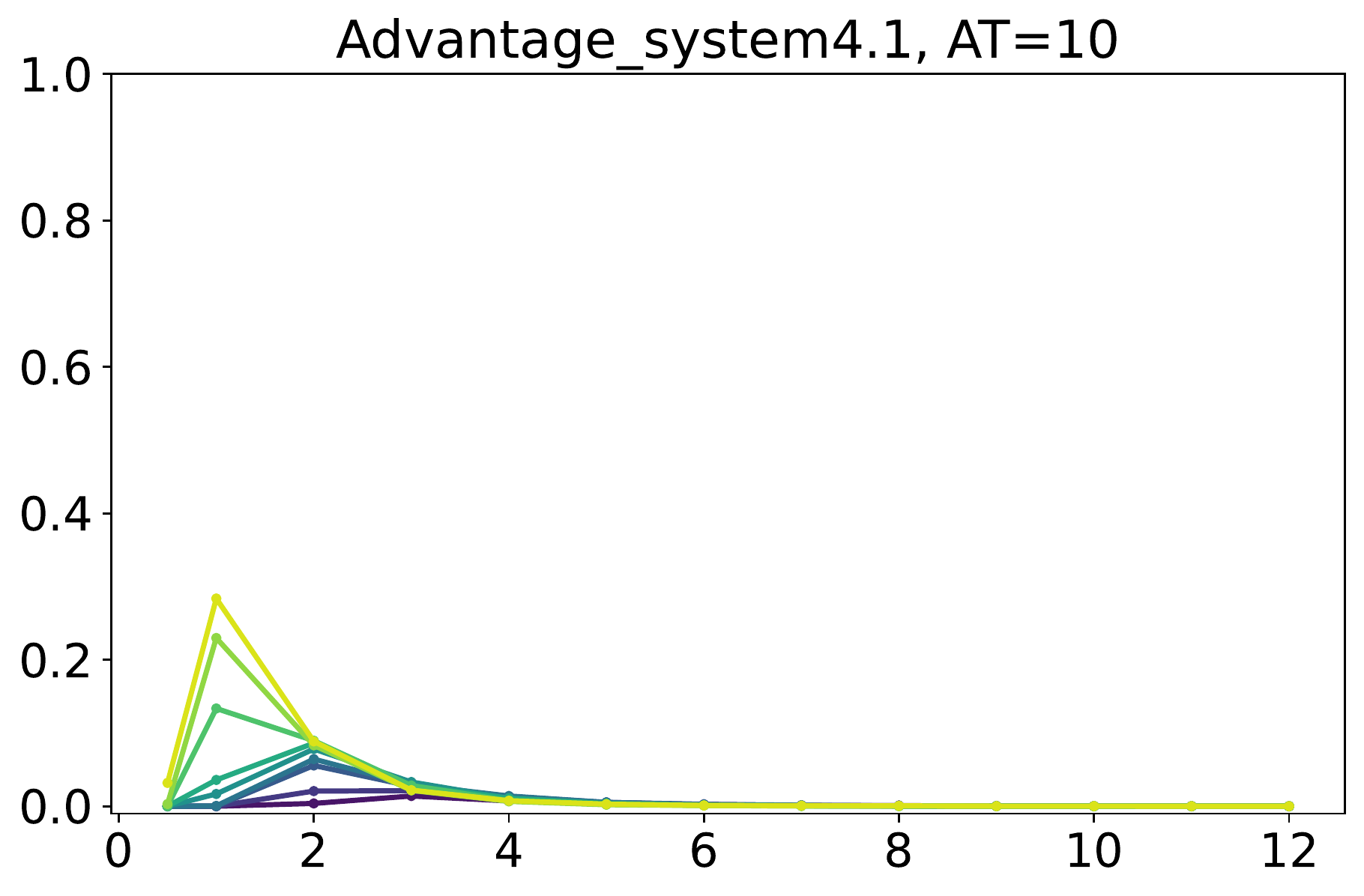}
    \includegraphics[width=0.24\textwidth]{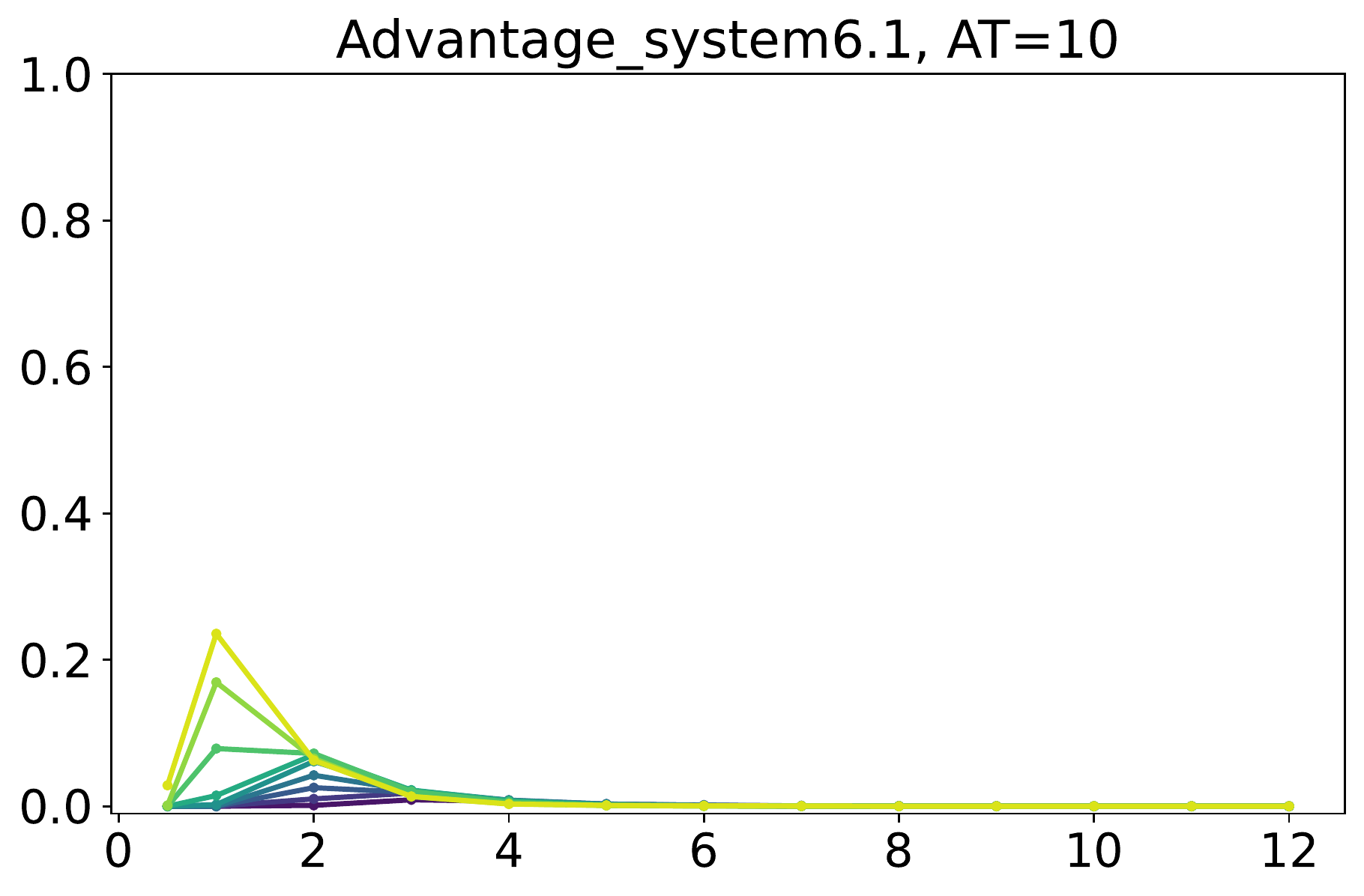}
    \includegraphics[width=0.24\textwidth]{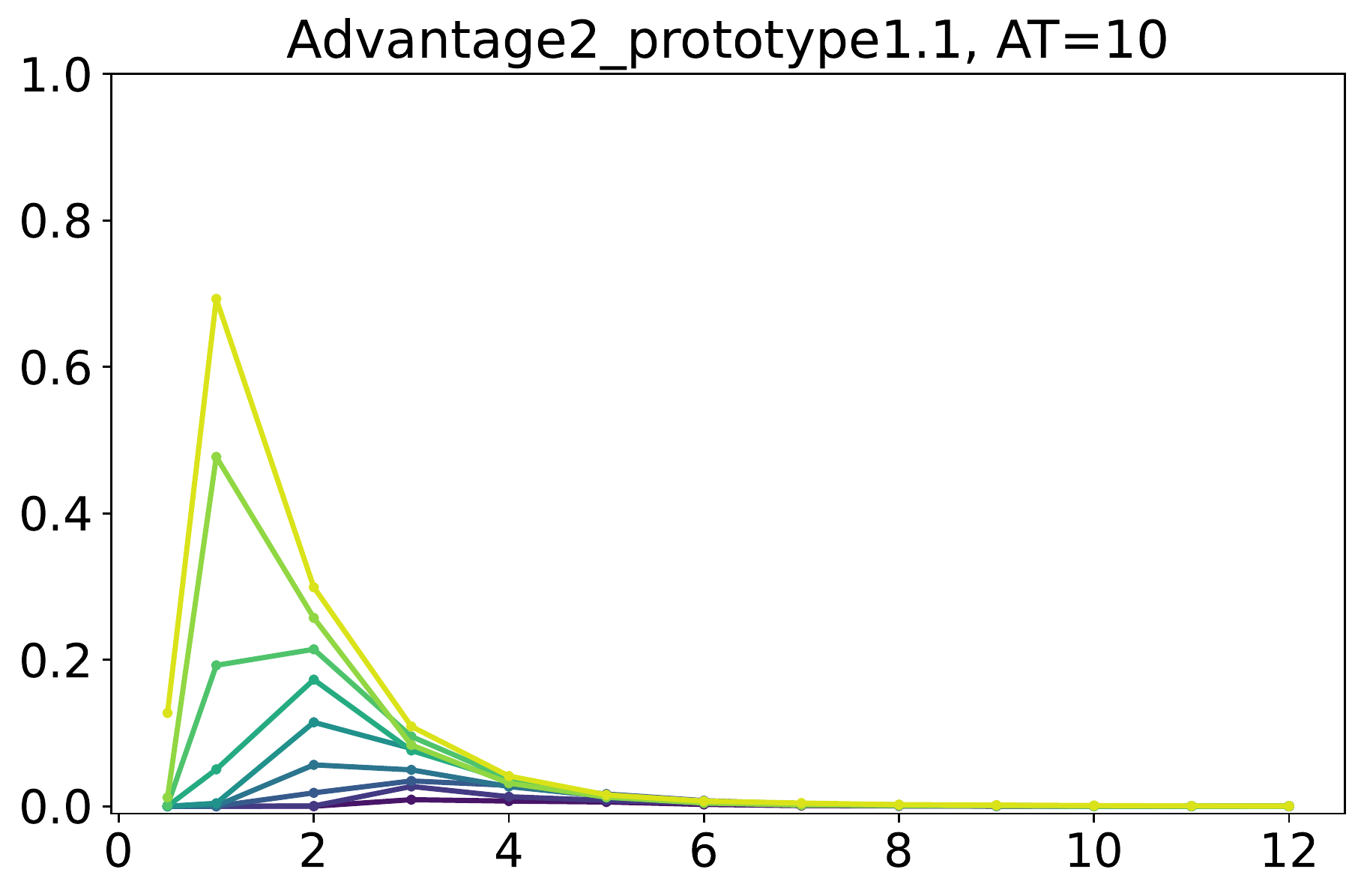}\\
    \includegraphics[width=0.24\textwidth]{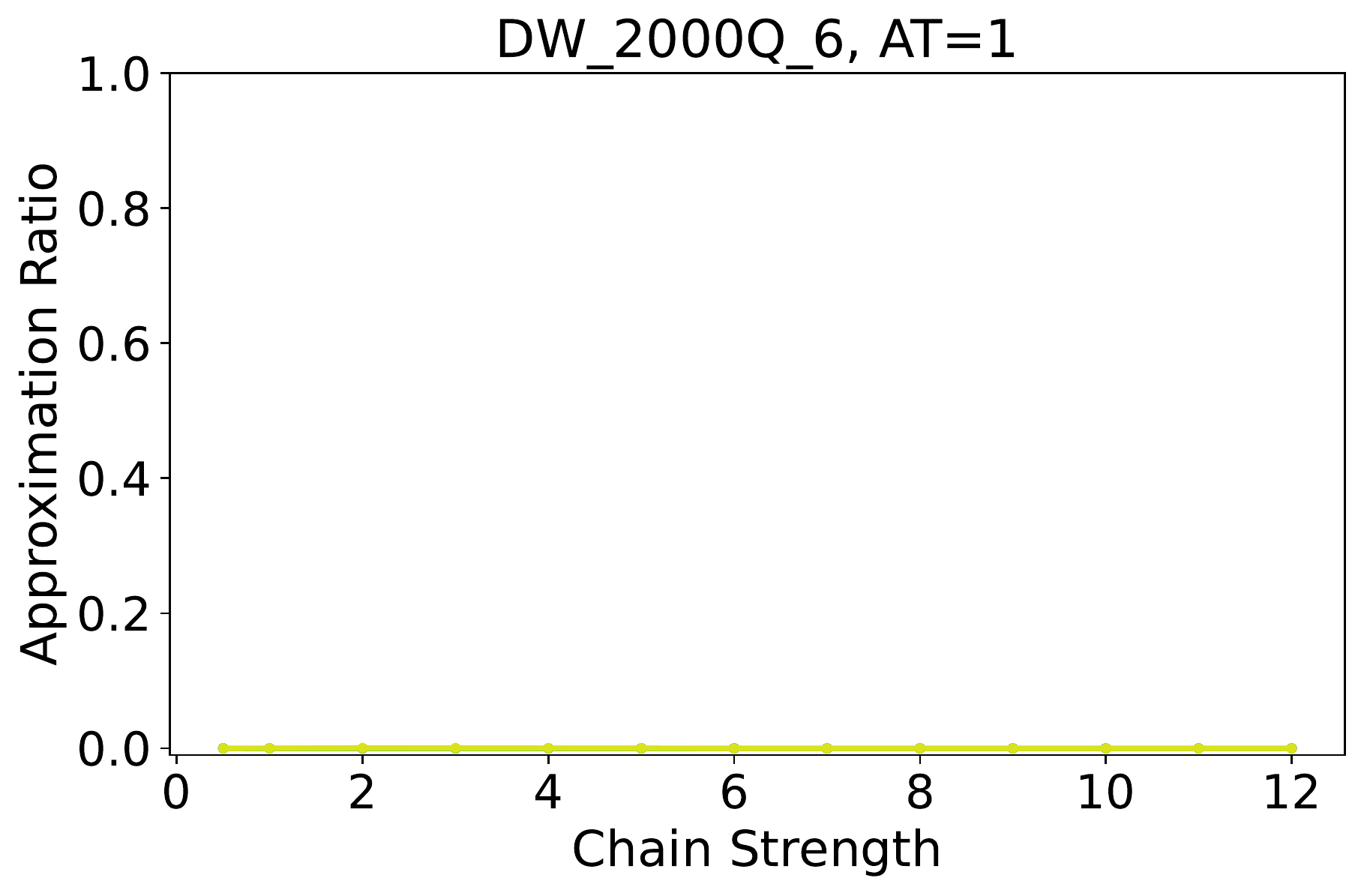}
    \includegraphics[width=0.24\textwidth]{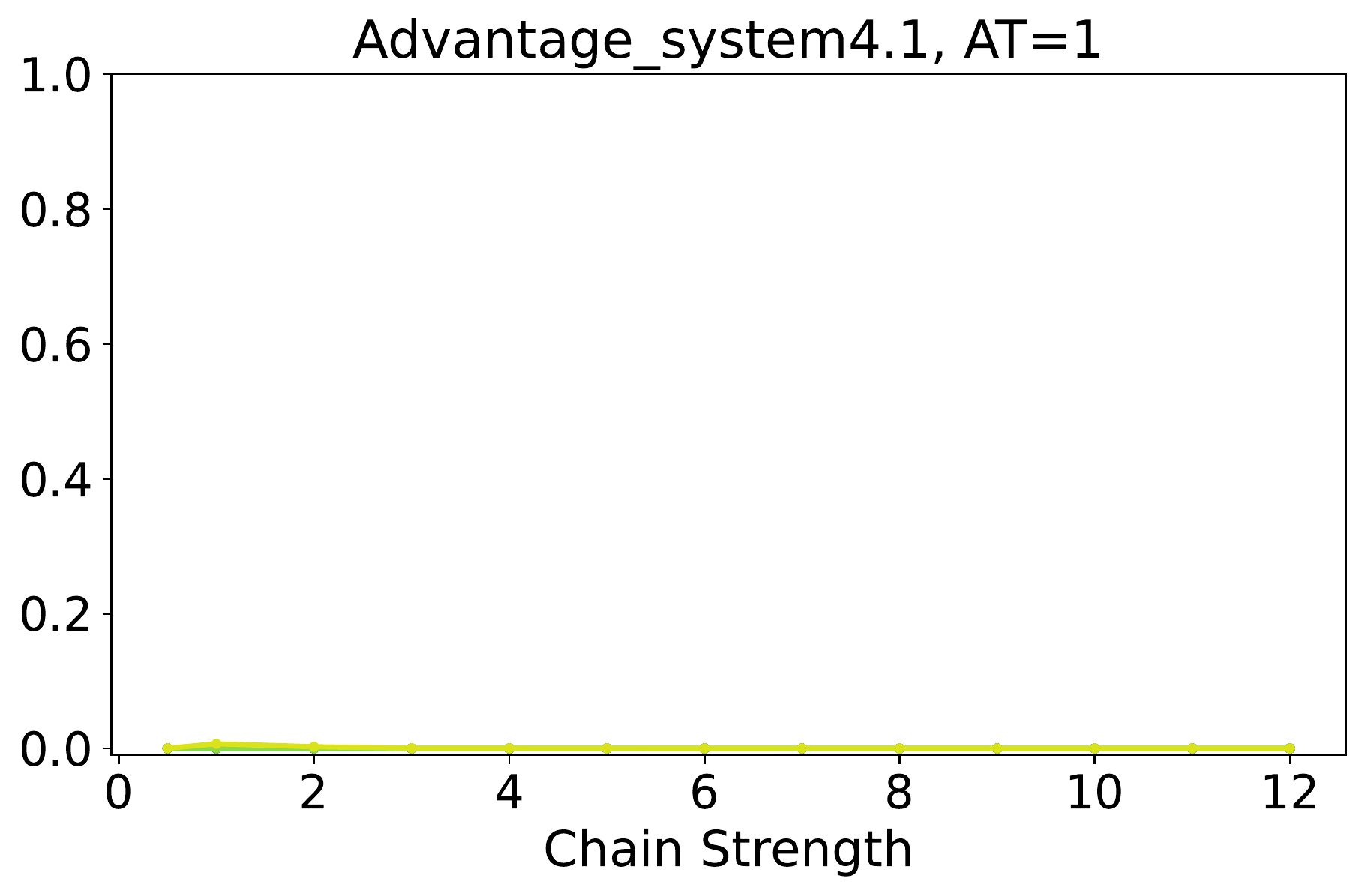}
    \includegraphics[width=0.24\textwidth]{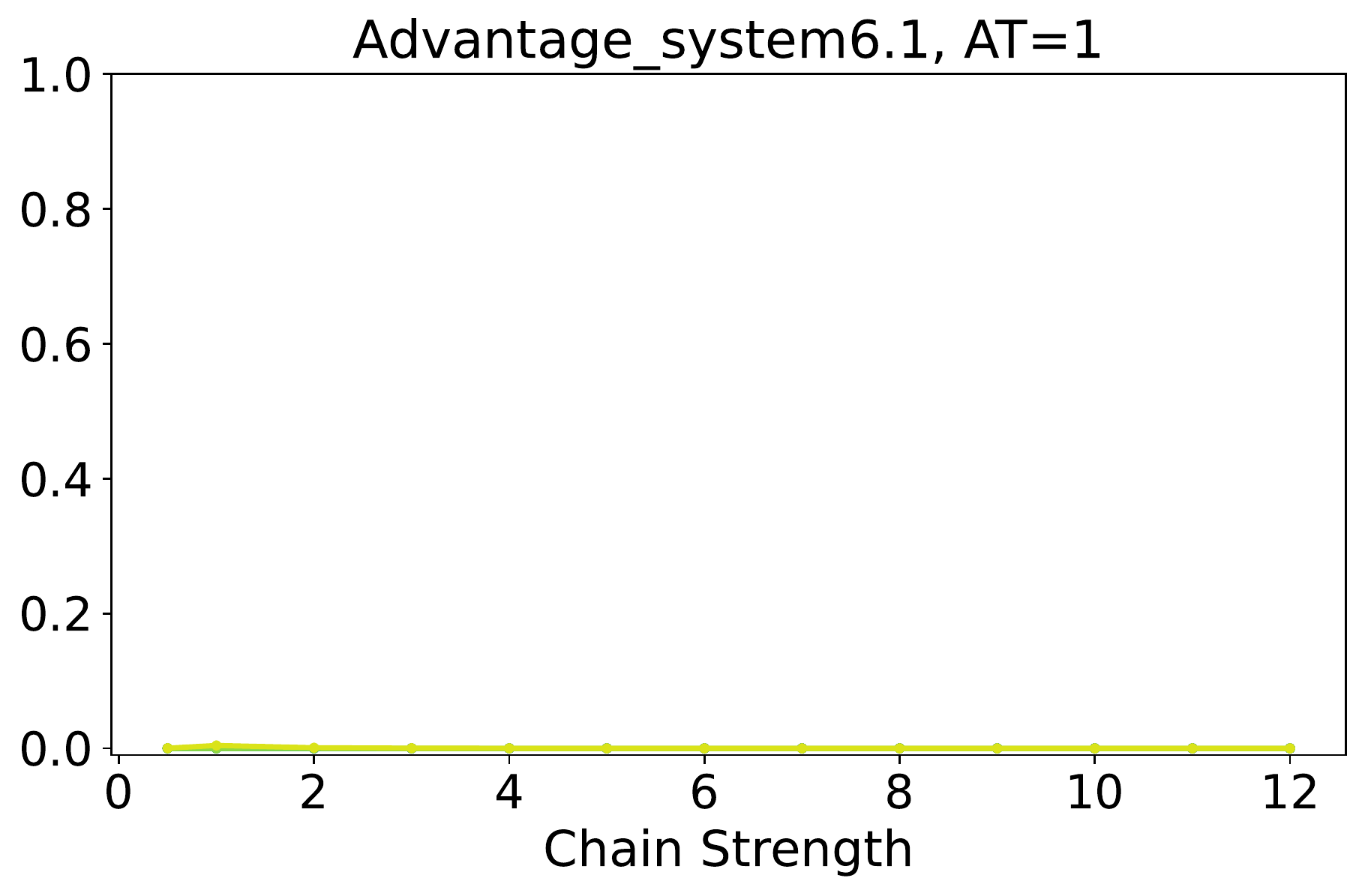}
    \includegraphics[width=0.24\textwidth]{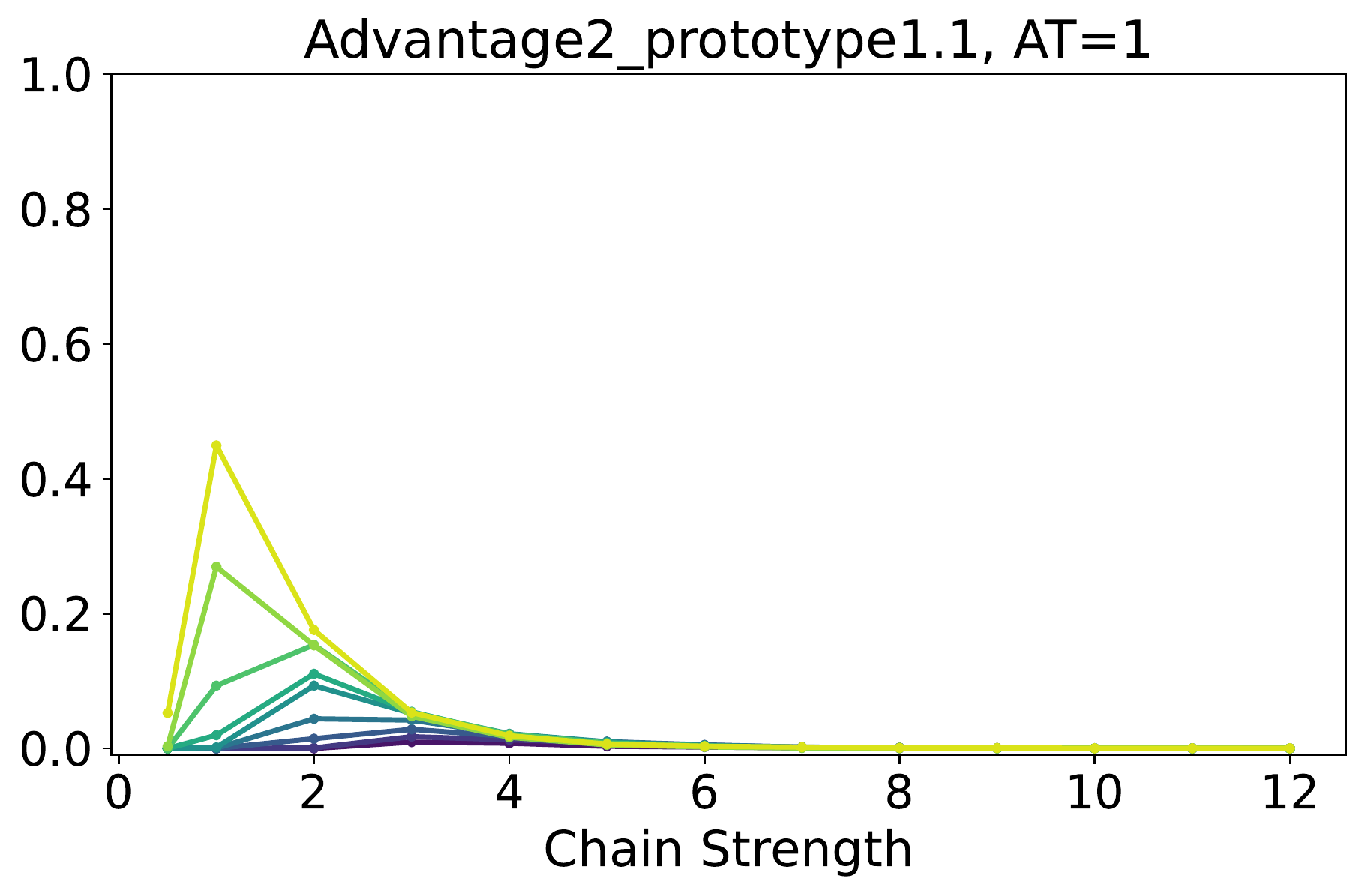}
    \includegraphics[width=0.71\textwidth]{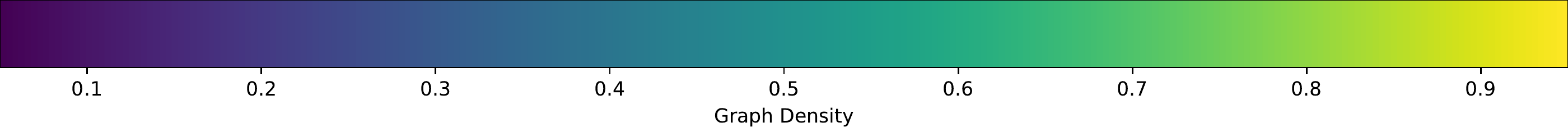}
    \caption{Maximum clique mean approximation ratios (y-axis) vs chain strength (x-axis) for the $200$ $G(n,p)$ random graphs. The aggregated results are shown in the form of $10$ lines, per plot, representing the mean approximation ratio for $10$ linearly spaced graph density intervals from $0.05$ to $0.95$, where the color of each line encodes the mean graph density for that interval. The color coding is shown in the colorbar below the plots. Problem QUBOs executed on \texttt{DW\_2000Q\_6} (left column), \texttt{Advantage\_system4.1} (center-left column), \texttt{Advantage\_system6.1} (center-right column), and \texttt{Advantage2\_prototype1.1} (right column). The annealing time in microseconds are varied across $2000$ microseconds (top row), $1000$ microseconds (second row), $100$ microseconds (third row), $10$ microseconds (fourth row), and $1$ microsecond (bottom row). The x-axis and y-axis labels are consistent and shared between all of the sub-figures. }
    \label{fig:maximum_clique_approx_ratio}
\end{figure*}

\subsection{Maximum Clique results}
\label{section:results_maximum_clique}

\begin{figure*}[h!]
    \centering
    \includegraphics[width=0.24\textwidth]{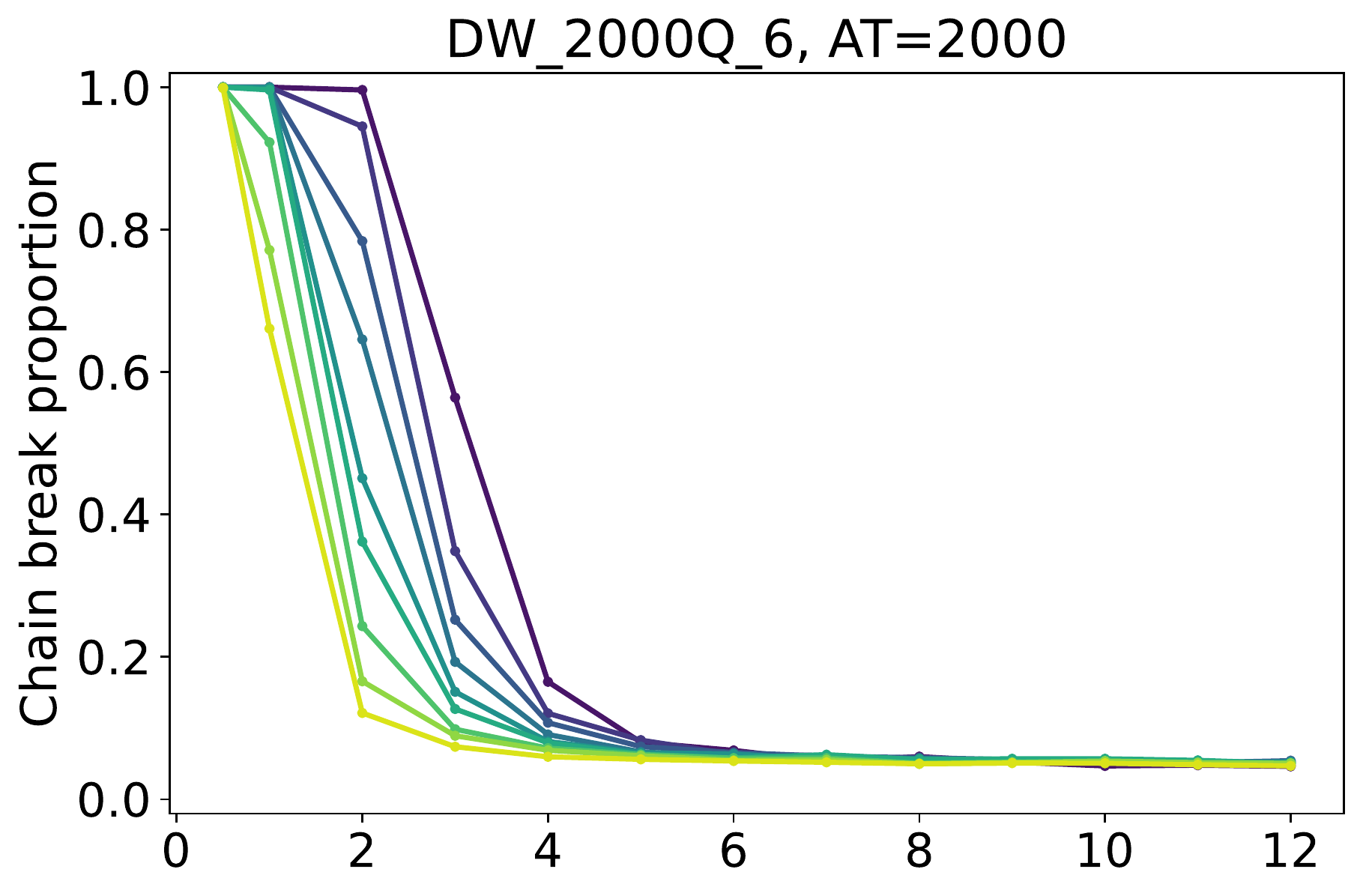}
    \includegraphics[width=0.24\textwidth]{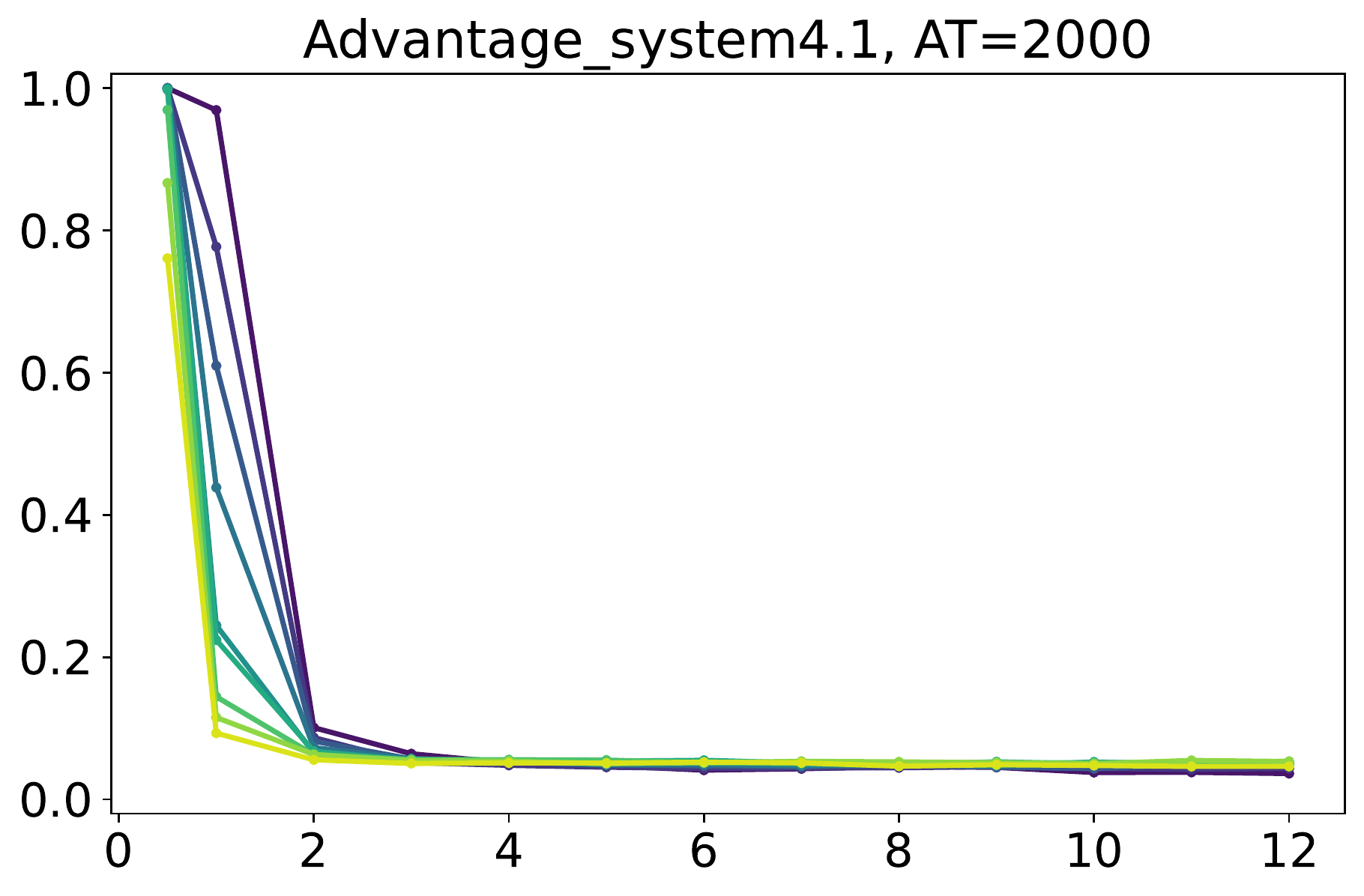}
    \includegraphics[width=0.24\textwidth]{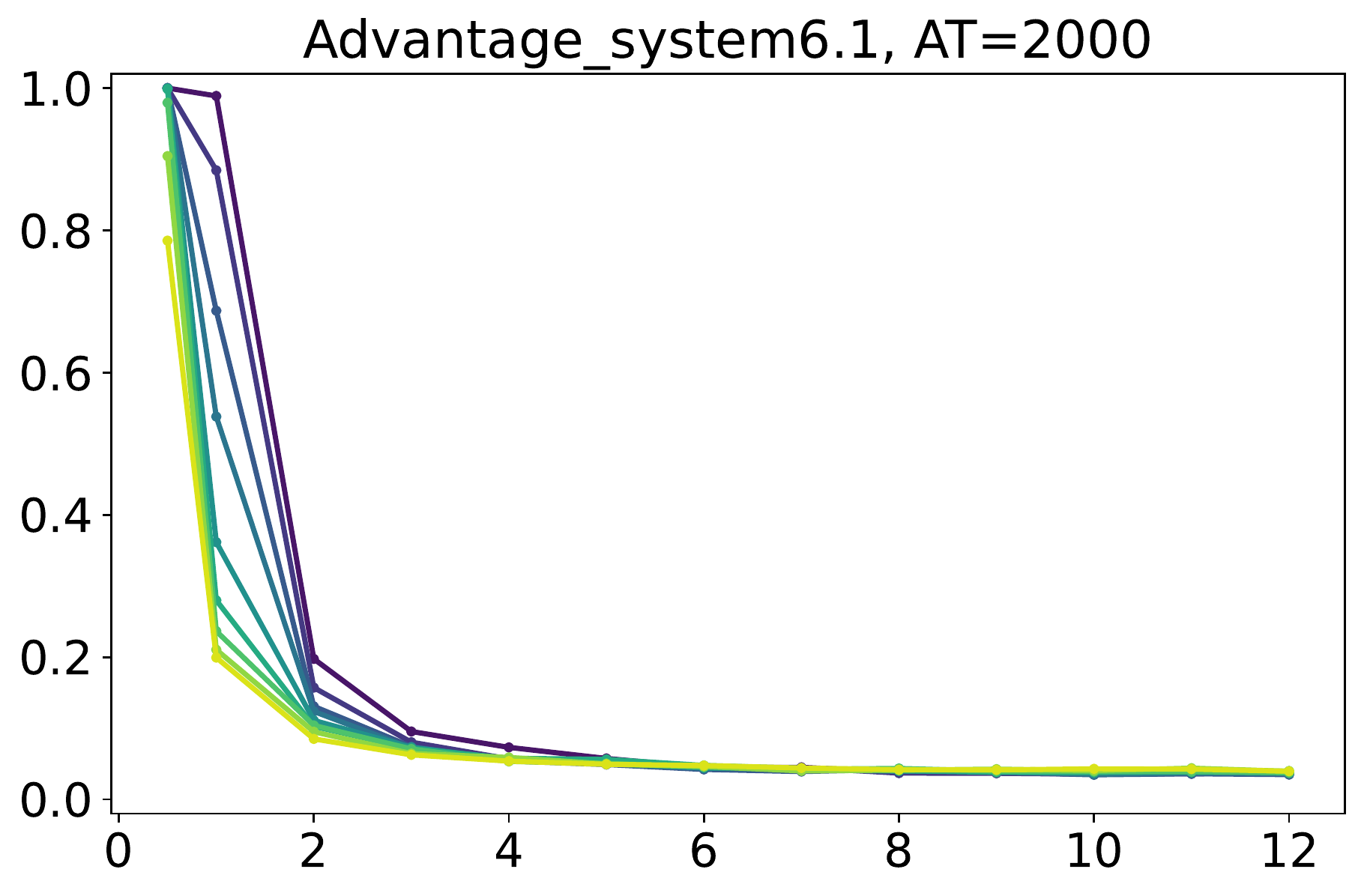}
    \includegraphics[width=0.24\textwidth]{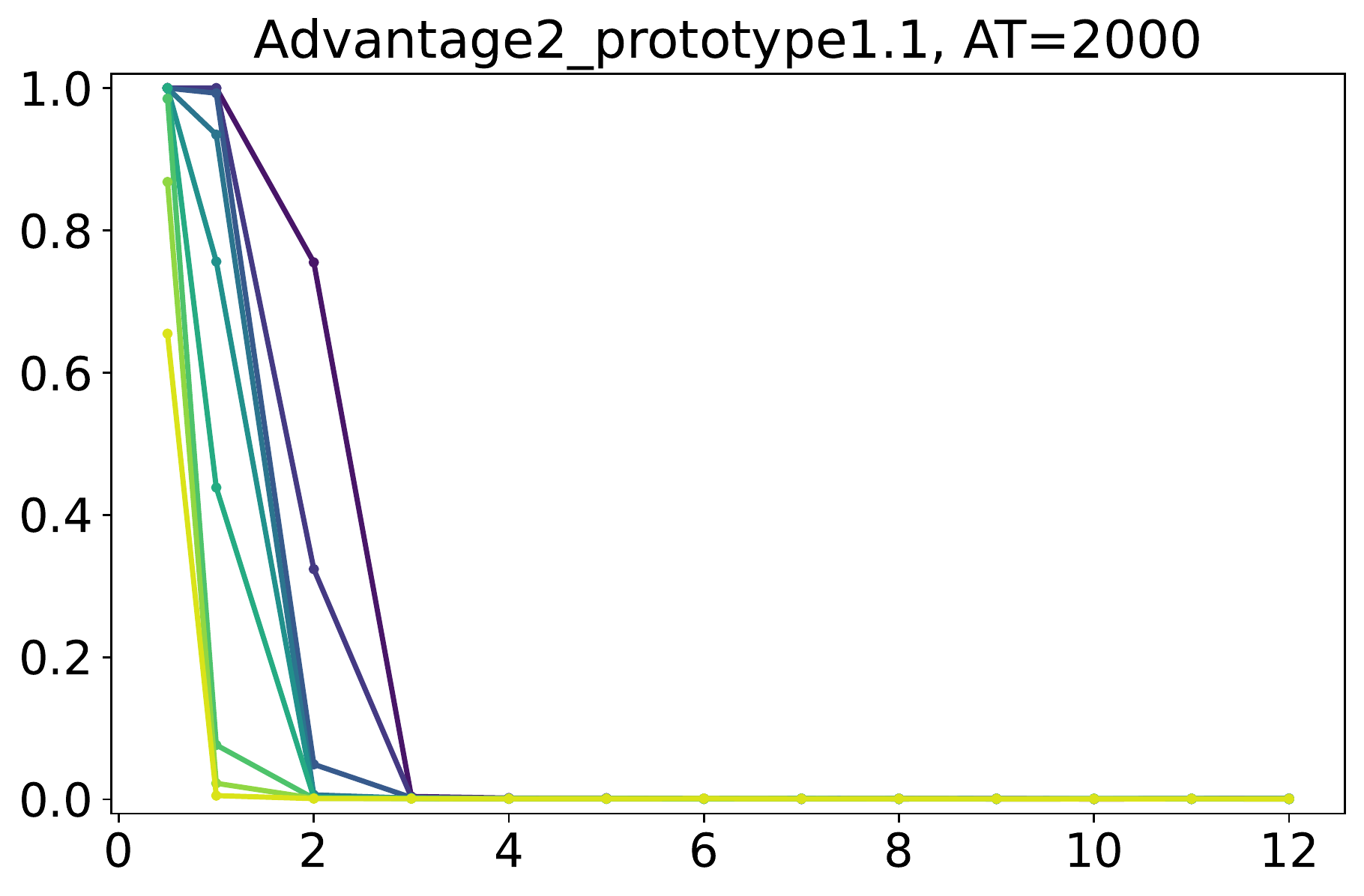}\\
    \includegraphics[width=0.24\textwidth]{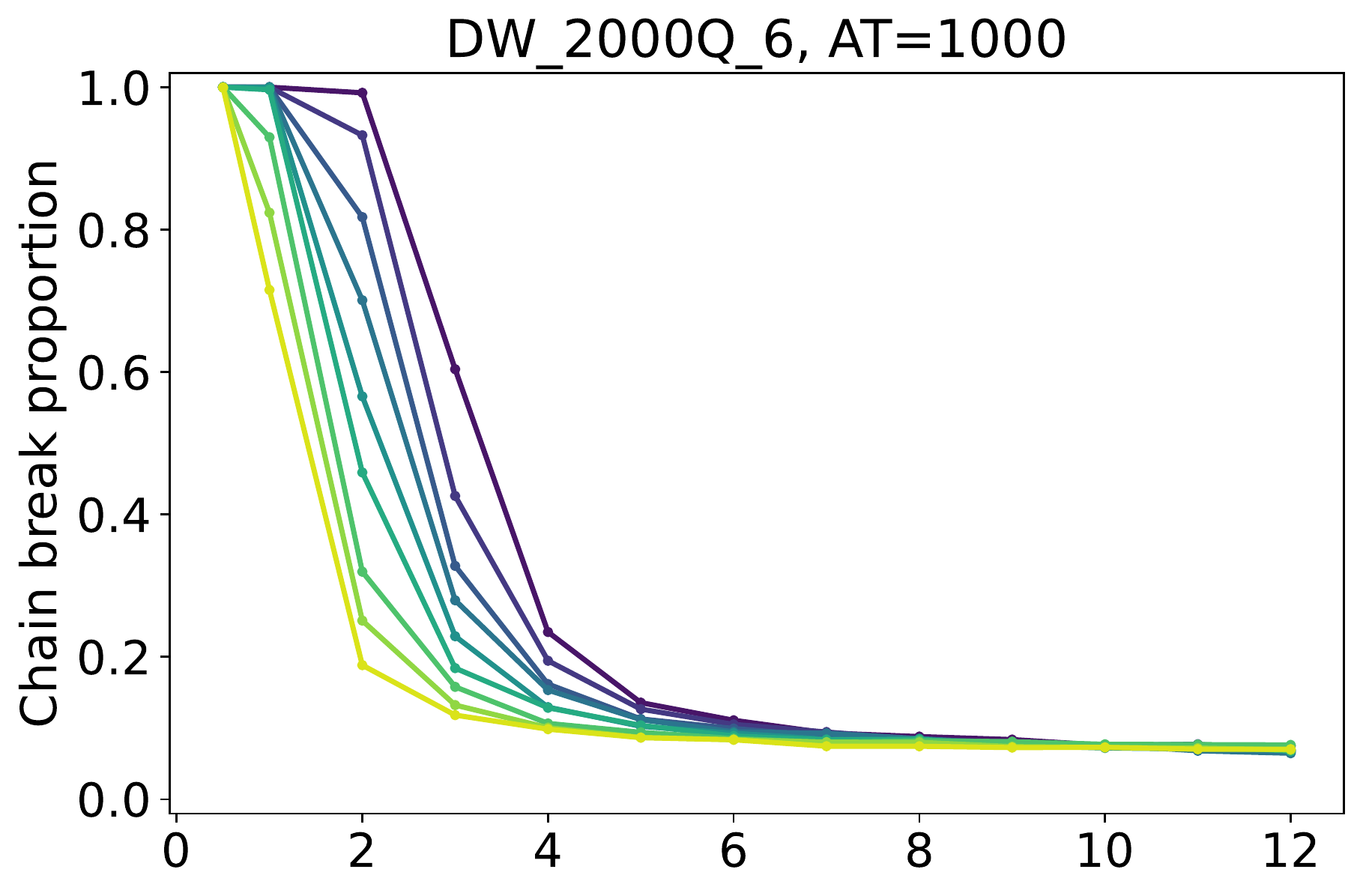}
    \includegraphics[width=0.24\textwidth]{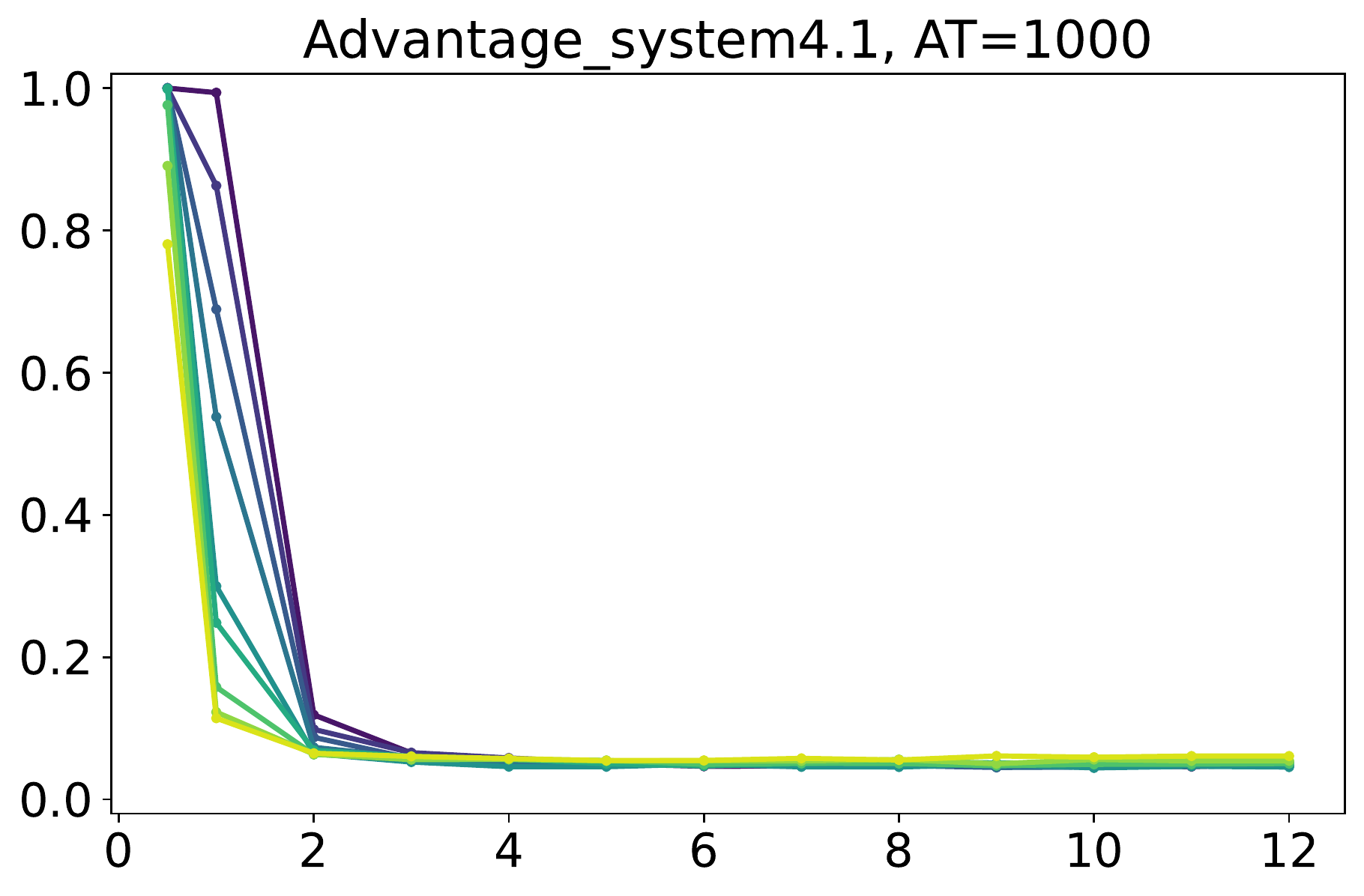}
    \includegraphics[width=0.24\textwidth]{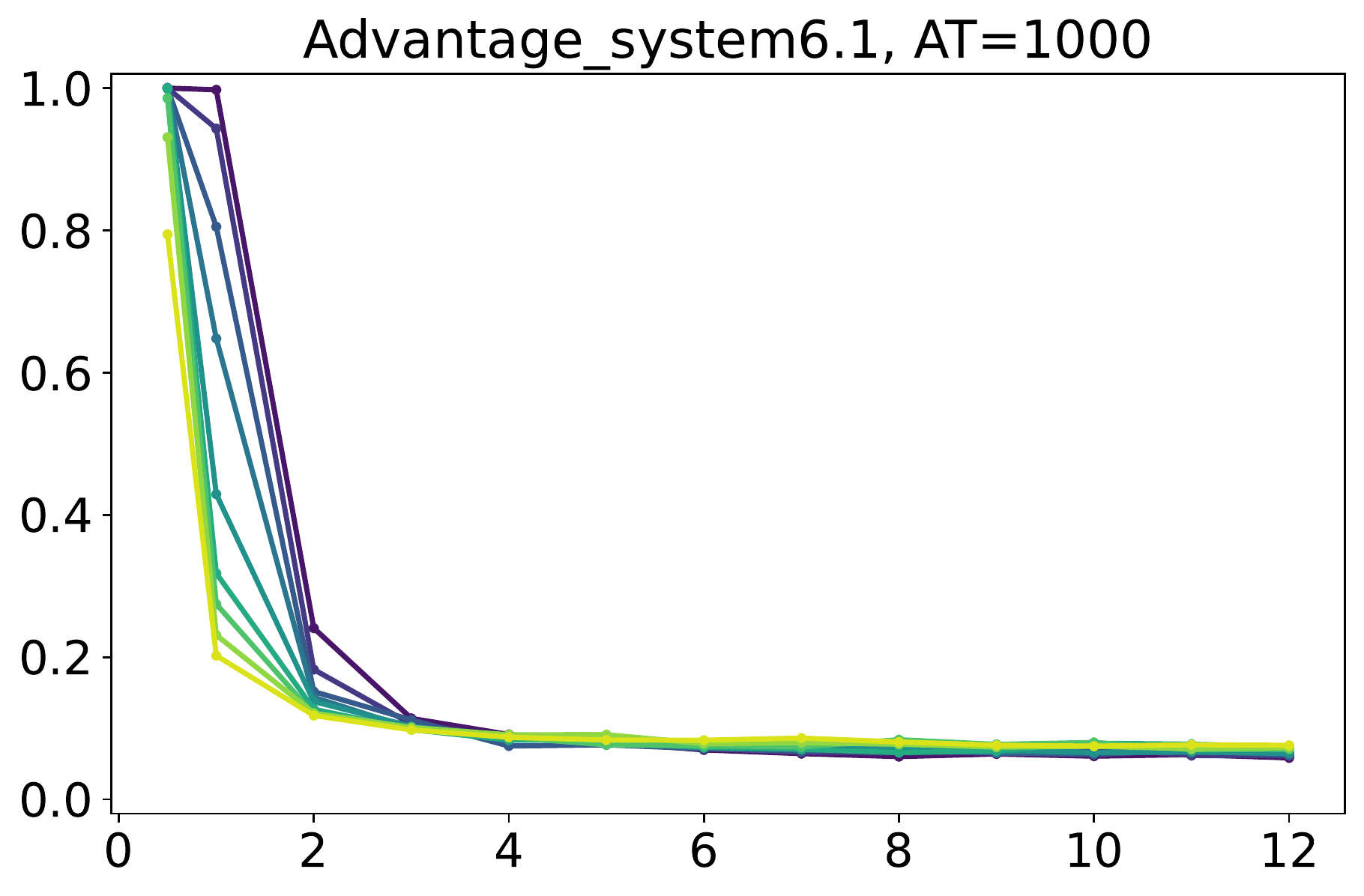}
    \includegraphics[width=0.24\textwidth]{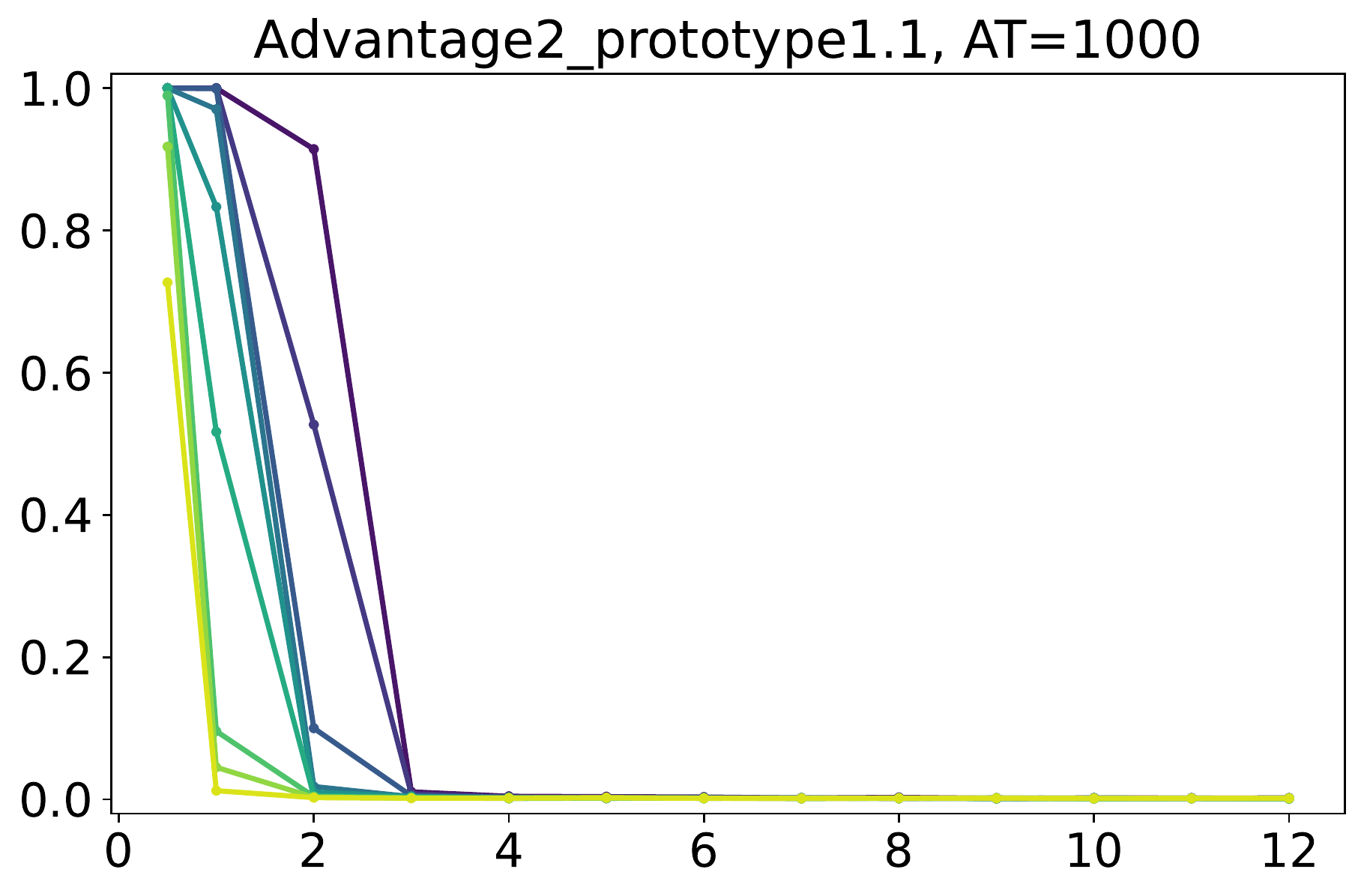}\\
    \includegraphics[width=0.24\textwidth]{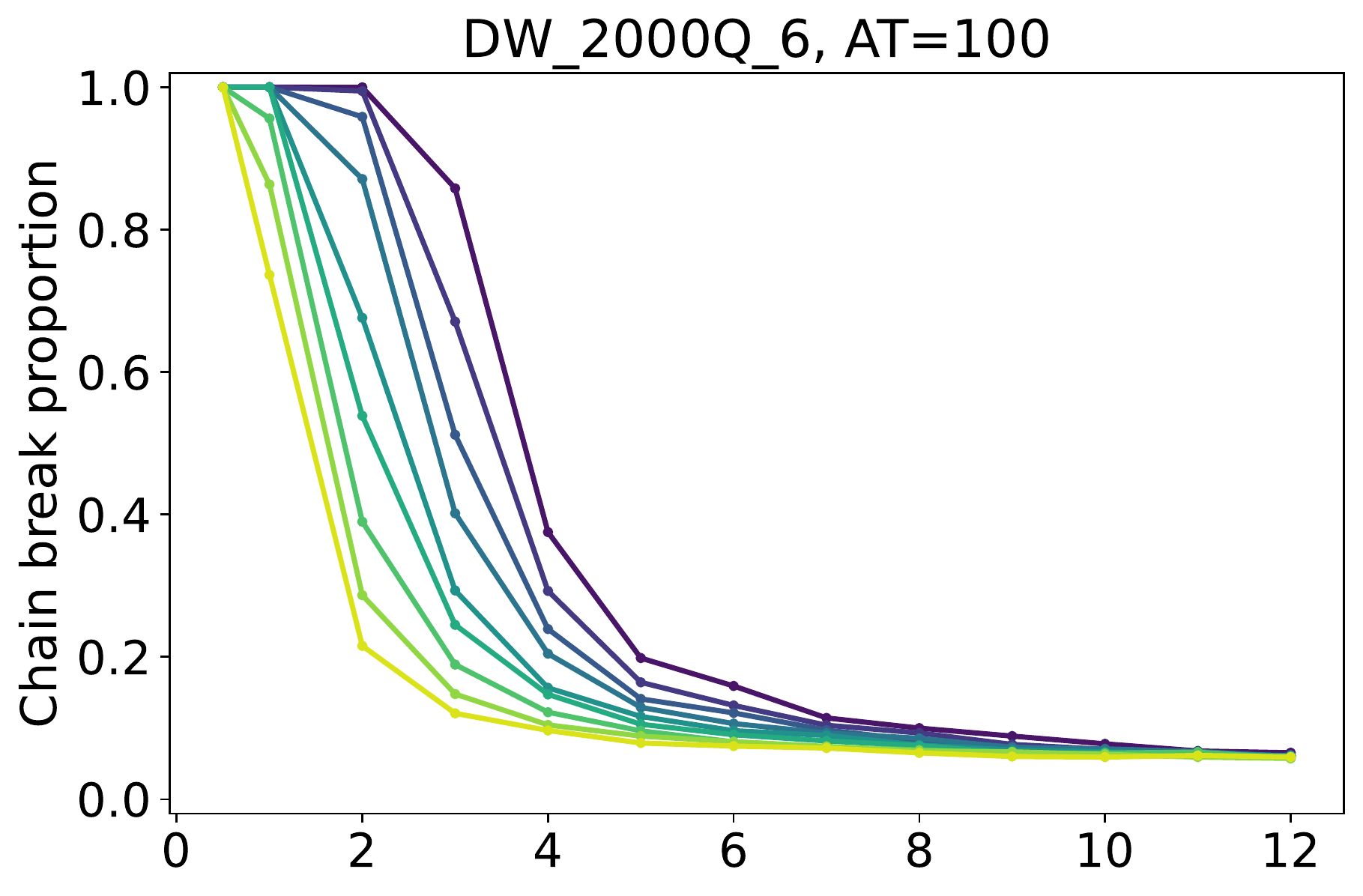}
    \includegraphics[width=0.24\textwidth]{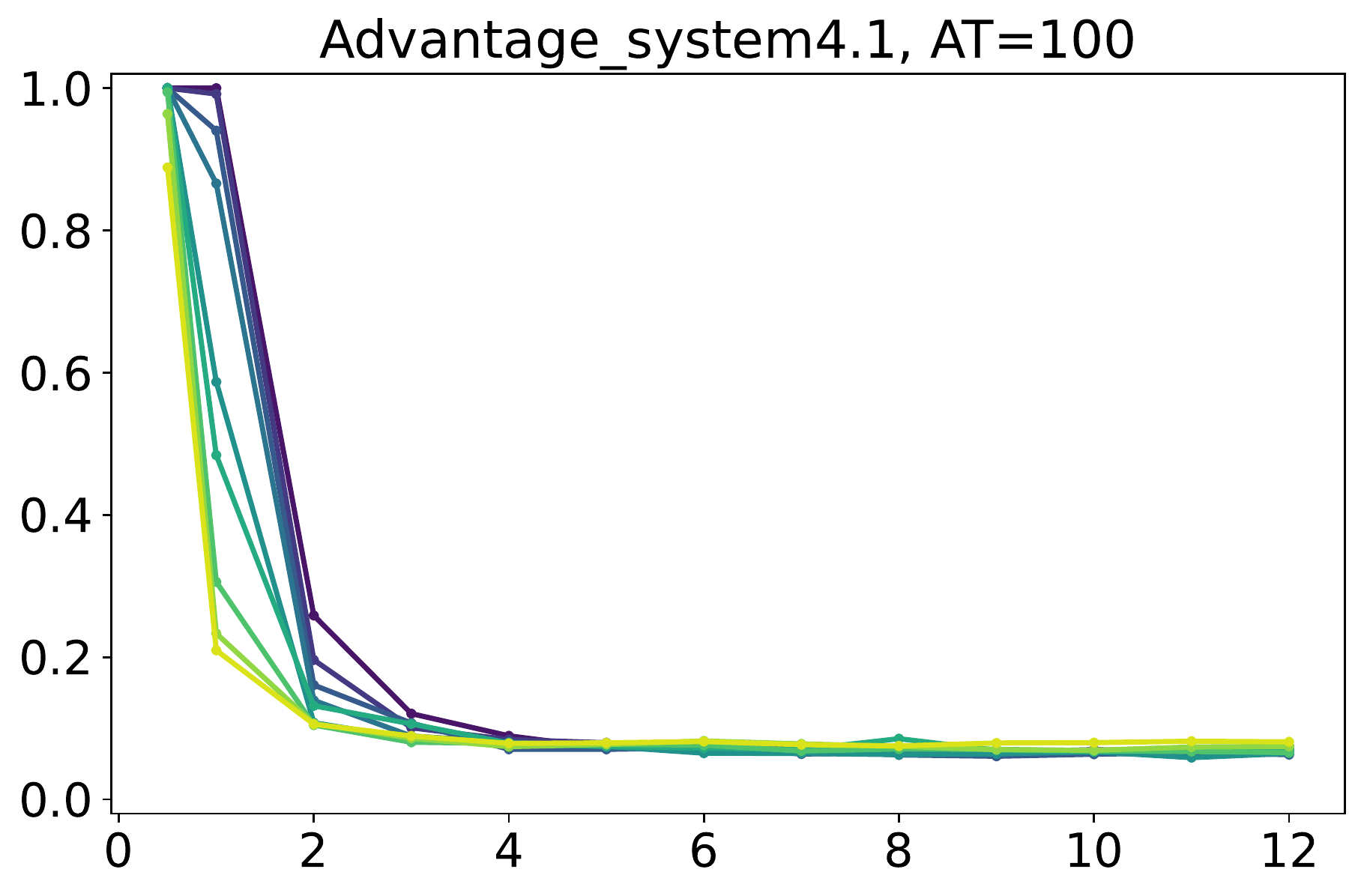}
    \includegraphics[width=0.24\textwidth]{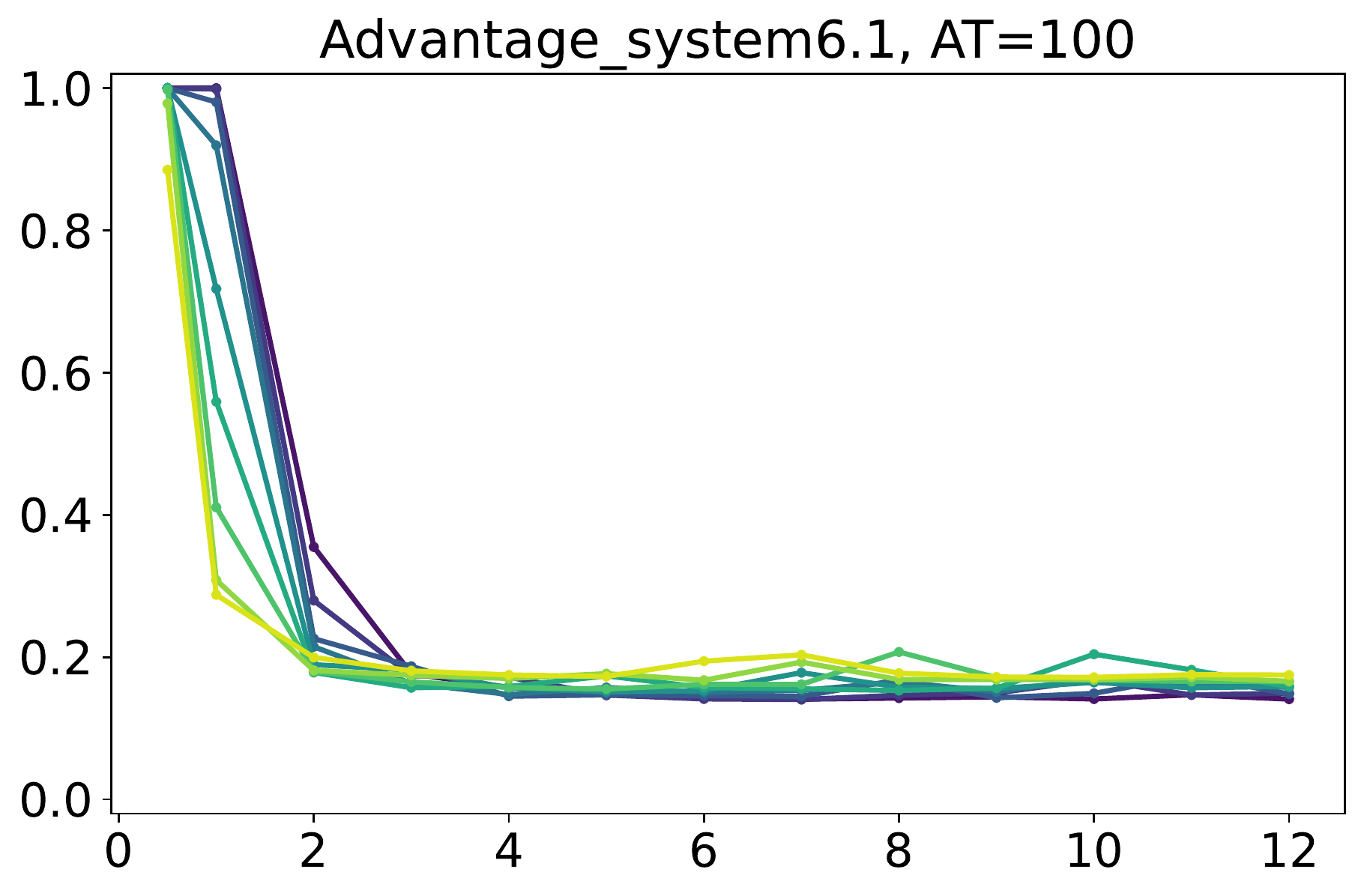}
    \includegraphics[width=0.24\textwidth]{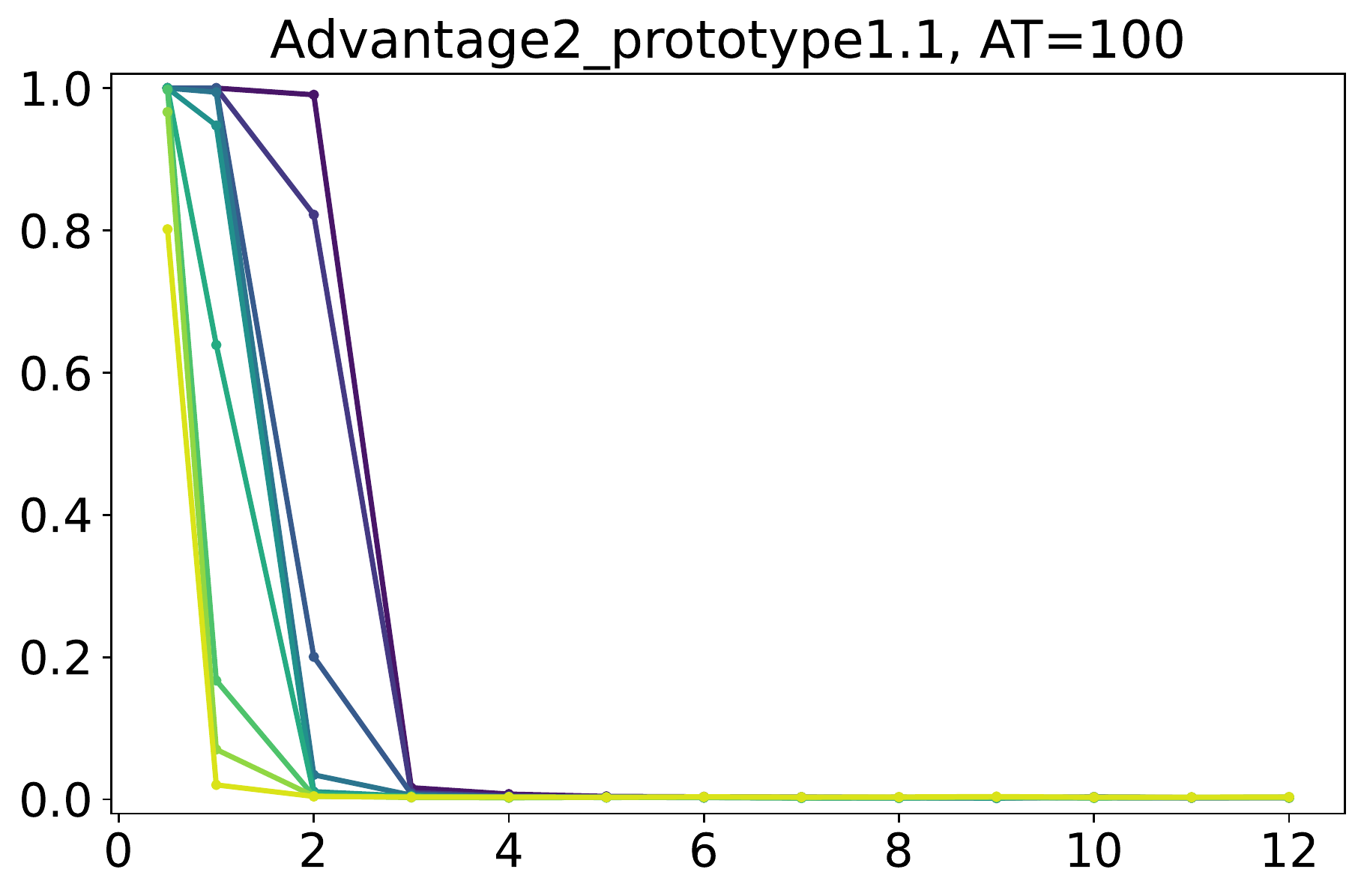}\\
    \includegraphics[width=0.24\textwidth]{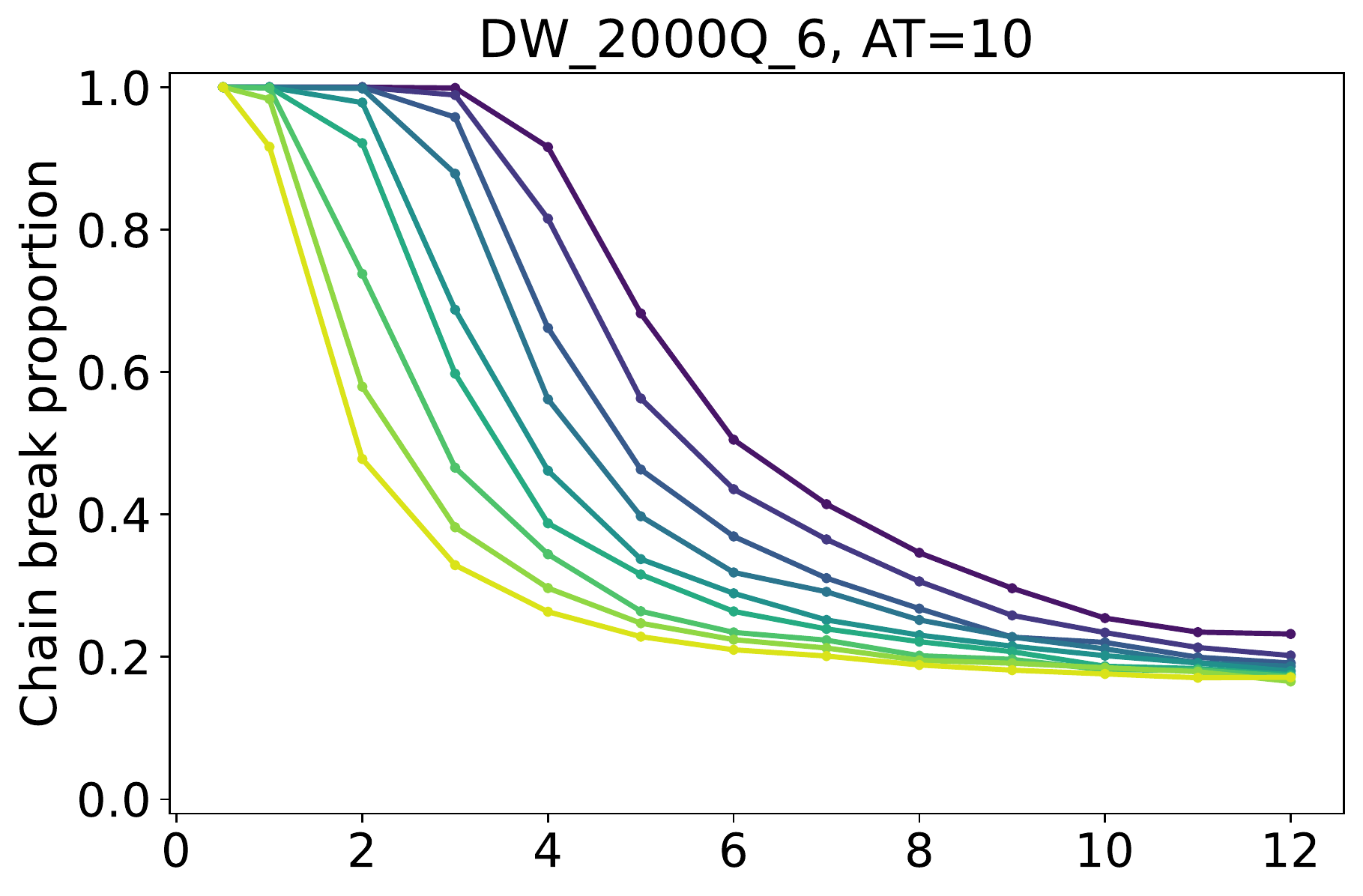}
    \includegraphics[width=0.24\textwidth]{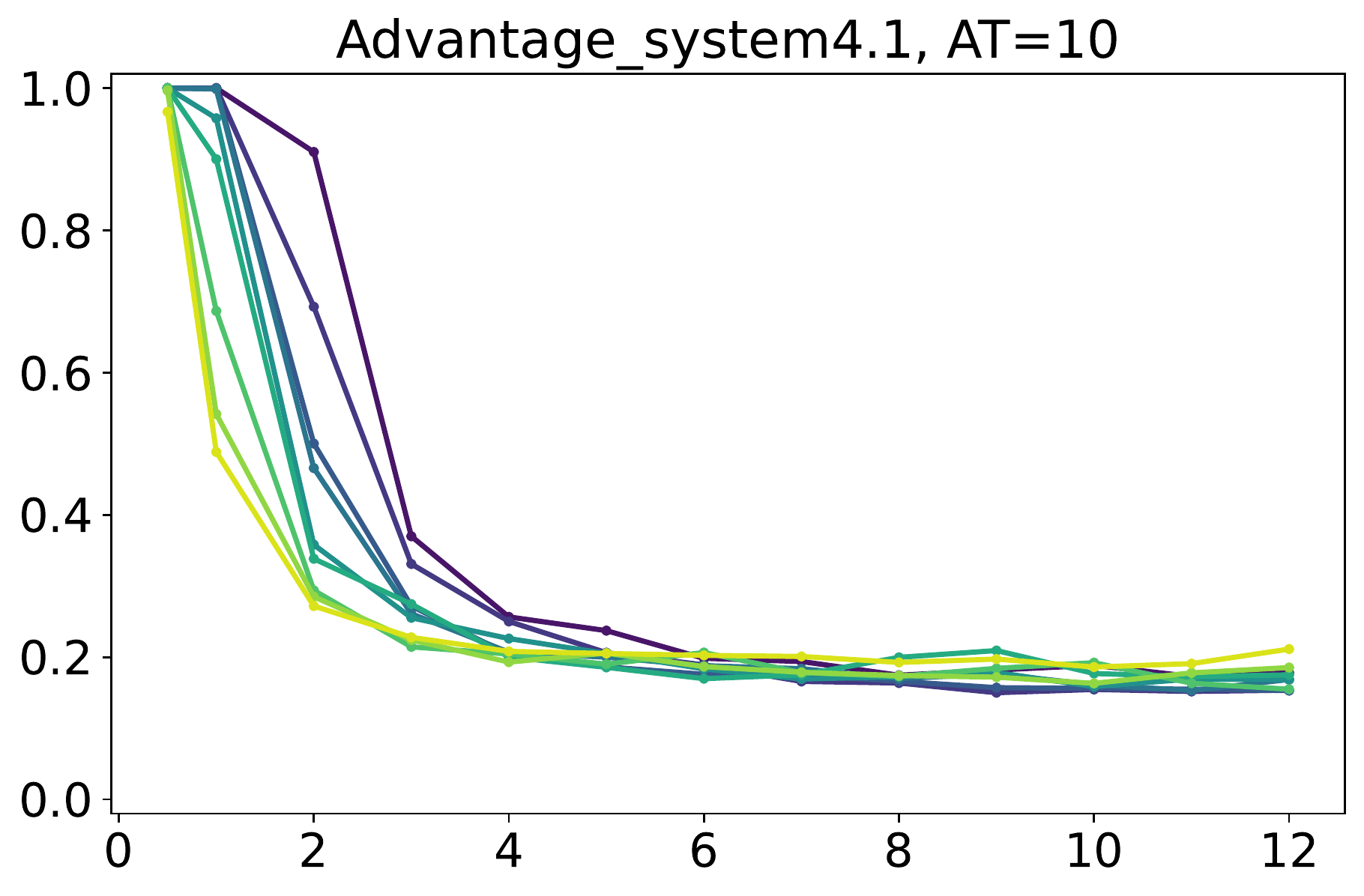}
    \includegraphics[width=0.24\textwidth]{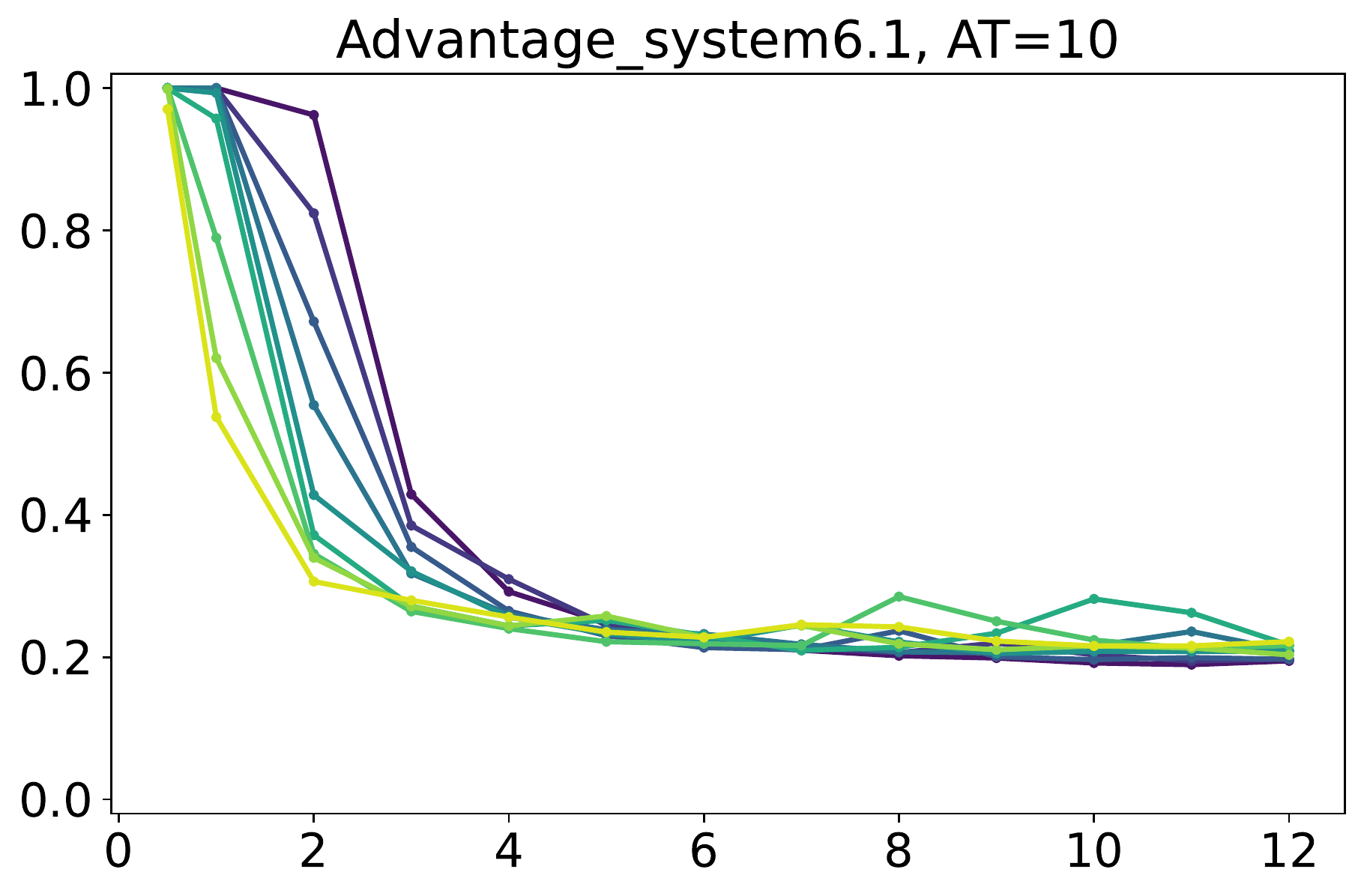}
    \includegraphics[width=0.24\textwidth]{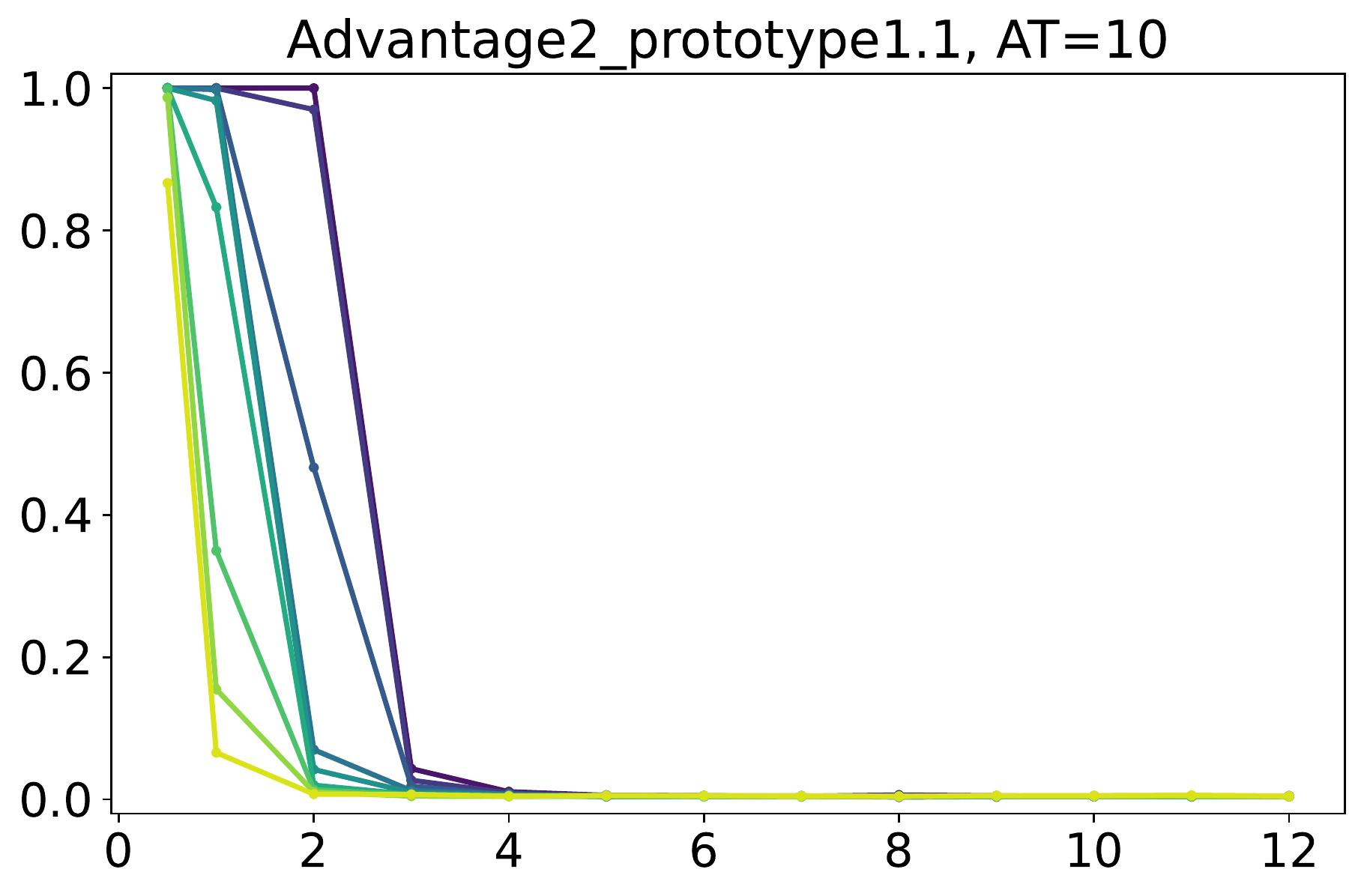}\\
    \includegraphics[width=0.24\textwidth]{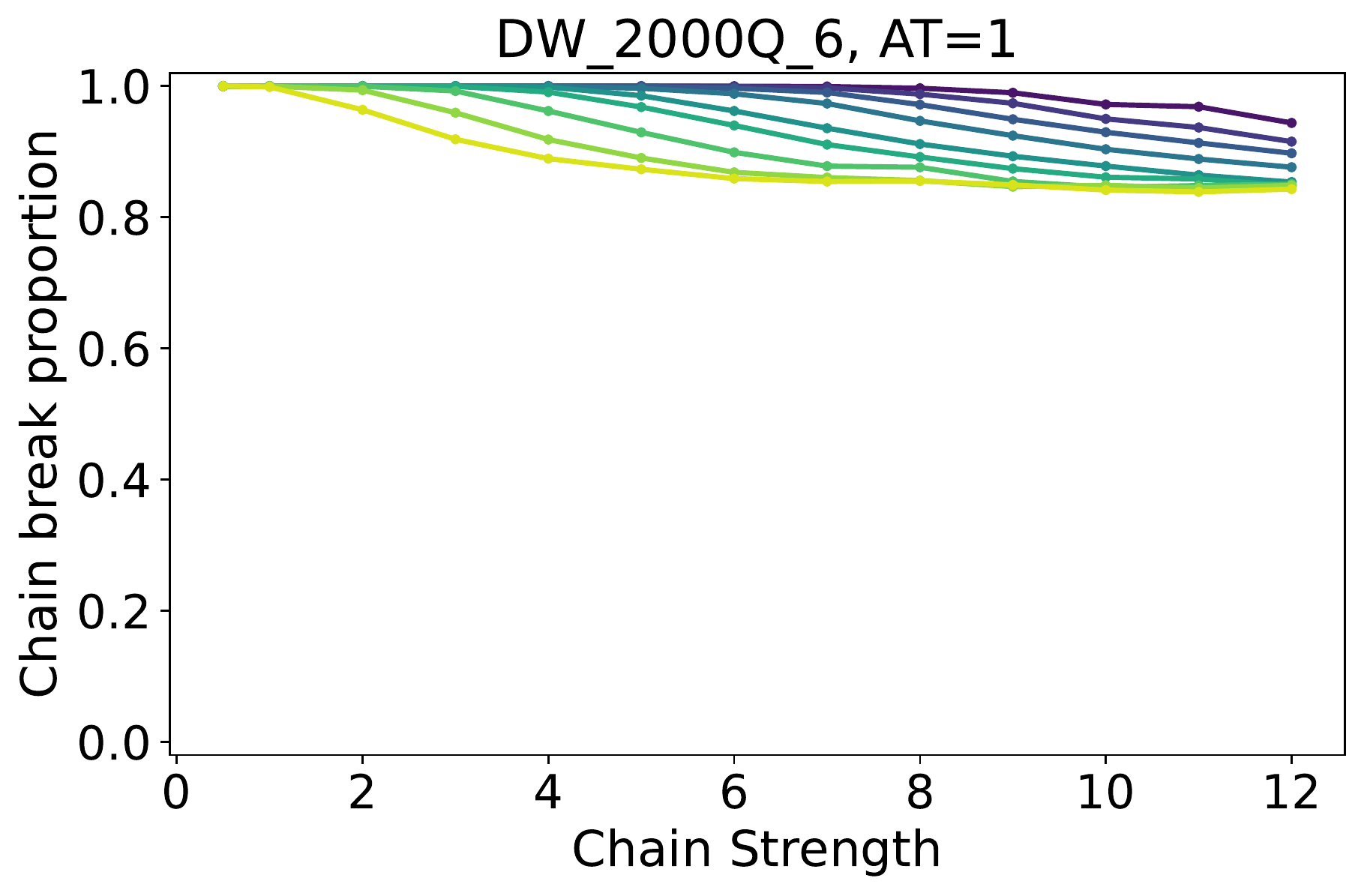}
    \includegraphics[width=0.24\textwidth]{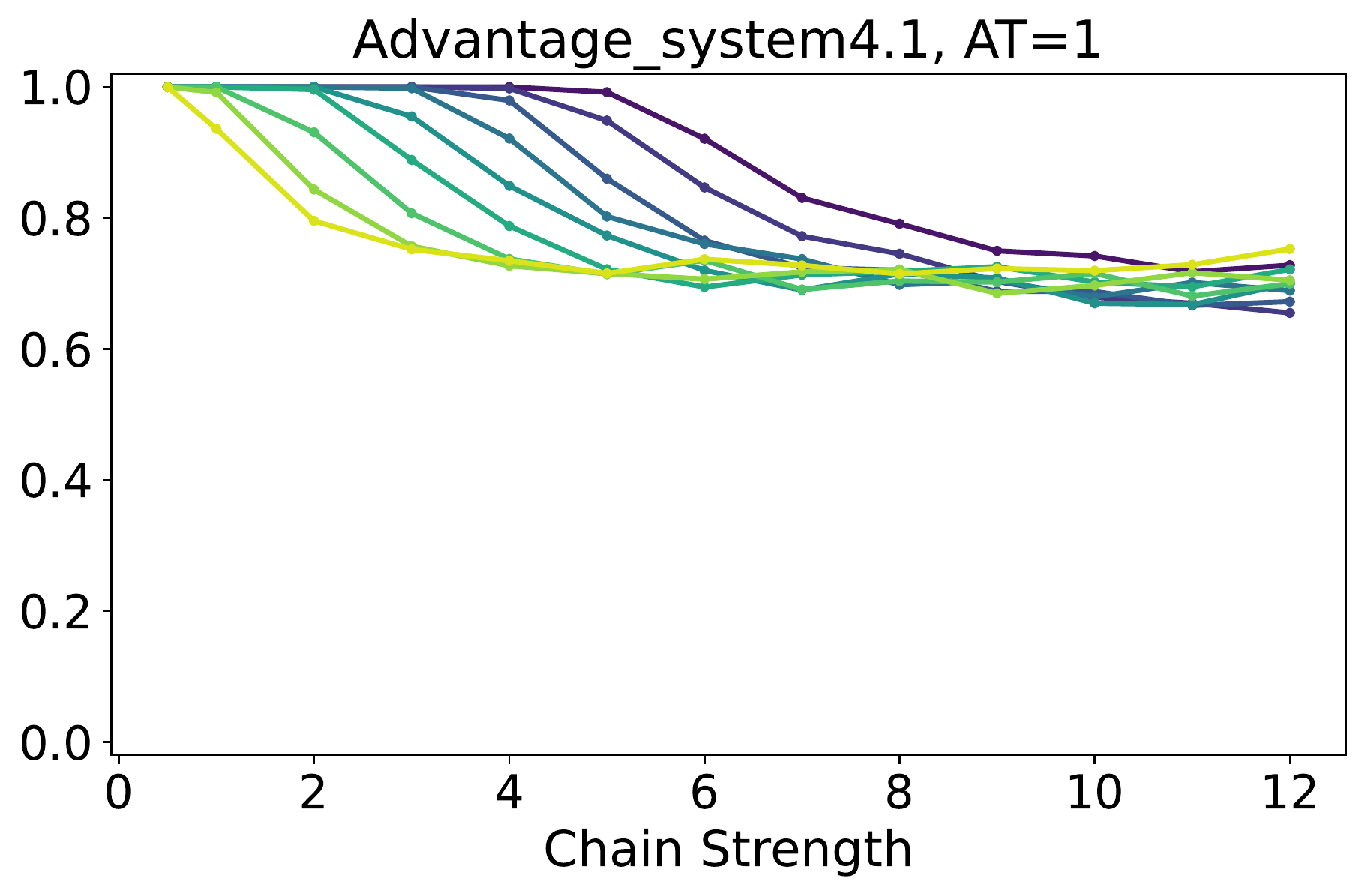}
    \includegraphics[width=0.24\textwidth]{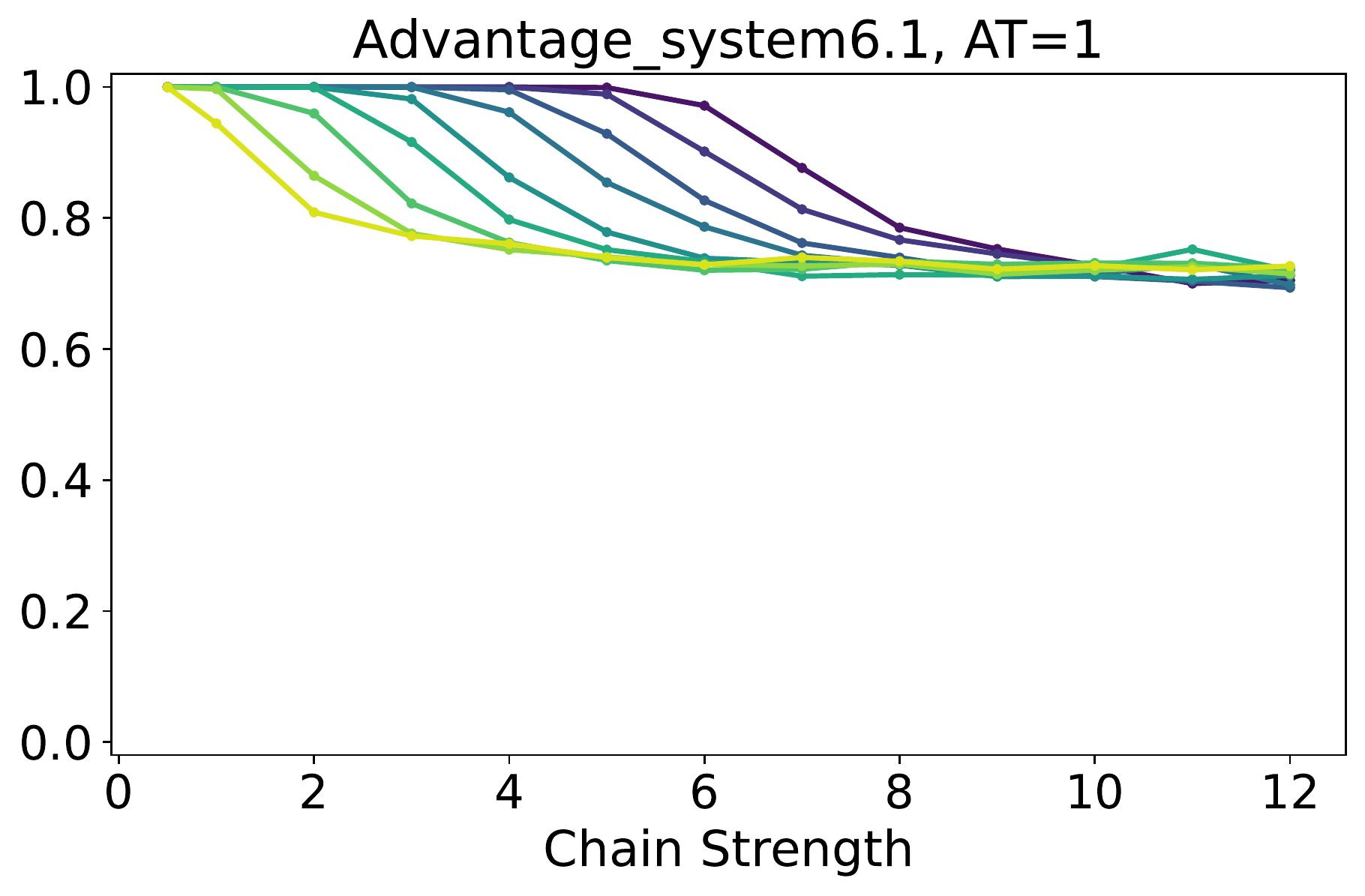}
    \includegraphics[width=0.24\textwidth]{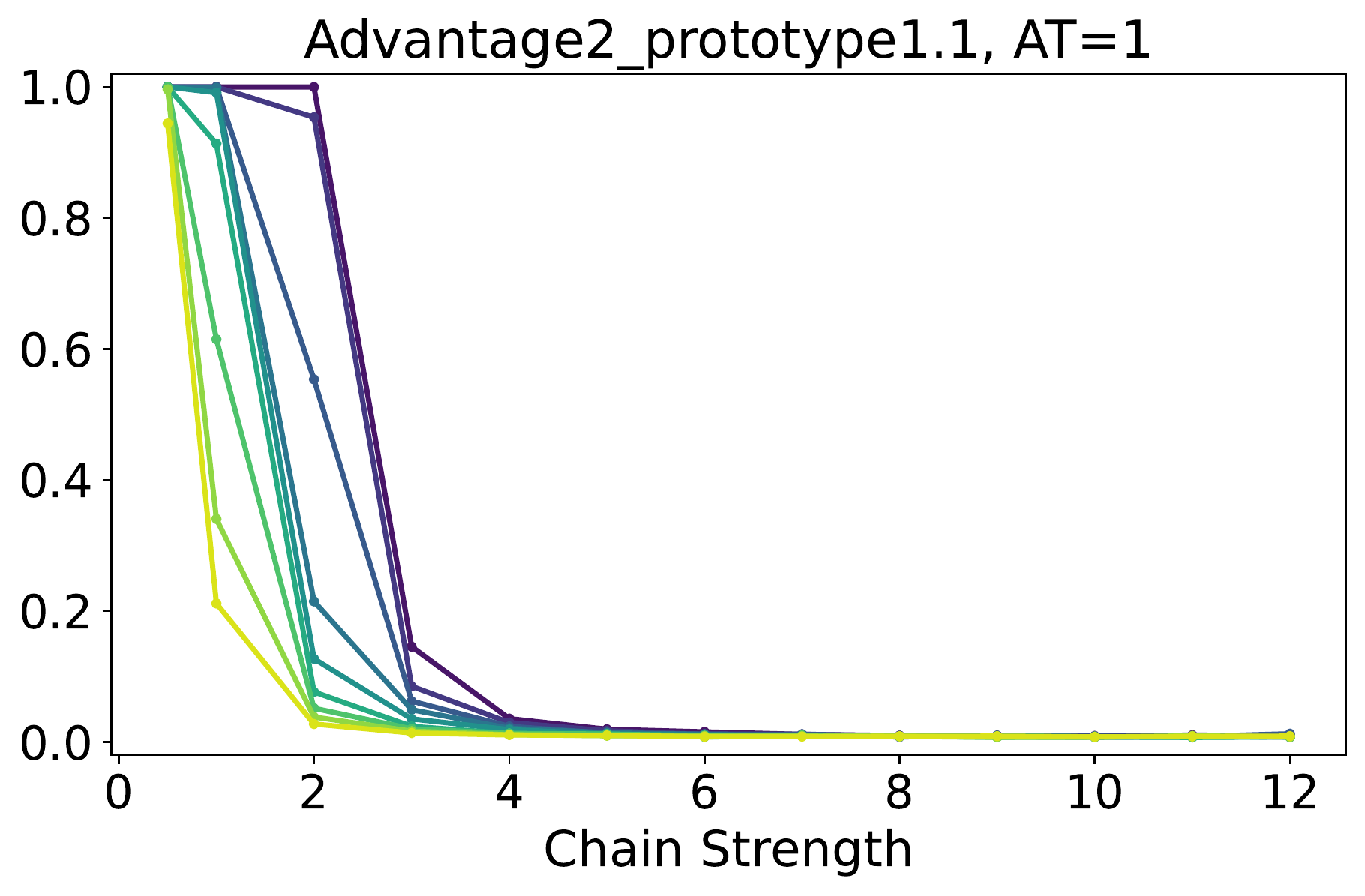}\\
    \includegraphics[width=0.71\textwidth]{figures/colorbar.pdf}
    \caption{Chain break proportions (y-axis) vs chain strength (x-axis) for each of the maximum clique instances on the $200$ $G(n,p)$ random graphs. The aggregated results are shown in the form of $10$ lines, per plot, representing the mean chain break proportion for $10$ linearly spaced graph density intervals from $0.05$ to $0.95$, where the color of each line encodes the mean graph density for that interval. The line colors encode the problem graph density using the same colorscale from Figure~\ref{fig:maximum_clique_approx_ratio}. The D-Wave devices are \texttt{DW\_2000Q\_6} (left column), \texttt{Advantage\_system4.1} (center-left column), \texttt{Advantage\_system6.1} (center-right column), and \texttt{Advantage2\_prototype1.1} (right column). The annealing time in microseconds are varied across $2000$ microseconds (top row), $1000$ microseconds (second row), $100$ microseconds (third row), $10$ microseconds (fourth row), and $1$ microsecond (bottom row). }
    \label{fig:chain_breaks_maximum_clique}
\end{figure*}

Figure~\ref{fig:maximum_clique_approx_ratio} shows the distribution of the mean maximum clique approximation ratios of the $200$ test graphs, binned into $10$ linearly spaced density intervals between $0.05$ and $0.95$, for different annealing times and chain strengths across the four D-Wave devices. Although it is difficult to discern if the devices are able to sample the optimal solutions with a non zero probability, as shown in Table~\ref{table:optimal_solution_table}, it is the case that the two Pegasus chip quantum annealers were able to find the maximum cliques of all $200$ of the test problems when the annealing times are sufficiently long, and the chain strengths are well tuned. The caveat that the devices only perform well when the annealing times are long and the chain strengths are properly tuned is of course important, but this shows that the quantum annealers have the capability to sample the optimal solutions for these moderately sized maximum clique QUBOs. Because Figure~\ref{fig:maximum_clique_approx_ratio} was computed with no un-embedding algorithm on the samples, something else that can be examined is the distribution of the chain breaks that occurred in those samples - this is shown in Figure~\ref{fig:chain_breaks_maximum_clique}. 

Figure~\ref{fig:maximum_clique_approx_ratio} shows that there is a trend of solution quality improvement across the three generations of D-Wave devices; \texttt{DW\_2000Q\_6} \footnote{Note that since this study was performed, \texttt{DW\_2000Q\_6} has been shut down and removed from the D-Wave Leap Cloud service. } (middle-right column) has the lowest overall mean approximation ratios, whereas the newer \texttt{Advantage2\_prototype1.1} performed the best. Although, it is notable that both \texttt{Advantage\_system4.1} and \texttt{Advantage\_system6.1} had higher approximation ratios for the lower density graphs (color coded as blue / purple) compared to \texttt{Advantage2\_prototype1.1}. \texttt{Advantage\_system4.1} appears to have performed slightly better than \texttt{Advantage\_system6.1}. Notably,~\texttt{Advantage2\_prototype1.1} was the only device of the four which was able to sample the optimal solutions reliably at the annealing time of $1$ microsecond, which suggests that this newer generation of hardware has less noise compared to the previous QPU generations. This result also suggests that the newer generation prototype could have longer coherence times than the 10's of nanoseconds qubit coherence times that have been measured on other D-Wave quantum annealers \cite{King_2022, King_2023_5000q}. 

Of the four devices in Figure~\ref{fig:maximum_clique_approx_ratio}, both~\texttt{Advantage\_system6.1} and \texttt{Advantage\_system4.1}, $2000$ and $1000$ microsecond annealing times have the lowest approximation ratios (among the $100$ problems). For example, the lowest approximation ratio among all $200$ Maximum Clique QUBOs sampled on \texttt{Advantage\_system4.1}, with an annealing time of $2000$ microseconds and a chain strength of $2$, is $0.2$ (top row, middle-left of Figure~\ref{fig:maximum_clique_approx_ratio}). This is in contrast to to both \texttt{Advantage2\_prototype1.1} and \texttt{DW\_2000Q\_6} where at the best performing chain strengths, the lowest approximation ratios are barely discernible from $0$. This shows that at least with respect to the worst approximation ratios, at the best parameter combinations, the two Pegasus topology devices performed the best compared to the other two devices. 

Figure~\ref{fig:maximum_clique_approx_ratio} shows that there is clear stratification in how graphs of different density are sampled on the quantum annealer. QA samples from higher density graphs have higher approximation ratios. This observation makes sense given the Maximum Clique QUBO formulation (eq.~\eqref{equation:maxclique_QUBO}) because the complement of higher density graphs result in fewer quadratic terms compared to lower density graphs, therefore the complexity of the QUBO is lower with higher density graphs. This observation is interesting since in classical maximum clique algorithms dense graphs are harder \cite{OSTERGARD2002197, WU2015693}, which suggests that quantum annealing could provide a speedup as a heuristic for sampling the maximum clique of dense graphs when strictly considering classical maximum clique solvers - however this same QUBO formulation sparsity property can be utilized by classical solvers to solve QUBOs of maximum clique problems faster as well. 

With respect to parameter tuning of chain strength and annealing time, Figures~\ref{fig:maximum_clique_approx_ratio} and~\ref{fig:chain_breaks_maximum_clique} show that there is an optimal choice of chain strength, which varies slightly with the problem graph density and which hardware is being used. A chain strength of between $1$ and $2$ performs the best overall. Longer annealing times (e.g. $2000$ microseconds) perform better across all four devices, which is notable because this time at which the measurements of the spins happen is well outside of the coherent regime of the qubits \cite{King_2022, King_2023_5000q}. Thermally assisted quantum annealing can assist in producing good optimization results \cite{dickson2013thermally}, but it is notable that we only get marginally better sampling results as the annealing time is extended up to $2000$ microseconds. 

There is a clear tradeoff where larger chain strengths reduce chain break proportions, with the cost being reduced solution quality. Interestingly, the chain break proportions for all devices substantially increased as the annealing time decreased, especially at $1$ microsecond, with the exception of \texttt{Advantage2\_prototype1.1}; this again suggests that the newer generation of hardware has less noise than the other three devices. 

Time-to-Solution (TTS) is a related metric that can also be computed, at least for the Maximum Clique sampling where the optimal solution is known -- quantifying TTS requires knowledge of what the optimal solution is. As opposed to approximation ratio, TTS intends to measure the expected computation time to reach optimality, whereas approximation ratio computes only solution quality and does not include computation time. Appendix~\ref{section:appendix_TTS_Max_Clique} shows Time-to-Solution results for the Maximum Clique QUBOs in Figure~\ref{fig:TTS_Max_Clique}.

\begin{figure*}[h!]
    \centering
    \includegraphics[width=0.24\textwidth]{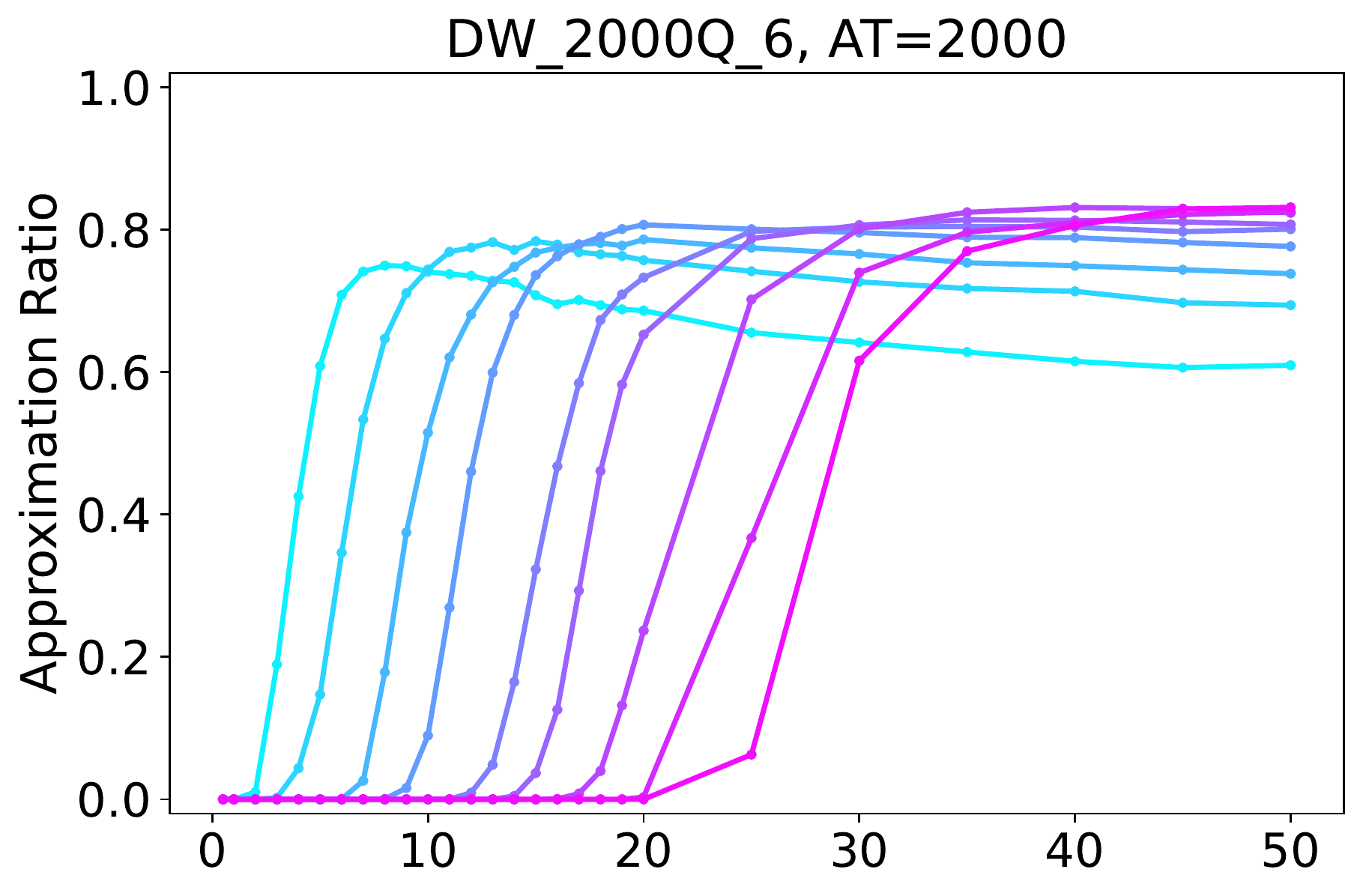}
    \includegraphics[width=0.24\textwidth]{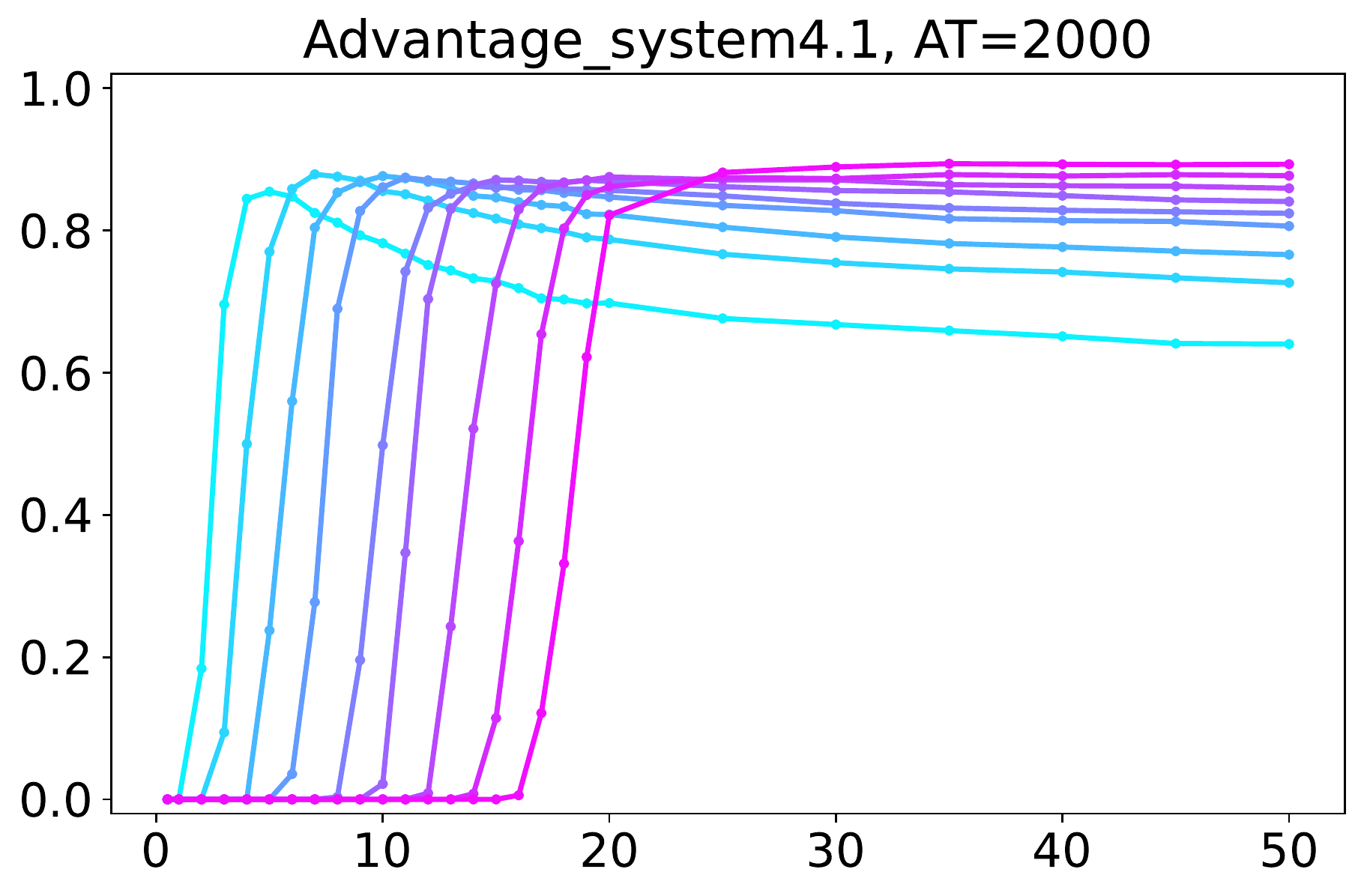}
    \includegraphics[width=0.24\textwidth]{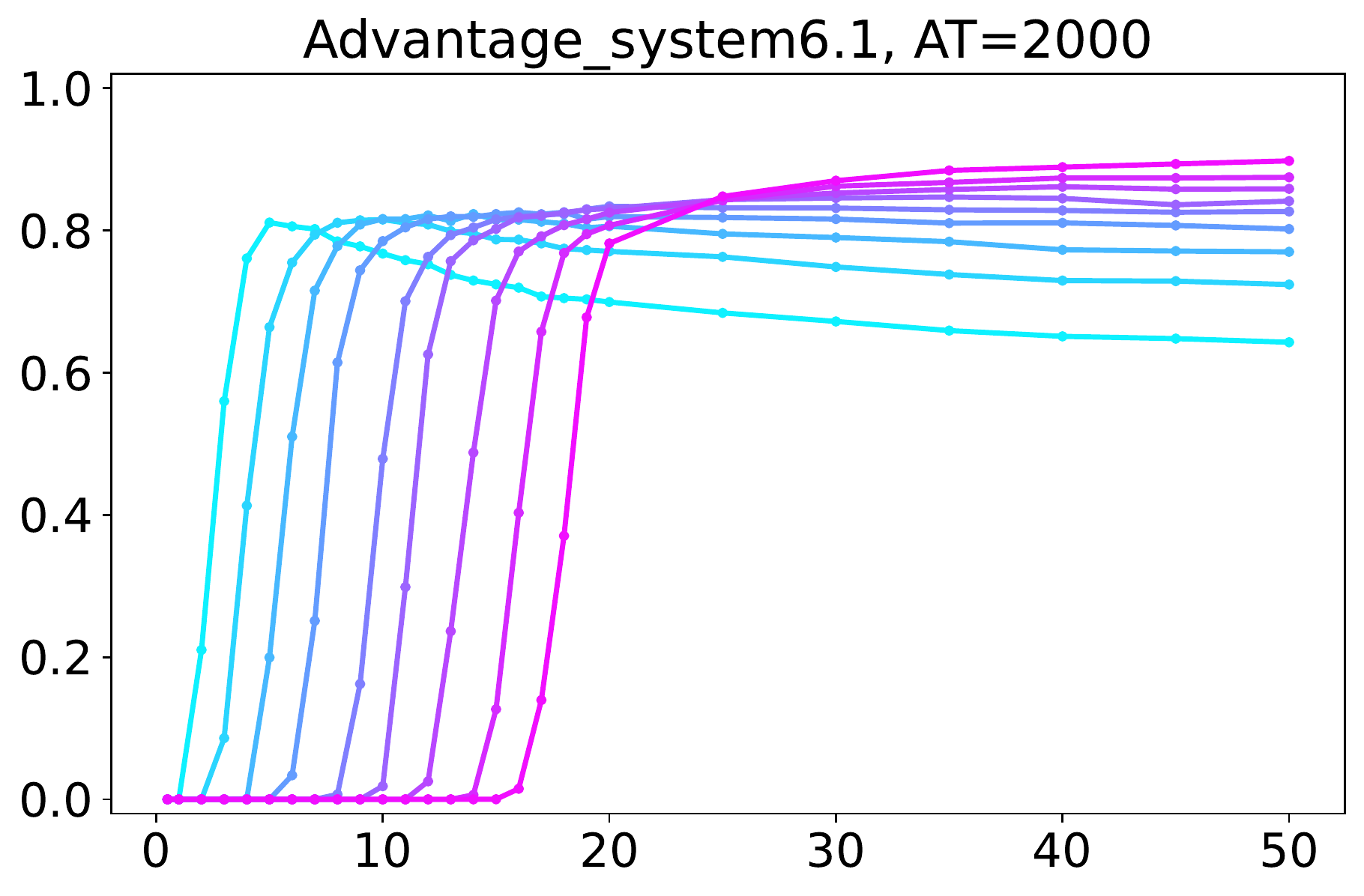}
    \includegraphics[width=0.24\textwidth]{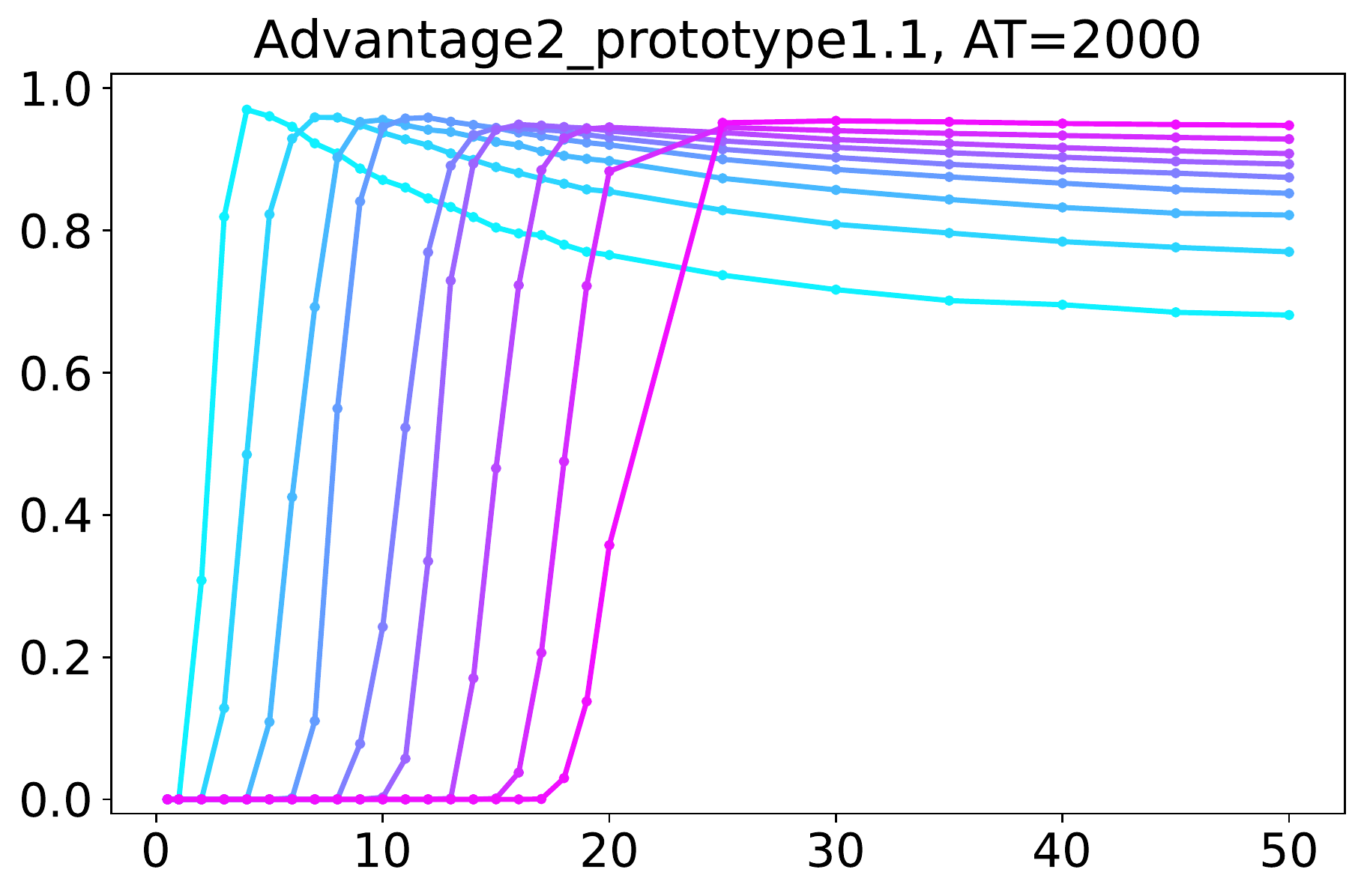}\\
    \includegraphics[width=0.24\textwidth]{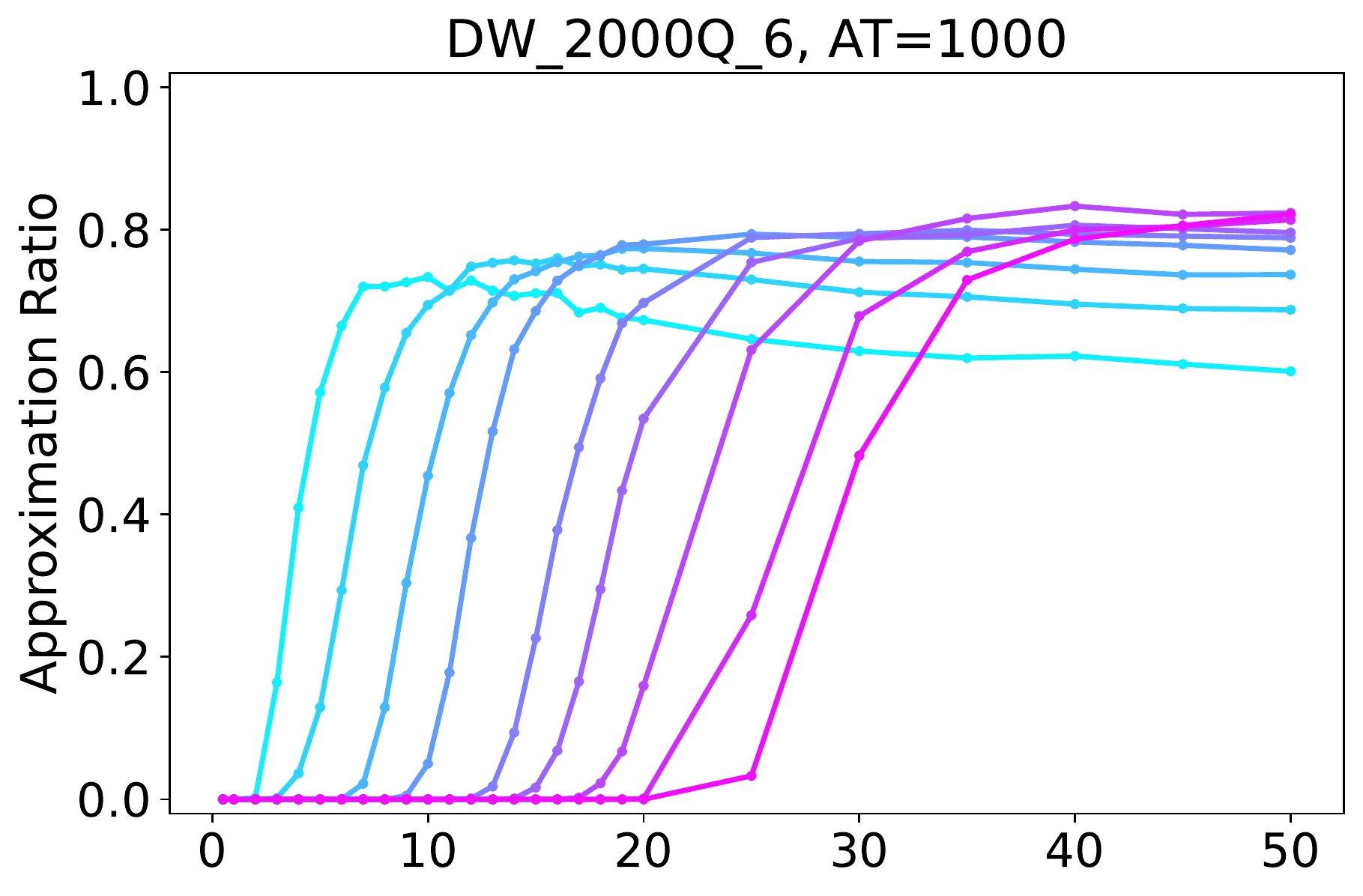}
    \includegraphics[width=0.24\textwidth]{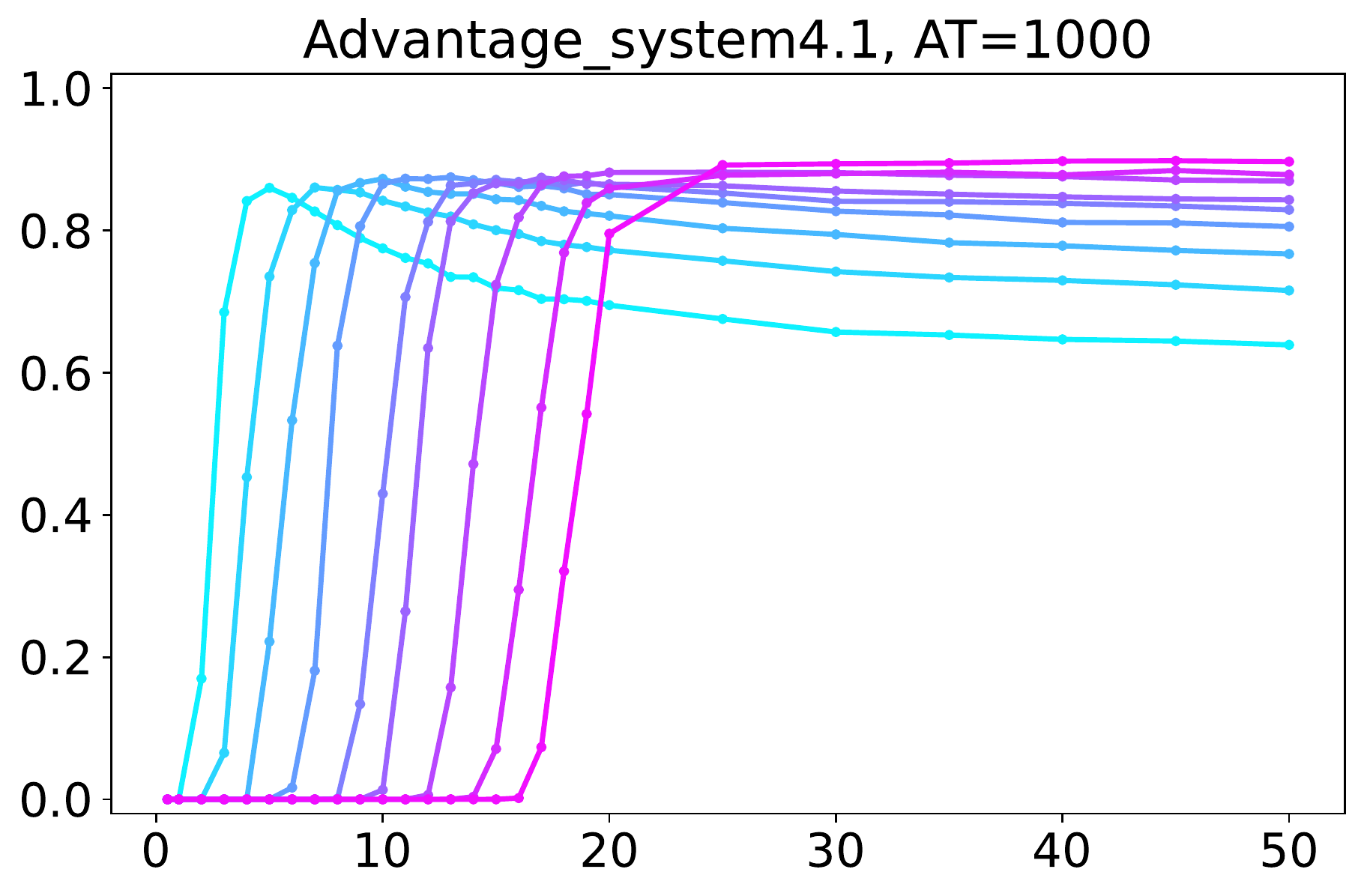}
    \includegraphics[width=0.24\textwidth]{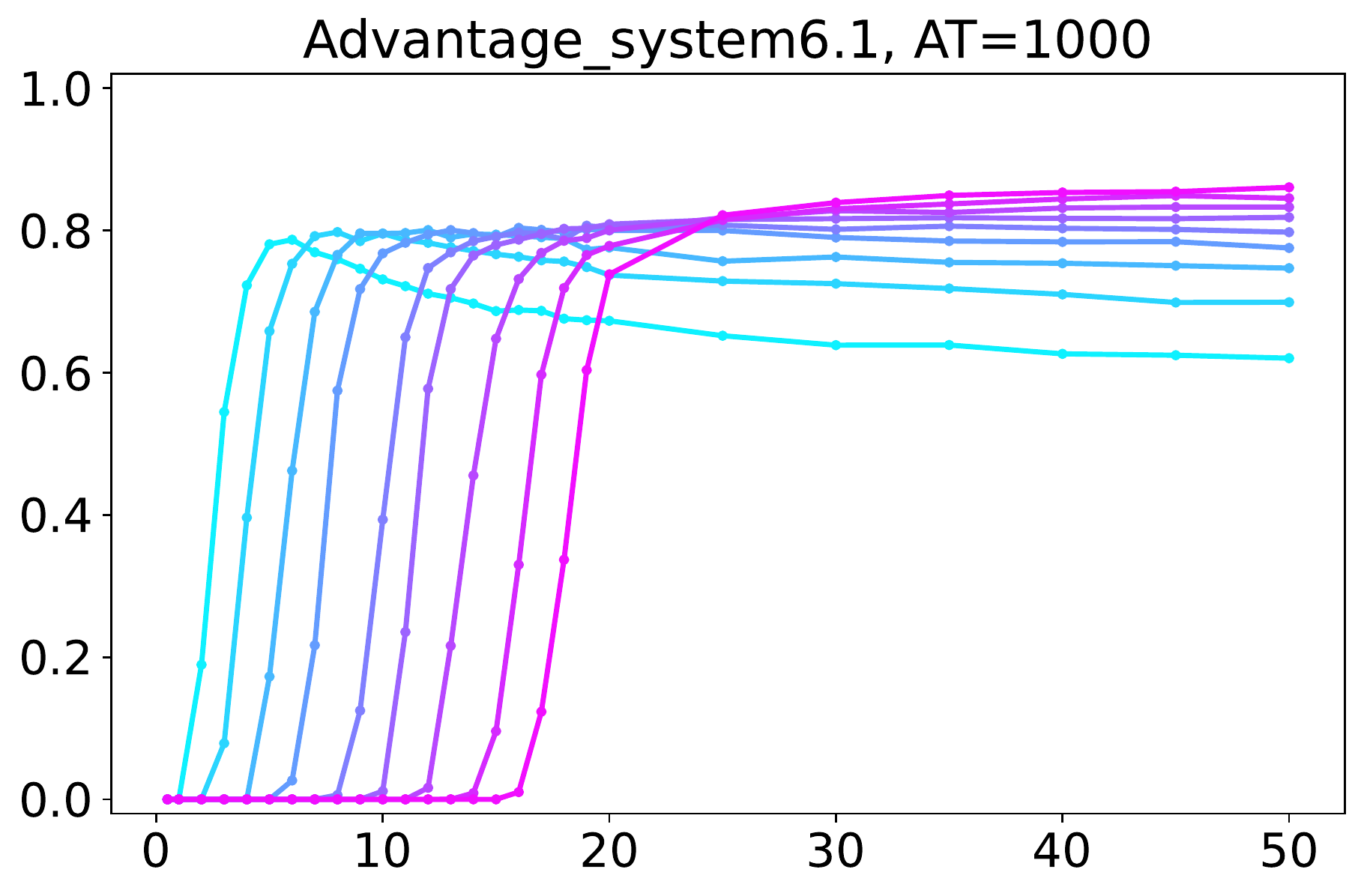}
    \includegraphics[width=0.24\textwidth]{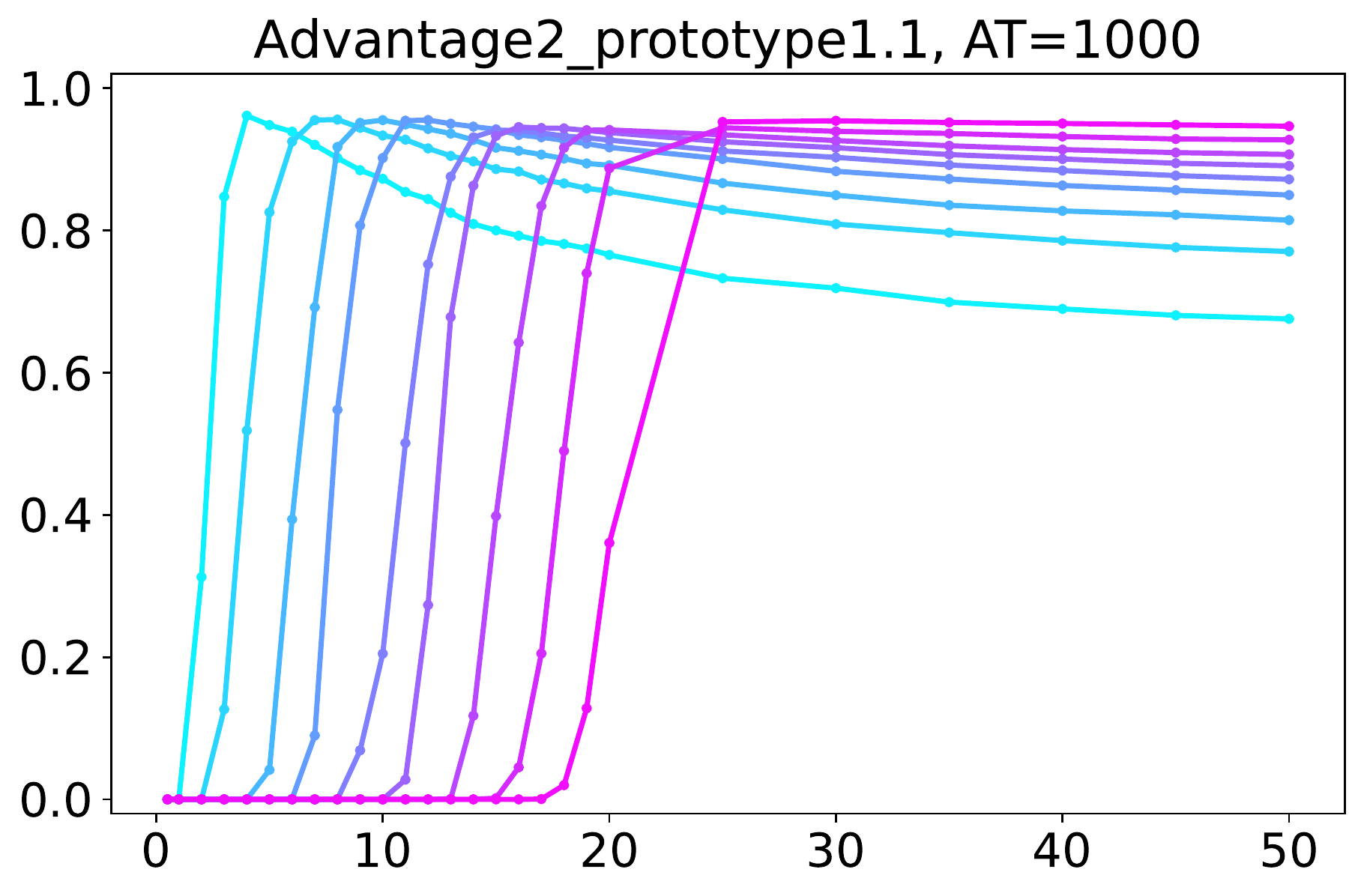}\\
    \includegraphics[width=0.24\textwidth]{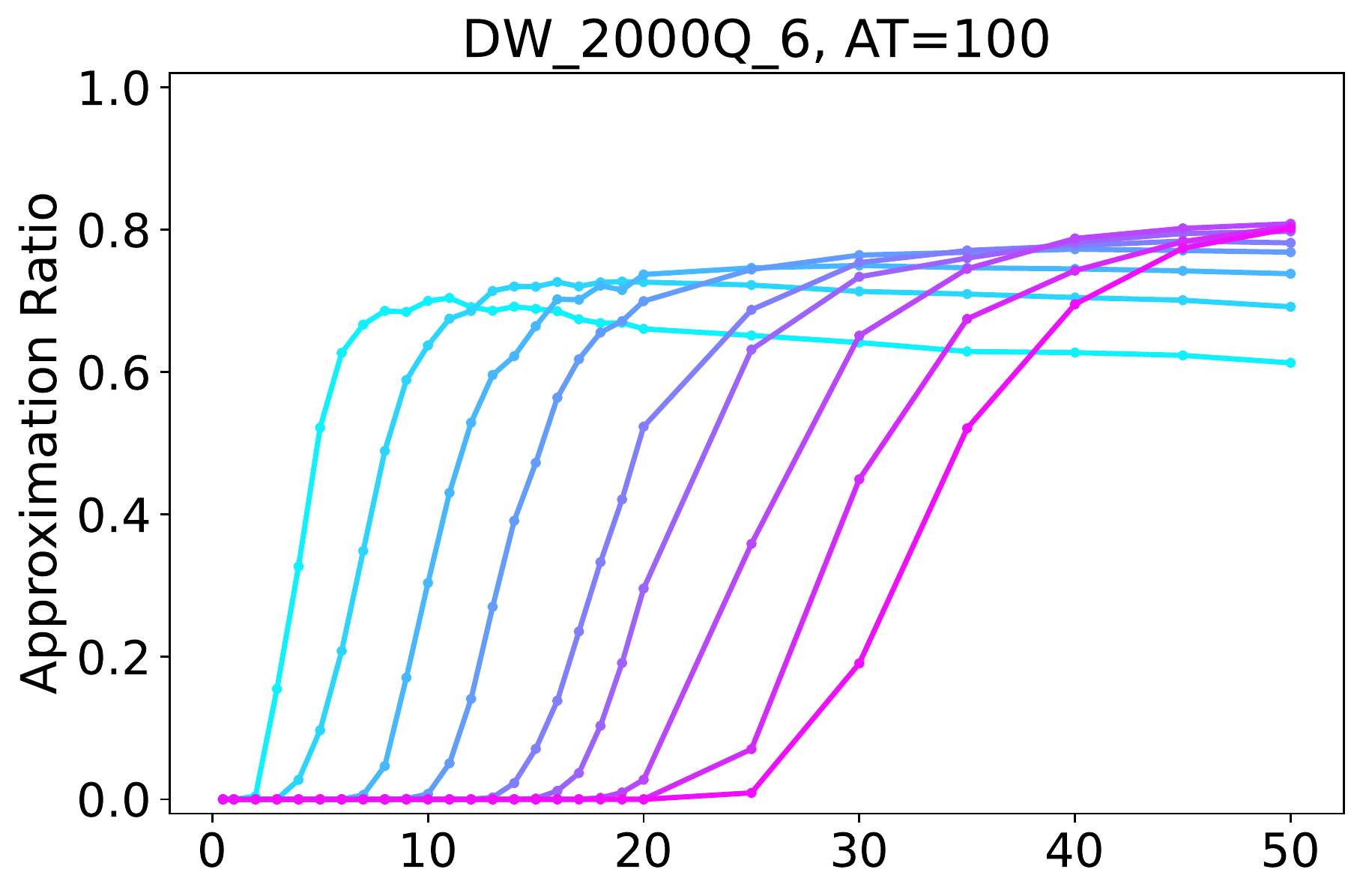}
    \includegraphics[width=0.24\textwidth]{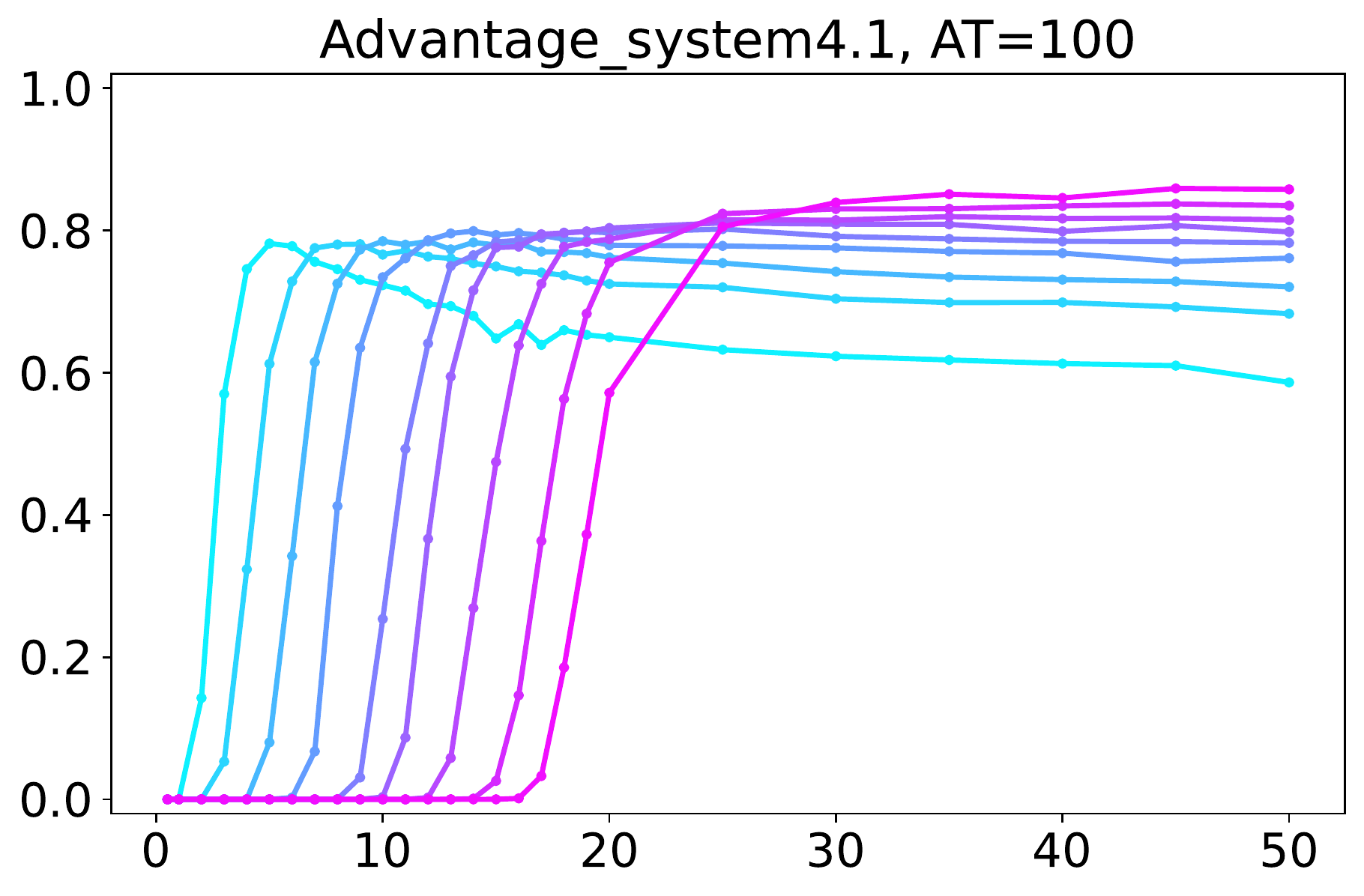}
    \includegraphics[width=0.24\textwidth]{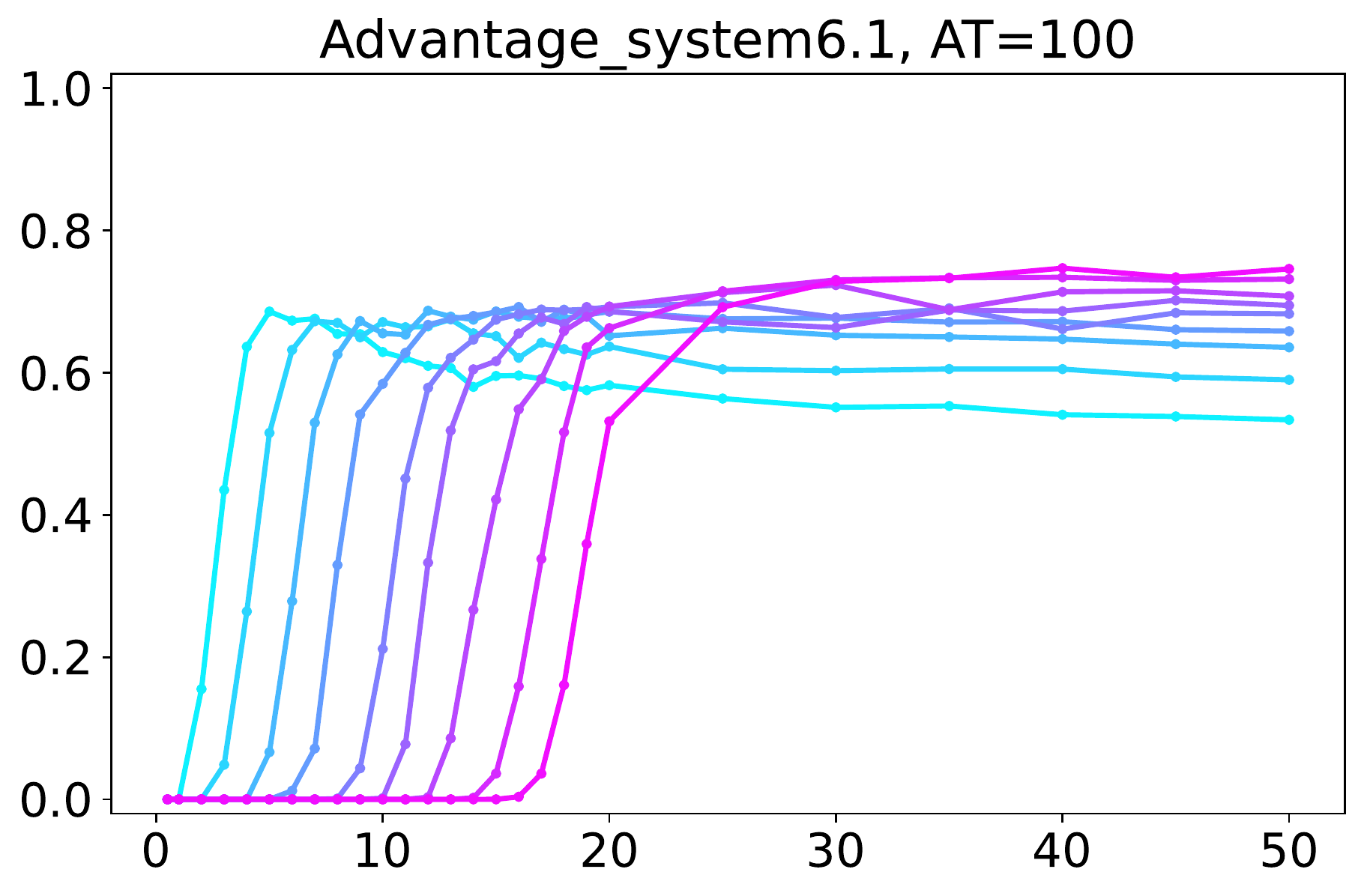}
    \includegraphics[width=0.24\textwidth]{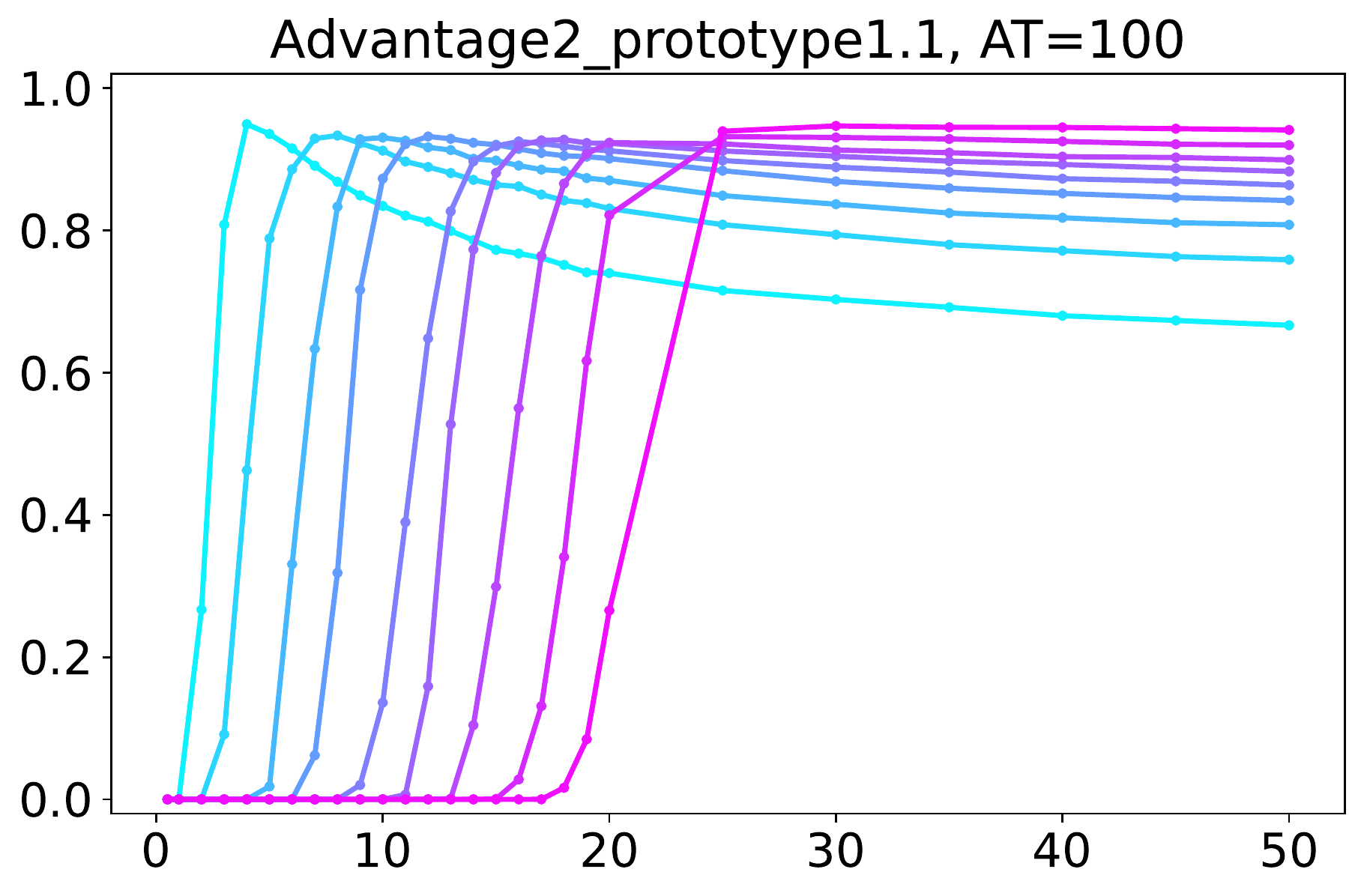}\\
    \includegraphics[width=0.24\textwidth]{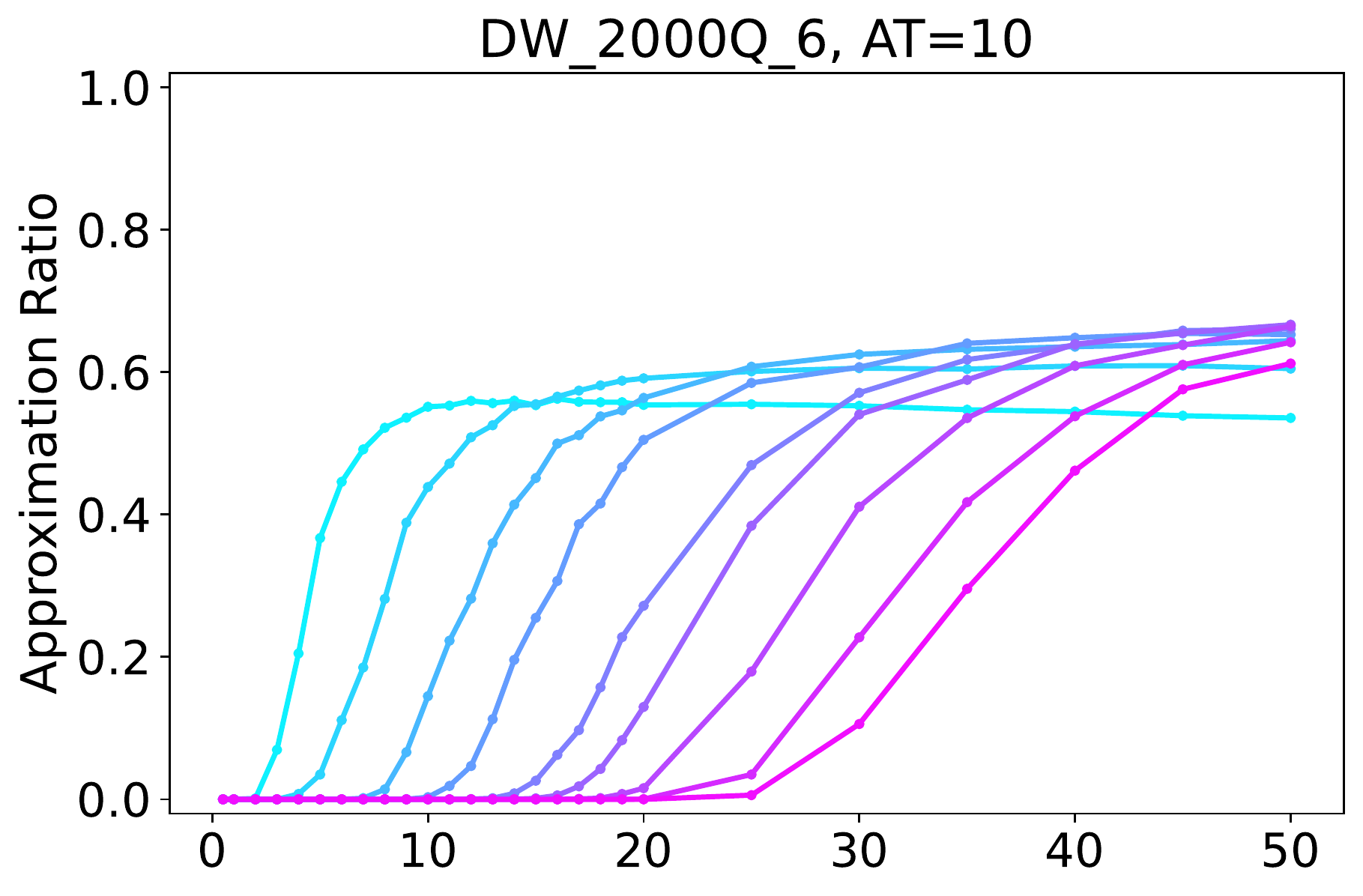}
    \includegraphics[width=0.24\textwidth]{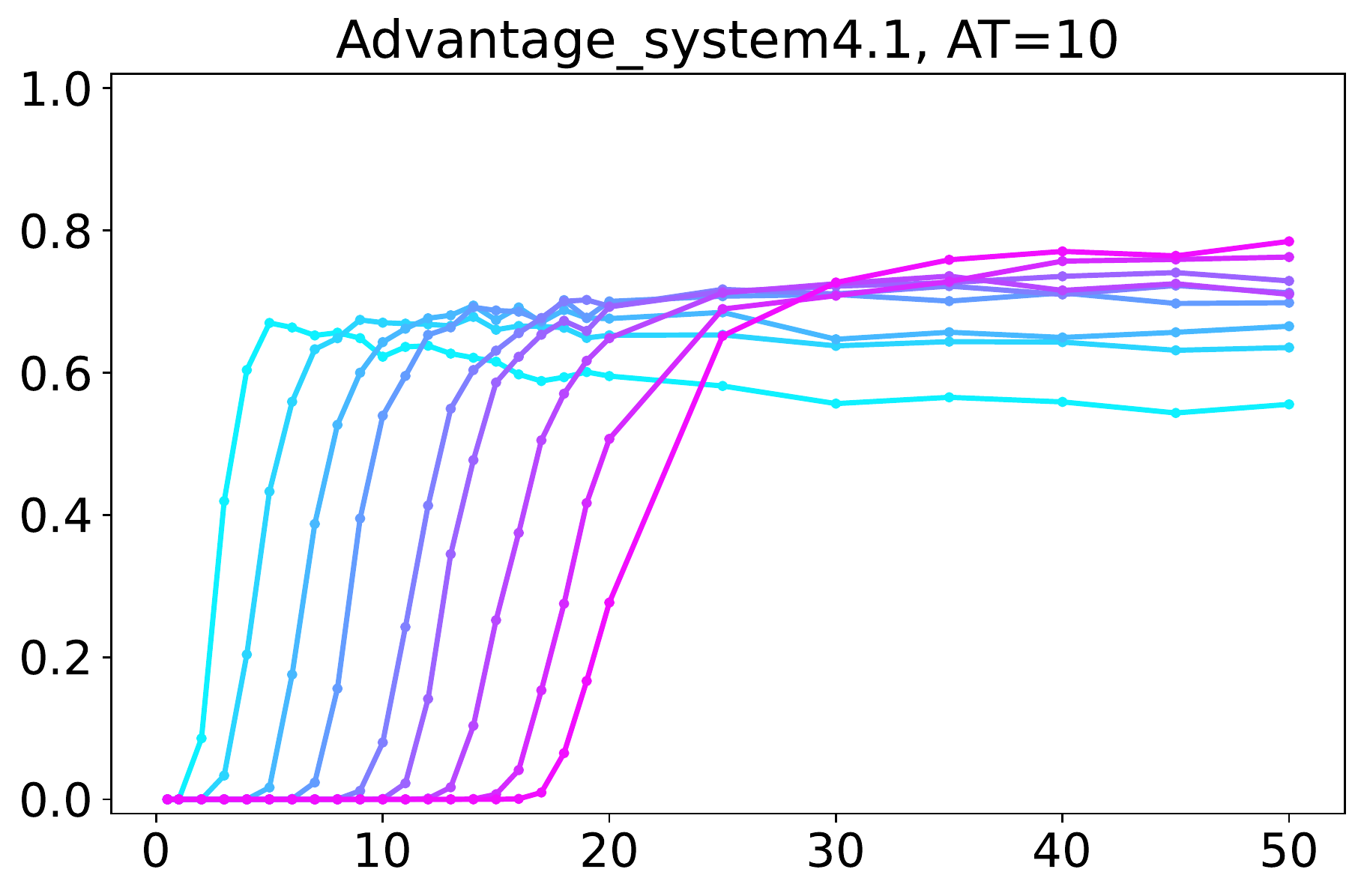}
    \includegraphics[width=0.24\textwidth]{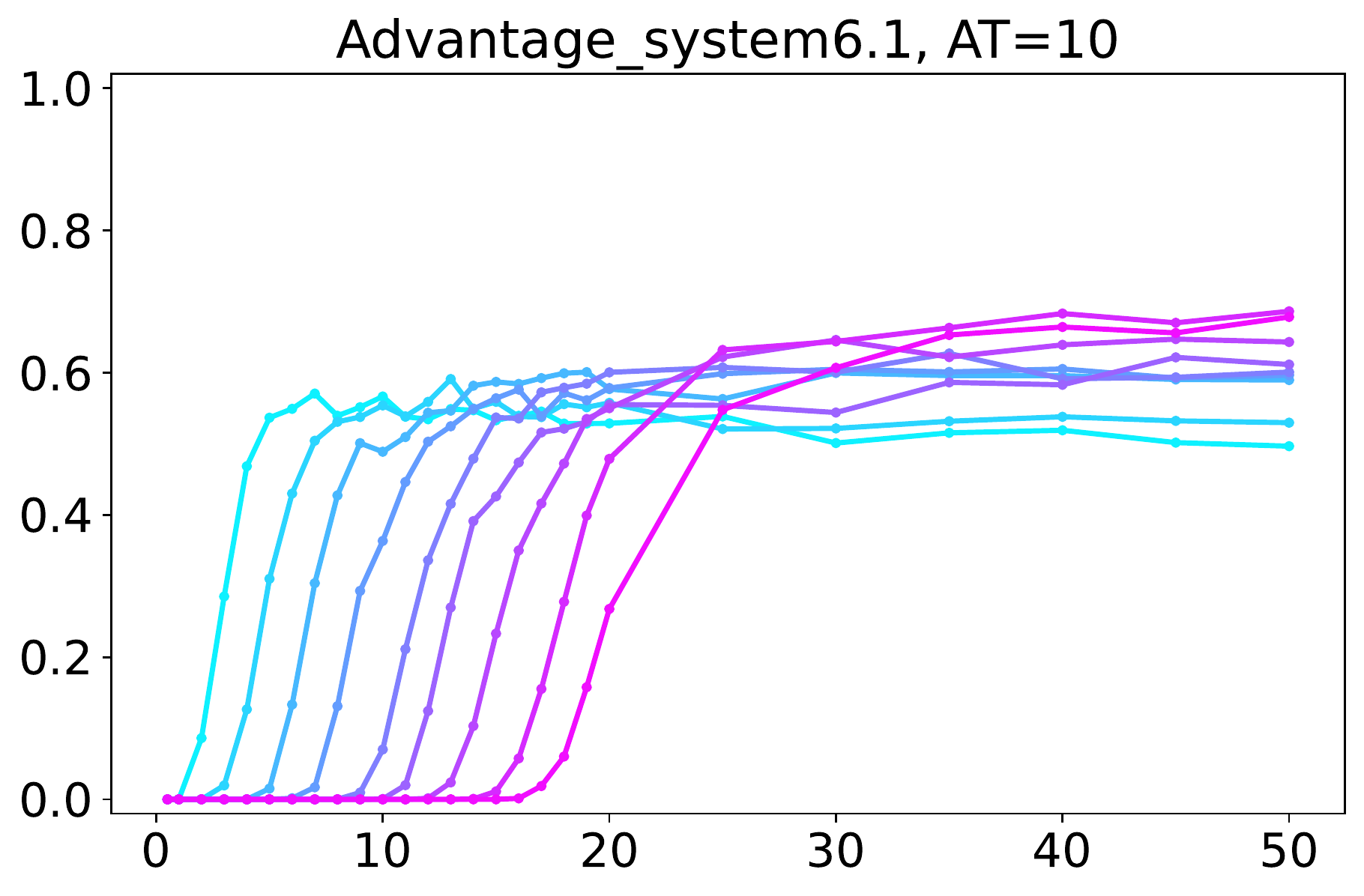}
    \includegraphics[width=0.24\textwidth]{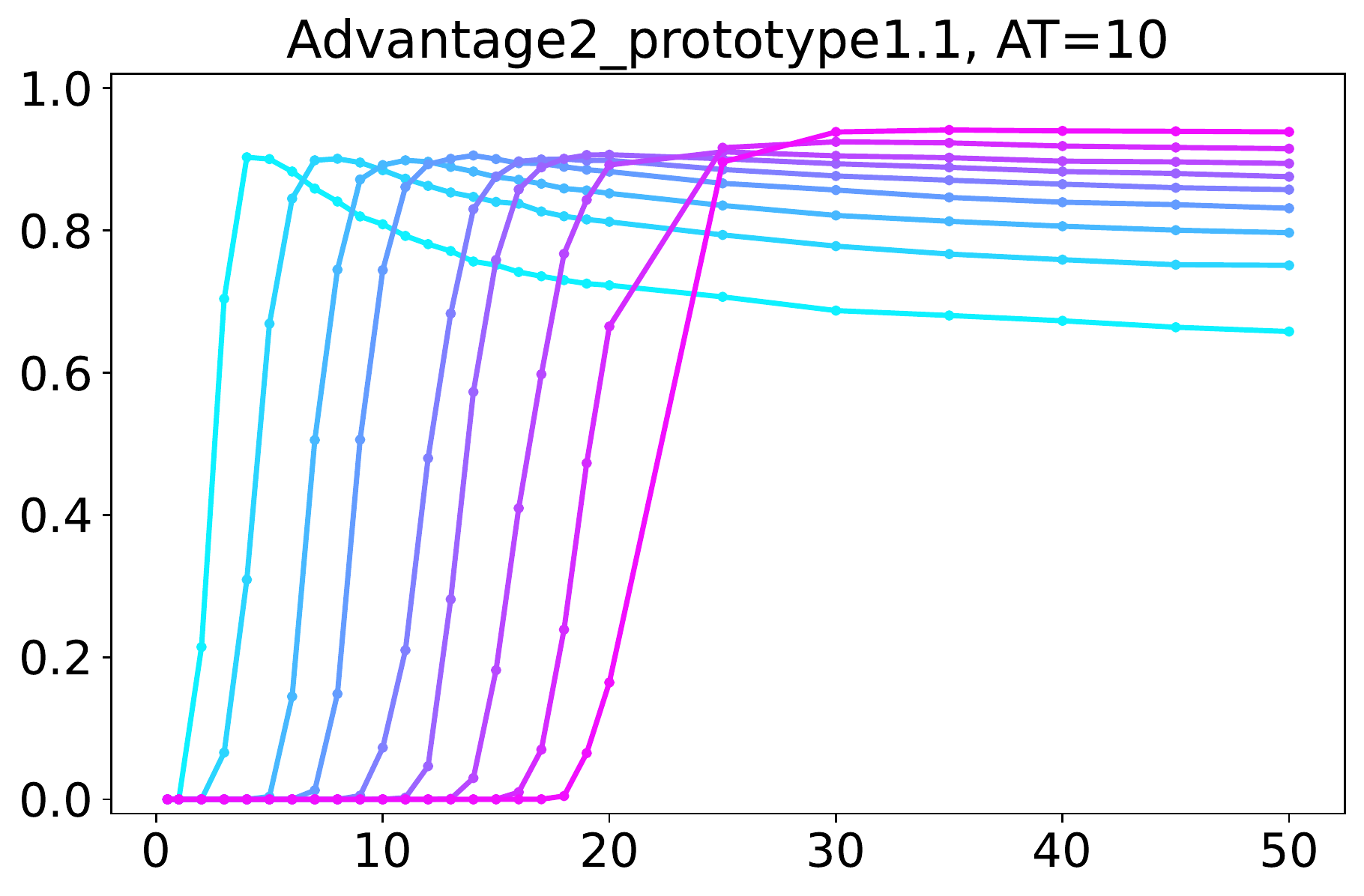}\\
    \includegraphics[width=0.24\textwidth]{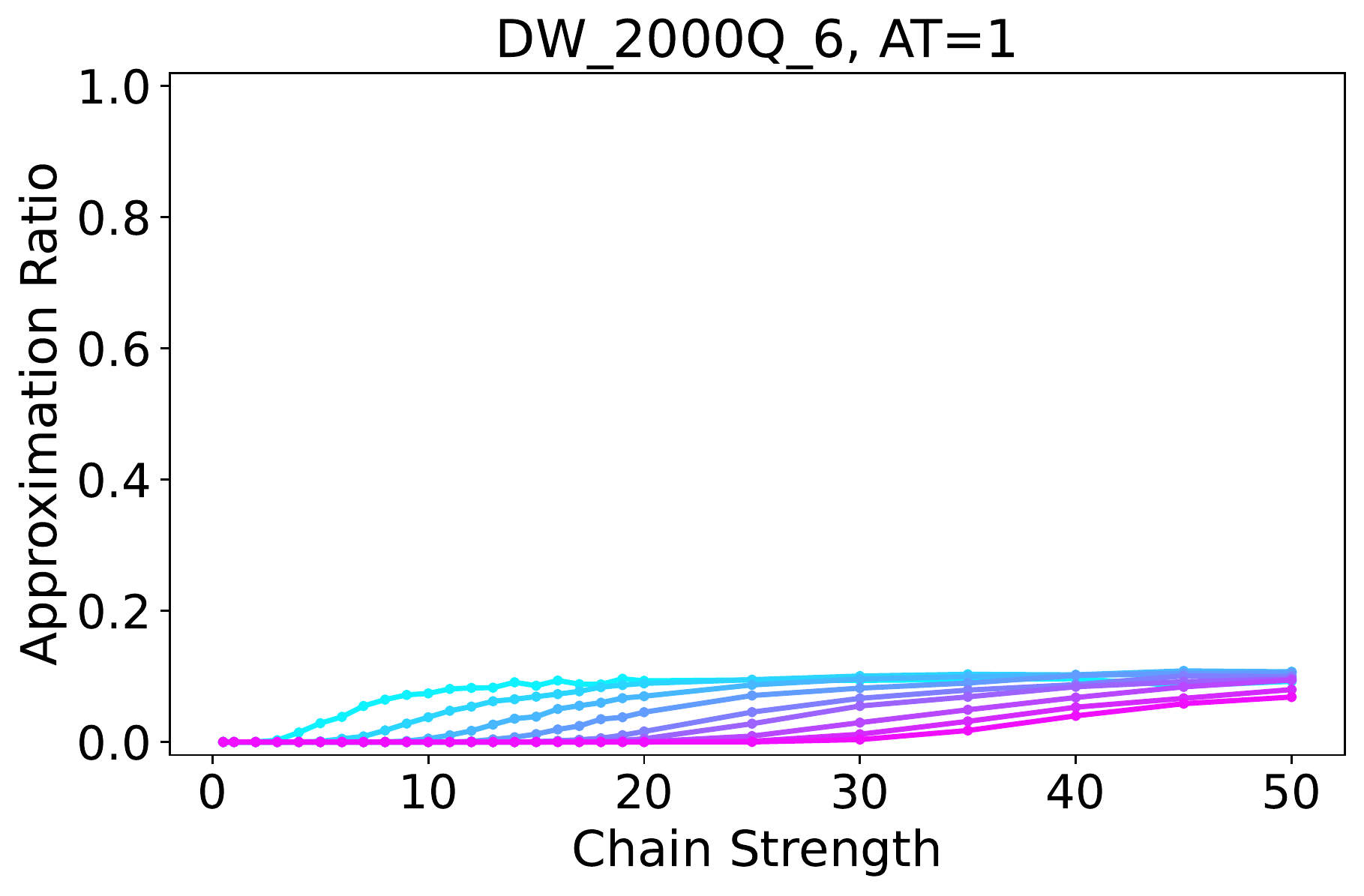}
    \includegraphics[width=0.24\textwidth]{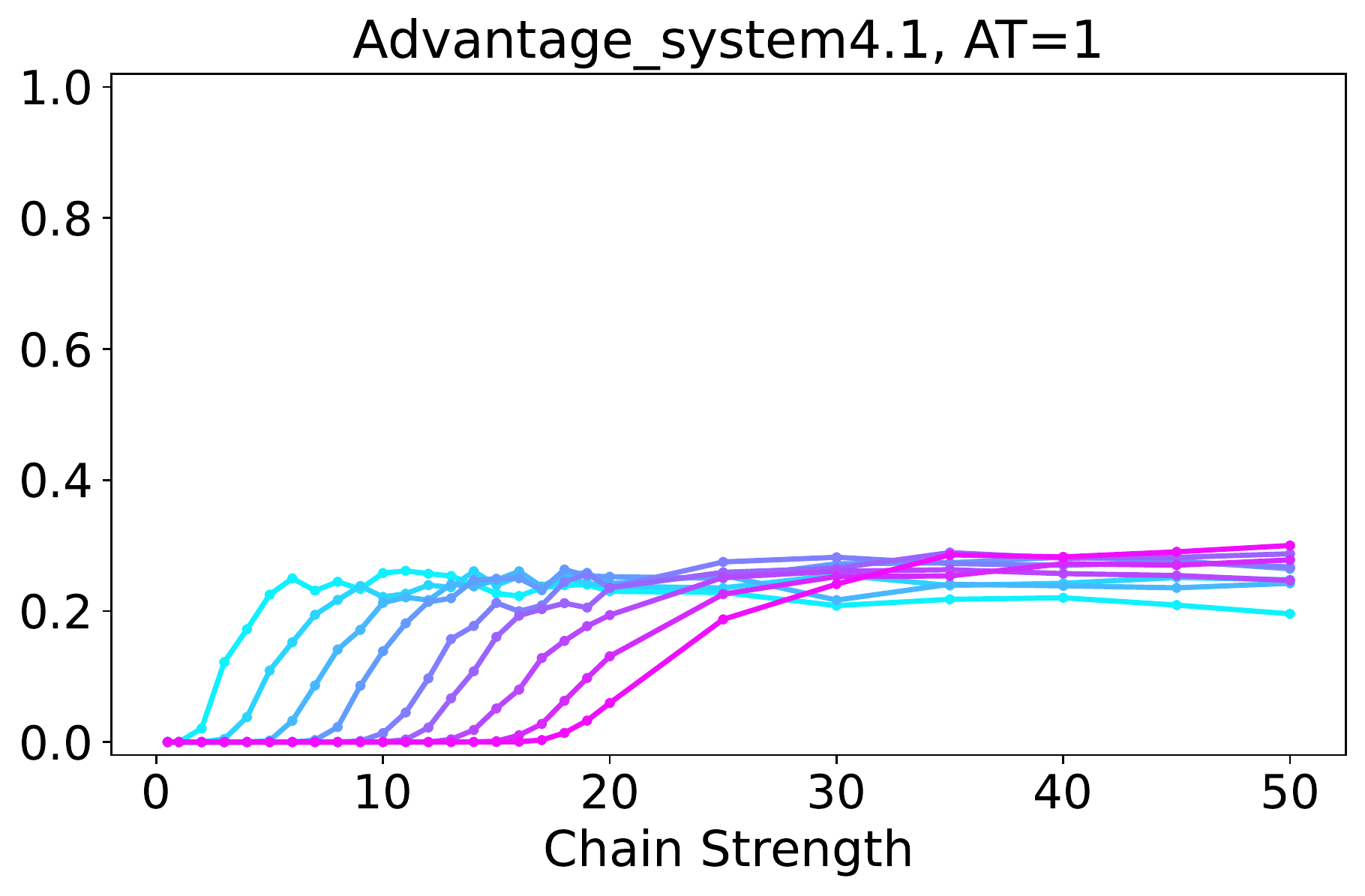}
    \includegraphics[width=0.24\textwidth]{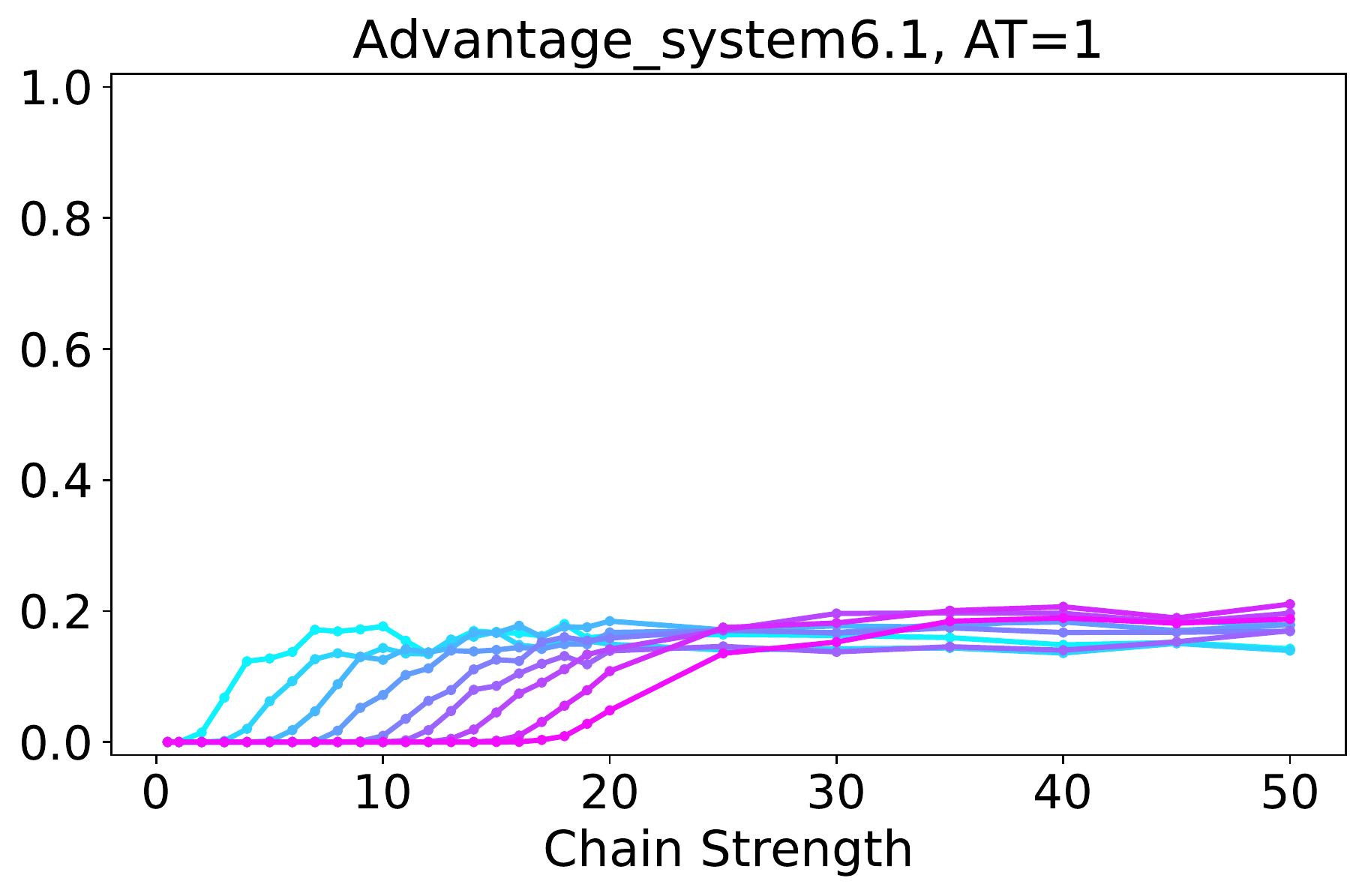}
    \includegraphics[width=0.24\textwidth]{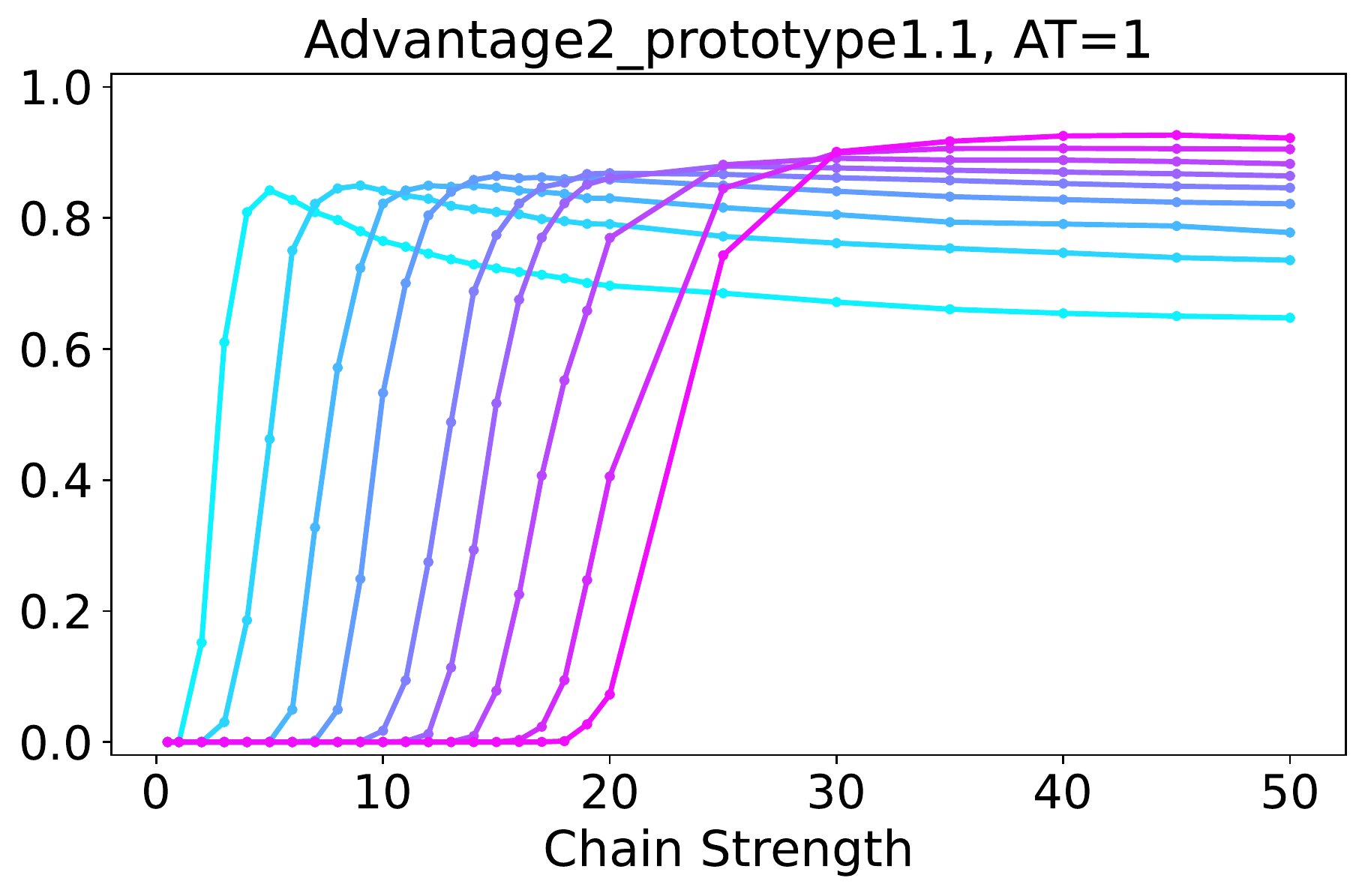}
    \includegraphics[width=0.71\textwidth]{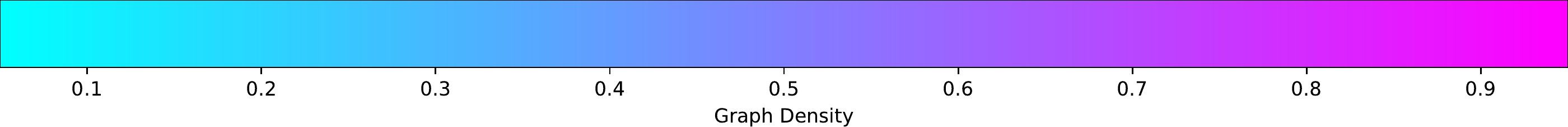}
    \caption{Maximum cut mean approximation ratios (y-axis) vs chain strength (x-axis) for each of the $150$ $G(n,p)$ random graphs. The aggregated results are shown in the form of $10$ lines per plot representing the mean approximation ratio for $10$ linearly spaced graph density intervals from $0.05$ to $0.95$, where the color of each line encodes the mean graph density for that interval. The color coding is shown in the colorbar below the plots. Problem QUBOs sampled using \texttt{DW\_2000Q\_6} (left column), \texttt{Advantage\_system4.1} (center-left column), \texttt{Advantage\_system6.1} (center-right column), and \texttt{Advantage2\_prototype1.1} (right column). The annealing time in microseconds are varied across $2000$ microseconds (first row), $1000$ microseconds (second row), $100$ microseconds (third row), $10$ microseconds (fourth row), and $1$ microsecond (bottom row). }
    \label{fig:maximum_cut_approx_ratio}
\end{figure*}

Approximation ratios and chain break proportions for an annealing time of $0.5$ microseconds are shown in Figure~\ref{fig:small_annealing_times_appendix_maxclique} in Section~\ref{section:appendix_small_annealing_times}. $0.5$ microseconds is currently the smallest annealing time that a user can program a D-Wave quantum annealer to execute; the two devices with Pegasus topology allow this capability. At $0.5$ microseconds annealing the chain break proportion is very high, and the approximation ratios are very near to zero.

\subsection{Maximum Cut results}
\label{section:results_maximum_cut}

\begin{figure*}[h!]
    \centering
    \includegraphics[width=0.24\textwidth]{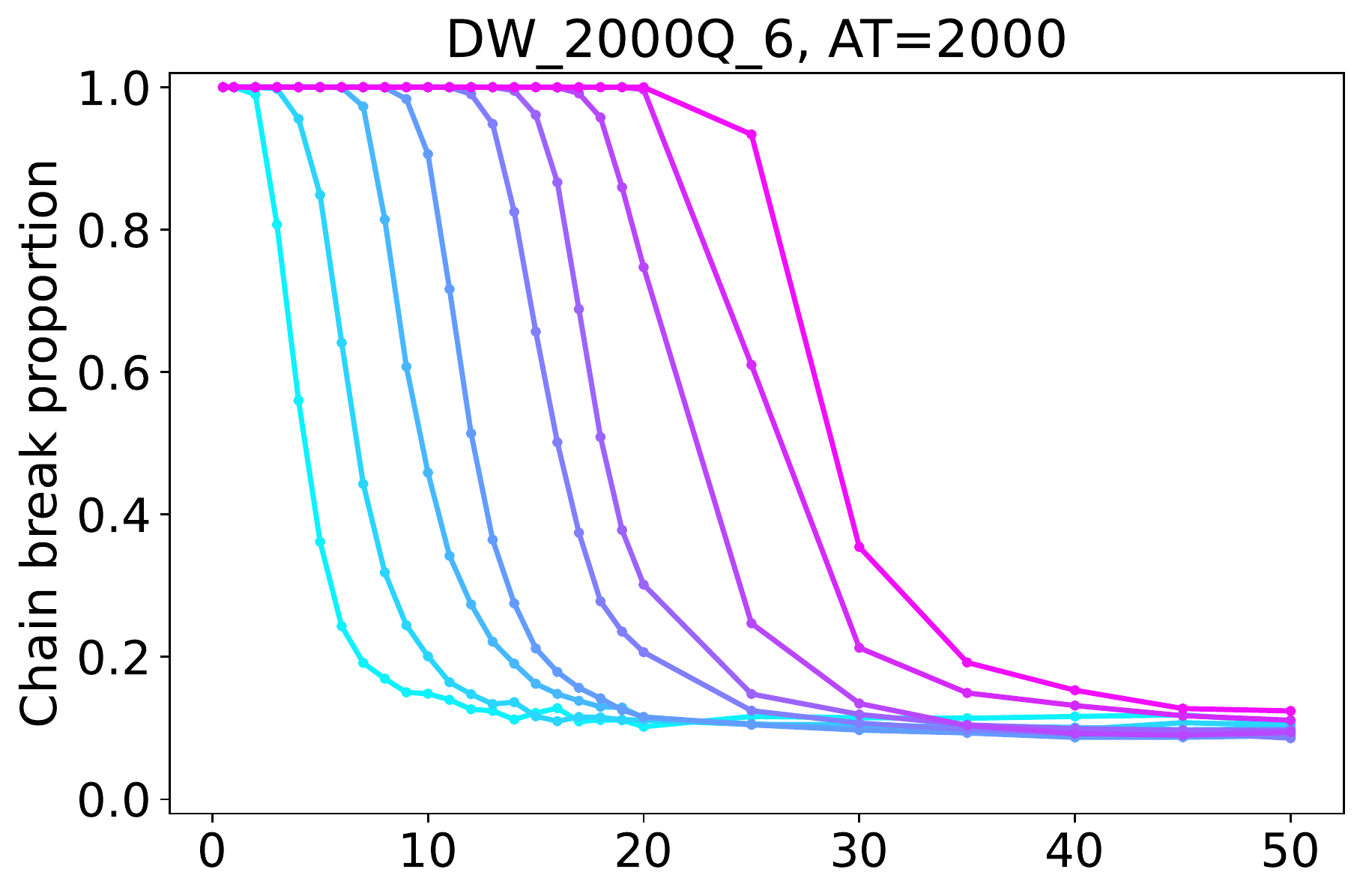}
    \includegraphics[width=0.24\textwidth]{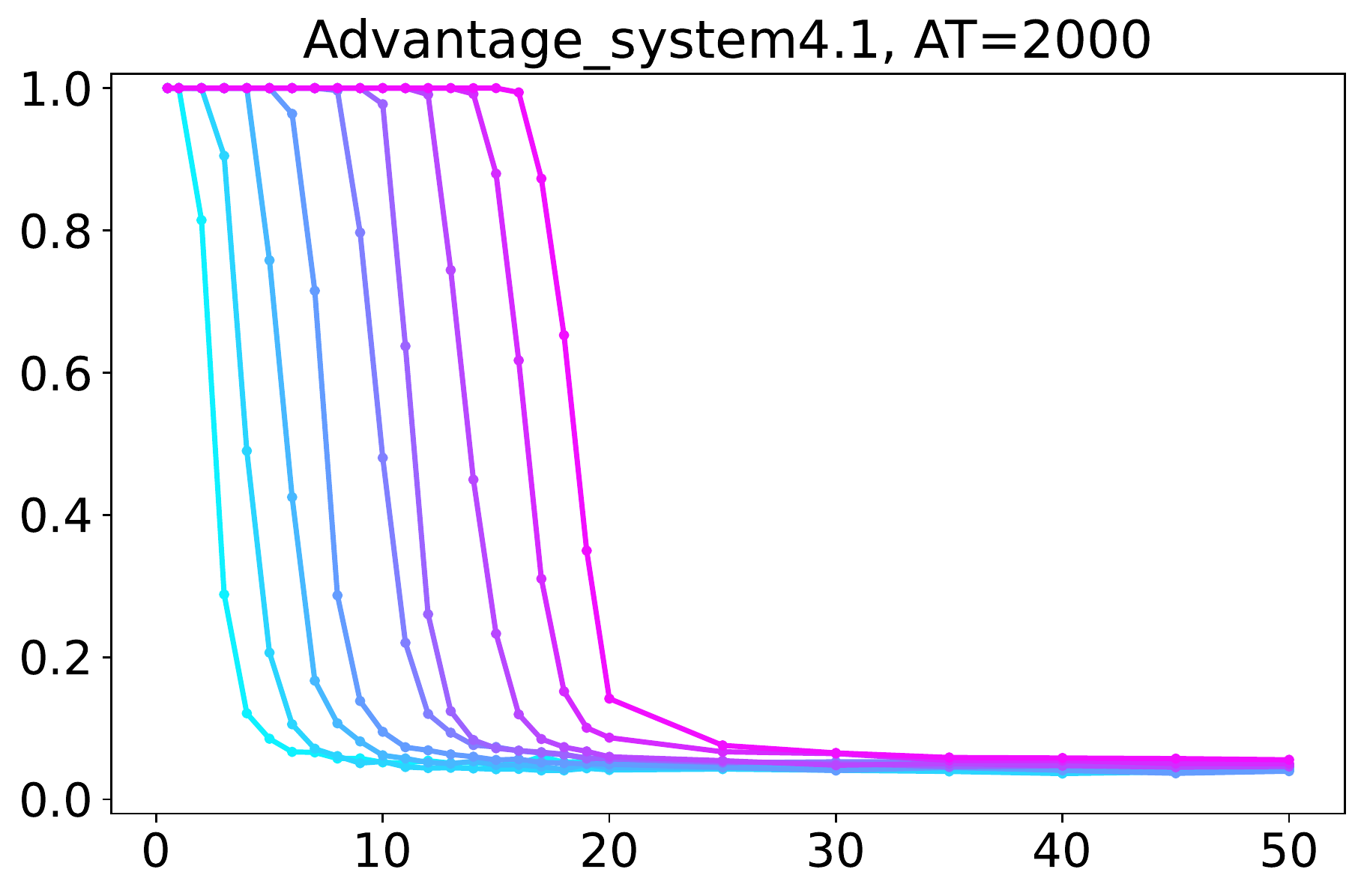}
    \includegraphics[width=0.24\textwidth]{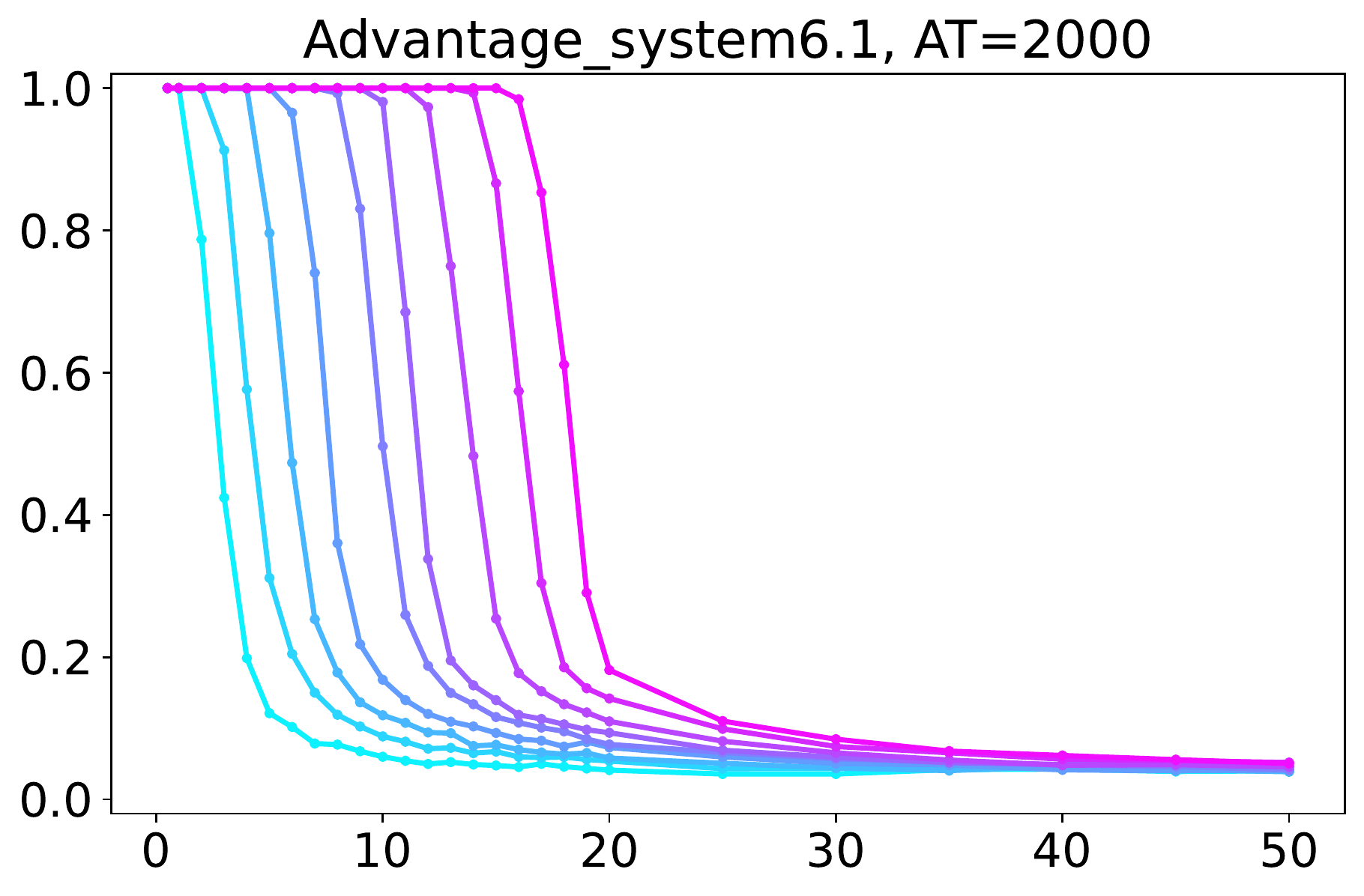}
    \includegraphics[width=0.24\textwidth]{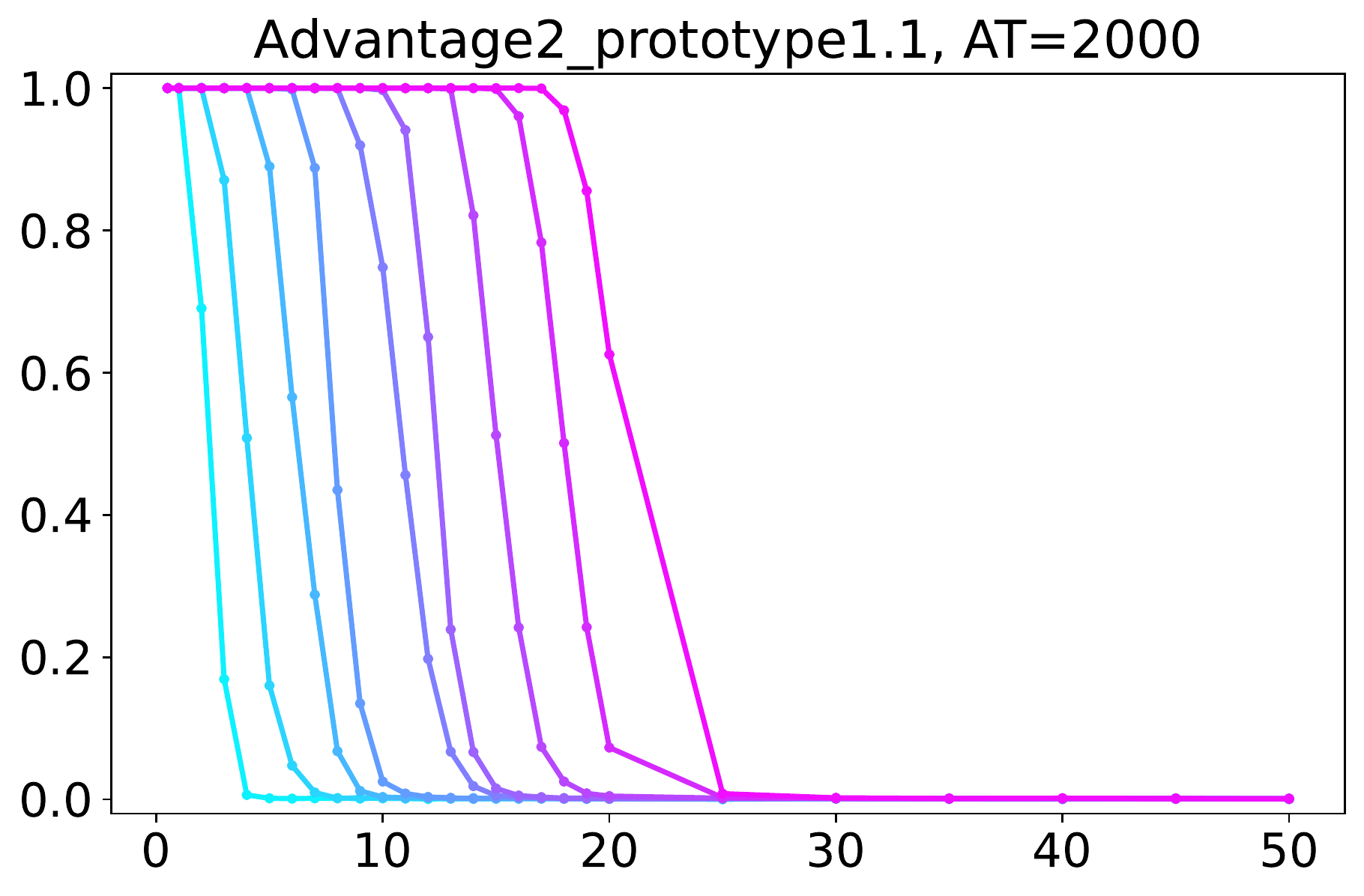}\\
    \includegraphics[width=0.24\textwidth]{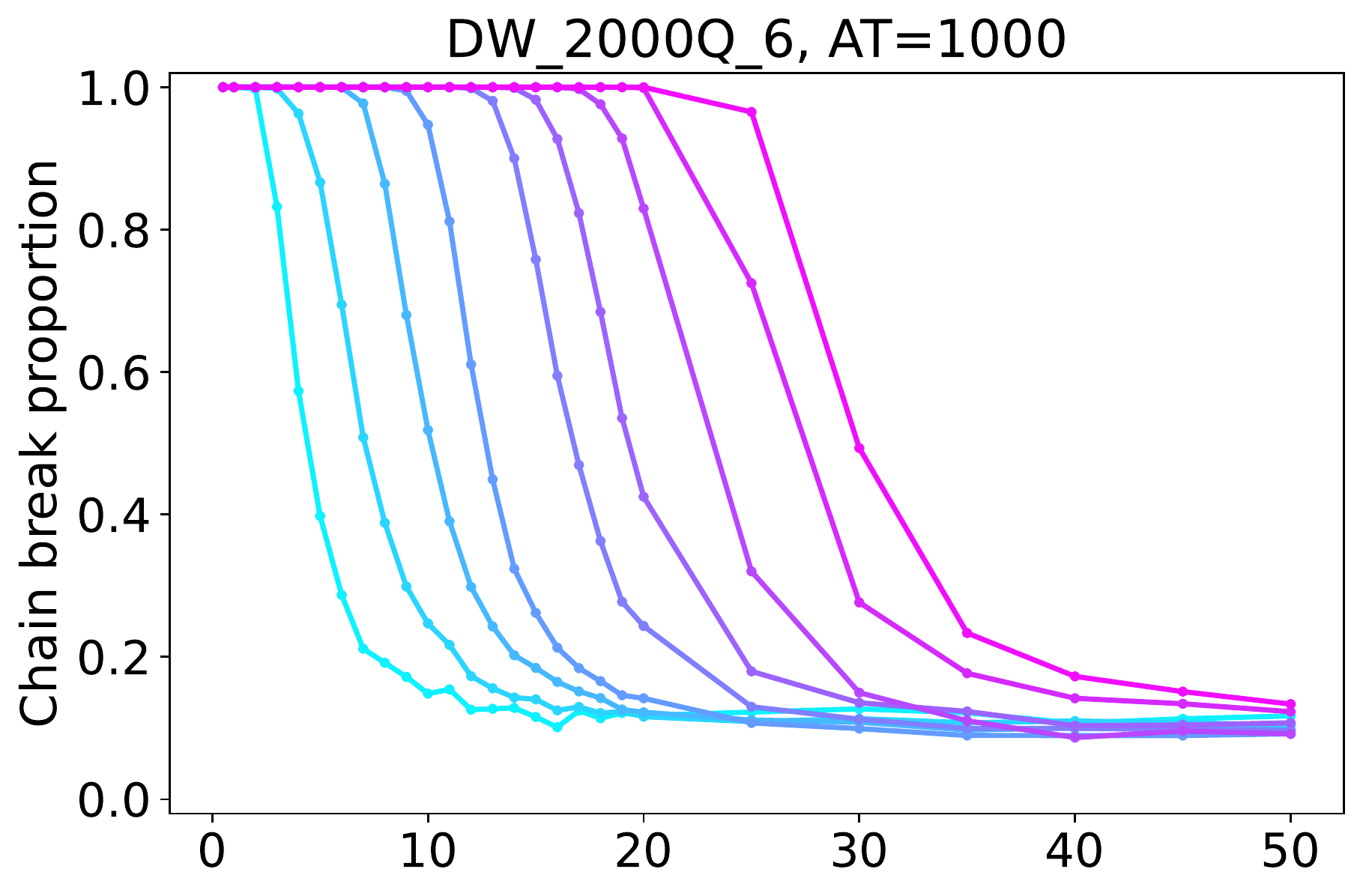}
    \includegraphics[width=0.24\textwidth]{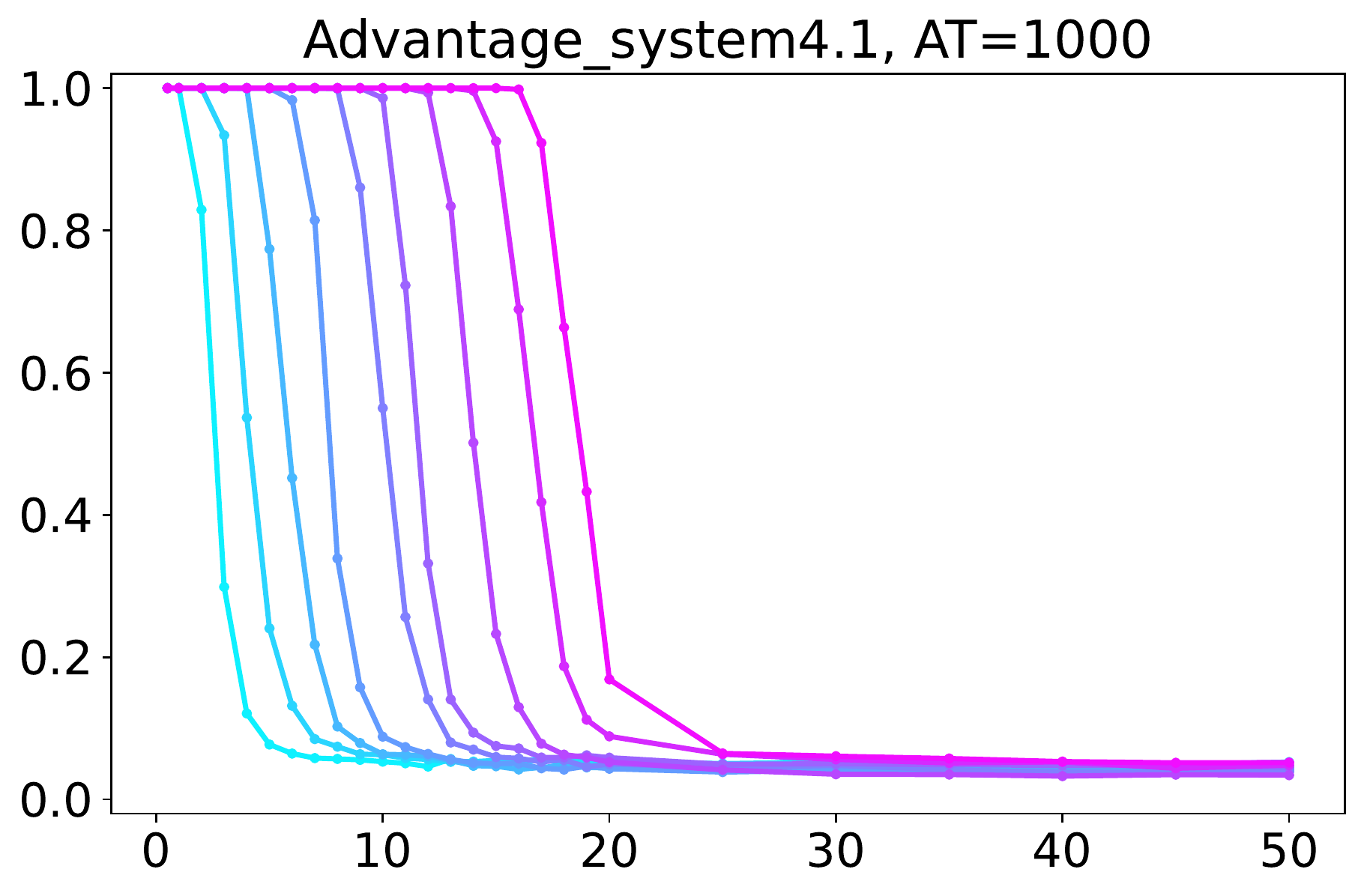}
    \includegraphics[width=0.24\textwidth]{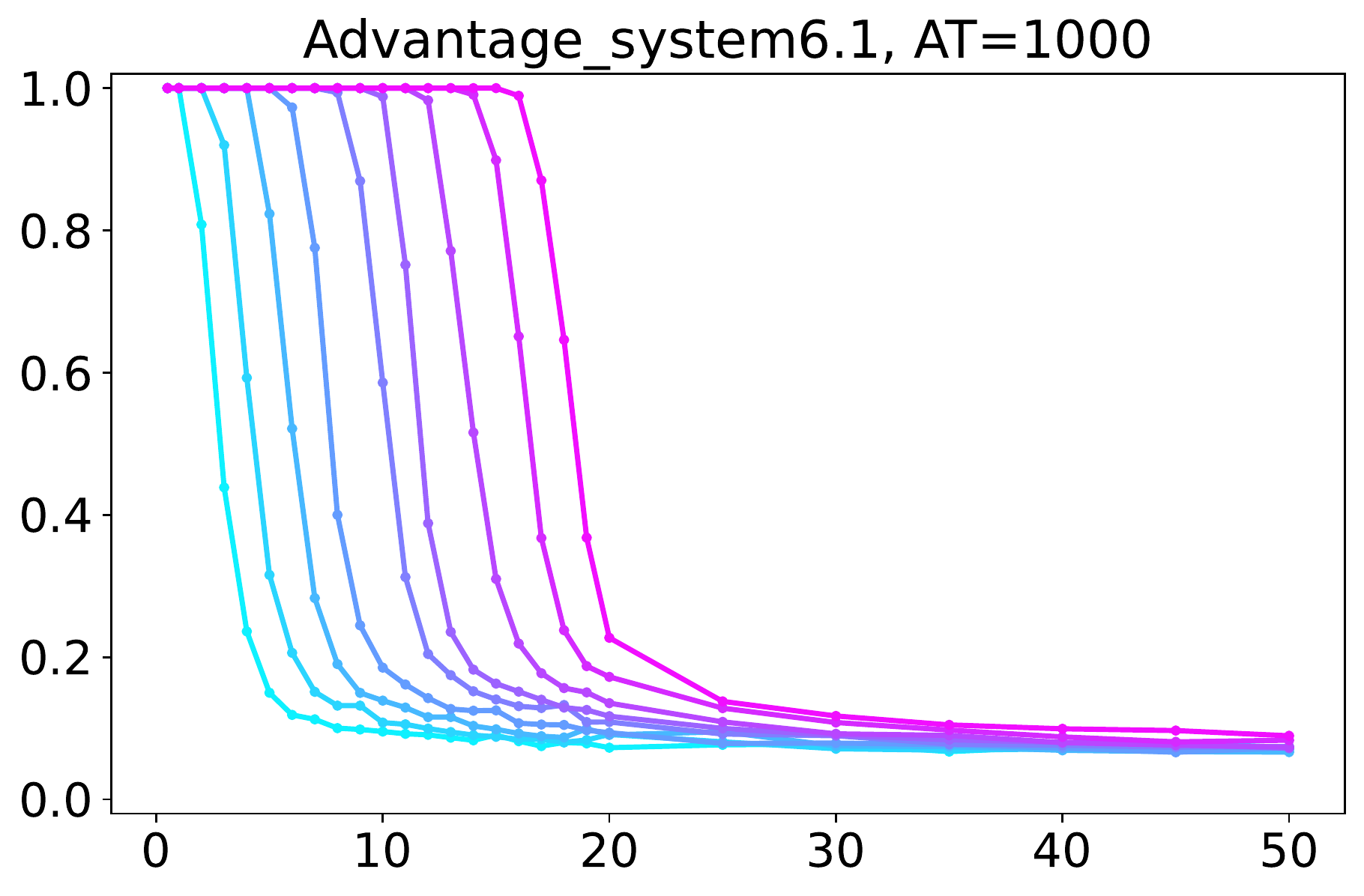}
    \includegraphics[width=0.24\textwidth]{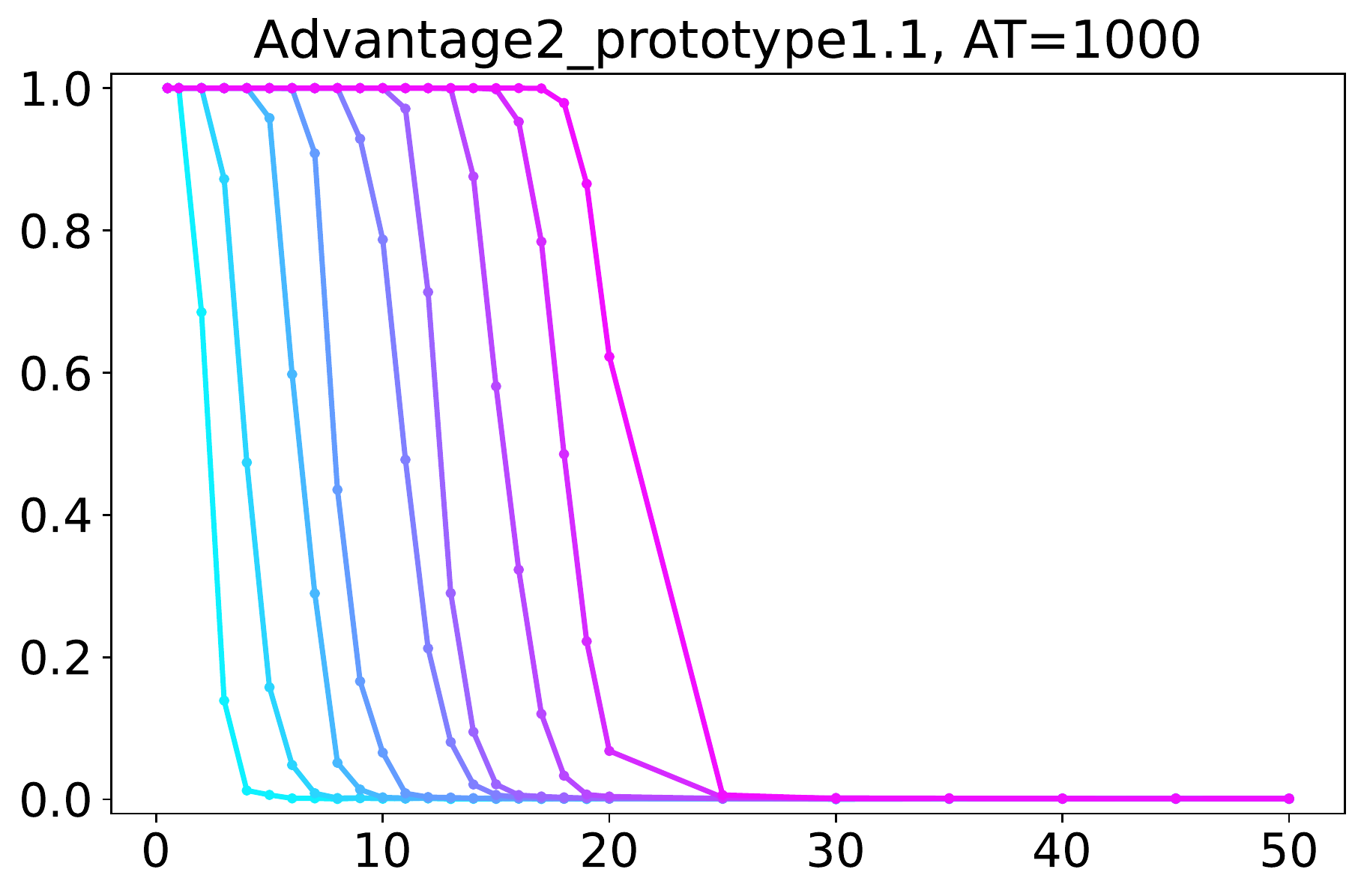}\\
    \includegraphics[width=0.24\textwidth]{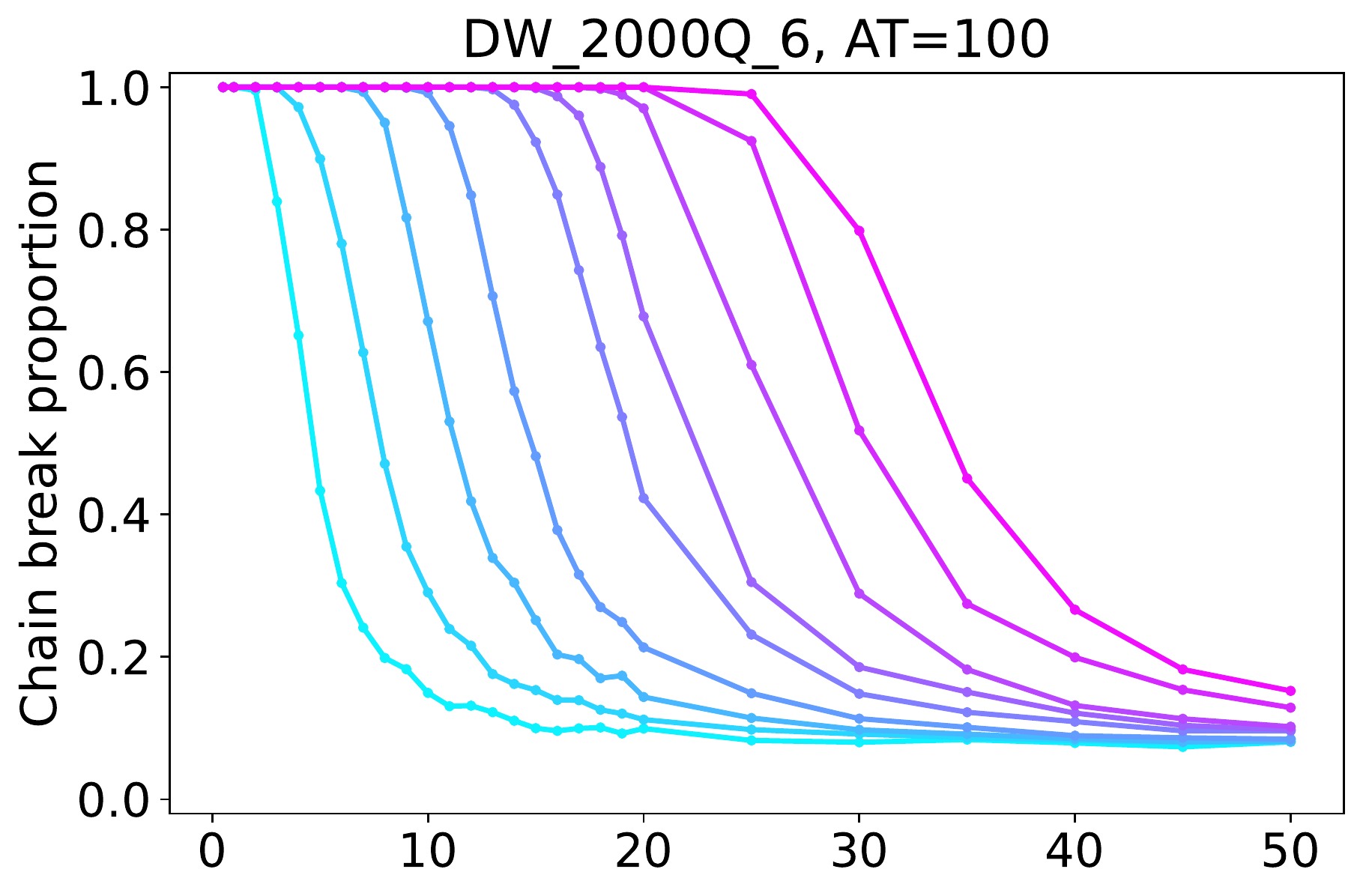}
    \includegraphics[width=0.24\textwidth]{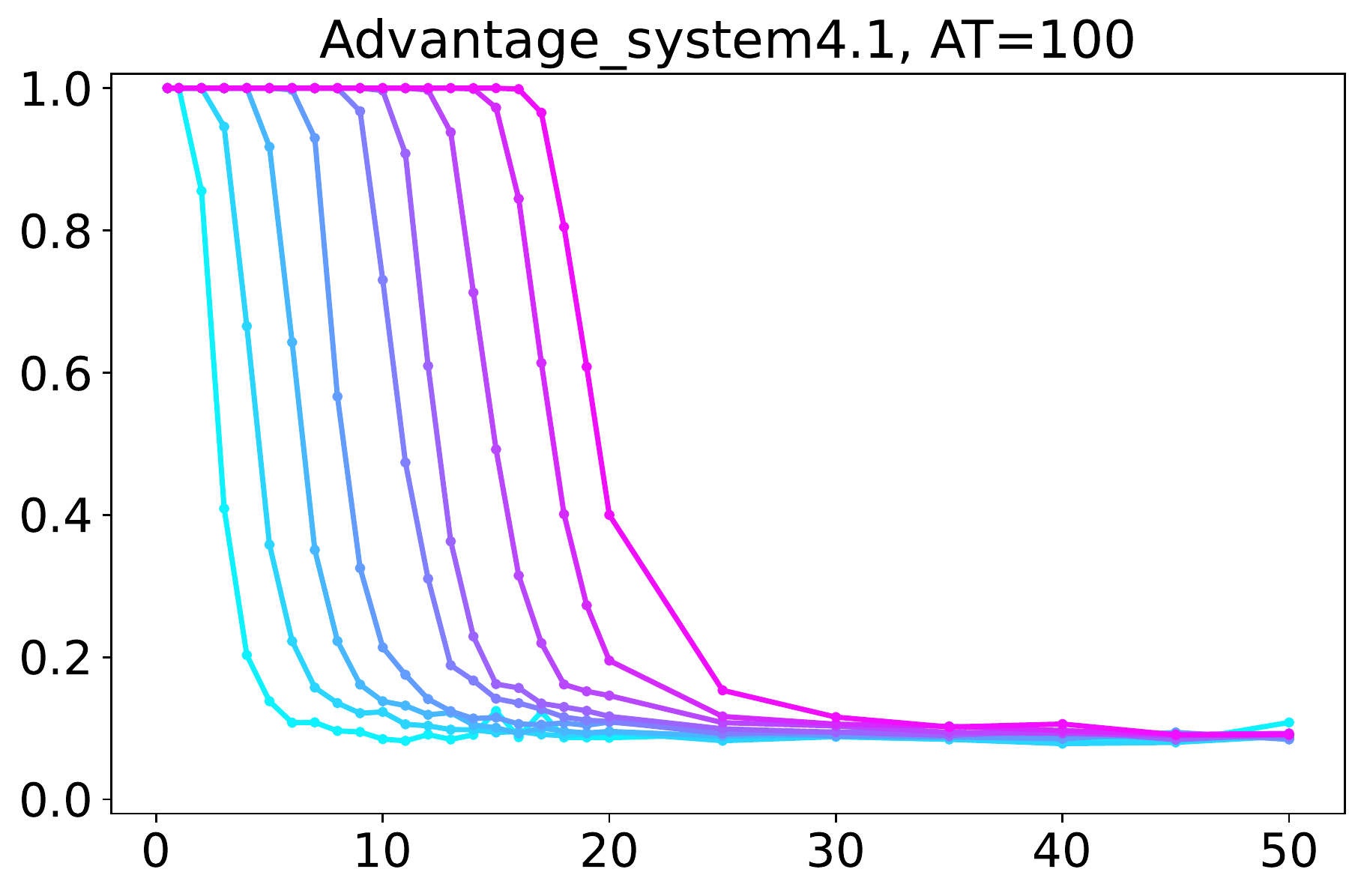}
    \includegraphics[width=0.24\textwidth]{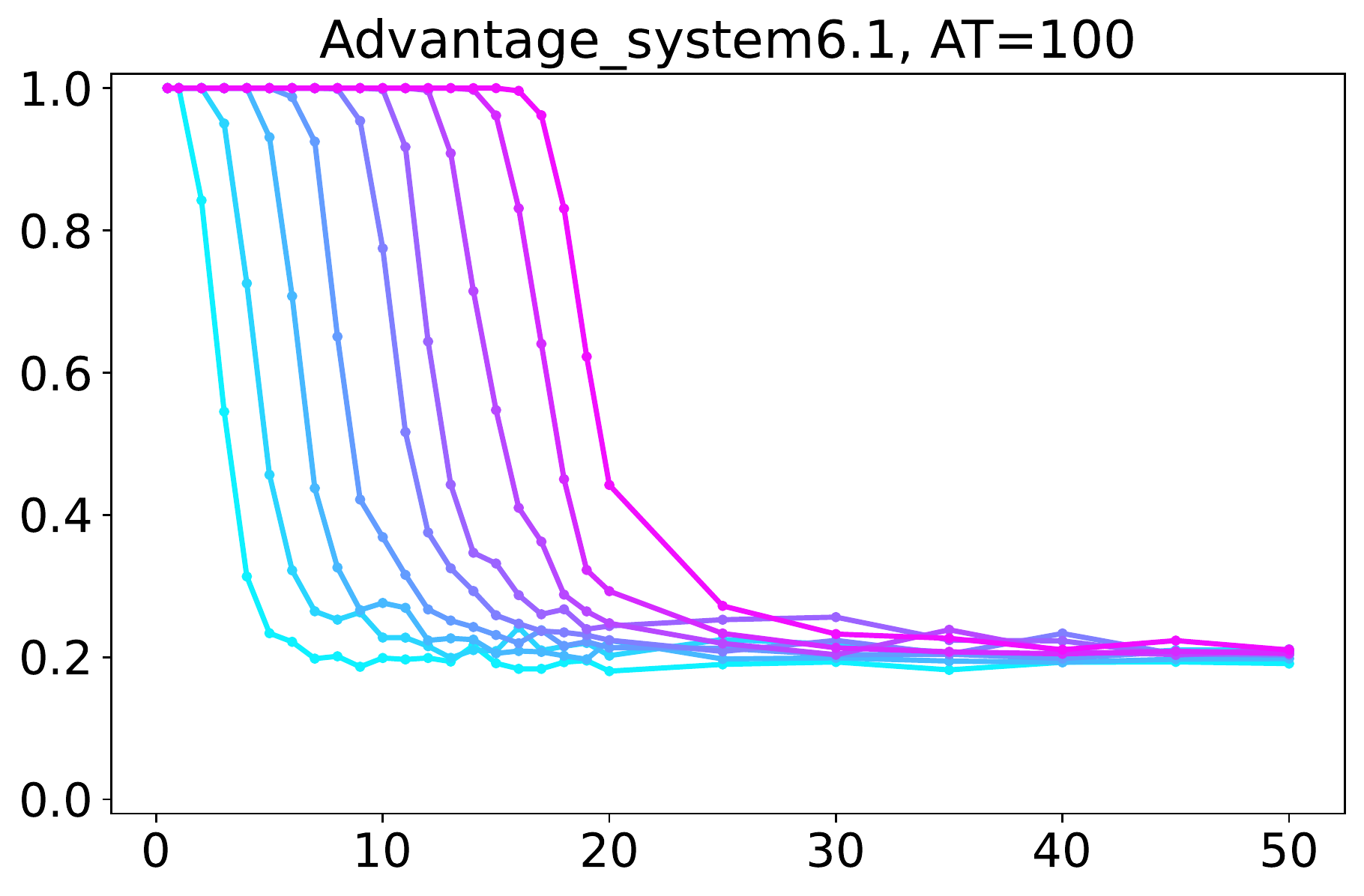}
    \includegraphics[width=0.24\textwidth]{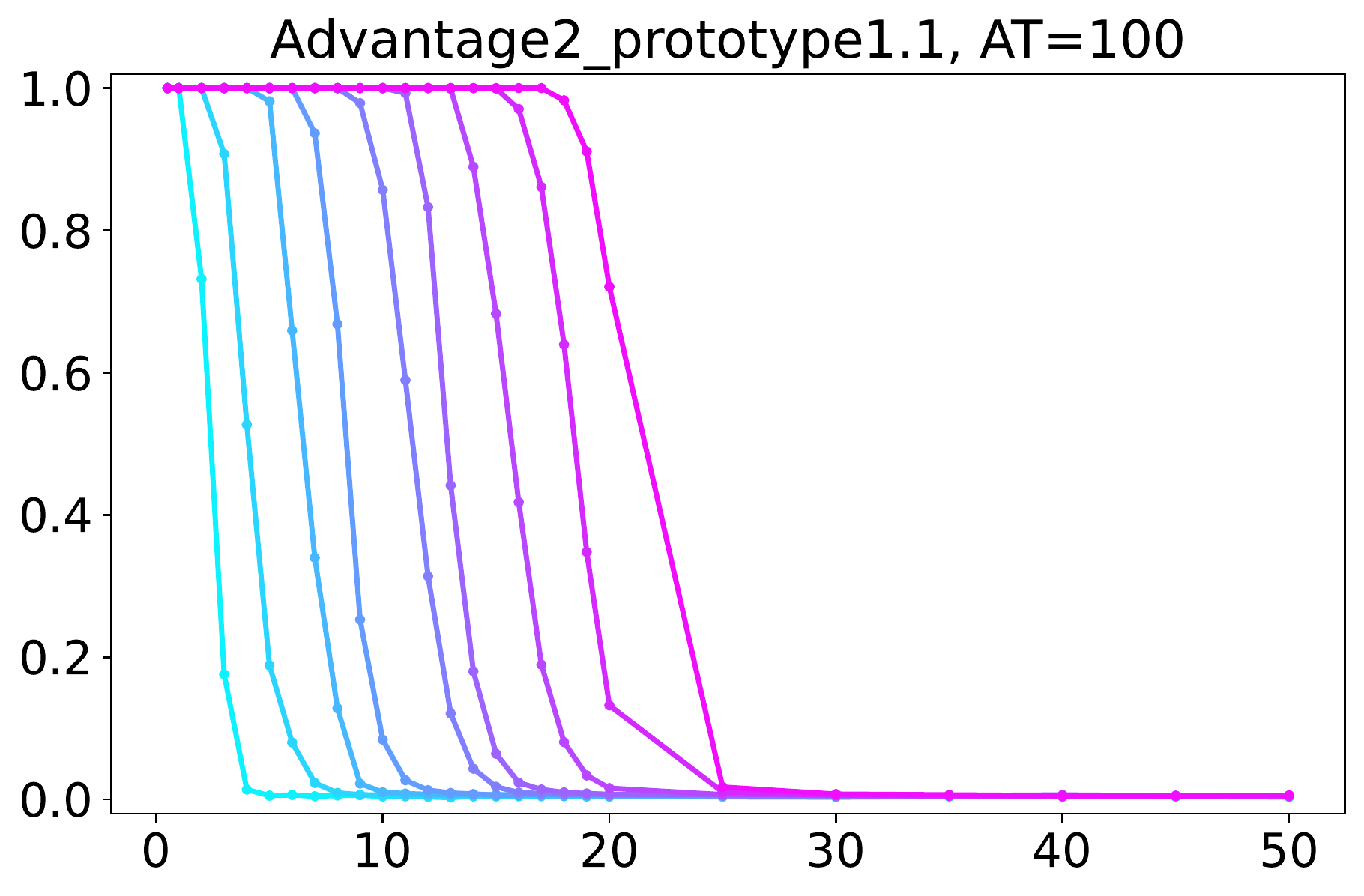}\\
    \includegraphics[width=0.24\textwidth]{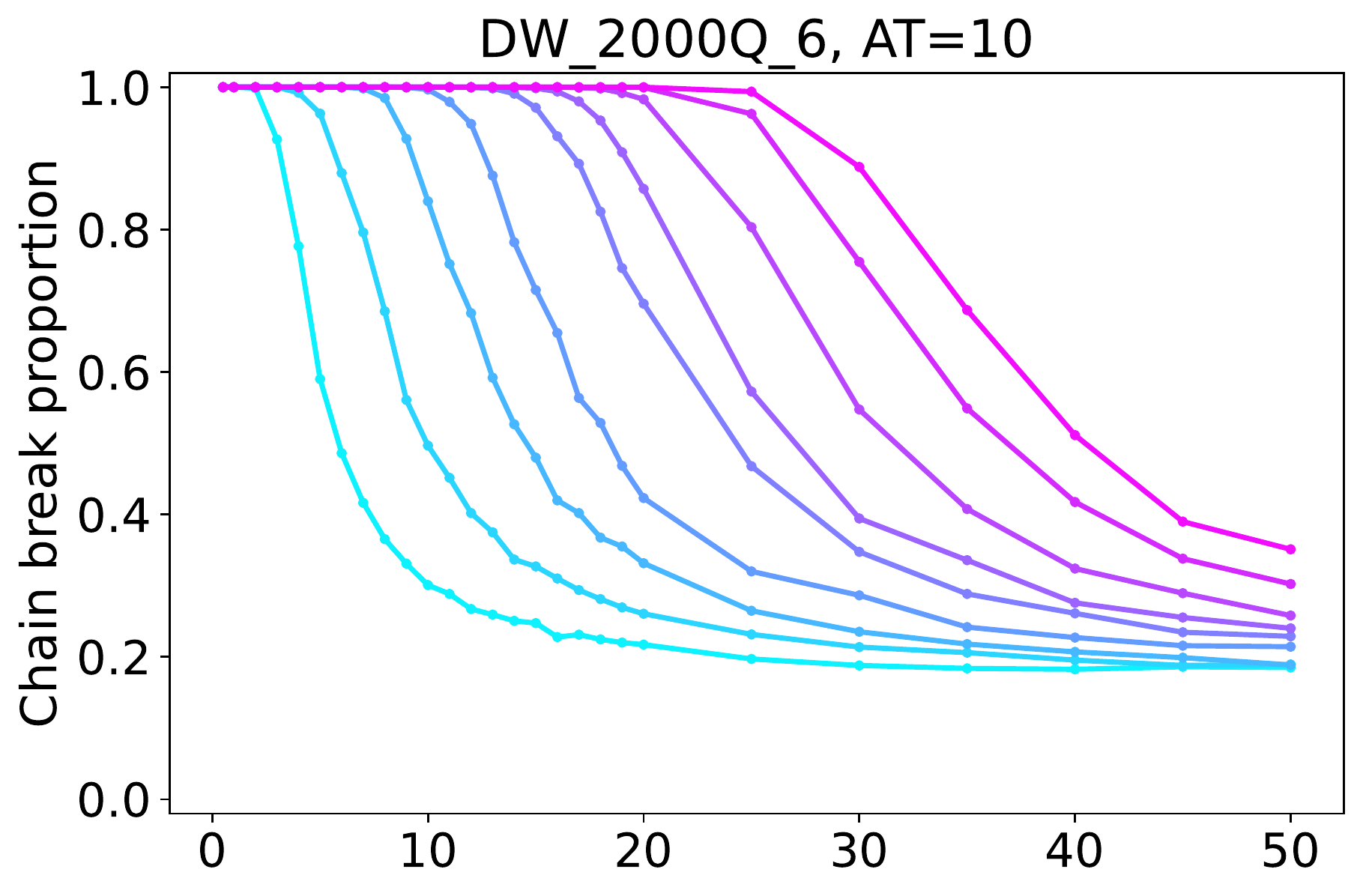}
    \includegraphics[width=0.24\textwidth]{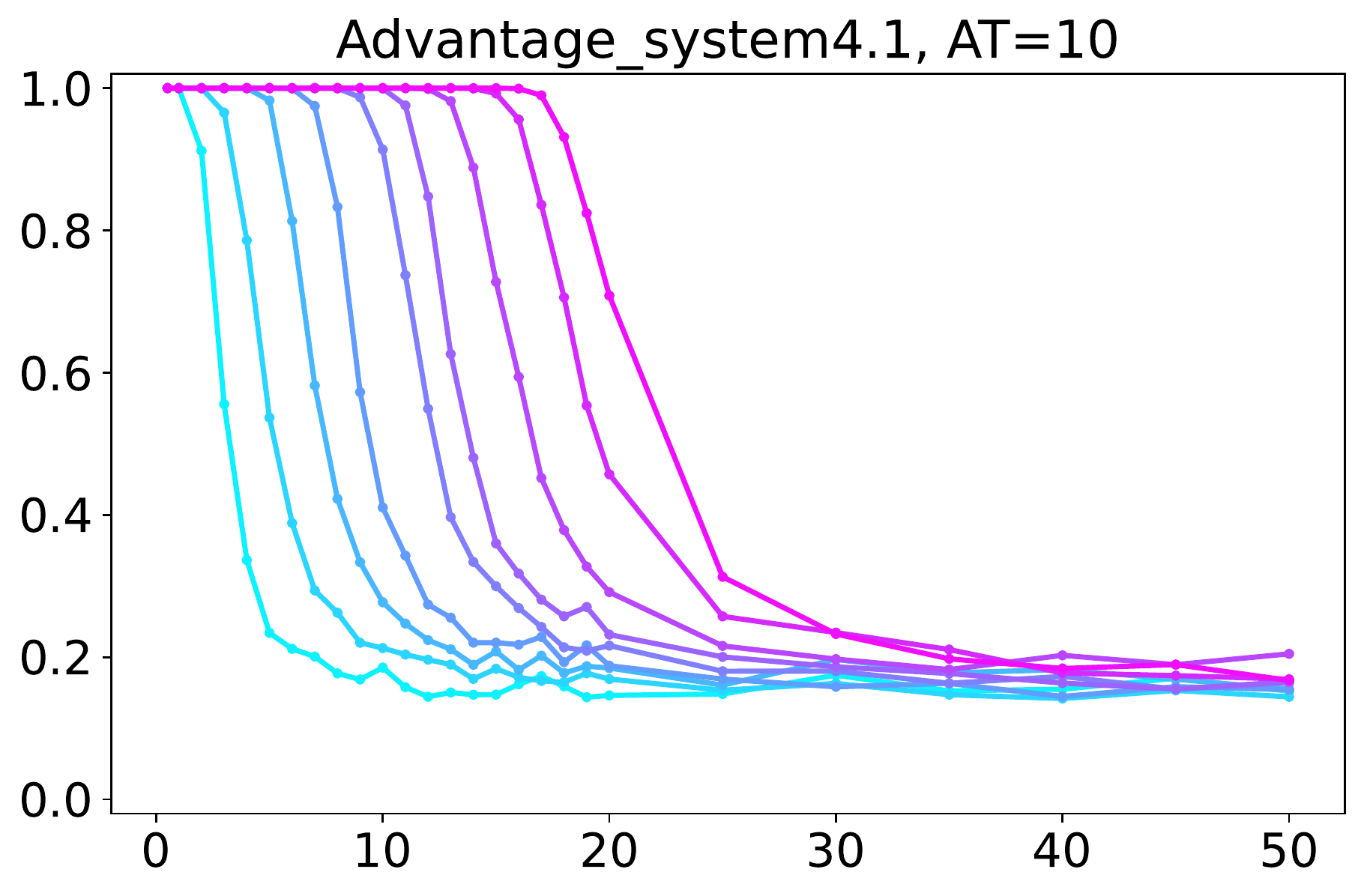}
    \includegraphics[width=0.24\textwidth]{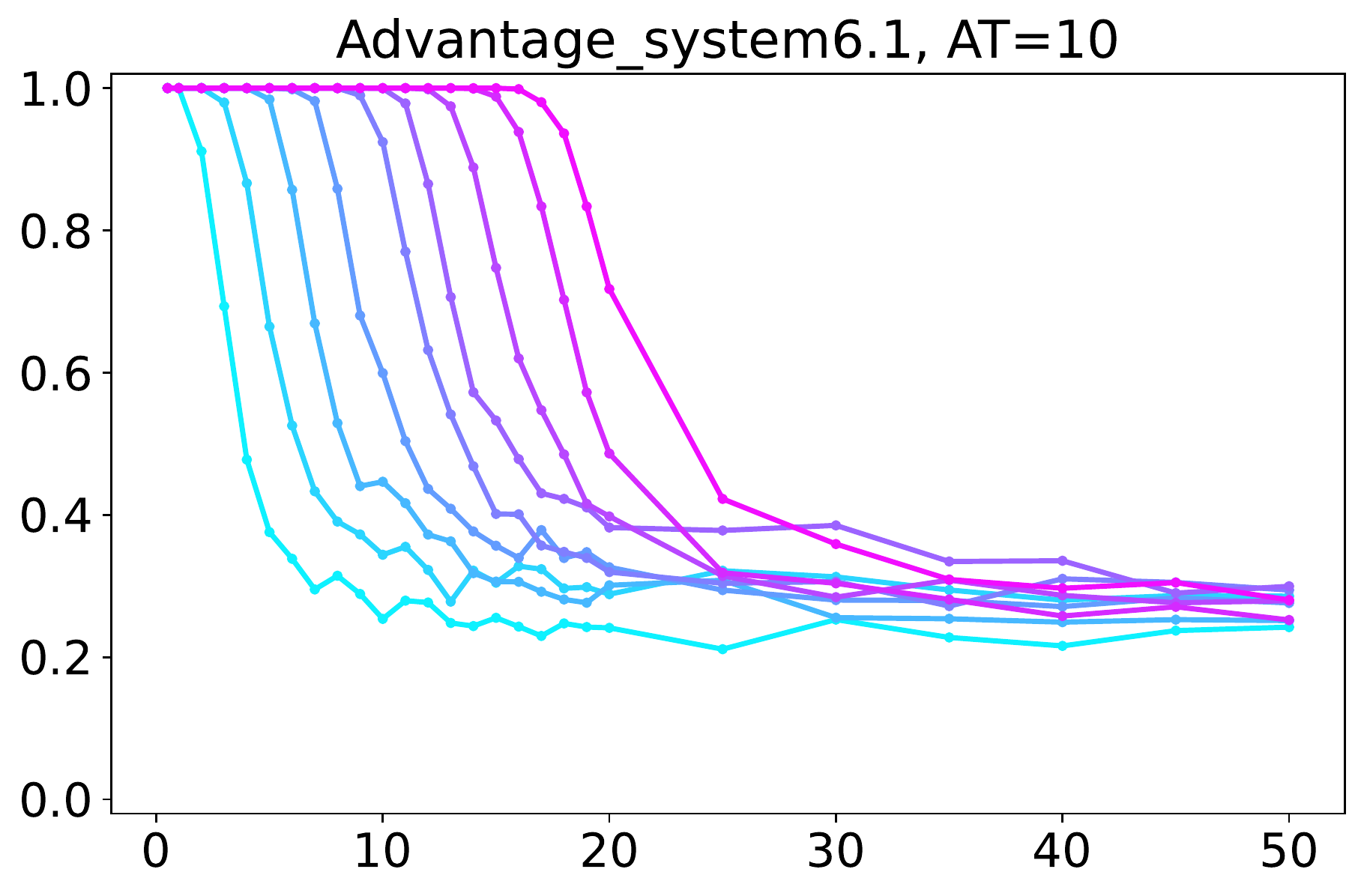}
    \includegraphics[width=0.24\textwidth]{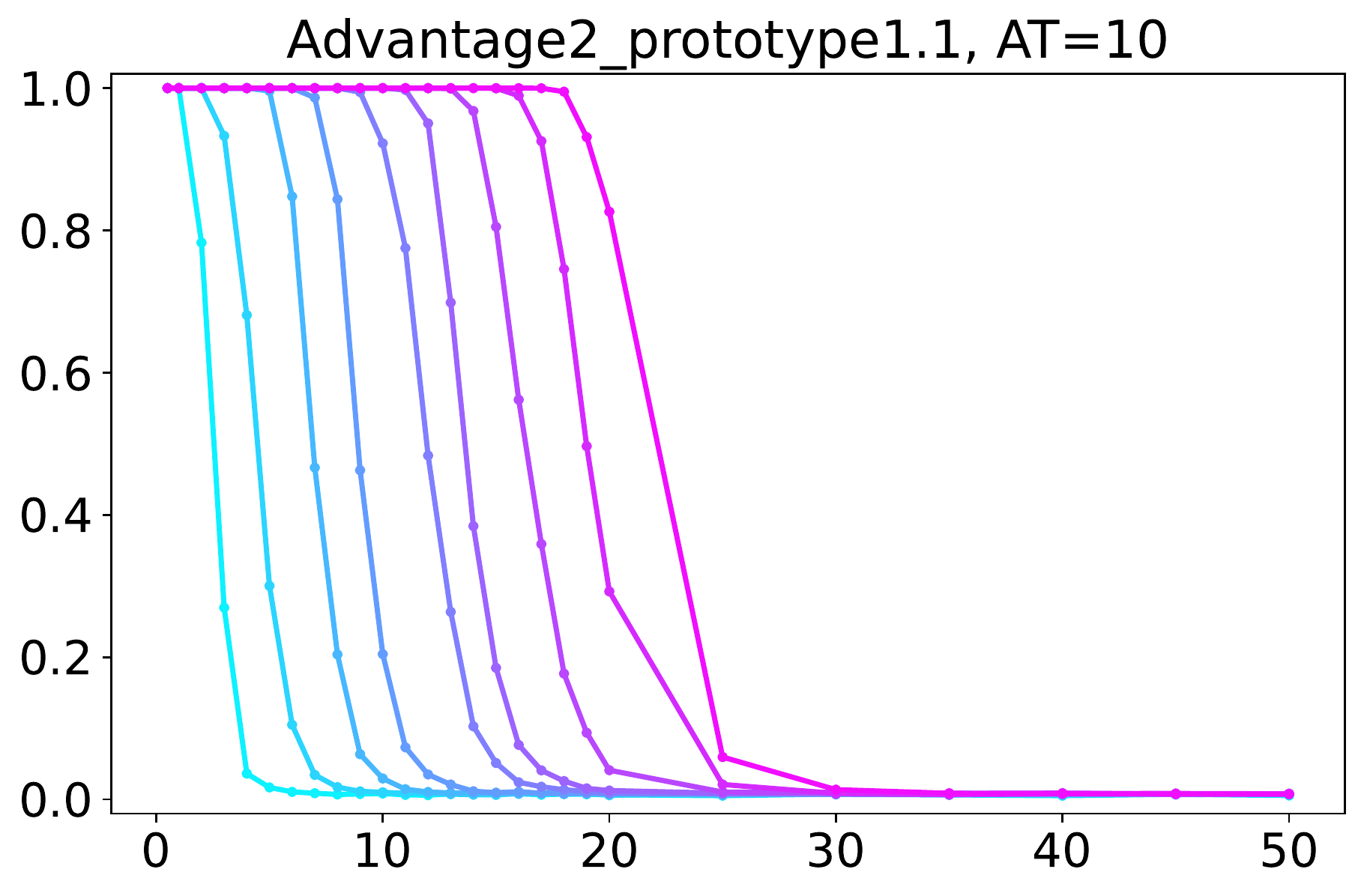}\\
    \includegraphics[width=0.24\textwidth]{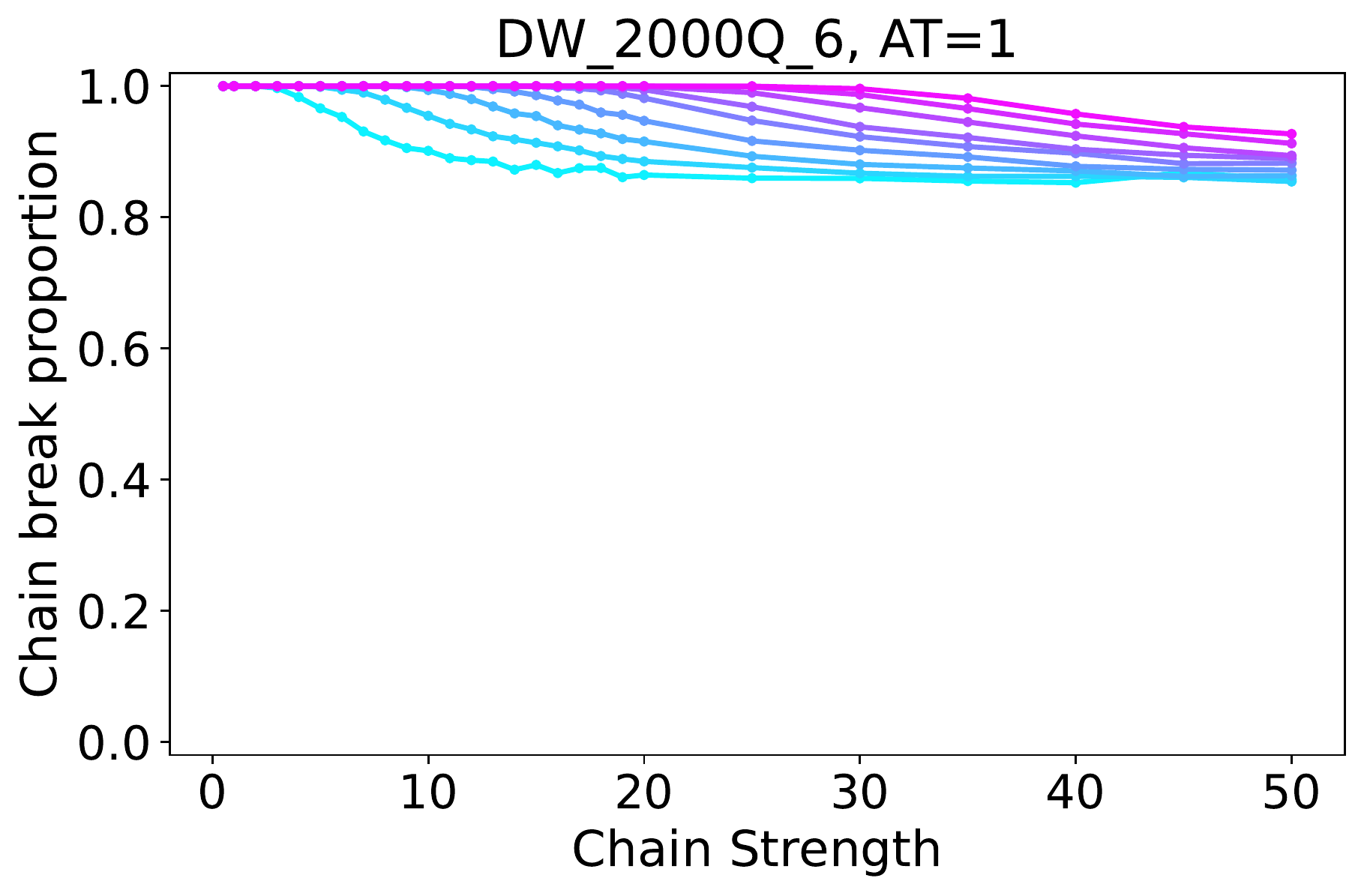}
    \includegraphics[width=0.24\textwidth]{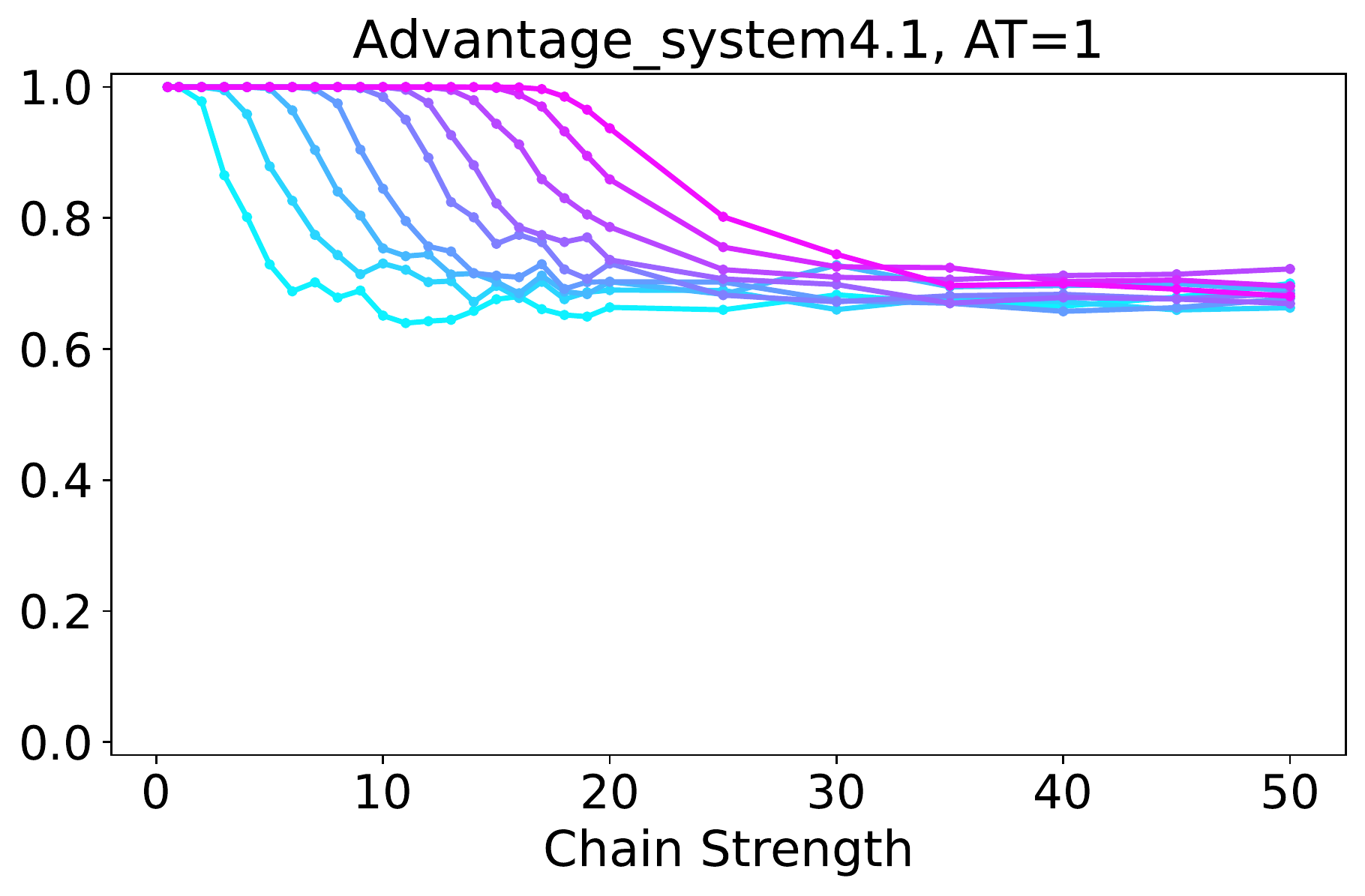}
    \includegraphics[width=0.24\textwidth]{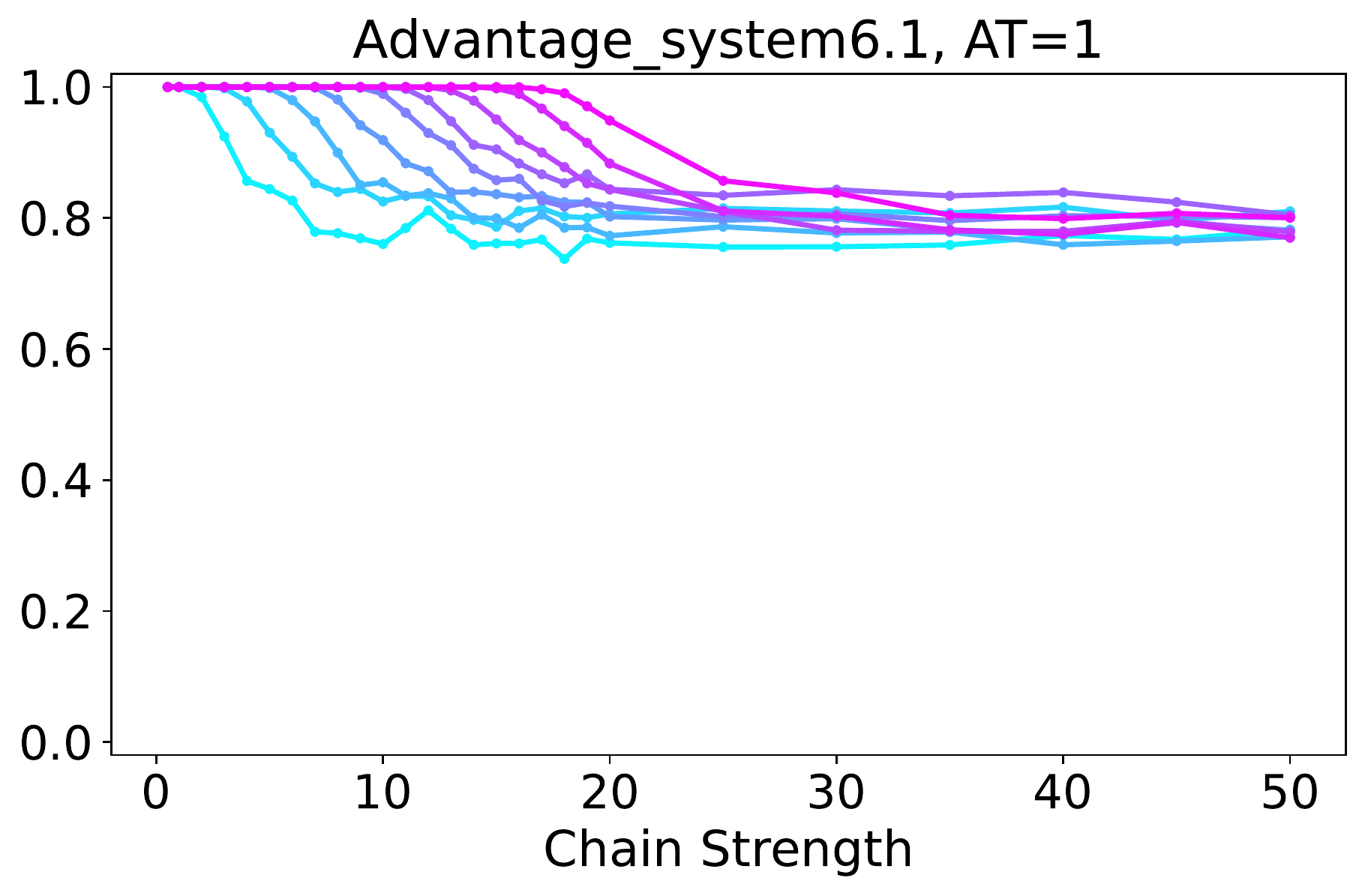}
    \includegraphics[width=0.24\textwidth]{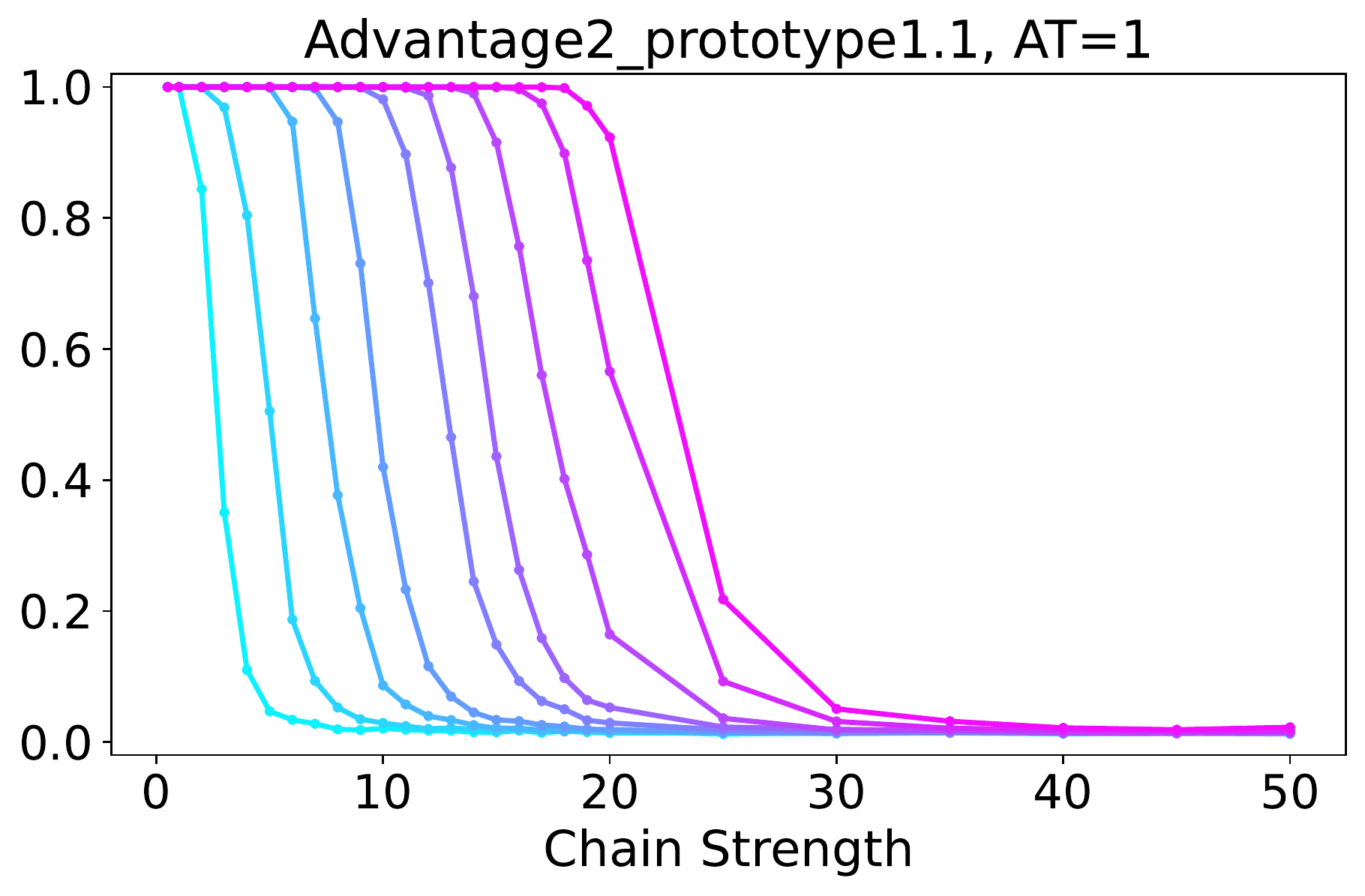}\\
    \includegraphics[width=0.71\textwidth]{figures/colorbar2.pdf}
    \caption{Chain break proportions (y-axis) vs chain strength (x-axis) for each of the maximum cut Isings from the $150$ $G(n,p)$ random graphs. The aggregated results are shown in the form of $10$ lines per plot representing the mean approximation ratio for $10$ linearly spaced graph density intervals from $0.05$ to $0.95$, where the color of each line encodes the mean graph density for that interval.  The line colors encode the problem graph density using the same colorscale from Figure~\ref{fig:maximum_cut_approx_ratio}. The D-Wave devices are \texttt{DW\_2000Q\_6} (left column), \texttt{Advantage\_system4.1} (center-left column), \texttt{Advantage\_system6.1} (center-right column), and \texttt{Advantage2\_prototype1.1} (right column). The annealing time in microseconds are varied across $2000$ microseconds (first row), $1000$ microseconds (second row), $100$ microseconds (third row), $10$ microseconds (fourth row), and $1$ microsecond (bottom row). }
    \label{fig:chain_breaks_max_cut}
\end{figure*}

Figure~\ref{fig:maximum_cut_approx_ratio} shows the computed Maximum Cut approximation ratios from the four D-Wave devices across different annealing times and chain strengths. At sufficient annealing times, in particular $2000$ and $1000$ microseconds, the approximation ratios are quite favorable across the four devices, with \texttt{Advantage2\_prototype1.1} having the best approximation ratios. As with the Maximum Clique results, the approximation ratios for the Maximum Cut problems sampled with \texttt{Advantage2\_prototype1.1} are the most stable across all annealing times. In particular, the approximation ratios for the other three D-Wave devices significantly fall at annealing times of $1$ microsecond. Interestingly, \texttt{Advantage\_system6.1} has the lowest approximation ratios across all four devices for $1$ microsecond annealing time. The reason for \texttt{Advantage\_system4.1} performing worse than \texttt{DW\_2000Q\_6} in terms of approximation ratios can not be entirely attributed to more chain breaks though; Figure~\ref{fig:chain_breaks_max_cut} shows that \texttt{DW\_2000Q\_6} has the highest chain breaks at $1$ microsecond annealing time. 

At annealing times of $1000$ and $2000$ microseconds, lower density graphs have a peak of approximation ratios at a chain strength of $5$, whereas for higher density graphs there is a plateau of approximation ratios approaching $1$ for all chain strengths between $30$ and $50$. The sharpness of the transition between lower approximation ratios and higher approximation ratios is very steep with \texttt{Advantage2\_prototype1.1}, whereas on \texttt{DW\_2000Q\_6} the slope of the transition is less steep. 

Of the four devices, it is clear that \texttt{Advantage2\_prototype1.1} performs the best on this ensemble of random graphs; at $1000$ and $2000$ microsecond annealing time, and for sufficiently large chain strengths, the approximation ratios are very nearly $1$ (Figure~\ref{fig:maximum_cut_approx_ratio}). Figure~\ref{fig:chain_breaks_max_cut} shows that in this regime, the chain break proportion for \texttt{Advantage2\_prototype1.1} is also effectively $0$, whereas all of the other three devices have a noticeably higher chain break proportion even at very high chain strengths. This shows that clearly the sampling capability of the quantum annealers is getting better, and for moderately sized problems is already quite good. However, the important caveat is that Table~\ref{table:optimal_solution_table} shows that even in this favorable regime of parameter choices, \texttt{Advantage2\_prototype1.1} does not always find the maximum cut of the graphs. 

An interesting trend that occurs in the \texttt{DW\_2000Q\_6} data, but not in the other three devices, is that as the annealing time increases (Figure~\ref{fig:maximum_cut_approx_ratio}, left column), in particular for $1000$ and $2000$ microseconds, the range of chain strengths which result in a plateau of the approximation ratios for high density graphs gets smaller, and the transition from very low approximation ratio to high approximation ratio has a less pronounced slope. This is in contrast to the other three devices, where as the annealing time increases the larger chain strengths cause a sharp increase and then plateau of high approximation ratios. What this means is that \texttt{DW\_2000Q\_6} requires more chain strength compared to the other three quantum annealers in order to attain a good approximation ratio. This is to be expected, primarily because of the longer chain lengths required for the sparser 2000Q connectivity (see Table~\ref{table:hardware_summary}). However, the unexpected part is that this trend reverses for longer annealing times; for example at $100$ microsecond annealing time for \texttt{DW\_2000Q\_6} in Figure~\ref{fig:maximum_cut_approx_ratio}, the approximation ratio steep climb and plateau for high density graphs is very similar to the other devices. At $1000$ and $2000$ microsecond annealing time this trend reverses. This could be indicative of higher noise on \texttt{DW\_2000Q\_6} compounding at longer annealing times and thereby degrading the computation, especially for the longer chains, compared to the other three devices.

\subsection{Fair sampling of Maximum Cliques}
\label{section:results_fair_sampling}

\begin{figure*}[h!]
    \centering
    \includegraphics[width=0.24\textwidth]{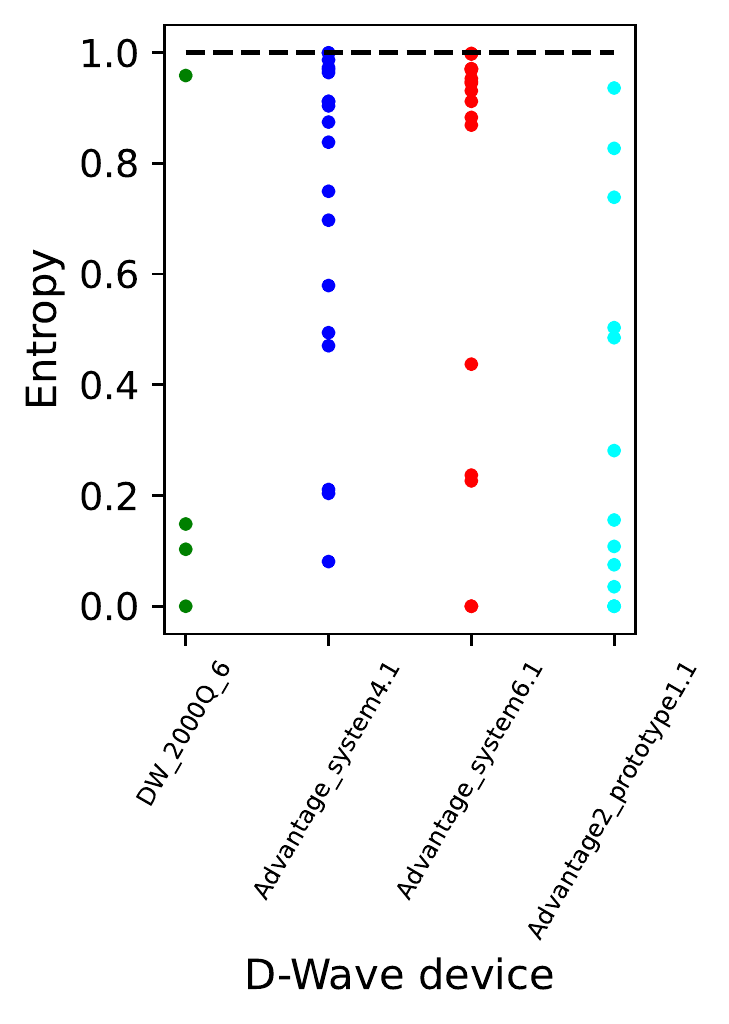}
    \includegraphics[width=0.24\textwidth]{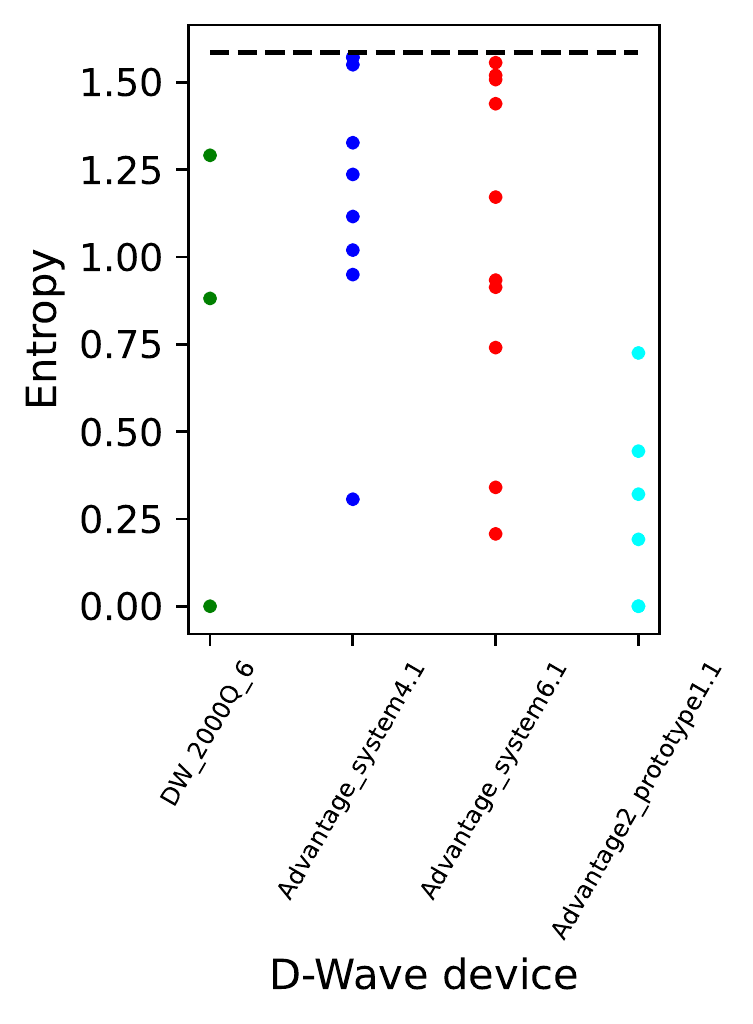}
    \includegraphics[width=0.24\textwidth]{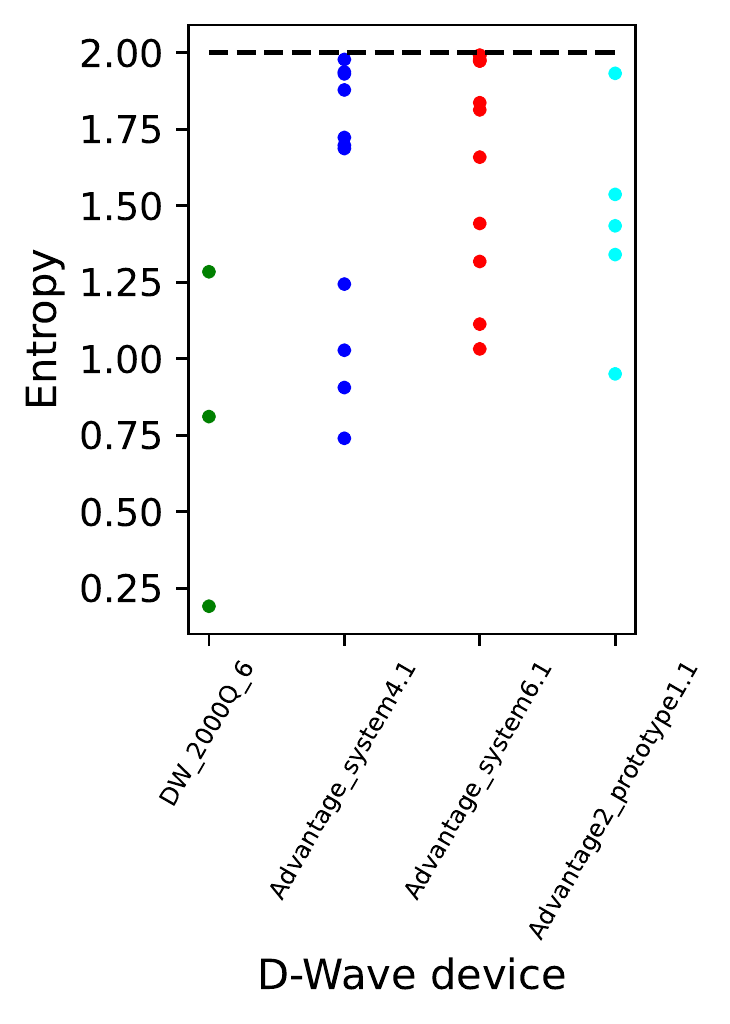}
    \includegraphics[width=0.24\textwidth]{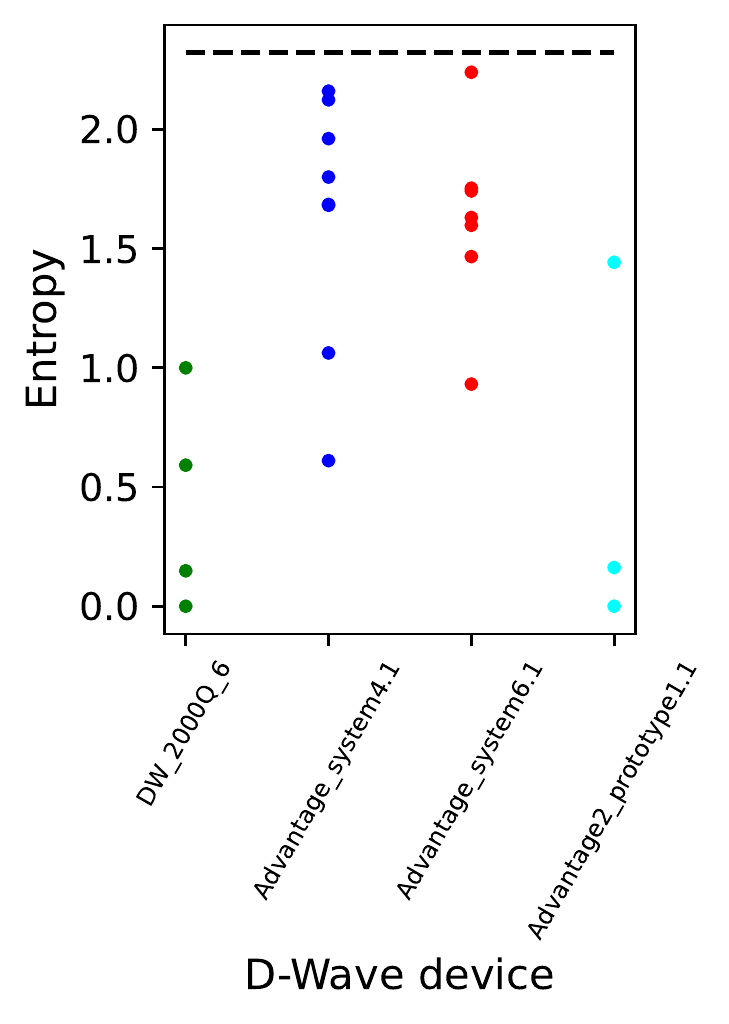}
    \caption{Fair sampling analysis, measured by the entropy of the sampled ground state distribution (y-axis), across the four D-Wave devices (x-axis) colored green for \texttt{DW\_2000Q\_6}, blue for \texttt{Advantage\_system4.1}, red for \texttt{Advantage\_system6.1}, and cyan for \texttt{Advantage2\_prototype1.1}. \textbf{Left:} Sampling of graphs with $2$ maximum cliques, \textbf{middle-left}: sampling of graphs with $3$ maximum cliques, \textbf{middle-right}: sampling of graphs with $4$ maximum cliques, \textbf{right}: sampling of graphs with $5$ maximum cliques. The maximum entropy possible for each of the number of maximum cliques is plotted as a dashed horizontal black line; this line denotes the entropy that can be achieved for a uniform distribution of the ground states. The minimum entropy possible is $0$, which denotes that the ground state sampling distribution is maximally biased towards always selecting a single state out of $N$ ground states. }
    \label{fig:fair_sampling_degenerate_maximum_cliques}
\end{figure*}

An interesting question associated with combinatorial optimization, quantum annealing in particular, is in what proportions degenerate ground states are sampled. Here, these experiments used structured minor embeddings which means that we may expect the results are reasonably non-biased compared to a random minor embedding with highly variable chain lengths. Interestingly though, it is well known that in general quantum annealing in the Transverse field Ising model does not sample degenerate ground states fairly \cite{matsuda2009quantum, 9605329, PhysRevLett.118.070502, PhysRevA.100.030303, Matsuda_2009, PhysRevE.99.063314, https://doi.org/10.48550/arxiv.2110.10560, https://doi.org/10.48550/arxiv.2110.10196, https://doi.org/10.48550/arxiv.2110.10930, konz2019embedding}. It has been suggested this unfair sampling could be useful in certain sampling applications \cite{zhang2017advantages}, but there are several applications where uniformly sampling degenerate ground states is important \cite{douglass2015constructing, jerrum1986random, 10.21468/SciPostPhys.2.2.013, weaver2012satisfiability}. 

The random graphs which were generated for these tests were not constructed to specifically have any number of degenerate ground states in particular, but it turns out that many of them do have multiple Maximum Cliques and Maximum Cuts. The CPLEX solver method for finding the maximum cuts though does not enumerate through all optimal solutions, whereas the networkx solver for maximum cliques does. Therefore, we only have complete optimal solutions for the maximum clique's of the random graphs. The fact that many of these random graphs do have multiple Maximum Cliques means that we can examine how fairly the four different quantum annealers sampled the optimal solutions to these minor embedded optimization problems.

The measure of information entropy \cite{6773024} is used to quantify fair sampling. Given we found $N$ anneals which correctly found a maximum clique for the given graph $G$ ($N$ could also be referred to as the measured ground state probability), we can set the distribution of the ground state samples as $\frac{p_0}{N}, \frac{p_1}{N}, \dots, \frac{p_k}{N}$ where $k$ indexes, in any order, the total number of optimal solutions which exist for this graph $G$. Using entropy as a metric for fair sampling means is useful because entropy is maximized for a uniform distribution, and is minimized when the distribution entirely skewed towards a single state. Samples which contain any broken chains are not considered in this computation. This fair sampling computation is performed on the $1000$ anneals, from each of the four devices, which had the greatest measured ground state probability (GSP) across the parameter search of annealing time and chain strengths. For the purpose of obtaining a high confidence measure of the true underlying distribution of the sampled optimal solutions, a reasonably large number of samples is required. Additionally, since the number of anneals is fixed to $1000$, if the number of optimal solutions is too large then we will also not see a good representation of the underlying distribution because of the finite number of samples. Therefore, the number of optimal solutions is restricted to $2, 3, 4, 5$ and the total GSP for the $1000$ anneals must be at least $30$. The entropy of the degenerate ground state distribution is computed in scipy \cite{2020SciPy-NMeth} with base $2$. 

Figure~\ref{fig:fair_sampling_degenerate_maximum_cliques} shows the fair sampling distributions for graphs with $2, 3, 4$, and $5$ maximum cliques, respectively across the four quantum annealers. Note that the number of datapoints being plotted for each D-Wave device for each plot can be different between each plot and device; the reason is because some devices have better ground state sampling probability than others. The first notable pattern in this fair sampling analysis is that the two devices with Pegasus topology appear to have the best fair sampling capability, with some of the entropy scores reaching the maximum, whereas both \texttt{DW\_2000Q\_6} and \texttt{Advantage2\_prototype1.1} have worse fair sampling. The other relevant property to note is that there is no device, or number of optimal solutions, which provide consistently fair sampling; both \texttt{Advantage\_system6.1} and \texttt{Advantage\_system4.1} have entropy measures which range over the spectrum of fairness. It has been observed before that NISQ devices with higher noise tend to have reasonably fair samples, at the cost of low GSP, \cite{10.1145/3510857}. Therefore, it could be the case that \texttt{Advantage\_system6.1} and \texttt{Advantage\_system4.1} introduced more noise into the computation, perhaps as a result of the longer-chain minor embeddings used, compared to \texttt{DW\_2000Q\_6} and \texttt{Advantage2\_prototype1.1}. While there is a clear difference in the QA device fair sampling capability, there is no evident trend that exists across the different ground state number plots.

\begin{table}[h!]
\begin{center}
\begin{tabular}{ |c| } 
 \hline
 Example broken chains \\ 
 \hline
 \hline
 \texttt{[0, 0, 0, 0, 0, 0, 1, 1, 1, 1, 1, 1, 1, 1]} \\ 
 \hline
 \texttt{[1, 1, 0, 0, 0, 0, 1, 1, 1, 1, 1, 1, 1, 1]} \\ 
 \hline
 \texttt{[1, 1, 1, 1, 0, 0, 0, 0, 0, 0, 0, 0, 0, 0]} \\
 \hline
 \texttt{[1, 1, 1, 1, 1, 1, 1, 0, 0, 0, 0, 1, 1, 1]} \\
 \hline
 \texttt{[0, 0, 0, 0, 0, 0, 0, 0, 0, 0, 1, 1, 1, 1]} \\
 \hline 
 \texttt{[1, 1, 1, 1, 0, 0, 0, 0, 0, 0, 1, 1, 1, 1]} \\
 \hline
 \texttt{[1, 1, 1, 1, 1, 1, 1, 1, 1, 1, 0, 0, 0, 0]} \\
 \hline
 \texttt{[0, 0, 0, 0, 0, 1, 1, 1, 1, 1, 1, 1, 0, 0]} \\
 \hline
 \texttt{[0, 0, 1, 1, 1, 1, 1, 1, 1, 1, 1, 1, 0, 0]} \\
 \hline
 \texttt{[1, 1, 1, 1, 1, 1, 1, 1, 1, 0, 0, 0, 0, 0]} \\
 \hline
\end{tabular}
\end{center}
    \caption{Table of $10$ arbitrarily chosen broken chain vectors (e.g. physically adjacent vectors of chains on the hardware) from Maximum Clique problems on \texttt{DW\_2000Q\_6} with a $1000$ microsecond annealing time. }
    \label{table:chain_break_examples}
\end{table}

\subsection{Broken chain properties}
\label{section:results_broken_chain_properties}
One of the interesting properties of samples which have chain breaks are that the bit flips are not typically uniformly distributed across the chain. For example, Table~\ref{table:chain_break_examples} shows $10$ broken chains from an arbitrarily selected sample from the Maximum Clique $1000$ microsecond annealing time experiments on \texttt{DW\_2000Q\_6}. 

The chain breaks seem to be grouped together very often. This observation has been made before in regards to quantum annealing chain breaks \cite{10.1145/3075564.3075575}. This observation on the characteristics of how the chains break could provide insight for understanding why chains have specific break patterns, which could be used to develop better heuristics for repairing chains similarly to developing heuristics for repairing chains from specific types of combinatorial optimization problems \cite{9325394}.

%%%%%%%%%%%%%%%%%%%%%%%%%%%%%%%%%%%%%%%%%%%%%%%%%%%%%%%%%%%%%%%%%
%%%%%%%%%%%%%%%%%%%%%%%%%%%%%%%%%%%%%%%%%%%%%%%%%%%%%%%%%%%%%%%%%
%%%%%%%%%%%%%%%%%%%%%%%%%%%%%%%%%%%%%%%%%%%%%%%%%%%%%%%%%%%%%%%%%
\section{Conclusion}
\label{section:conclusions}

Overall the four quantum annealers were able to sample reasonably high approximation ratios, even with no chain break resolution algorithms. These results are consistent with previous data on D-Wave quantum annealers, and also show that the newer devices have improved sampling capabilities for minor embedded combinatorial optimization problems. For example, the approximation ratios as a function of chain strength data is consistent with previous quantum annealing data including experiments on the D-Wave 2X \cite{Venturelli_2015}. Across the four devices there are reasonably consistent patterns, such as the optimal chain strength typically being a median value that is not too large and also not too small. These patterns can be used for parameter tuning of other problem instances in the future, and on future quantum annealing hardware. There are several clear findings from the data:

\begin{enumerate}[noitemsep]
    \item Longer annealing times are better, but with diminishing returns at longer anneal times. This is to be expected for typical combinatorial optimization problems and has been observed empirically in many other quantum annealing results \cite{9186612}. 
    \item The newest device \texttt{Advantage2\_prototype1.1} performs the best, in particular the approximation ratios for maximum clique and maximum cut were best for \texttt{Advantage2\_prototype1.1}. This can in part be attributed to a denser hardware graph, resulting in smaller chains being required for the minor embedding. This result is encouraging because it shows a progression of the quantum annealing technology over a relatively short time span of development. 
    \item At moderately sized problem instances, and with tuned chain strengths and annealing times, the D-Wave quantum annealers are able to serve as reasonable heuristics (meaning that they are able to approach approximation ratios of $1$) for Maximum Clique and Maximum Cut. 
    \item The optimal chain strengths varied slightly depending on the problem graph density and the D-Wave device. For Maximum Clique there is a clear peak of optimal chain strengths (see Figure~\ref{fig:maximum_clique_approx_ratio}) at $1.5$ to $2$ depending on the graph density. For Maximum Cut, the optimal chain strength for low density graphs is approximately $5$, whereas for higher density graphs the optimal chain strength is around $25$ to $50$. Note that these chain strength ranges are specific to the problem formulations of Maximum Clique and Maximum Cut given in Sections~\ref{section:Maximum_cut_Ising} and~\ref{section:Maximum_clique_QUBO}. 
    \item Although the four quantum annealers, as expected, did not consistently sample degenerate ground states of the random Maximum Clique QUBO's fairly (uniformly), both~\texttt{Advantage\_system4.1} and \texttt{Advantage\_system6.1} were able to sample the optimal solutions more fairly compared to \texttt{Advantage2\_prototype1.1} and \texttt{DW\_2000Q\_6}. 
\end{enumerate}

%%%%%%%%%%%%%%%%%%%%%%%%%%%%%%%%%%%%%%%%%%%%%%%%%%%%%%%%%%%%%%%%%
%%%%%%%%%%%%%%%%%%%%%%%%%%%%%%%%%%%%%%%%%%%%%%%%%%%%%%%%%%%%%%%%%
%%%%%%%%%%%%%%%%%%%%%%%%%%%%%%%%%%%%%%%%%%%%%%%%%%%%%%%%%%%%%%%%%
\section{Acknowledgments}
\label{sec:acknowledgments}
This work was supported by the U.S. Department of Energy through the Los Alamos National Laboratory. Los Alamos National Laboratory is operated by Triad National Security, LLC, for the National Nuclear Security Administration of U.S. Department of Energy (Contract No. 89233218CNA000001). The research presented in this article was supported by the Laboratory Directed Research and Development program of Los Alamos National Laboratory under project number 20220656ER. This research used resources provided by the Los Alamos National Laboratory Institutional Computing Program. Research presented in this article was supported by the NNSA's Advanced Simulation and Computing Beyond Moore's Law Program at Los Alamos National Laboratory. This work has been assigned the LANL technical report number LA-UR-23-20101.

\setlength\bibitemsep{0pt}
\printbibliography

\appendix

\section{$500$ nanosecond annealing times}
\label{section:appendix_small_annealing_times}
\begin{figure*}[h!]
    \centering
    \includegraphics[width=0.38\textwidth]{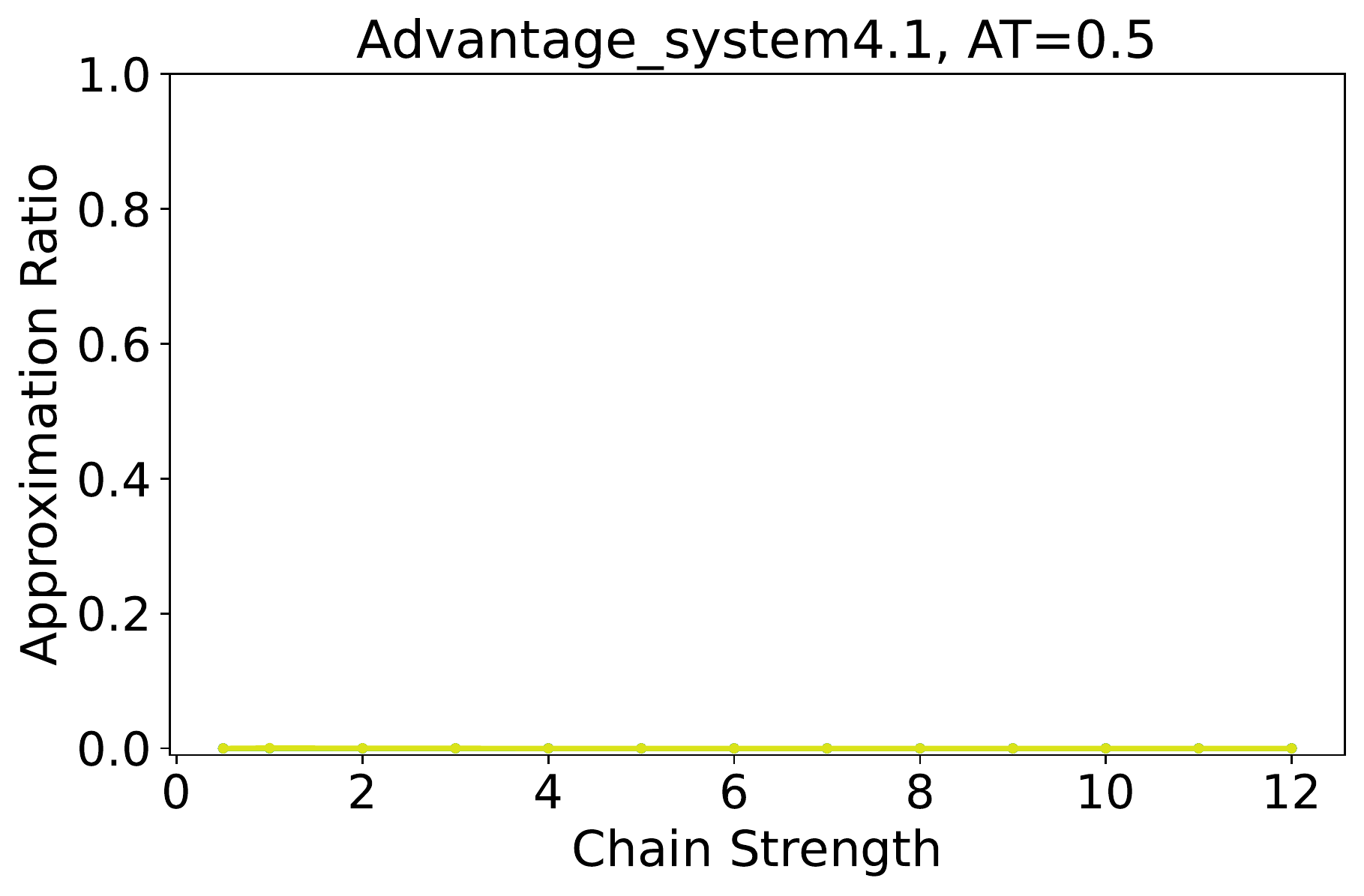}
    \includegraphics[width=0.38\textwidth]{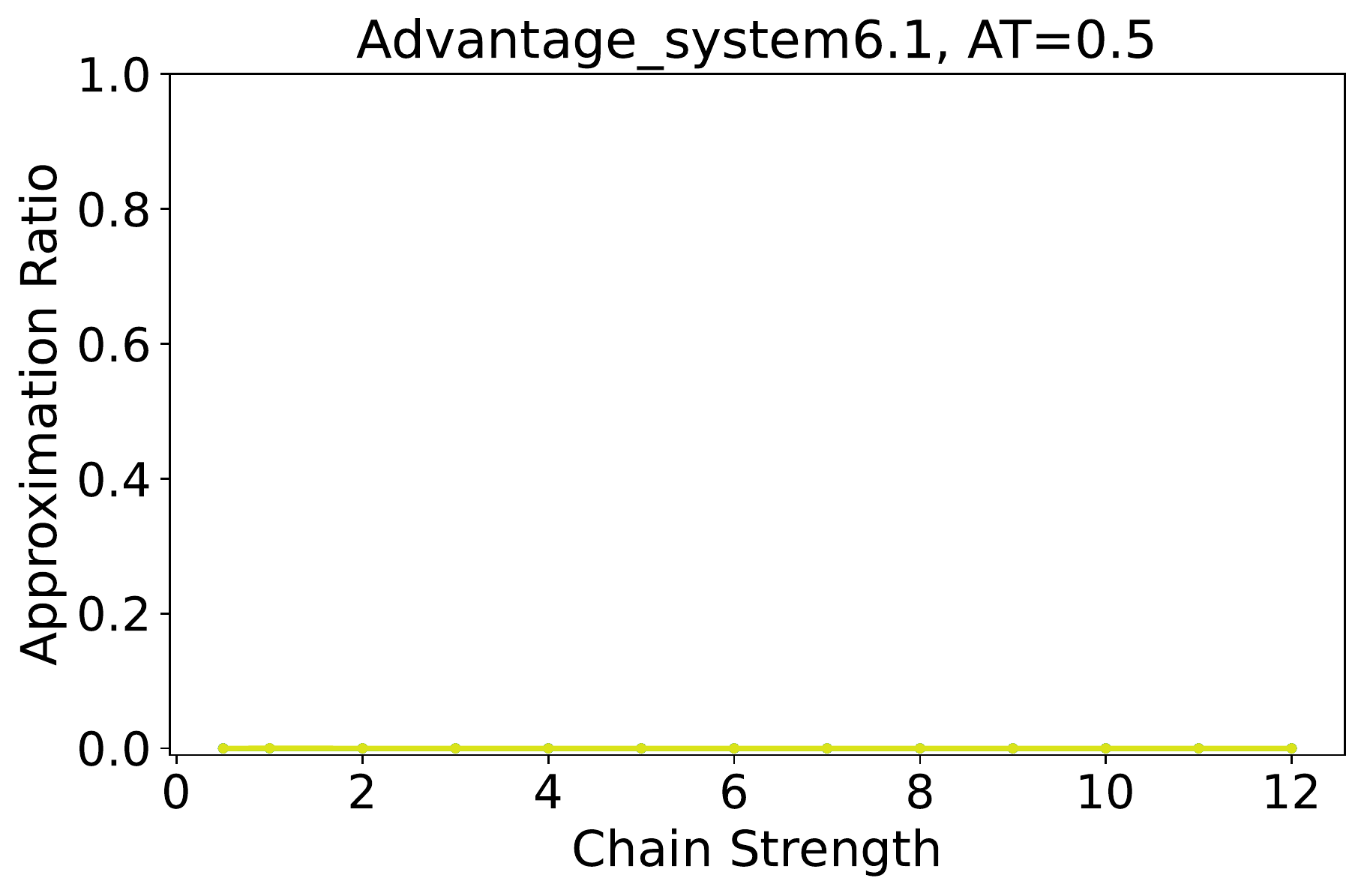}
    \includegraphics[width=0.38\textwidth]{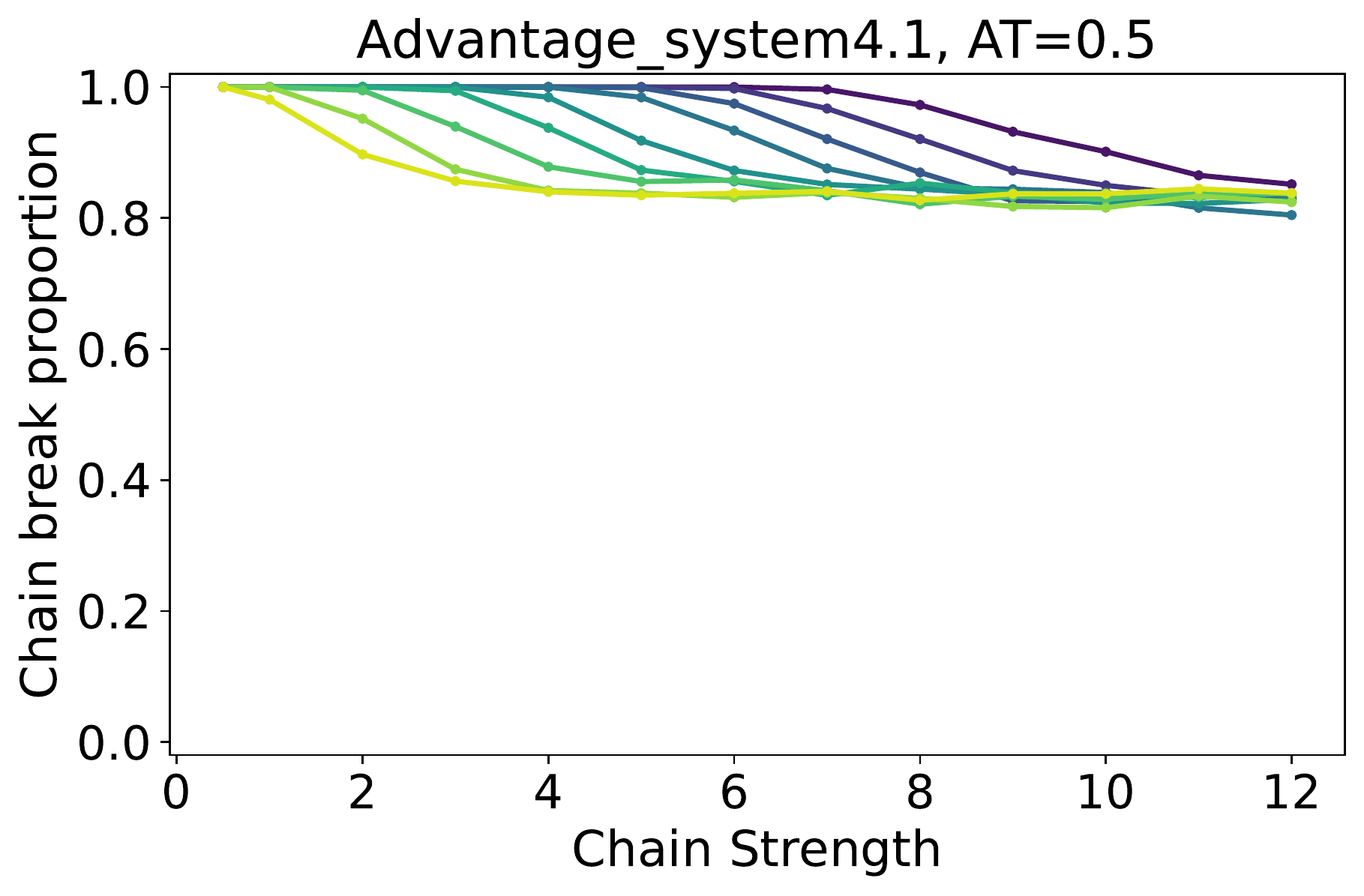}
    \includegraphics[width=0.38\textwidth]{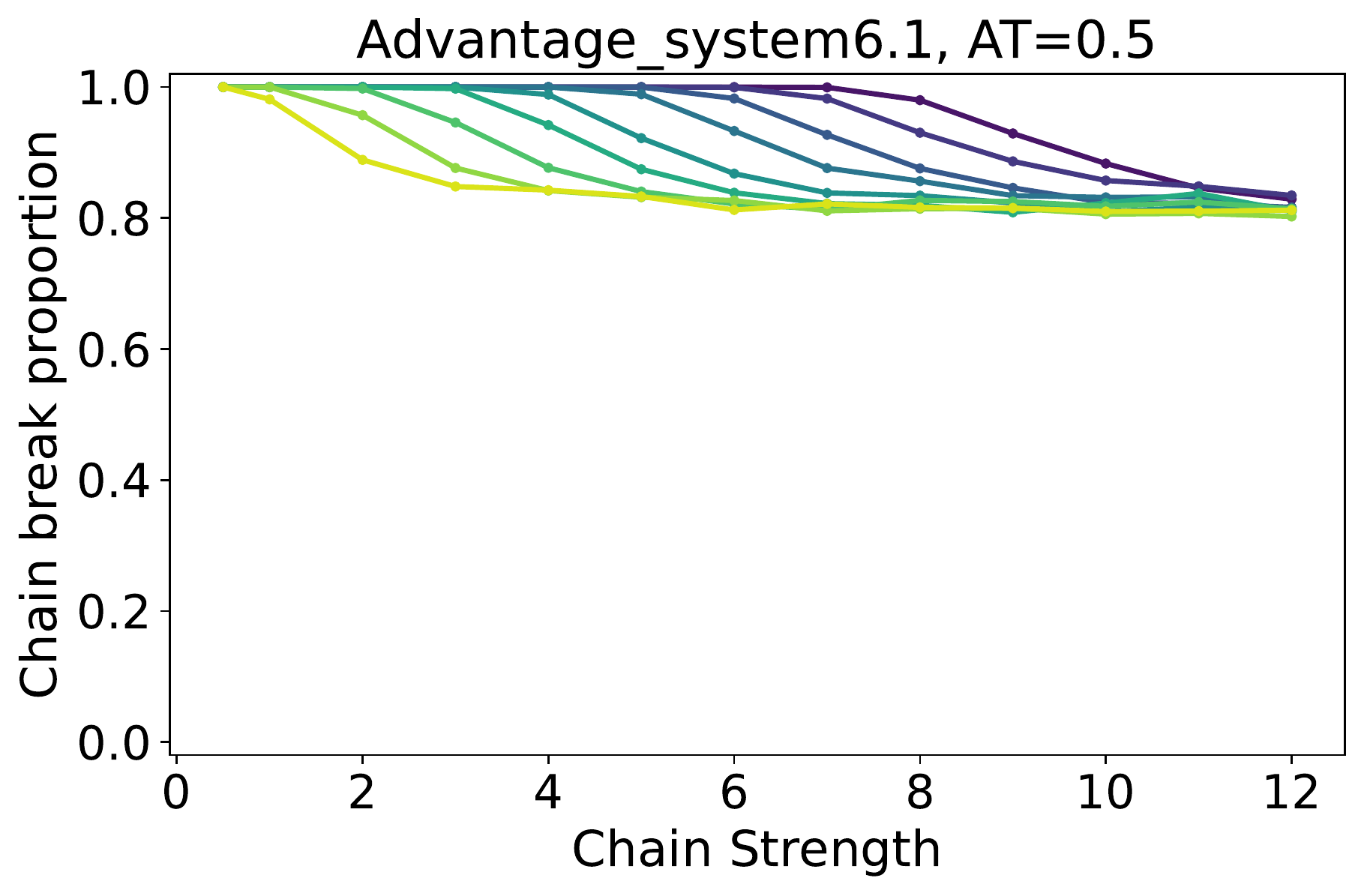}
    \includegraphics[width=0.65\textwidth]{figures/colorbar.pdf}
    \caption{Maximum Clique sampling approximation ratios (y-axis) vs chain strength (x-axis) in the top row. Chain break proportions (y-axis) vs chain strength (x-axis) in the bottom row. Annealing time is set to $500$ nanoseconds ($0.5$ microseconds). \texttt{Advantage\_system4.1} (left column) and \texttt{Advantage\_system6.1} (right column). These aggregated results are shown in the form of $10$ lines per plot representing the mean approximation ratio for $10$ linearly spaced graph density intervals from $0.05$ to $0.95$, where the color of each line encodes the mean graph density for that interval. The color coding is shown in the colorbar below the plots. Notably the chain break proportions are very high even at the highest chain strengths tested, and the approximation ratios are effectively $0$ across all chain strengths. }
    \label{fig:small_annealing_times_appendix_maxclique}
\end{figure*}

\begin{figure*}[h!]
    \centering
    \includegraphics[width=0.38\textwidth]{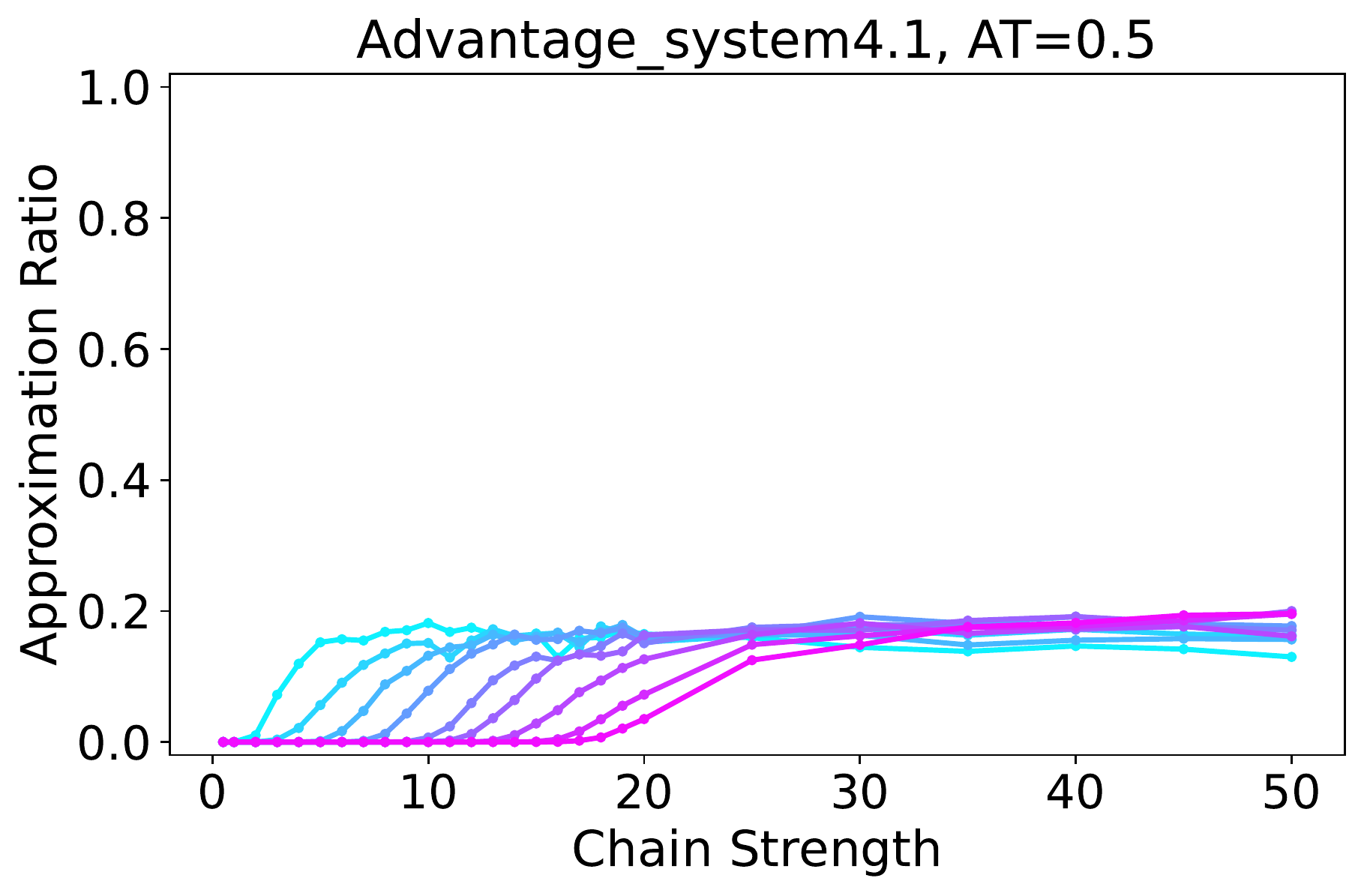}
    \includegraphics[width=0.38\textwidth]{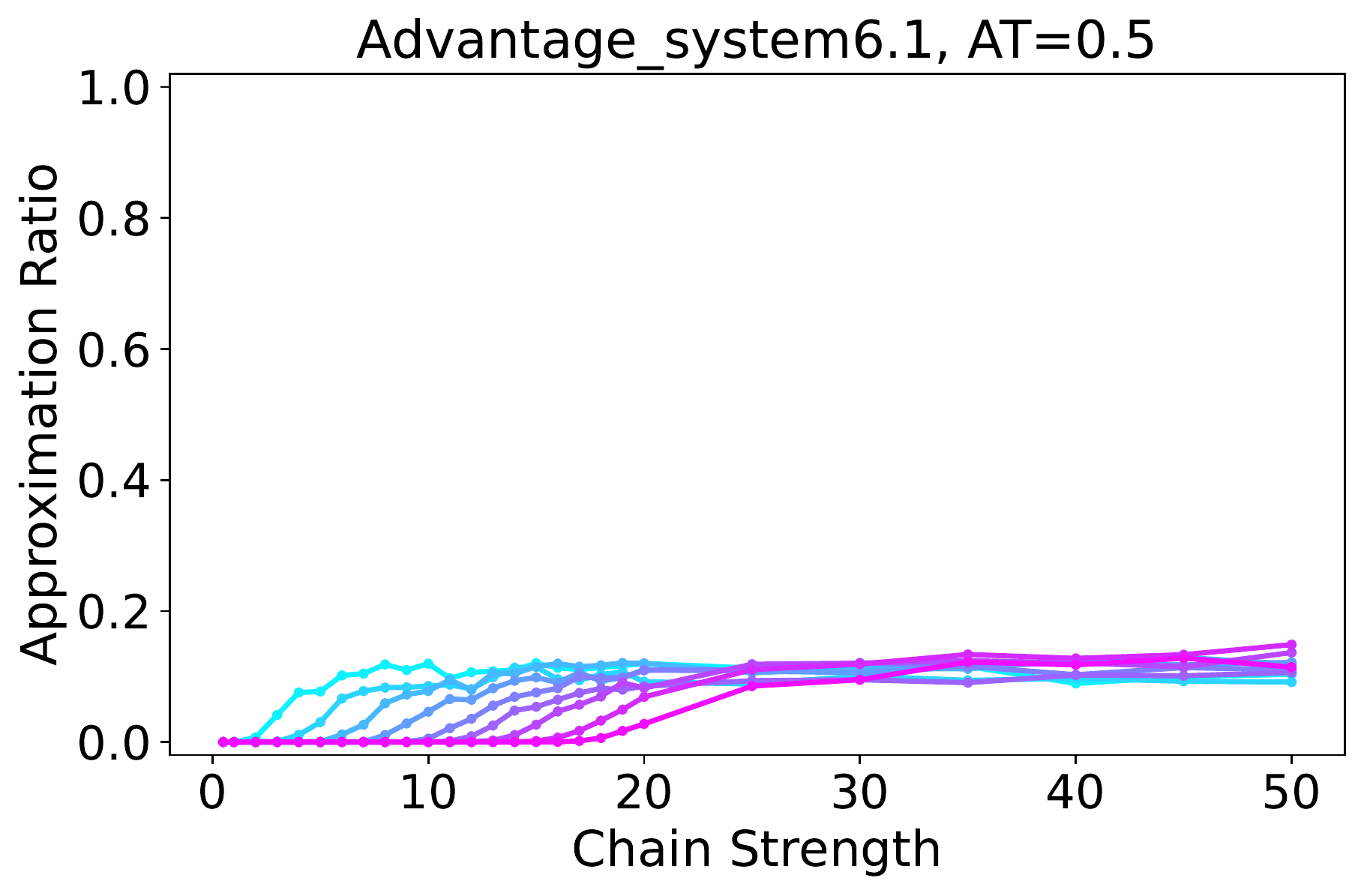}
    \includegraphics[width=0.38\textwidth]{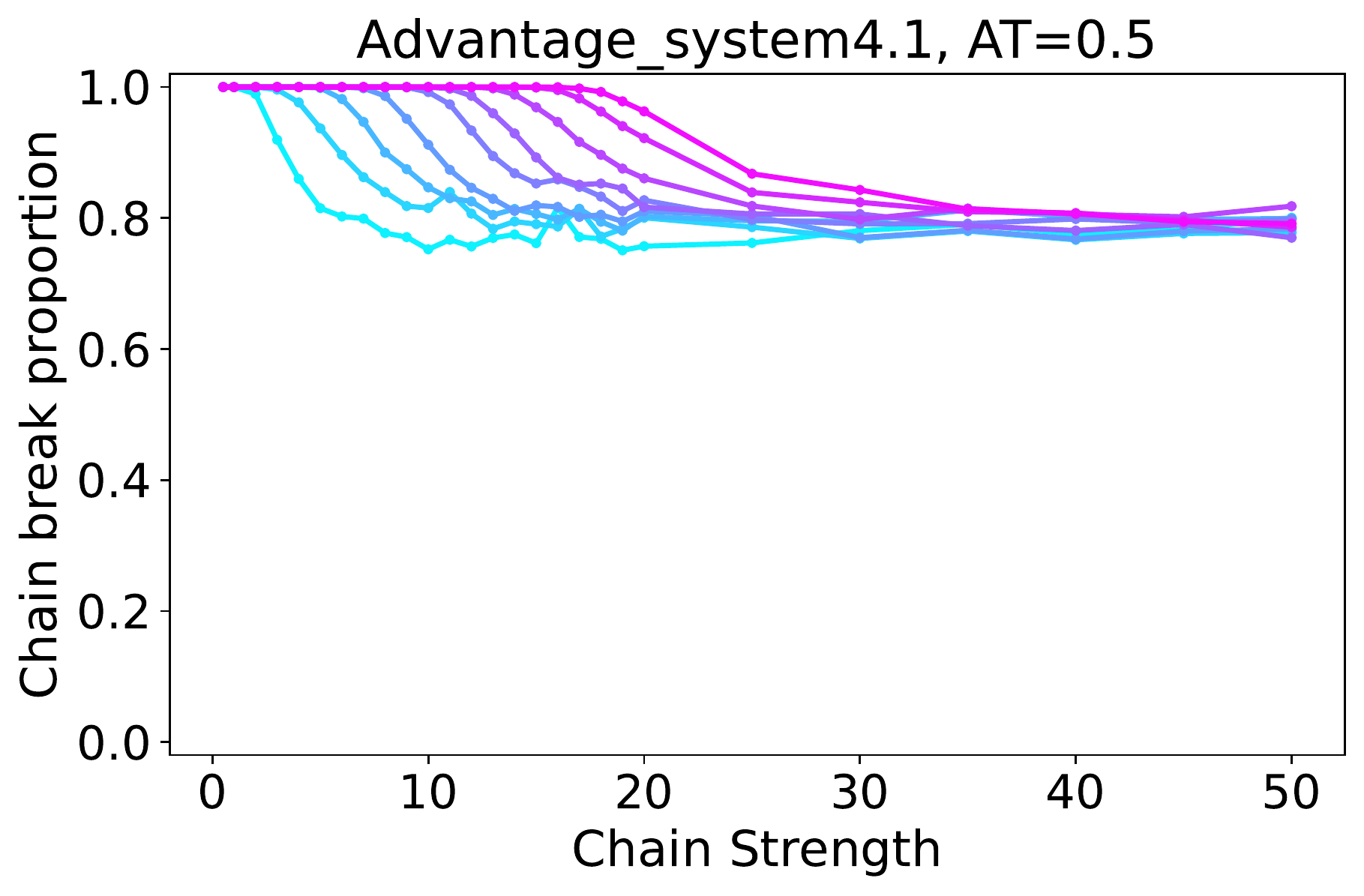}
    \includegraphics[width=0.38\textwidth]{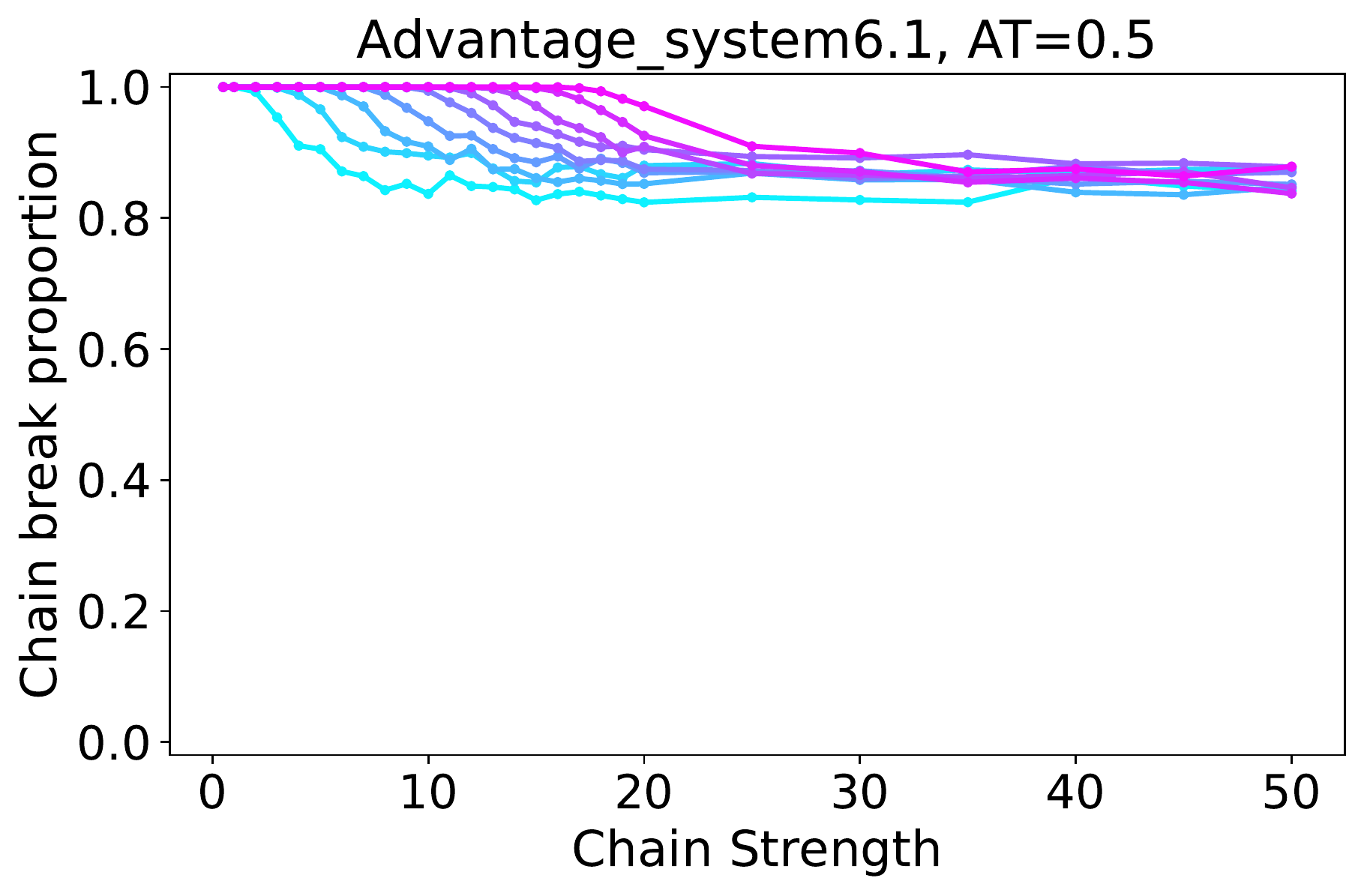}
    \includegraphics[width=0.71\textwidth]{figures/colorbar2.pdf}
    \caption{Maximum Cut sampling approximation ratios (y-axis) vs chain strength (x-axis) in the top row. Chain break proportions (y-axis) vs chain strength (x-axis) in the bottom row. Annealing time is set to $500$ nanoseconds ($0.5$ microseconds). \texttt{Advantage\_system4.1} (left column) and \texttt{Advantage\_system6.1} (right column). Although the approximation ratios are not exactly at $0$, similar to Figure~\ref{fig:small_annealing_times_appendix_maxclique} they do not reach above $0.4$, and \texttt{Advantage\_system6.1} has slightly smaller approximation ratios compared to \texttt{Advantage\_system4.1}. These aggregated results are shown in the form of $10$ lines per plot representing the mean approximation ratio for $10$ linearly spaced graph density intervals from $0.05$ to $0.95$, where the color of each line encodes the mean graph density for that interval. The color coding is shown in the colorbar below the plots. }
    \label{fig:small_annealing_times_appendix_maxcut}
\end{figure*}

Figure~\ref{fig:small_annealing_times_appendix_maxclique} shows the approximation ratios and chain break proportion plots for the Maximum Clique instances when sampled using $500$ nanosecond annealing times on \texttt{Advantage\_system4.1} and \texttt{Advantage\_system6.1}. $500$ nanoseconds is currently the smallest annealing time available on D-Wave quantum annealers and is only available for users to program on these two devices. Figure~\ref{fig:small_annealing_times_appendix_maxcut} shows the same $500$ nanosecond annealing time approximation ratios and chain strengths, except for the Maximum Cut problem instances.

\section{Time-to-Solution for Maximum Clique Instances}
\label{section:appendix_TTS_Max_Clique}

As seen in Table~\ref{table:optimal_solution_table}, the Maximum Clique QUBO's were always sampled to optimally under at least one parameter combination and D-Wave quantum annealer. This means that we can compute, at least for some parameter combinations and devices, the Time-to-Solution (TTS) for finding the optimal solution, with 99\% confidence. TTS \cite{ronnow2014defining, pearson2019analog, pokharel2022demonstration} is a well defined metric that can help quantify the computational quality of probabilistic algorithms, such as quantum annealing. Eq.~\eqref{equation:TTS} defines the TTS formula, given the \text{QPU-access-time} in seconds (which is the D-Wave QPU access time for a single device job, the number of anneals $A$ used in that job, and the proportion of those anneals that correctly found the optimal solution $p$. The prefactor of $\frac{\text{QPU-access-time}}{A}$ computes the amount of QPU time used for a single anneal-readout cycle within the job. Note that when $p=1$, TTS is computed as $\frac{\text{QPU-access-time}}{A}$, and when $p=0$ the TTS can not be computed and is therefore not plotted. 

\begin{equation}
    \text{TTS} = \frac{\text{QPU-access-time}}{A} \cdot  \frac{log(1-0.99)}{log(1-p)}
    \label{equation:TTS}
\end{equation}

\begin{figure*}[h!]
    \centering
    \includegraphics[width=0.24\textwidth]{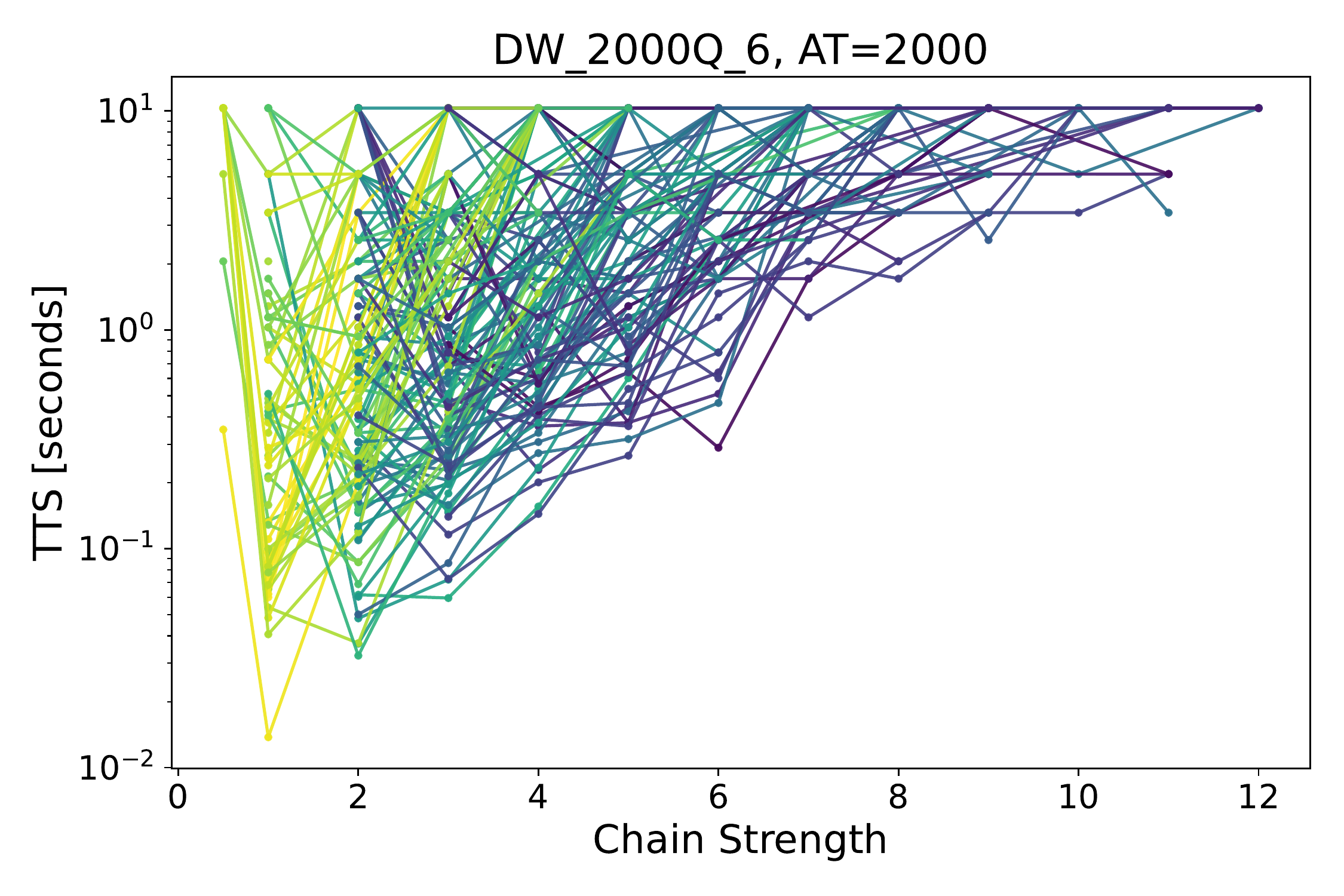}
    \includegraphics[width=0.24\textwidth]{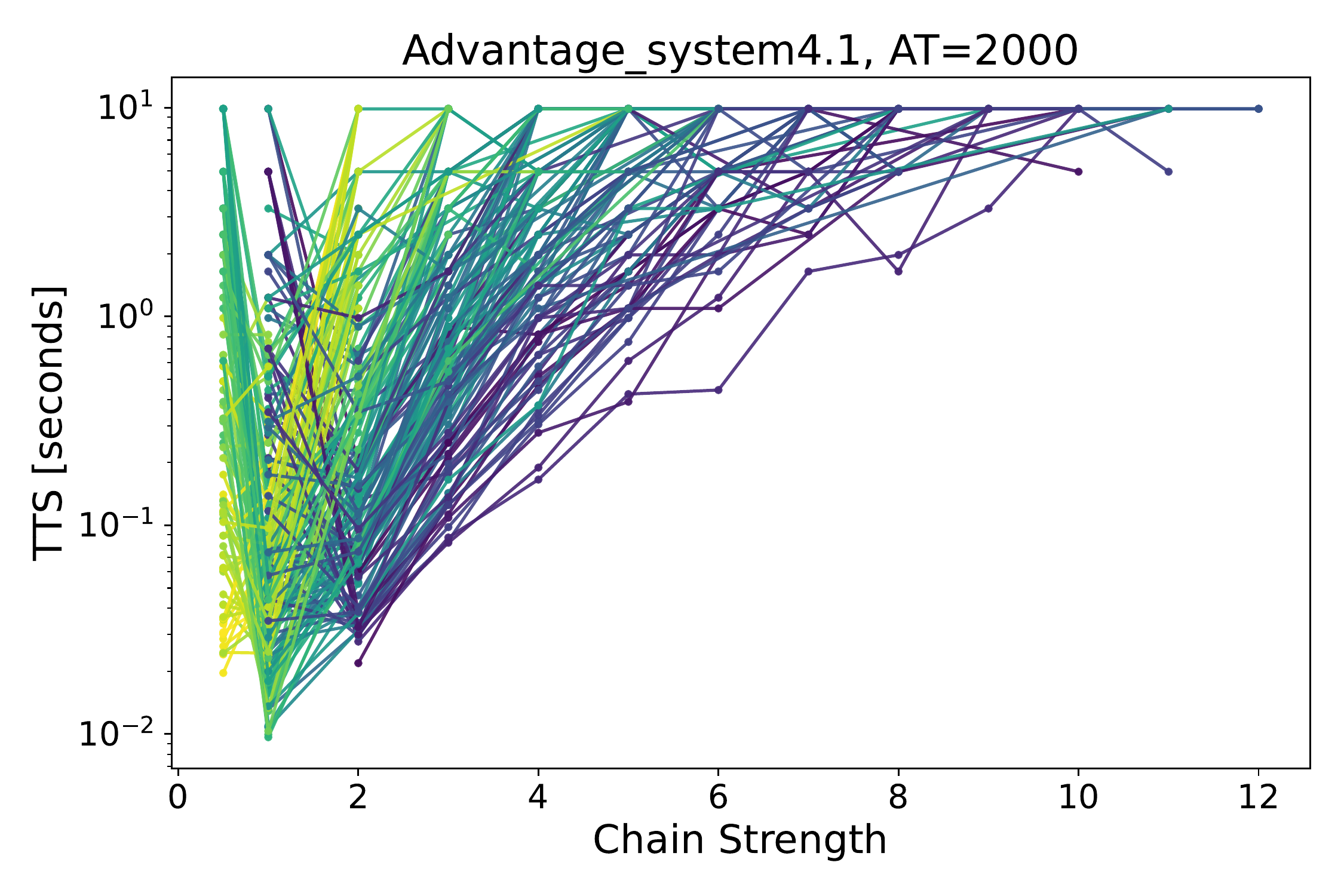}
    \includegraphics[width=0.24\textwidth]{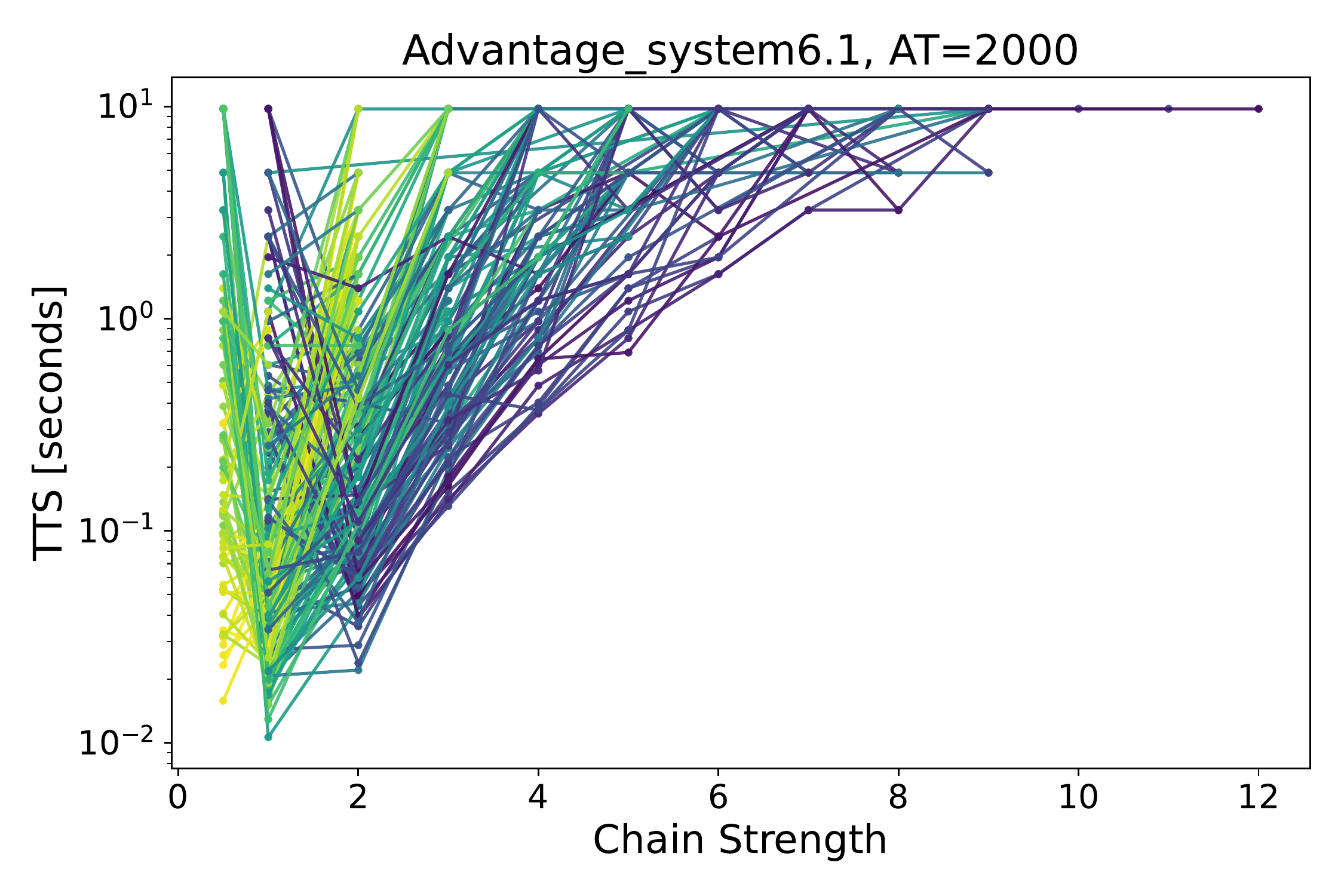}
    \includegraphics[width=0.24\textwidth]{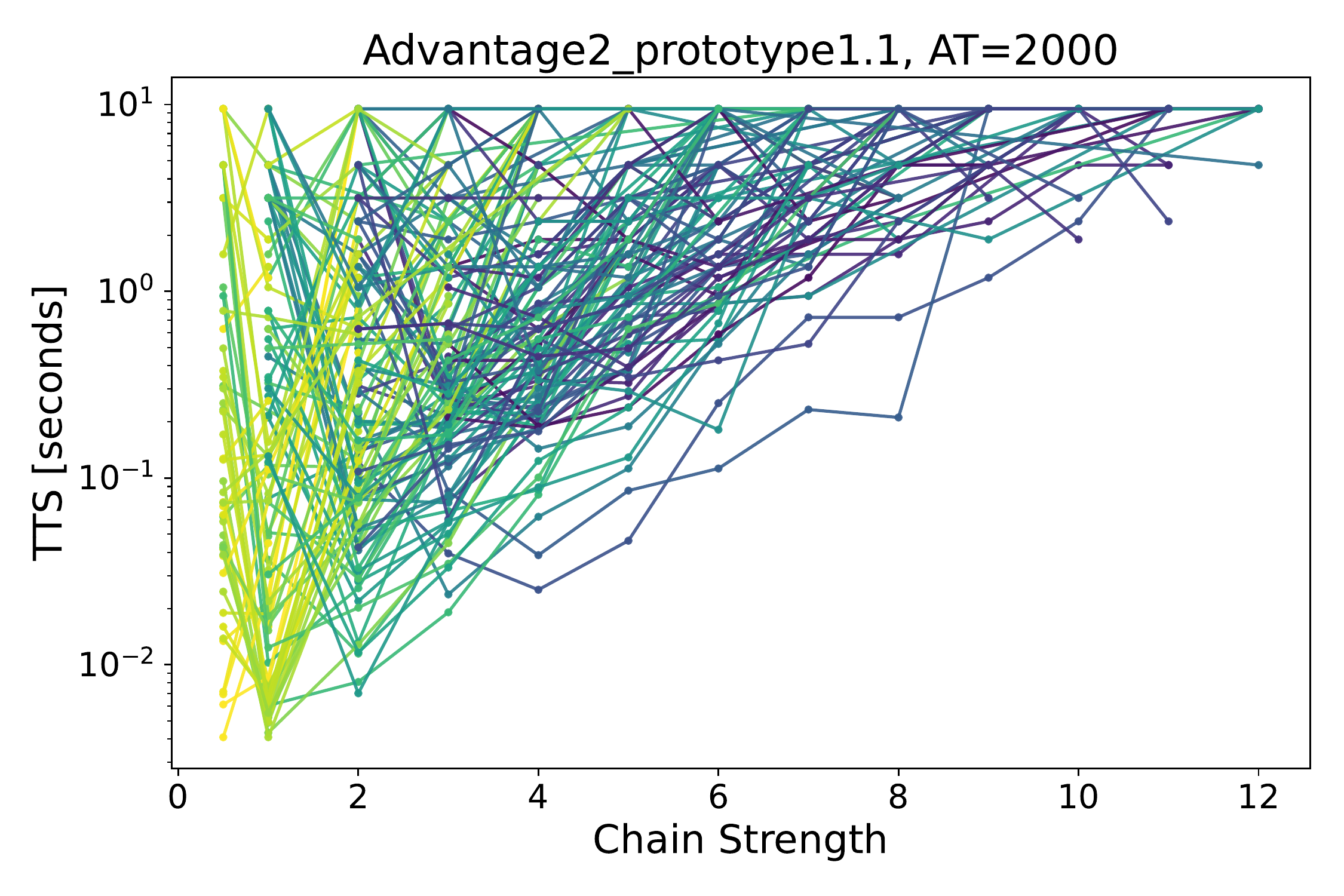}\\
    \includegraphics[width=0.24\textwidth]{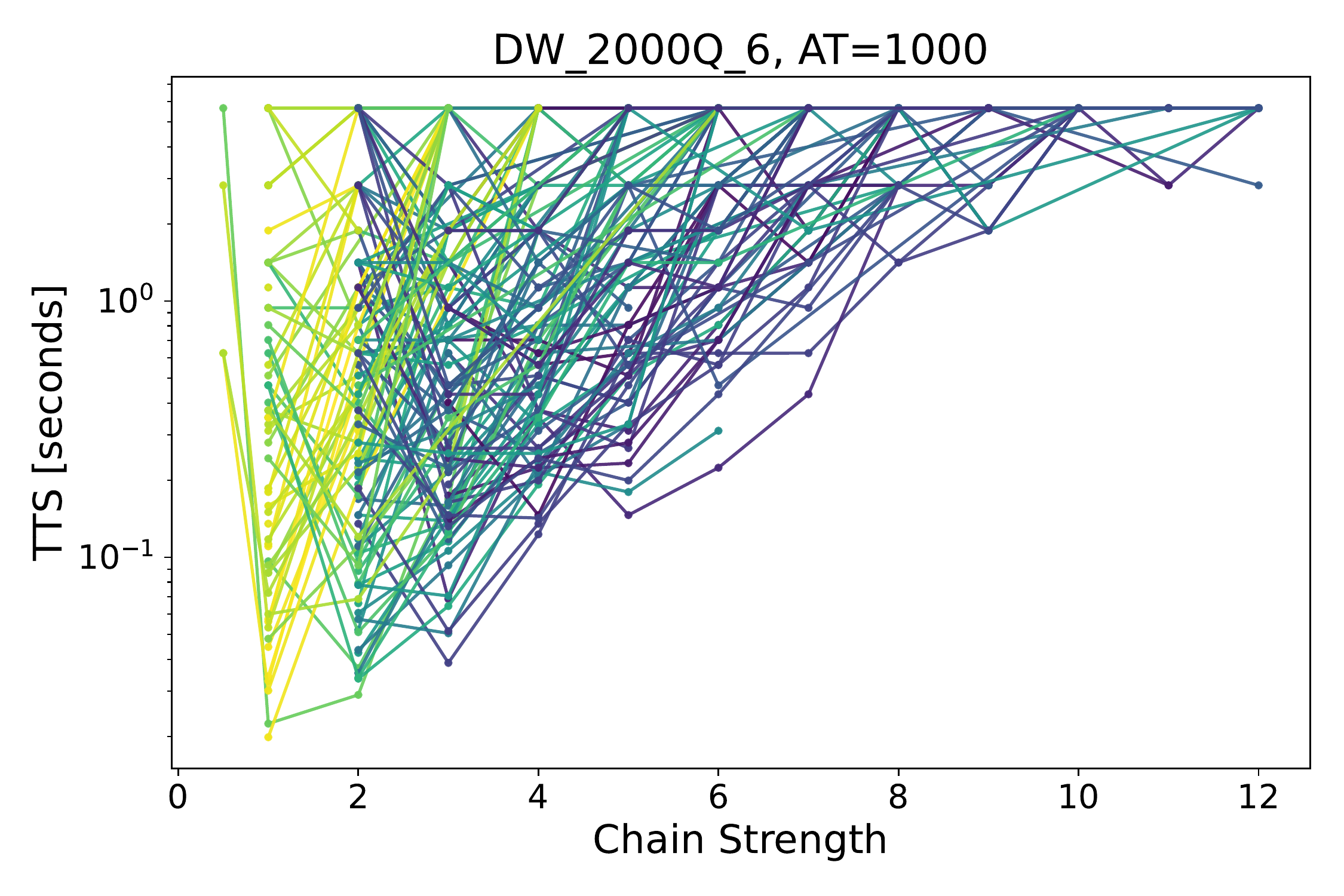}
    \includegraphics[width=0.24\textwidth]{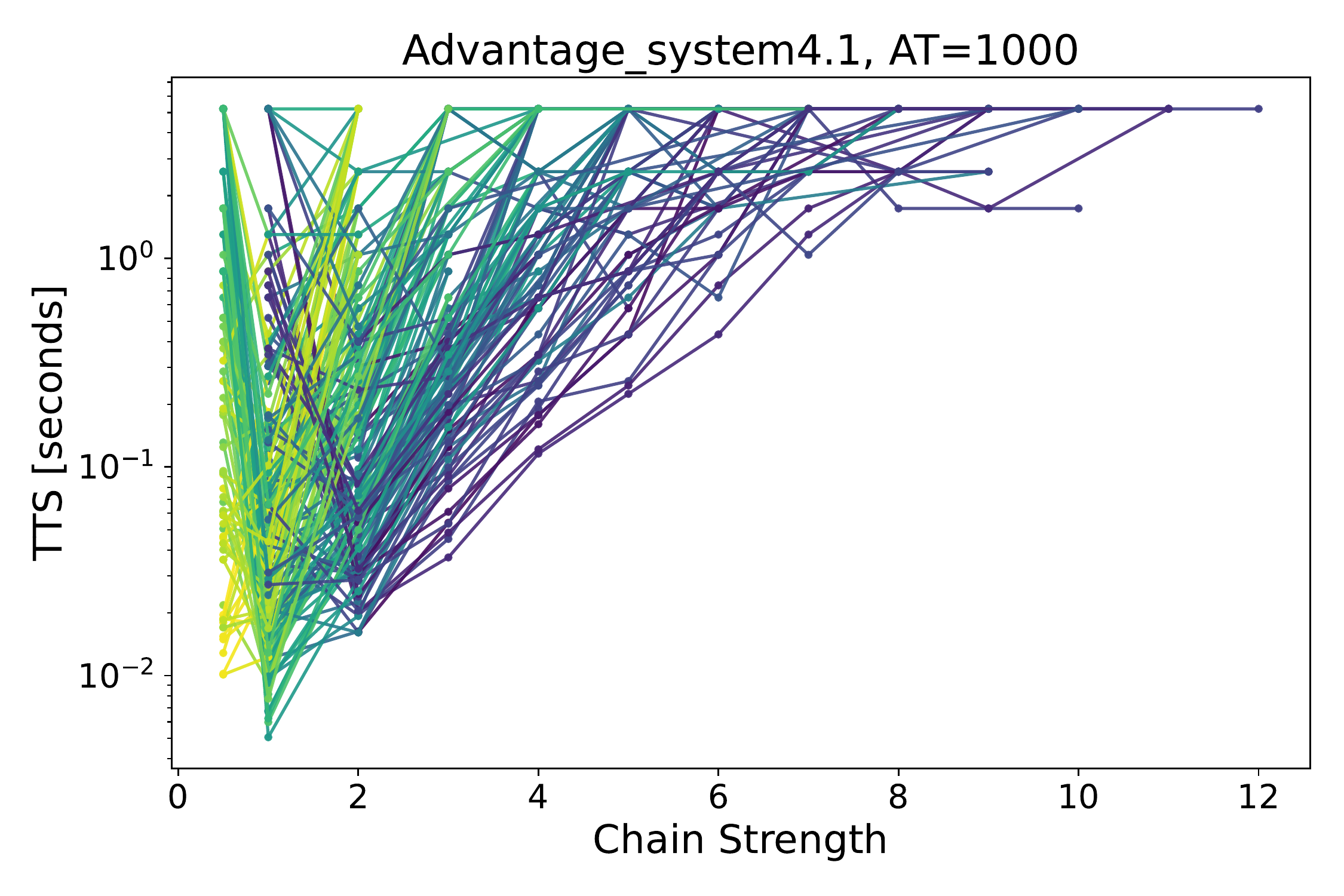}
    \includegraphics[width=0.24\textwidth]{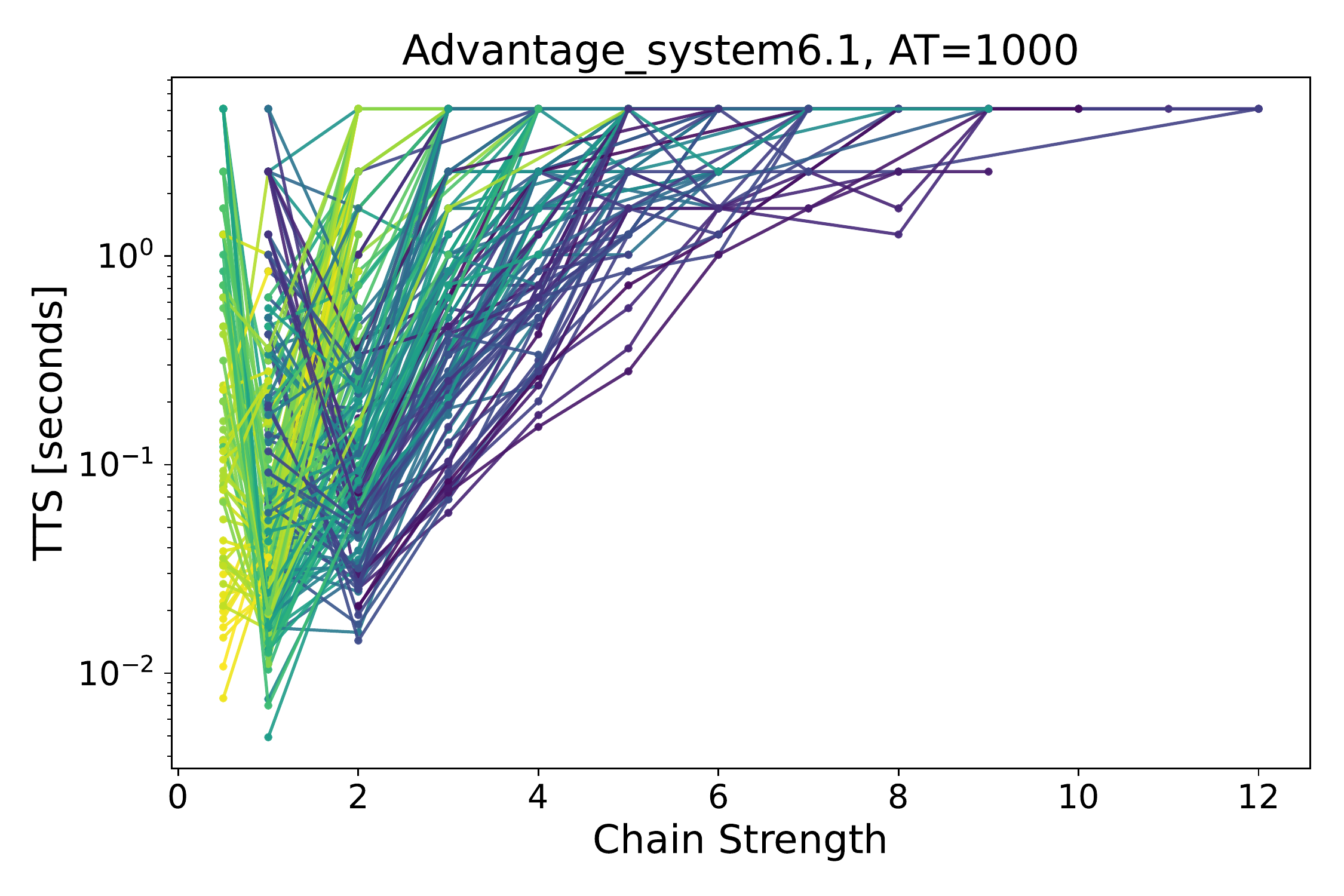}
    \includegraphics[width=0.24\textwidth]{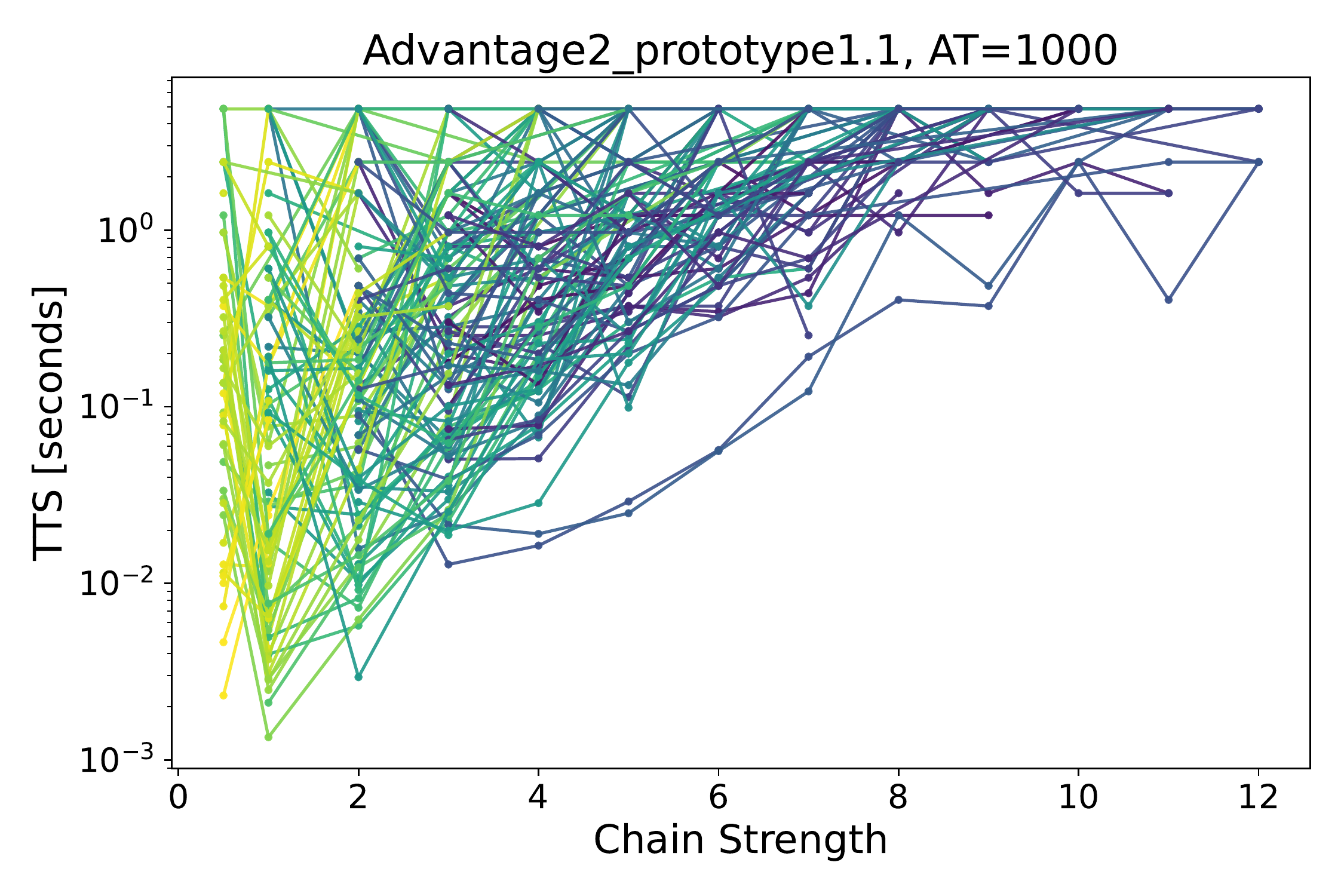}
    \includegraphics[width=0.24\textwidth]{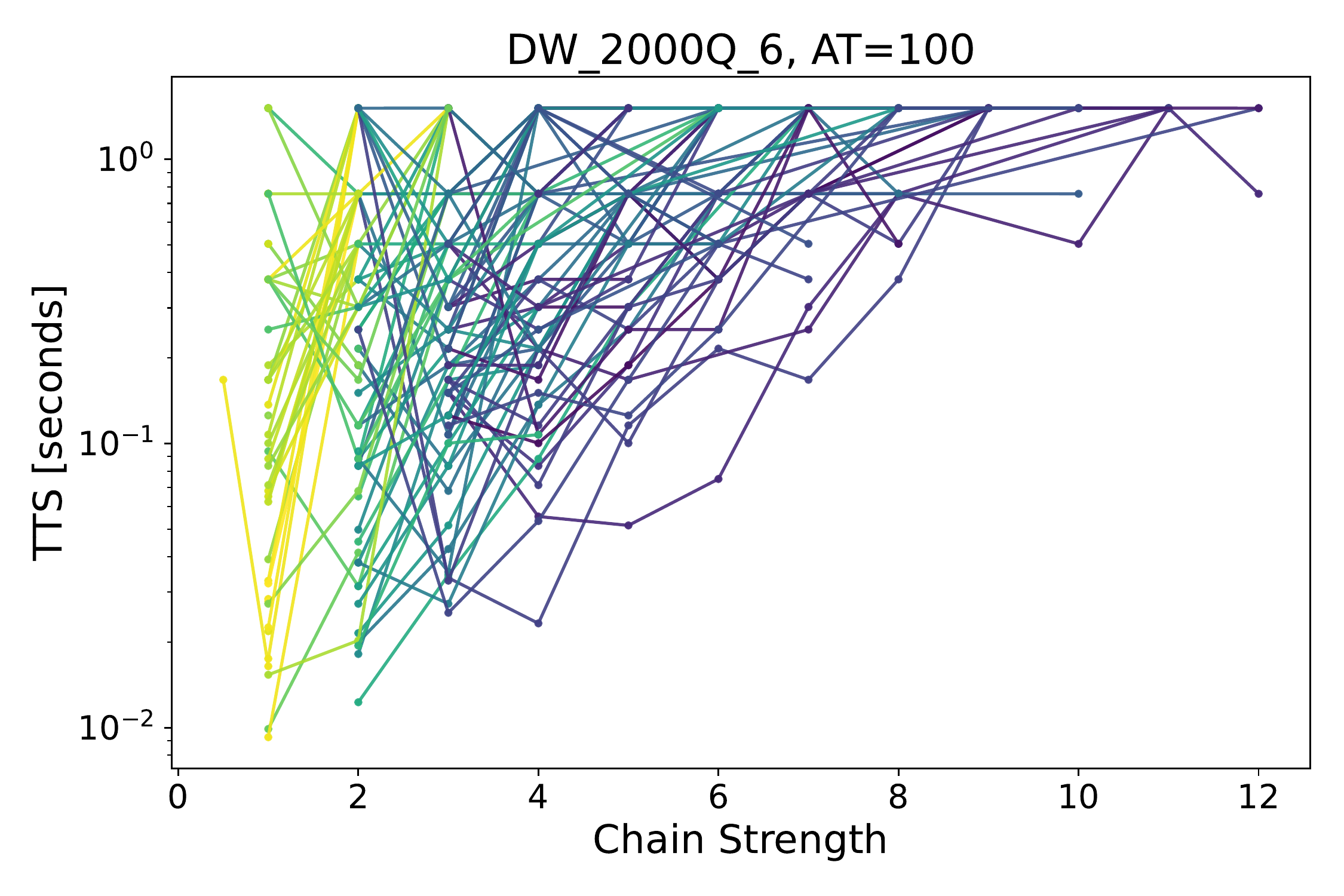}
    \includegraphics[width=0.24\textwidth]{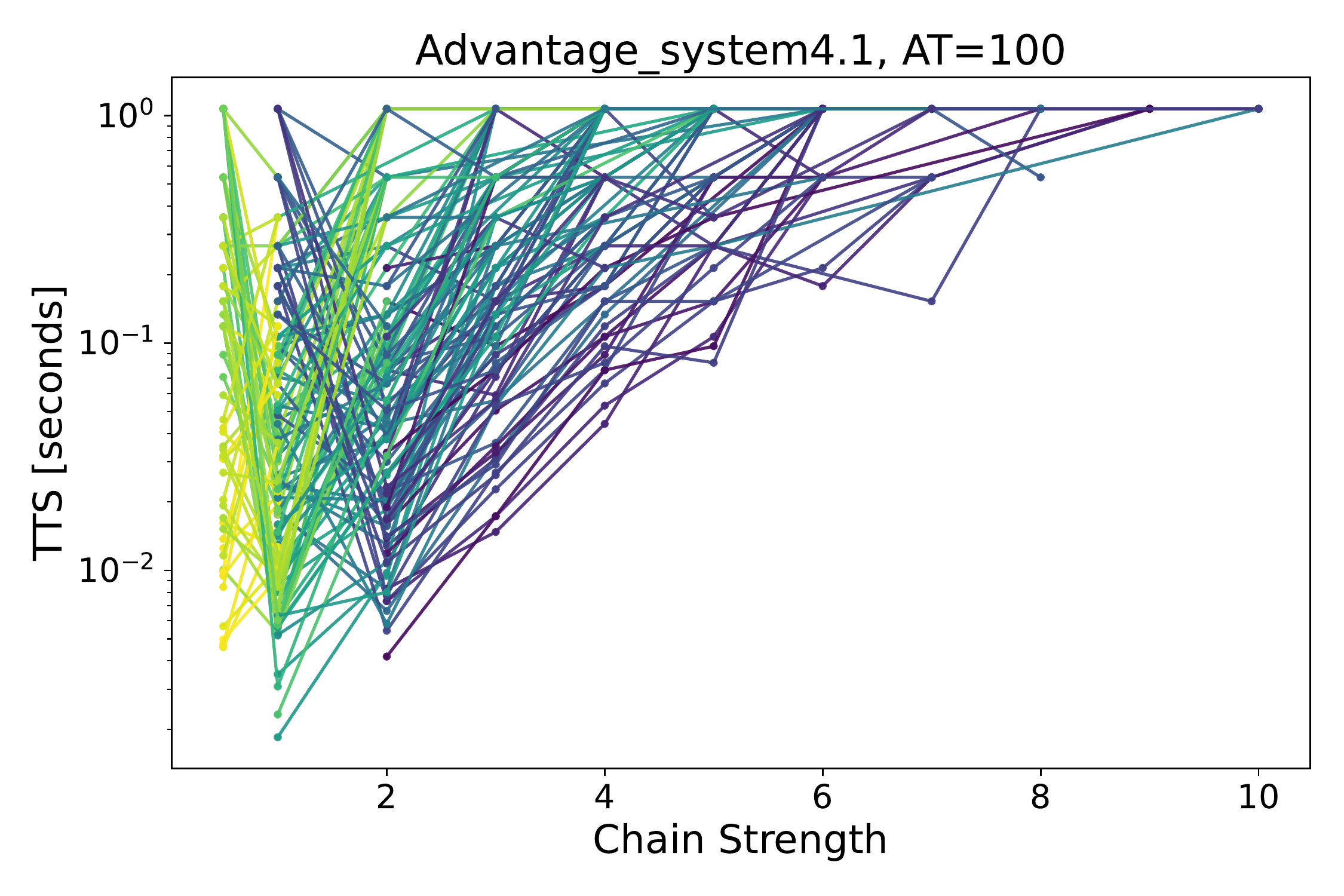}
    \includegraphics[width=0.24\textwidth]{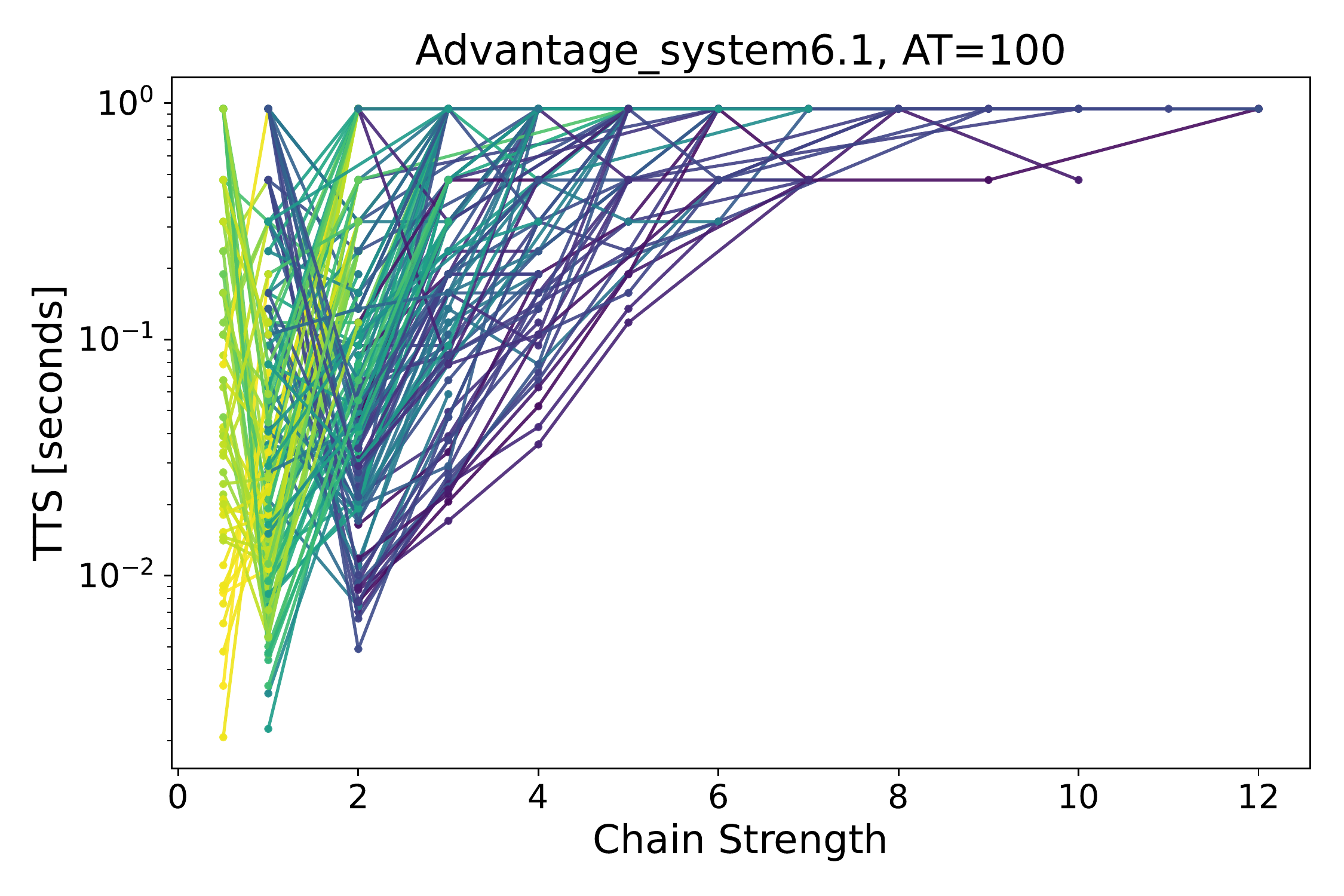}
    \includegraphics[width=0.24\textwidth]{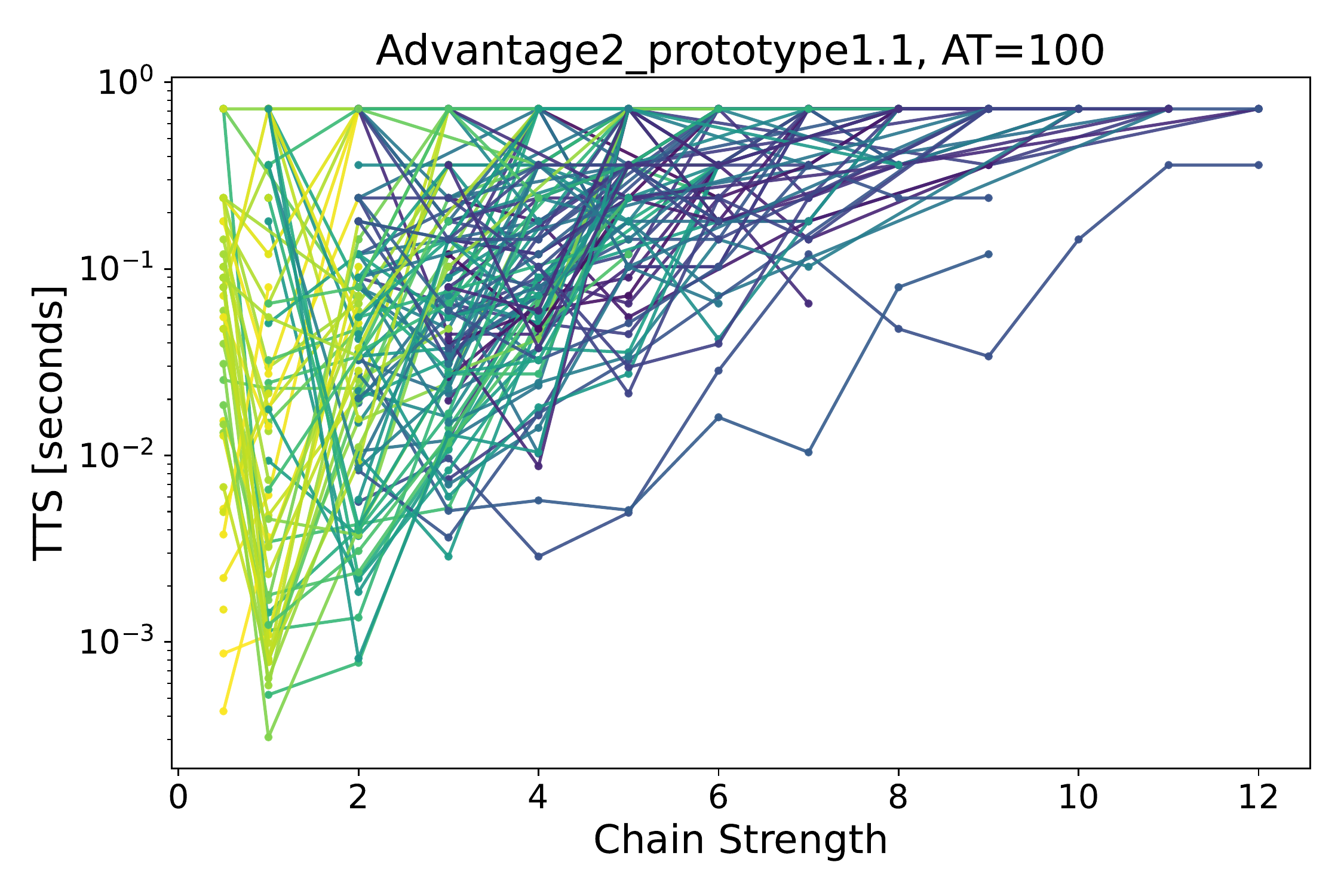}\\
    \includegraphics[width=0.24\textwidth]{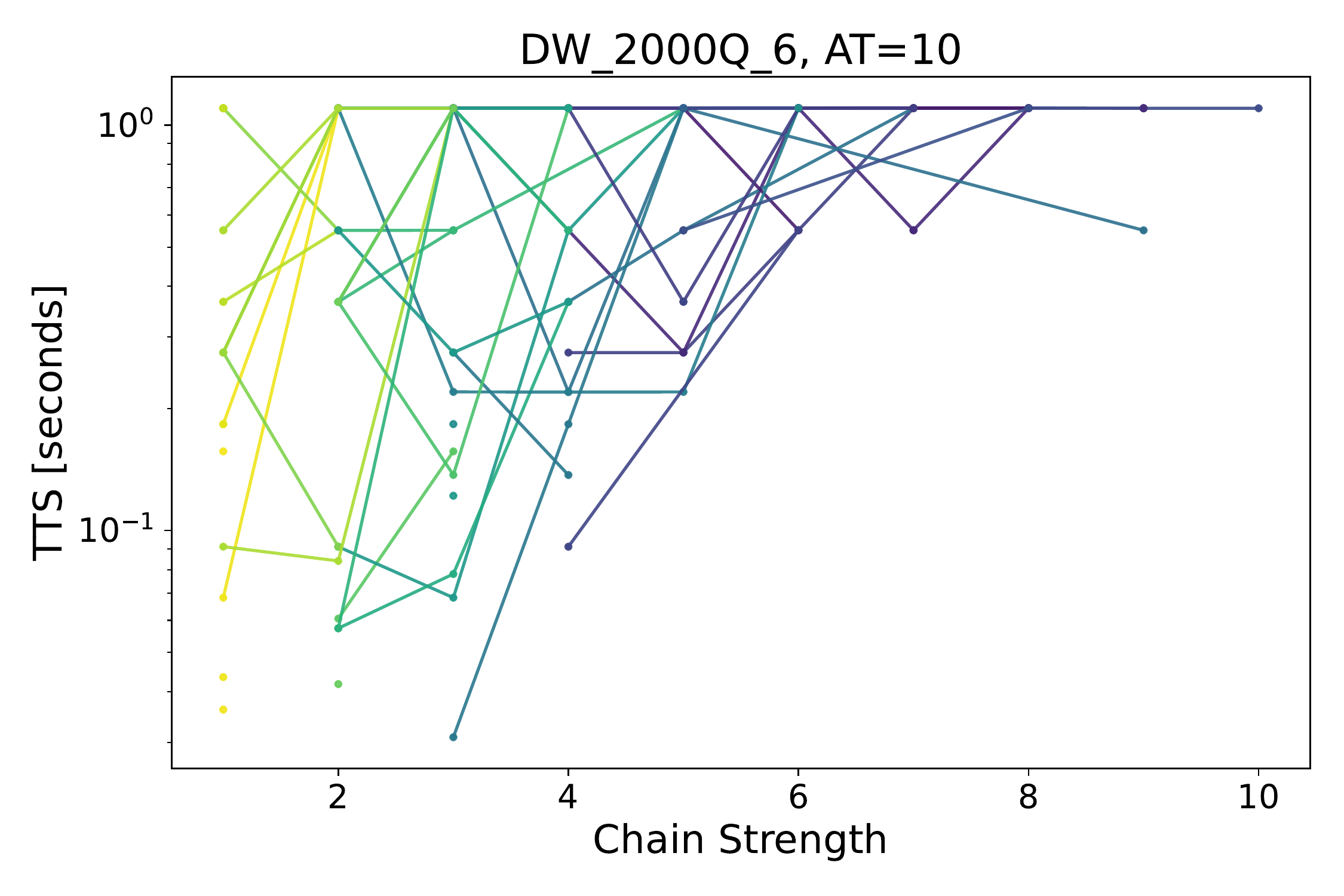}
    \includegraphics[width=0.24\textwidth]{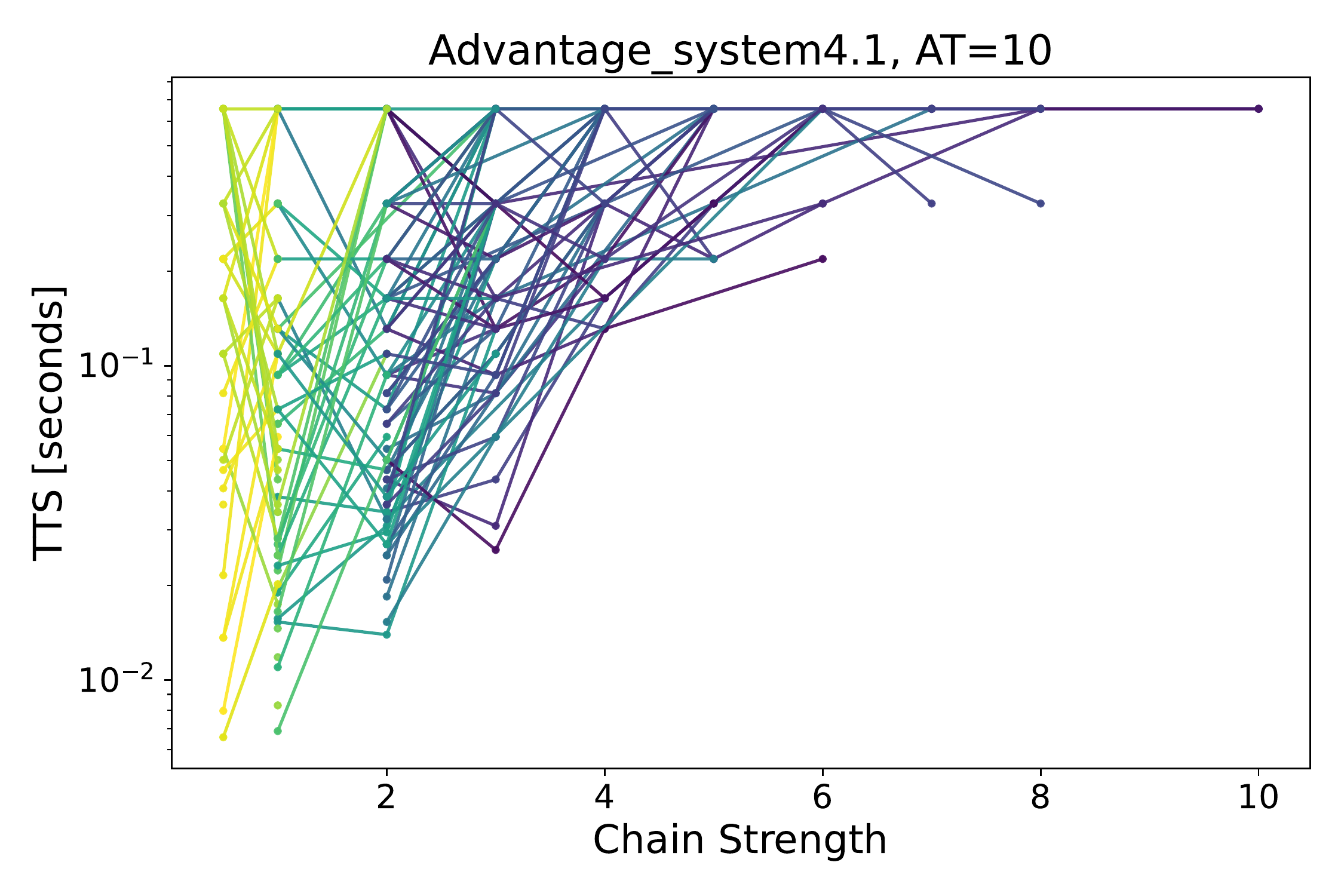}
    \includegraphics[width=0.24\textwidth]{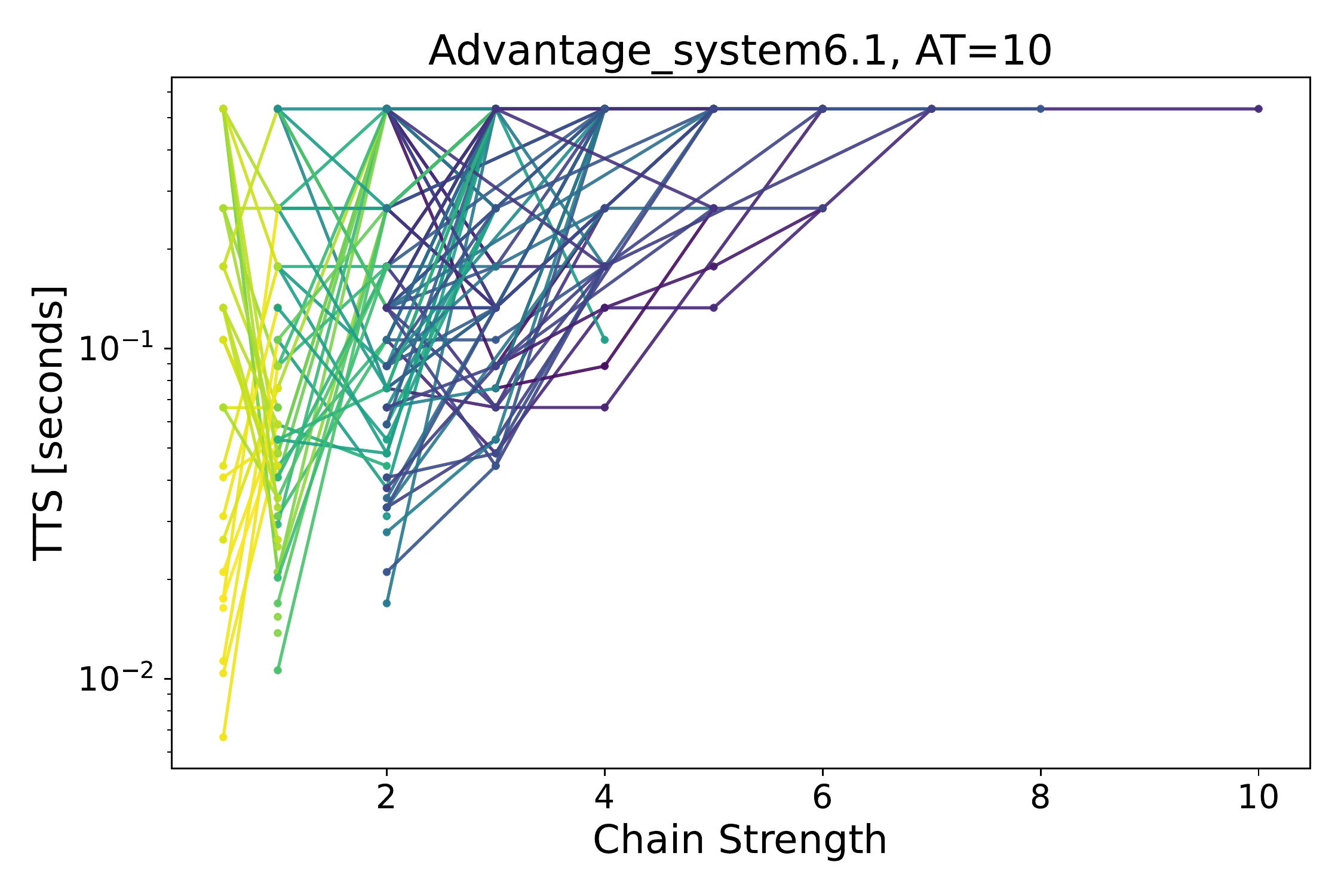}
    \includegraphics[width=0.24\textwidth]{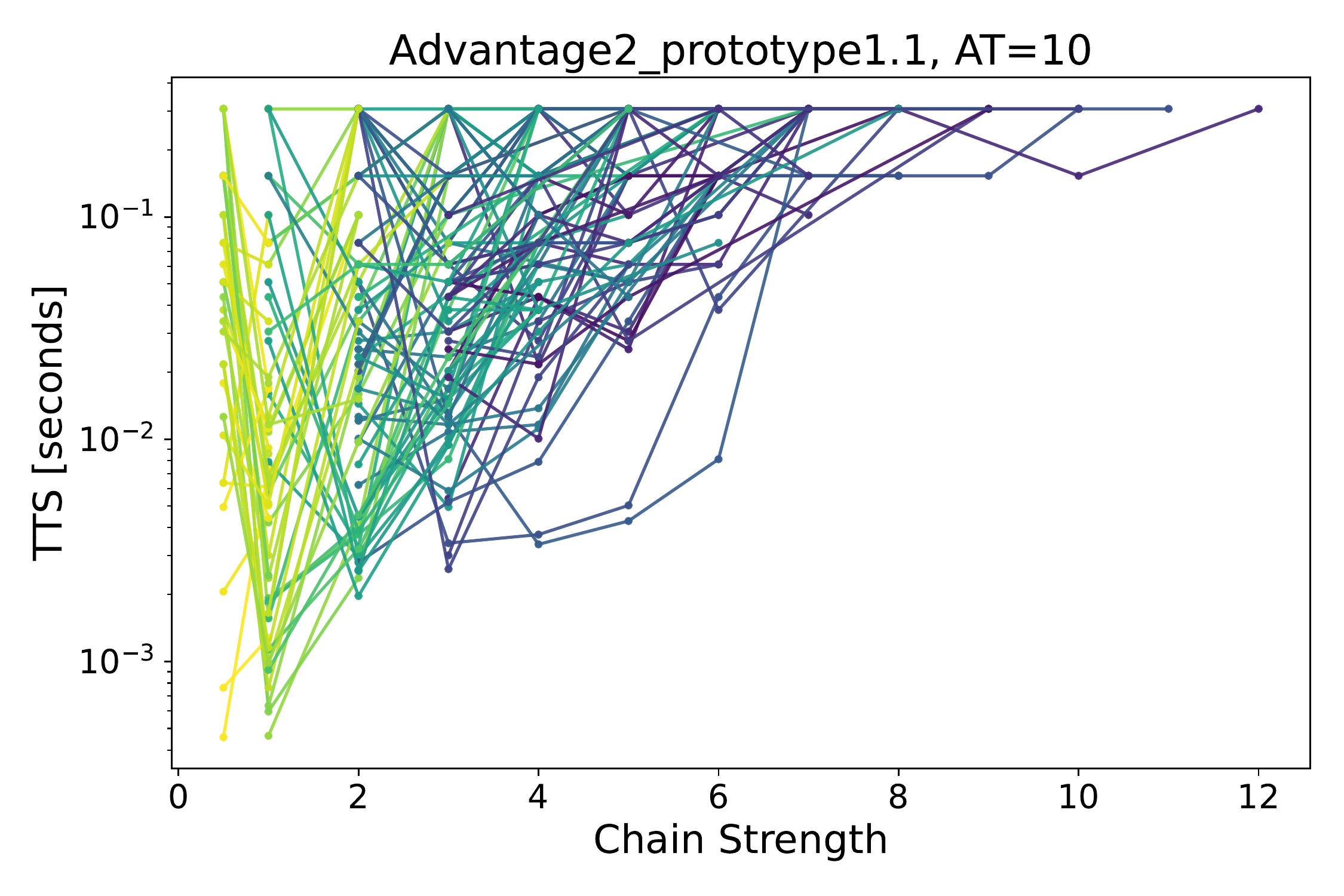}
    \includegraphics[width=0.71\textwidth]{figures/colorbar.pdf}
    \caption{Time-to-Solution (TTS) in seconds (log scale y-axis) vs chain strength (x-axis) for each of the Maximum Clique QUBOs from the $200$ $G(n,p)$ random graphs. The TTS for each of the $200$ graphs are plotted individually as a separate line for each graph - however if there are missing datapoints due to $p = 0$, then that data point is not plotted. The line colors encode the problem graph density. The D-Wave devices are \texttt{DW\_2000Q\_6} (left column), \texttt{Advantage\_system4.1} (center-left column), \texttt{Advantage\_system6.1} (center-right column), and \texttt{Advantage2\_prototype1.1} (right column). The annealing time in microseconds are varied across $2000$ microseconds (top row), $1000$ microseconds (second row), $100$ microseconds (third row), $10$ microseconds (bottom row). }
    \label{fig:TTS_Max_Clique}
\end{figure*}

Figure~\ref{fig:TTS_Max_Clique} plots QPU time based TTS computations for each of the QA QPU's and for varying annealing times. Note that the annealing time of $1$ microsecond is not present in these plots; this is because the success rates are very low at $1$ microsecond, and therefore the plots are quite sparse and not very meaningful. Also not that these TTS computations are based on the QPU access time only. This is because the classical un-embedding computation was simply to only consider non-broken chains, which does not introduce a large amount of classical computation overhead. However, for much larger system sizes classical data processing will become more important to consider in these metrics. 

Figure~\ref{fig:TTS_Max_Clique} shows that \texttt{Advantage2\_prototype1.1} gives the lowest measured TTS across all annealing times. Both \texttt{Advantage\_system4.1} and \texttt{Advantage\_system6.1} give the next lowest measured TTS, although in some instances \texttt{Advantage\_system6.1} provides slightly lower TTS compared to \texttt{Advantage\_system4.1}. \texttt{DW\_2000Q\_6} had the highest TTS compared to the other three devices, and also had the sparser measured TTS datapoints for smaller annealing times, which shows that the device had a lower solution quality compared to the other devices. This is notable because on average, the Maximum Clique sampling TTS matches the progression of the D-Wave device generation, thereby showing that there is an improvement as a function of D-Wave device generations. These plots show that there is some dependence of TTS on chain strength, in particular that the optimal chain strength regions (in the range of 1-2) have the most complete TTS data, with higher chain strengths resulting in higher TTS and sparser computable data. 

\begin{figure*}[h!]
    \centering
    \includegraphics[width=0.24\textwidth]{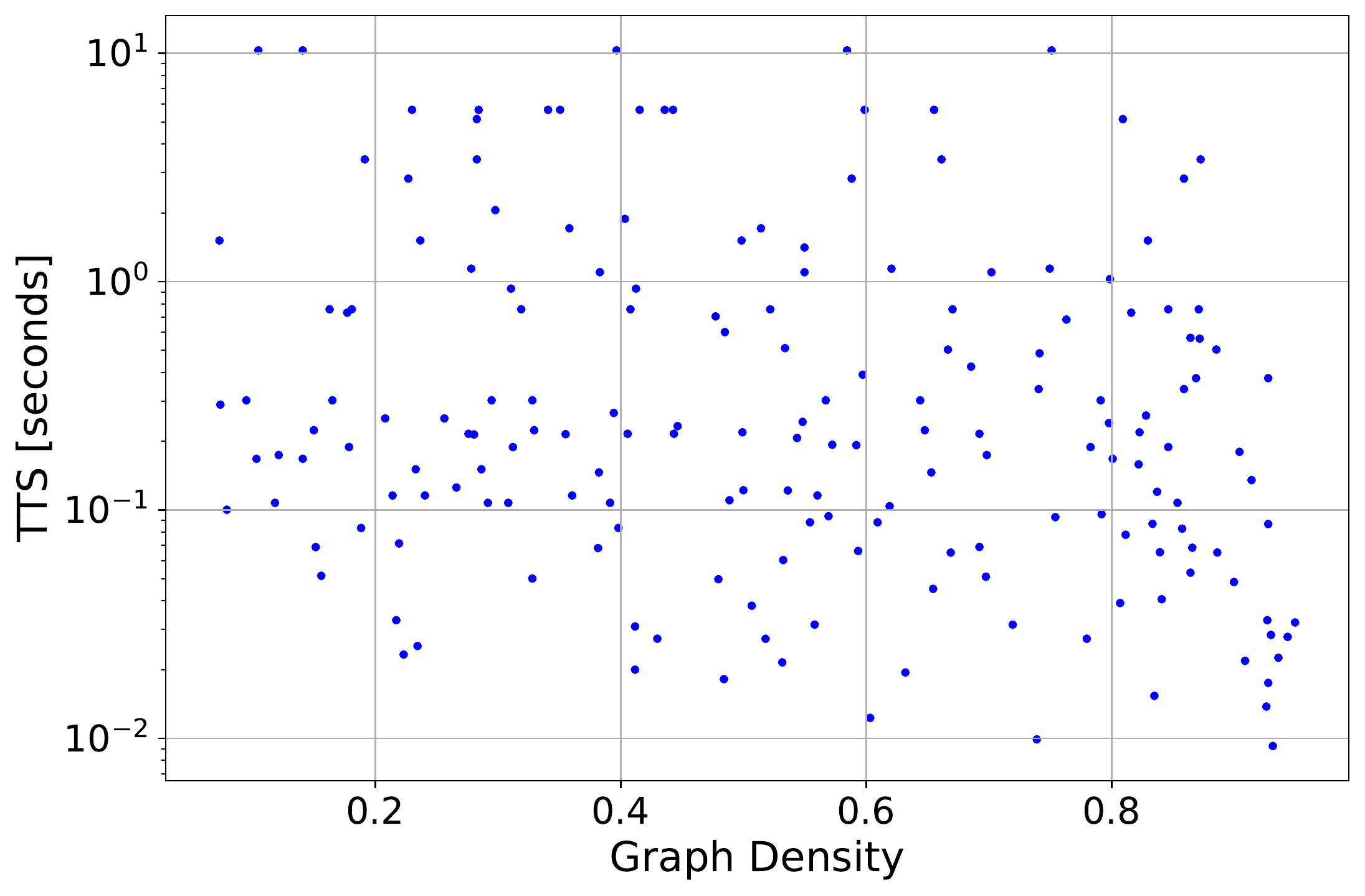}
    \includegraphics[width=0.24\textwidth]{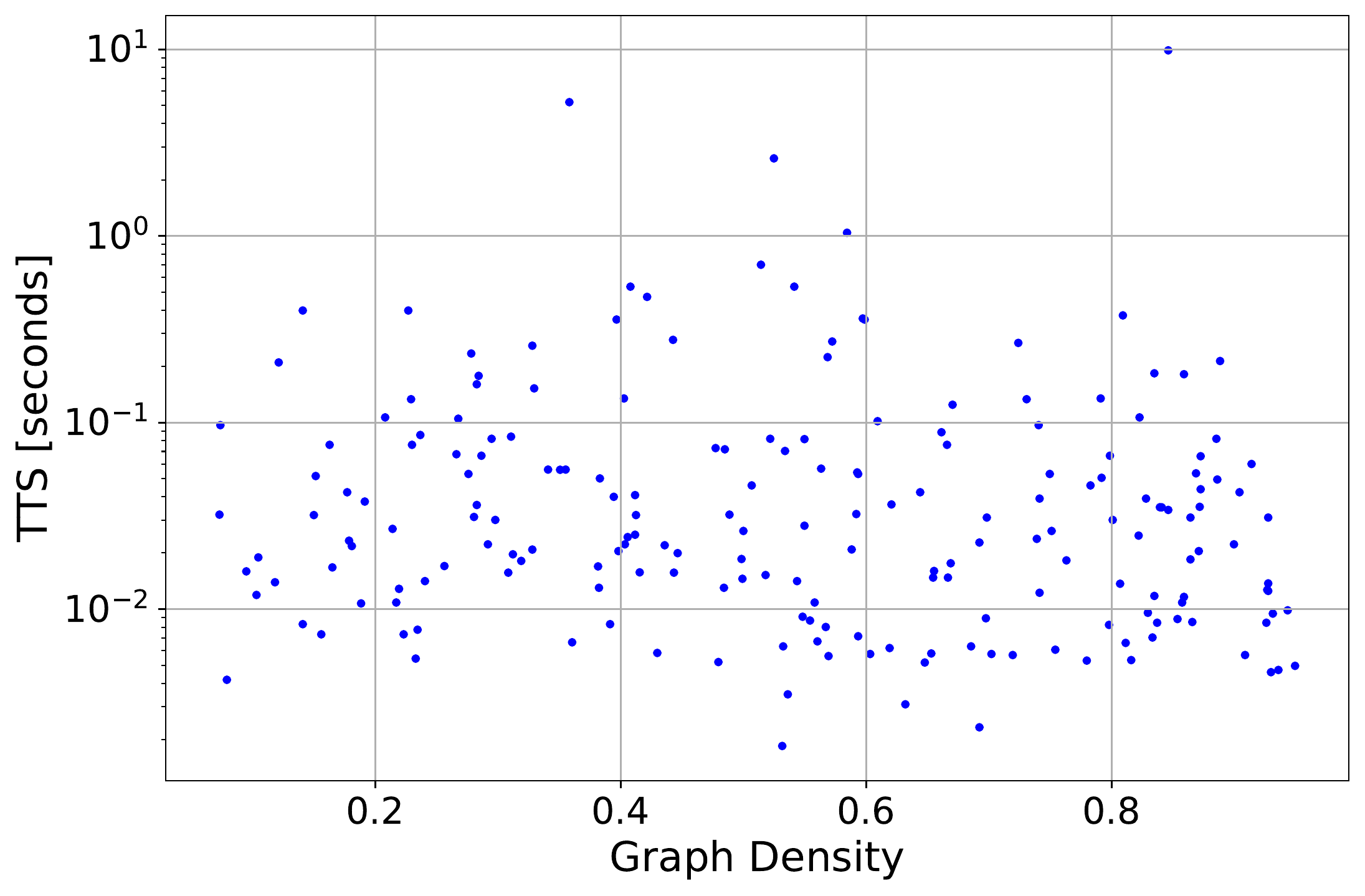}
    \includegraphics[width=0.24\textwidth]{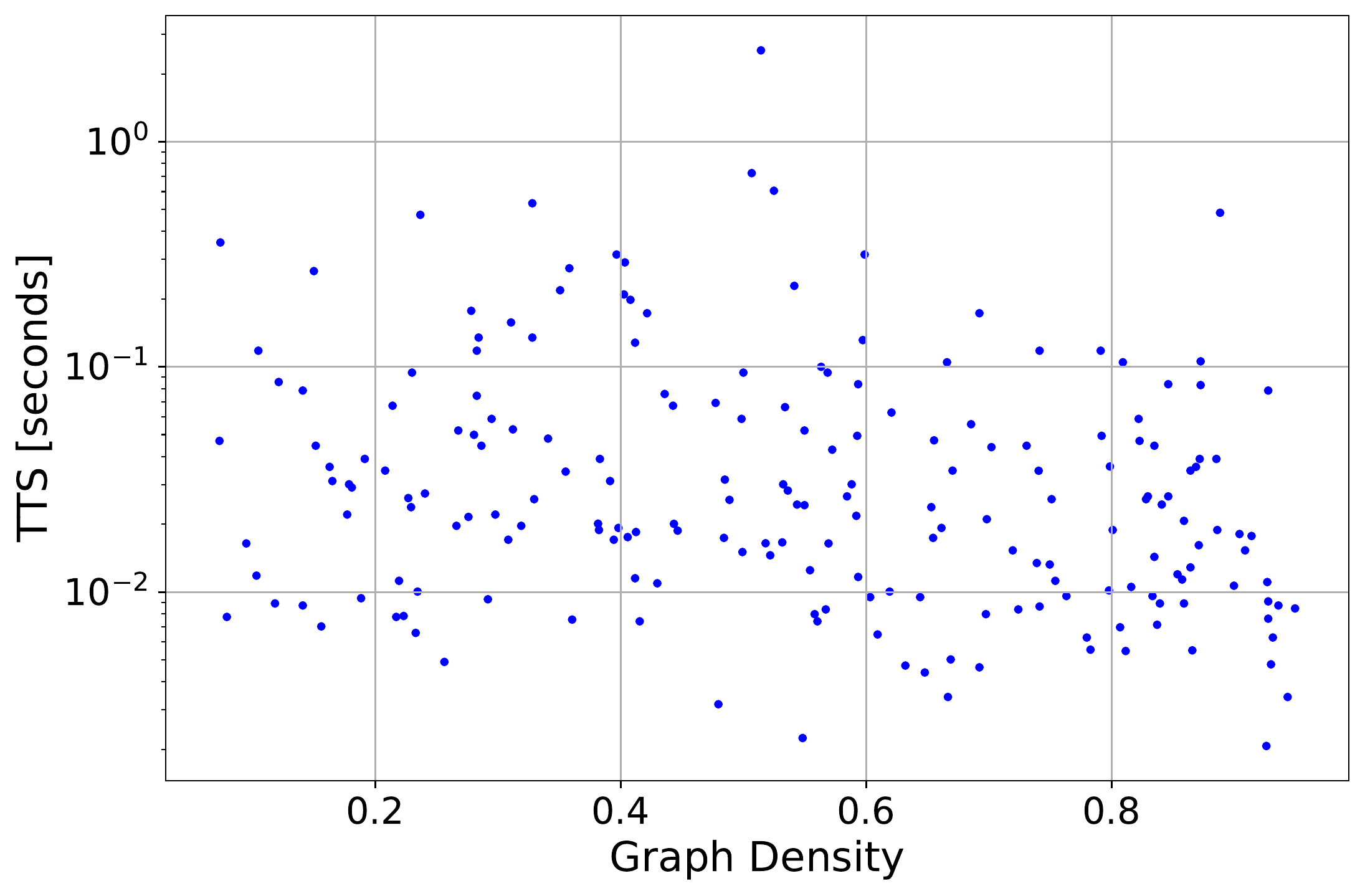}
    \includegraphics[width=0.24\textwidth]{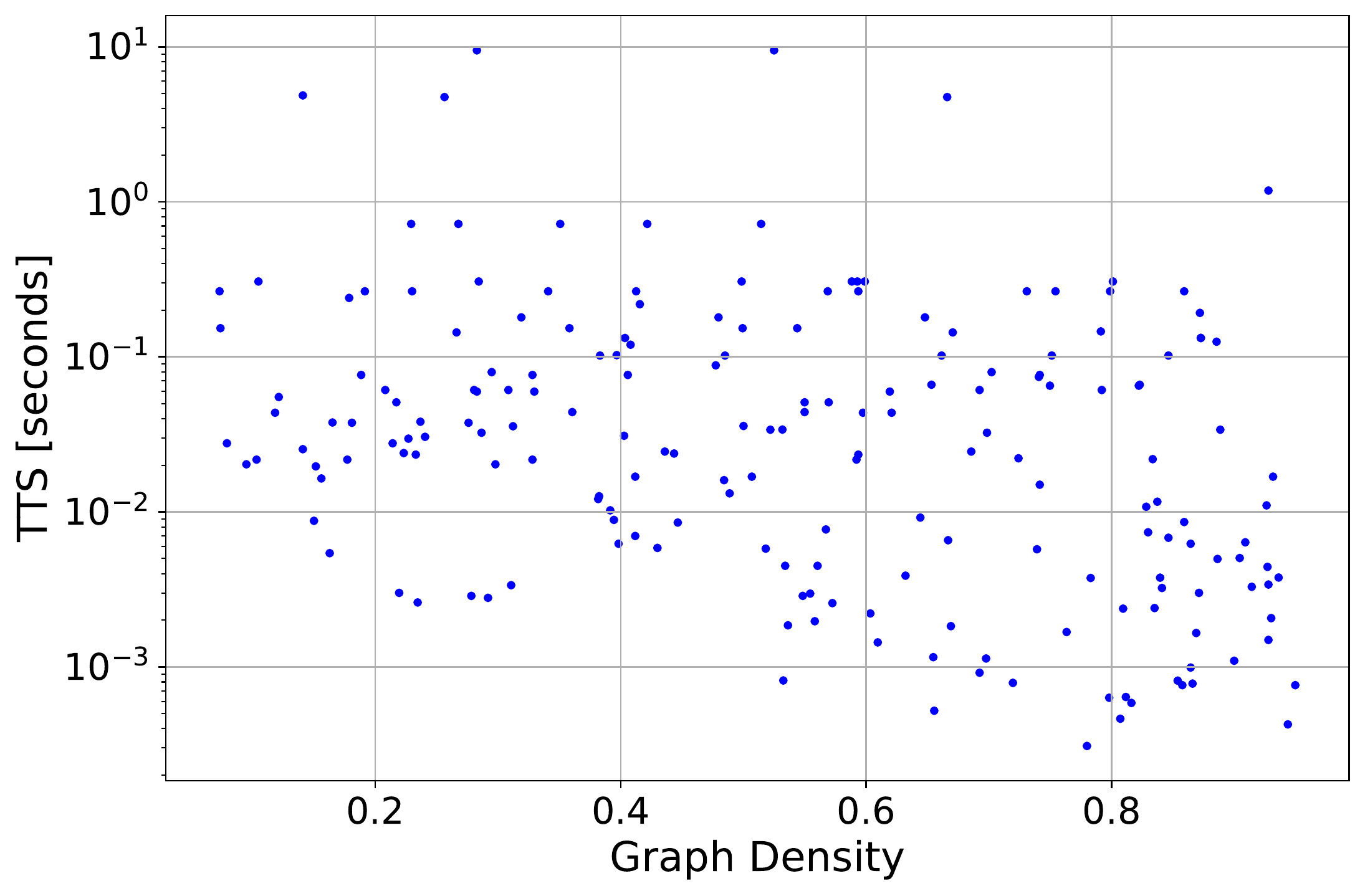}
    \caption{Optimal parameter Time-to-Solution (TTS) (log scale y-axix) vs graph density (x-axis) for the Maximum Clique problem instances, sampled on \texttt{DW\_2000Q\_6} (left) \texttt{Advantage\_system4.1} (center-left), \texttt{Advantage\_system6.1} (center-right), and \texttt{Advantage2\_prototype1.1} (right).  }
    \label{fig:TTS_optimal_Max_Clique}
\end{figure*}

Figure~\ref{fig:TTS_optimal_Max_Clique} plots the best (smallest) Time-to-Solution found across the parameter gridsearch (over annealing time and chain strengths) for each of the $4$ D-Wave devices for sampling the Maximum Clique problem instances. These optimal TTS values are plotted vs the problem instance density, which shows no clear dependence on the optimal parameter TTS as a function of graph density. No datapoints are plotted for the instances where the device did not sample the optimal solution at least once (see Table~\ref{table:optimal_solution_table}). 

\section{Minor Embeddings}
\label{section:appendix_minor_embeddings}

\begin{figure*}[h!]
    \centering
    \includegraphics[width=0.33\textwidth]{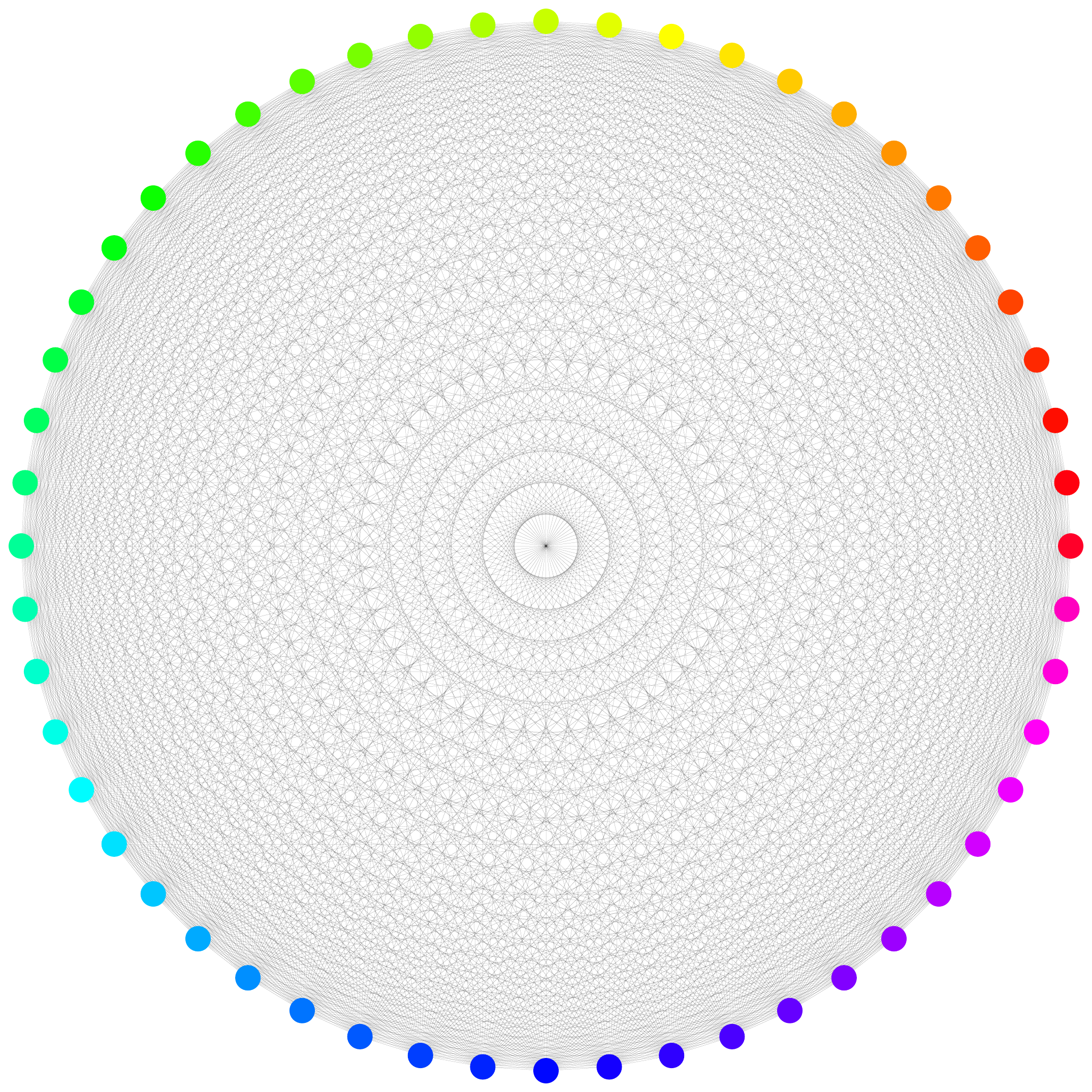}\\
    \includegraphics[width=0.41\textwidth]{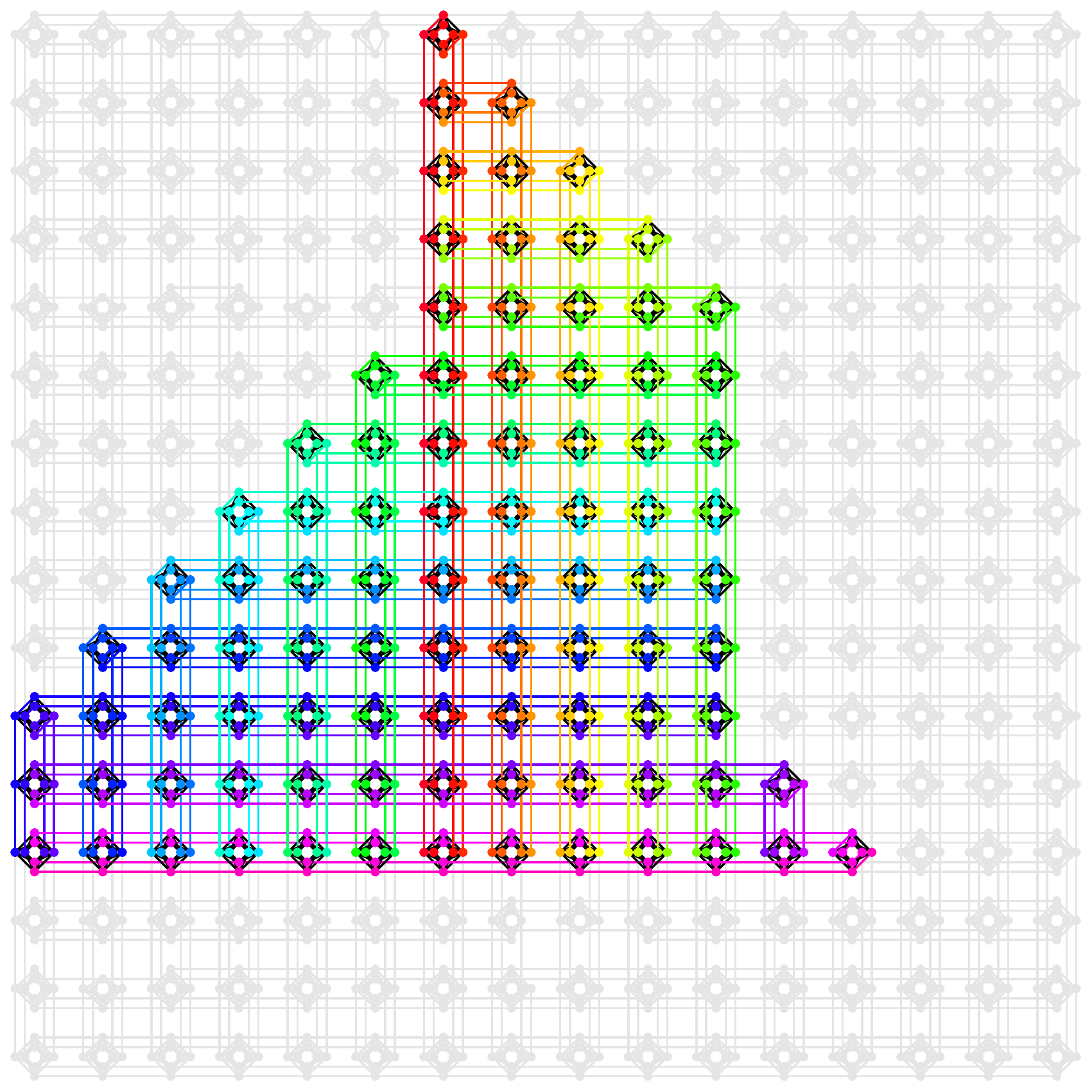}
    \includegraphics[width=0.41\textwidth]{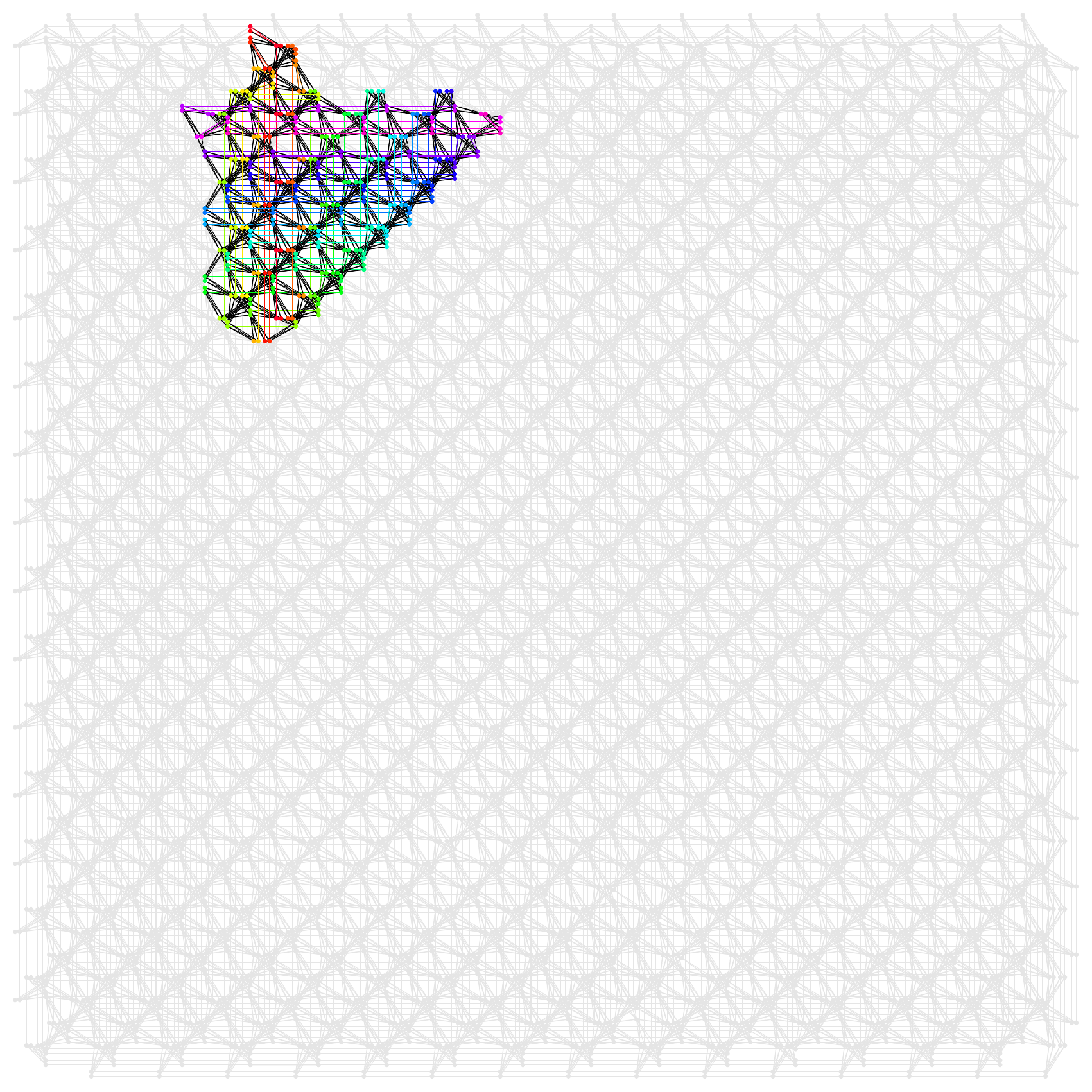}
    \includegraphics[width=0.41\textwidth]{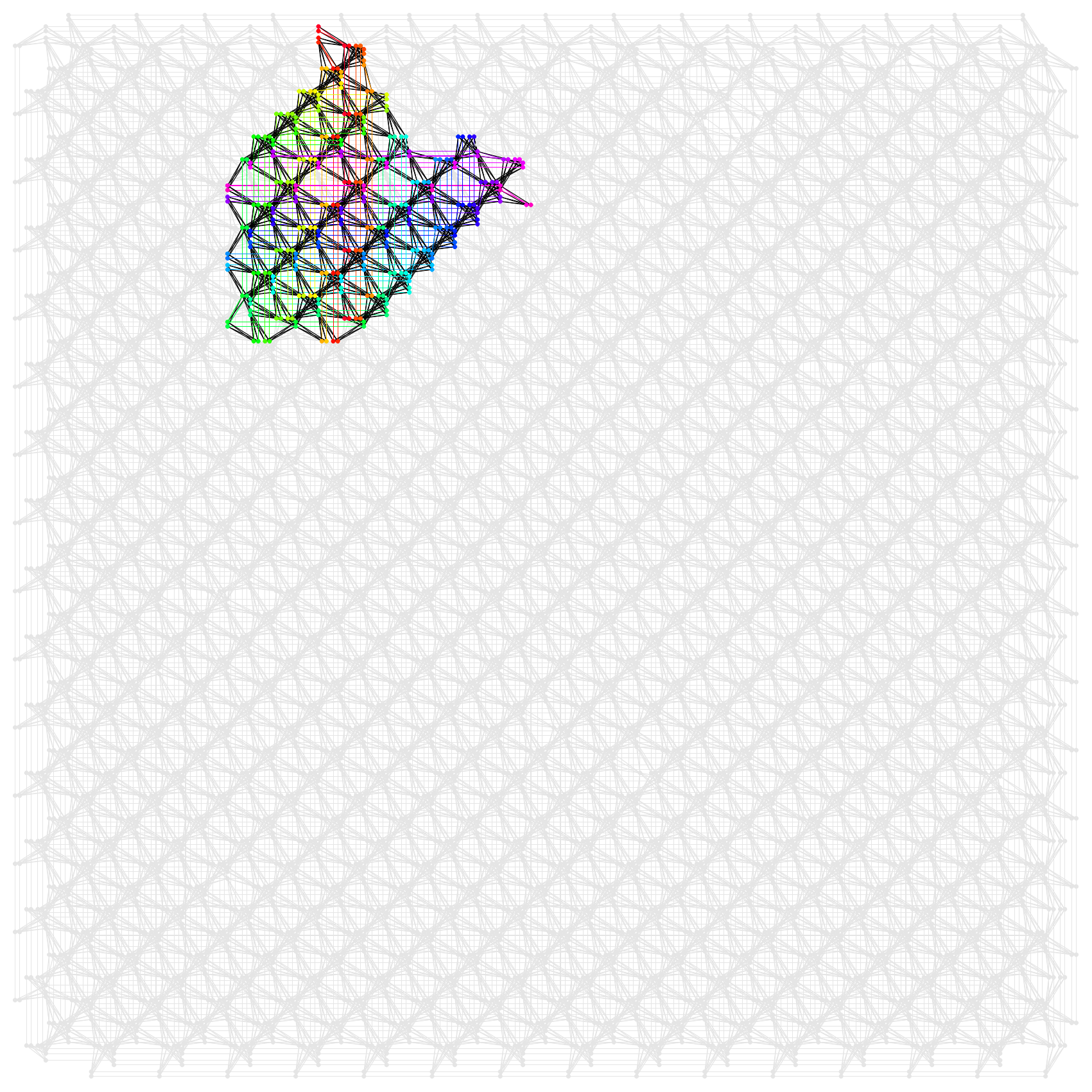}
    \includegraphics[width=0.41\textwidth]{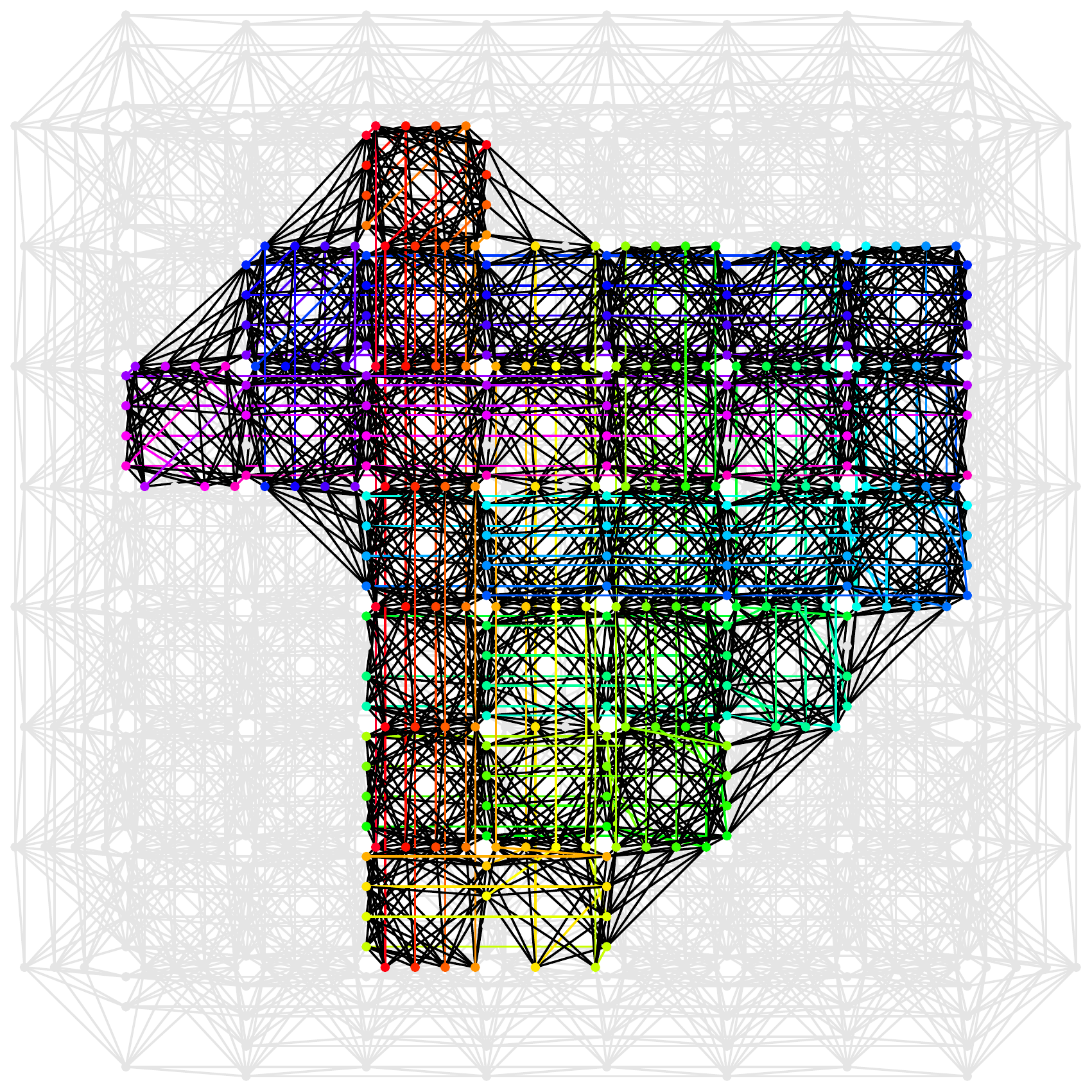}
    \caption{$52$ node clique graph ($K_{52}$ graph) (top single figure) which is minor embedded onto the four different D-Wave quantum annealing device topologies. 
    \texttt{DW\_2000Q\_6} (top left), \texttt{Advantage\_system4.1} (top right), \texttt{Advantage\_system6.1} (bottom left), \texttt{Advantage2\_prototype1.1} (bottom right). Each chain is uniquely color-coded to visually show the differentiation in the chains which are used. Each chain is physically a linear-nearest-neighbors sequence of qubits which are linked together by strong ferromagnetic couplers in order to represent the logical variable state of the problem Ising while providing much higher connectivity than the native device topology. The full hardware graphs are drawn, but the unused hardware sections are colored high transparency grey to highlight the embeddings. Notice that the embeddings for the Pegasus chips use a very small region of the entire chip compared to the much smaller Zephyr and Chimera chips. }
    \label{fig:minor_embedding}
\end{figure*}

Figure~\ref{fig:minor_embedding} shows the exact $N=52$ uniform chain length minor embeddings used in all experiments. 

\end{document}